\documentclass[a4paper,11pt,twoside]{book}

\pdfoutput=1

\usepackage{amsmath} 
\usepackage[FIGTOPCAP]{subfigure}
\usepackage{amsmath}
\usepackage{lipsum}
\usepackage{setspace}
\usepackage{enumerate}
\usepackage{xcolor,colortbl}
\usepackage{amsfonts}
\usepackage{epigraph}
\usepackage{hyperref}
\usepackage{slashed}
\usepackage{amssymb}
\usepackage{tensor}
\usepackage{booktabs}
\usepackage{amsmath}
\usepackage{amssymb}
\usepackage{amsthm}
\usepackage{psfrag}
\usepackage{graphicx}
\usepackage{color}
\usepackage{bbm}
\usepackage{color}
\usepackage{hyperref}
\usepackage{eso-pic}
\usepackage{graphicx}  
\usepackage{epstopdf}
\usepackage{fancyhdr}  
\usepackage{hyperref}
\usepackage{mathrsfs}
\usepackage{amsbsy}
\usepackage{amssymb}    
\usepackage{amsmath}    
\usepackage[english]{babel}
\usepackage{amsfonts}
\usepackage{pstricks}
\usepackage{color}
\usepackage{url}
\usepackage{cases}
\usepackage[framemethod=TikZ]{mdframed}
\usepackage{cite}
\usepackage{float}
\usepackage{multicol}
\usepackage{emptypage}
\usepackage{mdframed}
\usepackage{lipsum} 
\usepackage[nohyperlinks]{acronym}
\usepackage{lmodern}
\usepackage{ragged2e}
\usepackage[framemethod=TikZ]{mdframed}
\usepackage{lipsum}
\mdfdefinestyle{MyFrame}{%
    linecolor=blue,
    outerlinewidth=1pt,
    roundcorner=20pt,
    innertopmargin=\baselineskip,
    innerbottommargin=\baselineskip,
    innerrightmargin=20pt,
    innerleftmargin=20pt,
    backgroundcolor=gray!50!white}

\definecolor{Gray}{gray}{0.85}
\definecolor{LightCyan}{rgb}{0.88,1,1}

\newcolumntype{a}{>{\columncolor{Gray}}c}
\newcolumntype{b}{>{\columncolor{white}}c}

\newenvironment{nscenter}
 {\parskip=0pt\par\nopagebreak\centering}
 {\par\noindent\ignorespacesafterend}

\newcommand*{\boxcolor}{orange}
\makeatletter
\renewcommand{\boxed}[1]{\textcolor{\boxcolor}{%
\tikz[baseline={([yshift=-1ex]current bounding box.center)}] \node [rectangle, minimum width=1ex,rounded corners,draw] {\normalcolor\m@th$\displaystyle#1$};}}
 \makeatother

\newcommand{\HRule}{\rule{\linewidth}{0.5mm}} 
\newcommand{\be}{\begin{equation}}
\newcommand{\ee}{\end{equation}}
\newcommand{\bea}{\begin{eqnarray}}
\newcommand{\eea}{\end{eqnarray}}

\newcommand{\beq}{\begin{equation}}
\newcommand{\eeq}{\end{equation}}
\long\def\/*#1*/{}

\newcommand{\x}{{\widetilde x}}

\def\d{\partial}
\def\l{\left(}

\usepackage{etoc}

\newcommand{\bg}{\begin{gather}}
\newcommand{\eg}{\end{gather}}
\newcommand{\bseq}{\begin{subequations}}
\newcommand{\eseq}{\end{subequations}}

\renewcommand{\ln}{\mathop{\rm ln}\nolimits}

\def\half{\frac{1}{2}}

\DeclareMathOperator{\tr}{tr}

\newcommand{\I}{\mathcal I}

\newcommand{\p}{\partial}

\newcommand{\junk}[1]{}

\def\siml{{\ \lower-1.2pt\vbox{\hbox{\rlap{$<$}\lower6pt\vbox{\hbox{$\sim$}}}}\ }}

\def\dsl{\,\raise.15ex\hbox{/}\mkern-13.5mu D}

\def\dsl{\,\raise.15ex\hbox{/}\mkern-13.5mu D}

\newcommand\BackgroundPic{%
\put(0,0){%
\parbox[b][\paperheight]{\paperwidth}{%
\vfill
\centering
\includegraphics[width=30cm,height=30cm,
keepaspectratio]{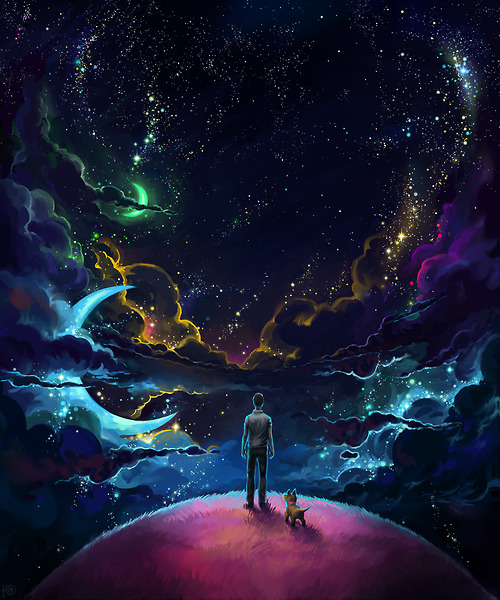}%
\vfill
}}}

\def\siml{{\ \lower-1.2pt\vbox{\hbox{\rlap{$<$}\lower6pt\vbox{\hbox{$\sim$}}}}\ }}

\allowdisplaybreaks[3]

\DeclareGraphicsExtensions{.eps,.pdf,.png,.jpg,.jpeg}

\usepackage[explicit]{titlesec}
\usepackage{titletoc}
\usepackage{tikz}
\usepackage{epigraph}
\usepackage{xpatch}
\usepackage{lmodern}
\usepackage{lipsum}

\newlength\ChapWd
\definecolor{myblue}{RGB}{0,0,122}

\titleformat{\chapter}[display]
  {\normalfont\filcenter\sffamily}
  {\tikz[remember picture,overlay]
    {
    \node[fill=myblue,font=\fontsize{70}{82}\selectfont\color{white},anchor=north east,minimum size=\ChapWd] 
      at ([xshift=-15pt,yshift=-15pt]current page.north east) 
      (numb) {\thechapter};
    \node[rotate=90,anchor=south,inner sep=0pt,font=\huge] at (numb.west) {\chaptertitlename};
    }
  }{0pt}{\fontsize{33}{40}\selectfont\color{myblue}#1}[\vskip10pt\Large***]
\titlespacing*{\chapter}
  {0pt}{50pt}{10pt}

\makeatletter
\xpatchcmd{\ttl@printlist}{\endgroup}{{\noindent\color{myblue}\rule{\textwidth}{1.5pt}}\vskip30pt\endgroup}{}{}
\makeatother

\usepackage{blindtext}

\usepackage{fix-cm}
\usepackage{titlesec}
\usepackage{tikz} 
\usetikzlibrary{calc} 
\usetikzlibrary{decorations,decorations.shapes,shapes} %

\begin{document}
\nocite*{}
\AddToShipoutPicture*{\BackgroundPic}


\pagenumbering{roman}
\setcounter{page}{1}

\begin{titlepage}
\begin{center}

\begin{figure}[h!]
\includegraphics[height=15mm]{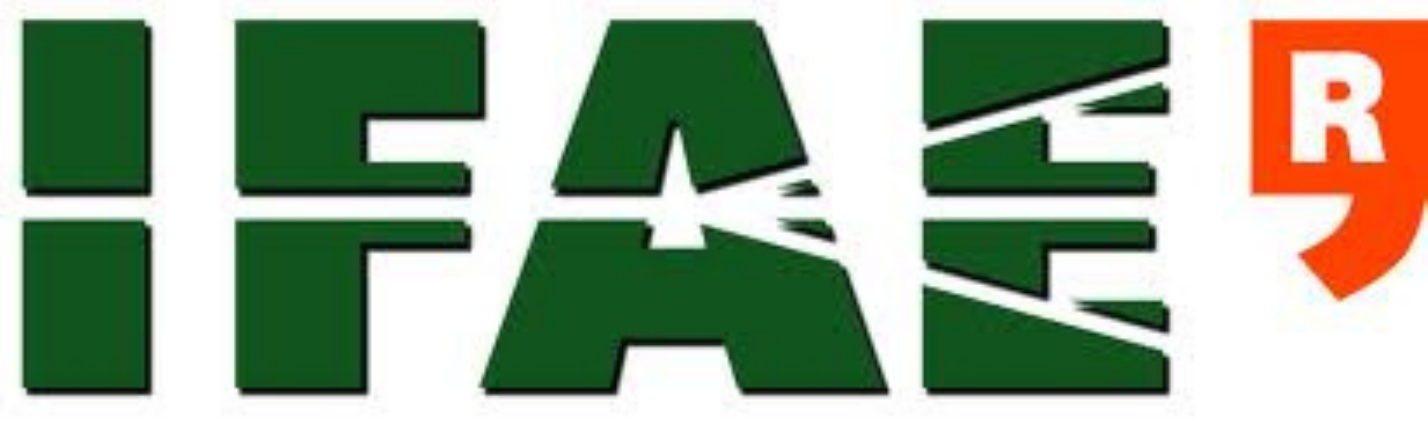}
\hfill
\includegraphics[height=15mm]{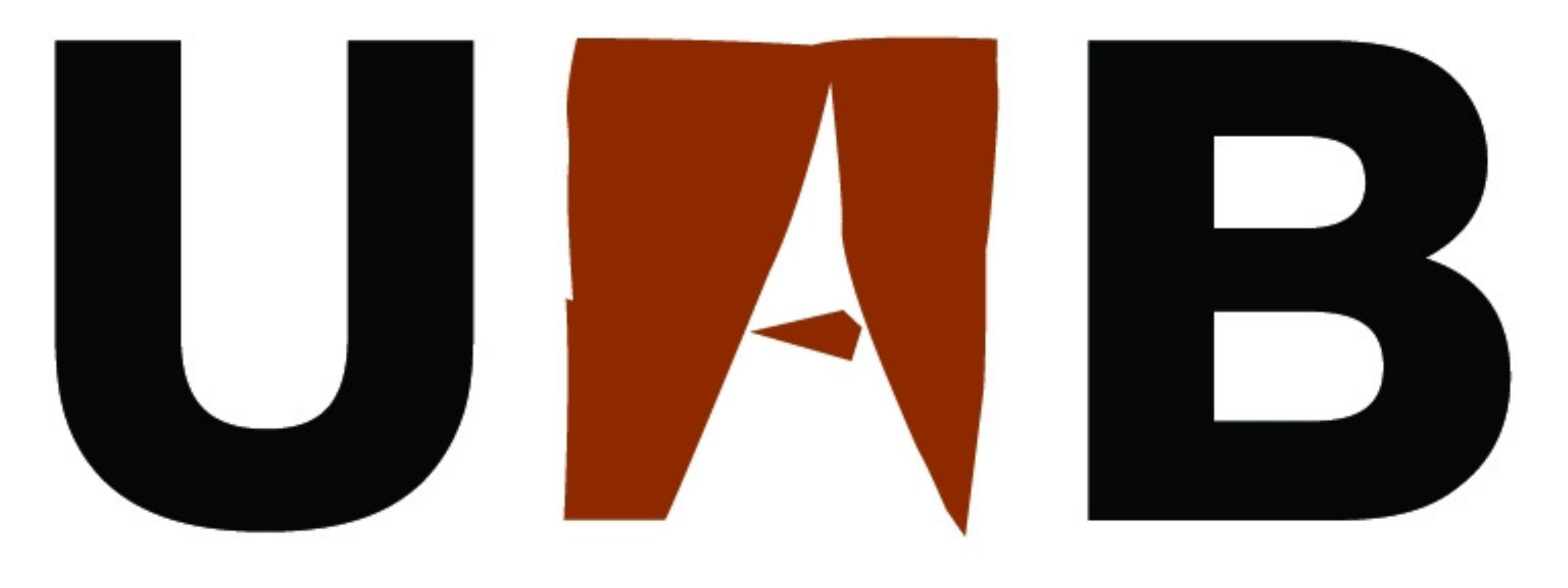}
\end{figure}

\vspace{1cm}
\HRule\\
\vspace{0.3cm}
{\Huge \bf \color{white}Gravity, Holography and \\[0.34cm] Applications to Condensed Matter}
\HRule\\
\par
\vspace{1.5in}

\par
\color{white}
\Large 
Based on a dissertation by\\
\vspace{0.5cm}
{\huge \bf Matteo Baggioli}
\vspace{0.8in}
\par
\Large aimed at the achievement of\\
\vspace{0.5cm}
{\huge \bf Ph.D. in physics}\\
\vspace{0.5cm}
\Large granted by\\
\Large Departament de F\'{i}sica\\
\Large Universitat Aut\`{o}noma de Barcelona\\
\par
\vspace{0.5cm}

\today
\par
\vspace{0.5in}

\emph{\Large Supervised by:}\\
{\bf \Large Prof. Oriol Pujol\'as Boix}\\

\color{black}
\end{center}
\end{titlepage}

\cleardoublepage
\thispagestyle{empty}
\chapter*{Acknowledgements}
\label{chapter:acknowledgements}
\addcontentsline{toc}{chapter}{Acknowledgements}
\textit{I have been planning for a long time to make a huge and very funny list of acknowledgements to start with a laugh the reading of my thesis and with the convictions that physicists could be social and amusing animals but I suddenly realized that maybe it is a good occasion to be serious (once in life) and maybe... we are really the nerds people think we are.\\
That said, the first thank goes, without any doubt, to my advisor Oriol Pujol\`as for bearing with me all these 3 years. It must have been a hard task!!! Thanks for sharing your time and your knowledge with me and for teaching me how to behave in this scientific world.\\
A sincere thank goes also to all my collaborators and colleagues (too many to list them all), who enlightened my lazy brain with deep discussions and intriguing problems. I owe you much of what I know. \\
There are not enough time nor enough meaningful words to show my family the gratitude they deserve. I am grateful to let me be who I am and to encourage me unconditionally in everything I did and I have been doing. You a big part of my strength.\\
Lot of people shared their time, knowledges, energies, thoughts and fears with me during these past three years in Spain (Oops, Catalunya!). I would like to thank you one by one but I do not trust much words, I just prefer to remember the moments spent together without any ''italianadas''. I just want to make sure that you know that without you this would have not been possible (and probably nobody would have felt the difference!).\\
Everybody of you gave me something (and someone of you, the lucky ones lol, are still giving it) and most importantly taught me something, which I will bring with me forever and everywhere together with the tons of ink I have on my skin.\\
Part of these 200 pages of ''very useful'' scientific developments are yours and more than that my survival after 3 years of intense theoretical physics activity has been possible just because all of you.\\
Thanks...}

\thispagestyle{empty}
\newpage
\thispagestyle{empty}
\vspace*{0.01\textheight}
\begin{figure}
\centering
\bf \Large '' BE WHO YOU ARE AND DO WHAT YOU WANT,\\[0.1cm] BECAUSE THOSE WHO MIND DONT MATTER\\[0.1cm]AND THOSE WHO MATTER DONT MIND AT ALL ... ''     \\[0.2cm](dedicated)
\end{figure}
\begin{quote}
\bf \large Non ce la farai mai ...
\end{quote}
\hfill La Dodo 
\begin{quote}
\bf \large Non importa chi sei ma chi vuoi diventare ...
\end{quote}
\hfill  \begin{flushright}Big Z\end{flushright}
\begin{quote}
\bf \large Que faena me das idioti ...
\end{quote}
\centering
\begin{mdframed}[style=MyFrame]
\centering
\Large \textbf{DEDICATION}
\flushleft{\textit{Dedicated to my sister who always told me to do something better remunerated\ldots\\[0.15cm]
Dedicated to my mom, now you know what I was thinking about when I was not replying you about choosing pasta or pizza\ldots\\[0.15cm] 
Dedicated to my dad, because you gave me the \textit{locura} I need for this mad world\ldots\\[0.15cm]
Dedicated to all the eyes I crossed during these years and all the people who made me smile when a smile was the last of my thoughts\ldots\\[0.15cm]
Dedicated to all the girls (and sometimes boys, huh!) who stole my time from physics, you could have done it way more, who cares about physics c'mon \ldots\\[0.15cm]
Dedicated to all the people I did not listen to, it looks like I did well\ldots}}
\end{mdframed}
\clearpage
\thispagestyle{empty}
\justify
\chapter*{Abstract}
\label{chapter:abstract}
\addcontentsline{toc}{chapter}{Abstract}
Strongly coupled systems are elusive and not suitable to be described by conventional and perturbative approaches. They are nevertheless ubiquitous in nature, especially in condensed matter physics. The AdS/CFT correspondence, born from the excitement of ideas and efforts employed in finding out a possible description of quantum gravity, introduced an unexpected and brandnew perspective for dealing with strongly coupled quantum field theories. In its more general formulation, known as gauge-gravity duality, it accounts for an effective and efficient tool to tackle these questions using a dual gravitational description which turns out to be much easier than the original one. In the last years, an increasing number of developments have been achieved in applying the duality to understand new strongly correlated phases of matter and their origin.\\
Momentum relaxation is an ever-present and unavoidable ingredient of any realistic condensed matter system. In real-world materials the presence of a lattice, impurities or disorder forces momentum to dissipate and leads to relevant physical effects such as the finiteness of the DC transport properties, i.e. conductivities. The main purpose of this thesis is the introduction of momentum dissipation and its consequent effects into the framework of AdS/CMT, namely the applications of the gauge-gravity duality to condensed matter.\\
A convenient and effective way of breaking the translational symmetry associated to such a conservation law is provided by massive gravity (MG) bulk theories. We consider generic massive gravity models embedded into asymptotically Anti de Sitter spacetime and we analyze them using holographic techniques. We study in detail their consistency and stability. We then focus our attention on the transport properties of the CFT duals. A big part of our work is devoted to the analysis of the electric conductivity in relation to possible universal bounds and the existence of holographic metal-insulator transitions. We moreover initiate the study of the viscoelastic response and we consider the possible violation of the well known KSS bound. We finally describe the effects of momentum relaxation on the well known holographic models for superconductivity.\color{black}

\thispagestyle{empty}
\cleardoublepage
\thispagestyle{empty}
\acresetall
\chapter*{List of works}
\label{papers}
\addcontentsline{toc}{chapter}{List of published works}

\begin{enumerate}
  \item \textbf{Electron-Phonon Interactions, Metal-Insulator Transitions, and Holographic Massive Gravity}
  
   M.~Baggioli, O.~Pujolas
  
  Published in Phys.Rev.Lett. 114 (2015) no.25, 251602,\\ \href{https://inspirehep.net/record/1326017}{arXiv:1411.1003 [hep-th,cond-mat.str-el]}

 \item \textbf{Phases of holographic superconductors with broken translational symmetry}
  
   M.~Baggioli, M.~Goykhman
  
  Published in JHEP 1507 (2015) 035 ,\\ \href{https://inspirehep.net/record/1362529}{arXiv: 1504.05561[hep-th,cond-mat.str-el,cond-mat.supr-con]}
  
  \item \textbf{Drag Phenomena from Holographic Massive Gravity }
  
   M.~Baggioli, D.~K.~Brattan
  
  Sent to Quantum and Classical Gravity,\\\href{https://inspirehep.net/record/1365318}{arXiv: 1504.07635[hep-th]}
  
  \item \textbf{Under The Dome: Doped holographic superconductors with broken translational symmetry}
  
   M.~Baggioli, M.~Goykhman
  
  Published in JHEP 1601 (2016) 011 ,\\ \href{https://inspirehep.net/record/1399219}{arXiv: 1510.06363[hep-th, cond-mat.str-el, cond-mat.supr-con]}
  
   \item \textbf{Solid Holography and Massive Gravity }
  
   L.~Alberte, M.~Baggioli,  A.~Khmelnitsky, O.~Pujolas
  
   Published in JHEP 1602 (2016) 114,\\ 
   \href{https://inspirehep.net/record/1402123}{arXiv: 1510.09089[hep-th, cond-mat.str-el]}
   
   \item \textbf{Viscosity bound violation in holographic solids and the viscoelastic response }
  
   L.~Alberte, M.~Baggioli, O.~Pujolas
   
   Published in JHEP 1607 (2016) 074,\\
   \href{https://inspirehep.net/record/1415150}{arXiv: 1601.03384[hep-th]}
    
      \item \textbf{	
On holographic disorder-driven metal-insulator transitions}
  
   M.~Baggioli, O.~Pujolas
   
   Sent to JHEP,\\\href{https://inspirehep.net/record/1418214}{arXiv: 1601.07897[hep-th,  cond-mat.str-el]}
    
    \item \textbf{Chasing the cuprates with dilatonic dyons	
}
  
  A.~Amoretti, M.~Baggioli, N.~Magnoli, D.~Musso
  
  Published in JHEP06(2016)113,\\
  \href{https://inspirehep.net/record/1426857}{arXiv: 1603.03029[hep-th,  cond-mat.str-el]}
    
    \item \textbf{On effective holographic Mott insulators	
}
  
  M.~Baggioli, O.~Pujolas
  
  Sent to JHEP,\\ \href{https://inspirehep.net/record/1454106}{arXiv: 1604.08915[hep-th,  cond-mat.str-el]}
\end{enumerate}

\thispagestyle{empty}
\cleardoublepage 
\thispagestyle{empty}

\tableofcontents
\cleardoublepage   

\chapter*{List of abbreviations}
\addcontentsline{toc}{chapter}{Abbreviations}
\begin{acronym}
\acro{QCD}{Quantum Cromodynamics}
\acro{QGP}{QUark Gluon Plasma}
\acro{FD}{Fermi-Dirac}
\acro{MB}{Maxwell-Boltzmann}
\acro{BH}{Black Hole}
\acro{GSL}{Generalized second law}
\acro{SC}{Superconductor}
\acro{RN}{Reissner Nordstrom}
\acro{BF}{Breitenlohner-Freedman}
\acro{MIT}{Metal Insulator transition}
\acro{KSS}{Kovtun, Son and Starinets}
\acro{NG}{Nambu Goto}
\acro{MG}{Massive Gravity}
\acro{NED}{Non Linear Electrodynamics}
\acro{DBI}{Dirac Born Infield}
\acro{EMD}{Einstein Maxwell Dilaton}
\acro{GPKW}{Gubser, Polyakov, Klebanov, Witten}
\acro{GR}{General Relativity}
\acro{AdS}{Anti De Sitter}
\acro{CFT}{Conformal Field Theory}
\acro{GGD}{Gauge Gravity Duality}
\acro{HMG}{Holographic Massive Gravity}
\acro{dRGT}{De Rham Gabadadze Tolley}
\acro{BB}{Black Brane}
\acro{CM}{Condensed Matter}
\acro{LVMG}{Lorentz Violating Massive Gravity}
\acro{UV}{Ultraviolet}
\acro{IR}{Infrared}
\acro{DC}{Direct Current}
\acro{AC}{Alternate Current}
\acro{VEV}{Vacuum Expectation Value}
\acro{QNM}{Quasinormal Mode}
\acro{GH}{Gibbons Hawking}
\acro{RG}{Renormalization Group}
\acro{TB}{Translation Breaking}
\acro{DDMIT}{Disorder Driven Metal Insulator Transition}
\acro{CCS}{Charge Conjugation Symmetric}
\acro{FP}{Fierz-Pauli}
\acro{vDVZ}{van Dam-Veltman-Zakharov}
\acro{EH}{Einstein-Hilbert}
\acro{ADM}{Arnowitt, Deser and Misner}
\acro{BD}{Boulware-Deser}
\end{acronym}
\cleardoublepage

\pagenumbering{arabic}
\setcounter{page}{1}


\part{Introduction}
\label{introduction}
\begin{figure}
\includegraphics[scale=0.3]{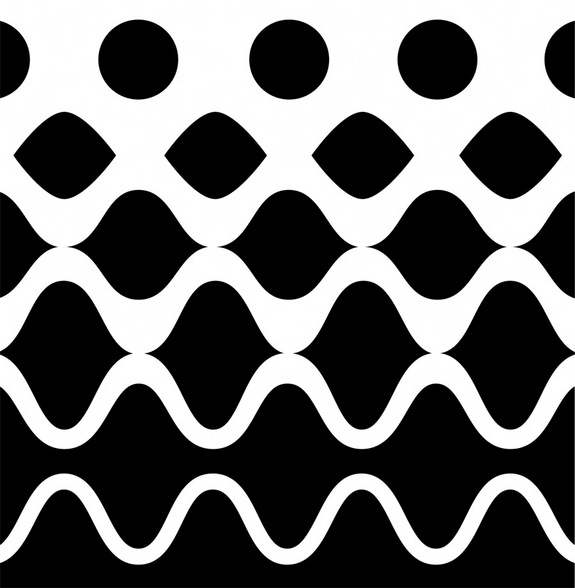}
\end{figure}
\epigraph{Duality in mathematics is not a theorem, but a ''principle''}{\textit{Michael Atiyah}}
The idea of \textbf{duality} is ubiquitous in fundamental sciences. It is very powerful and useful, and has a long history going back hundreds of years. Over time it has been adapted and modified and it has finally taken the stage in the modern scientific scenario. It appears in many subjects in mathematics (geometry, algebra, analysis) and in physics.
\textit{Fundamentally, a duality gives two different points of view of looking at the same object}. In theoretical physics one often says that a non-trivial equivalence between two models is a duality and that two very different looking physical	systems	can	nevertheless be	identical. The first example of such an idea in the context of physics goes back to the early history and it refers to the nature of light. Aristotle was one of the first to publicly hypothesize about the nature of light, proposing that light is a disturbance in the element aether (that is, it is a wave-like phenomenon). Democritus -the original atomist- argued that all things in the universe, including light, are composed of indivisible sub-components (light being some form of solar atom). This dicotomy formalized later on through the work of Max Planck, Albert Einstein, Louis de Broglie, Arthur Compton, Niels Bohr, and many others, takes the name of \textit{Wave-Particle duality} and it is nowadays phrased as: all particles also have a wave nature (and vice versa).\\
The idea that a particular problem can have more than one description and that depending on the situation one is more convenient than the other spread into several fields of physics and becomes a strong and robust computational tool. Early prototypes are the \textit{Electro-Magnetic duality} and the \textit{Kramers-Wannier duality}, which allows for example to solve the 2-dimensional Ising model exactly. Along with the formulation of Supersymmetry and String Theory a huge class of dualities has been discovered and analyzed: S-Duality, T-Duality, U-Duality, Mirror Symmetry, Montonen-Olive duality, etc\dots\\
The astonishing results following this program are that entities concerning the theoretical description of a system such as:
\begin{itemize}
\item the nature of the fundamental degrees of freedom;
\item the number of spacetime dimensions;
\item the spacetime's size and topology;
\item the couplings' strengths;
\end{itemize}
are not \textit{duality invariant} concepts and that despite the physics of a particular system is one and only its description can be absurdly different in different \textit{duality frames}.\\
All these ideas along with the brandnew openminded attitude lead to the birth of the so called \textbf{AdS-CFT correspondence}, first formulated by Juan Maldacena in 1997, which represents not only the single most important result in string theory in the last twenty years but also the most shining and deeply surprising example of duality. The original example of AdS/CFT linked two very special theories. The gravitational side involved a particular extension of gravity (type IIB supergravity) on a particular geometry (AdS$_5\,\times\,$S$_5$) whereas the QFT was the unique theory with the largest possible amount of supersymmetry ($\mathcal{N}=4 \,SYM$). There is a specific dictionary that translates between the theories. This relationship has no formal mathematical proof. However a very large number of checks have been performed. These checks involve two calculations, using different techniques and methods, of quantities related by the dictionary. Continual agreement of these calculations constitutes strong evidence for the correspondence.
The first example has by now been extended to many other cases, and AdS/CFT is more generally referred to as the \textbf{Gauge-Gravity duality} (GGD). Formally this is the statement that gravitational theories in (N+1) dimensions can be entirely and completely equivalent to non-gravitational quantum field theories in N dimensions.
The AdS/CFT correspondence has a very useful property. When the gravitational theory is hard to solve, the QFT is easy to solve, and vice-versa! This opens the door to previously intractable problems in QFT through simple calculations in gravity theories.
Moreover AdS/CFT allows a conceptual reworking of the classic problems of QFT. Indeed if a QFT can be equivalent to gravitational theory, then neither one is deeper than the other. Maybe, the non-perturbative definition of a QFT is not a QFT anymore but it takes the form of a gravitational one. Physicists can therefore use it to develop new intuitions for both QFT and Quantum Gravity in a symbiotic fashion.\\
The main feature of the GGD is that it qualifies as a \textit{Strong-Weak duality} in the sense that it relates a theory with a coupling constant g to an equivalent theory with coupling constant $1/g$. More in details, the \textit{dual} of a strongly coupled quantum field theory is represented by a weakly coupled and classical theory of gravity, i.e. General Relativity. Therefore, exploiting this connection GGD has become a very efficient (and sometimes the only one available) tool to attack strongly coupled problems in the context of:
\begin{enumerate}[(a)]
\item QCD and Quark Gluon Plasma (QGP) Physics
\item Condensed Matter and Quantum Phase Transitions
\item Non Equilibrium Physics
\item Information Theory
\end{enumerate}
In this work based on my PhD thesis we focus our attention on the applications of the Gauge-Gravity Duality towards the Condensed Matter world, which are usually referred as \textbf{AdS-CMT}, making use of the \textit{motto}:\\
\begin{nscenter}
\framebox{\underline{Strongly Coupled/Correlated} and \underline{Quantum} Condensed Matter Systems}\\[0.15cm]
$\updownarrow$\\[0.15cm]
\framebox{
\underline{Weakly Coupled} and \underline{Classical} Gravitational Theories (GR)}\vspace{0.5cm}
\end{nscenter}
It is the most brillant example of inter-disciplinarity.\\
Condensed Matter is a boiling pot of interesting questions and open problems which seems to conflict the old-known and well established paradigms of the field itself. The access and the study of strongly coupled and strongly correlated materials opened a completly new and misterious scenario where the single particle approximation and the perturbative methods are proved of no help.\\[0.2cm]
Through the chapters of this manuscript we will encounter the hottest open problems in CM such as:
\begin{enumerate}[(a)]
\item the nature of the \textit{Strange Metals}
\item the role of the \textit{High-Tc Superconductors}
\item the existence of \textit{Metal-Insulator transitions} (MIT)
\item the role of \textit{disorder} in CM systems and the appearance of \textit{Anderson Localization}
\end{enumerate}
and we will attack them using the new tool given us by the GGD.\\[0.15cm]
The novelty and the crucial point of the present work is the introduction of momentum dissipation effects into the GGD setup. The (explicit/spontaneous) breaking of translational symmetry is a mandatory ingredient to describe condensed matter system where the presence of \textit{lattice, impurities, disorder, etc...} is at the order of the day. In the spirit of \textit{effective field theories} (EFT) we mimick such a mechanism considering \textbf{Massive Gravity} (MG) theories where diffeomorphism invariance is (partially) broken. Such a modification of the usual GR picture will allow us to consider ''duals'' of metallic and insulating configurations and eventual transitions between them. This represents a step forward in realizing \textbf{holographic effective field theories} for condensed matter and in sharpening the GGD tool towards its concrete application to ''real world'' systems.\\[1cm]
\textbf{Organization of the manuscript:}\\[0.5cm]
Part \ref{introduction} is devoted to the theoretical background necessary in order to get through this work. It does not represent to any extent original material by the author. It appears in a schematic and coincise fashion in order to introduce in a comfortable and fast way the reader to the main body of the manuscript which constitute the original scientific contributions of the present work.\\[0.2cm]
In part \ref{results} we present the original results of this manuscript which contribute to the development of the AdS-CMT field and its ''Real-World'' applications.\\[0.2cm]
In part \ref{part:Final Remarks} we conclude with some final remarks, a brief summary and a list of ideas and homeworks for the future.\\[0.5cm]

This work is based on the following published papers:
\begin{enumerate}
\item \href{https://arxiv.org/abs/1411.1003}{''Electron-Phonon Interactions, Metal-Insulator Transitions, and Holographic Massive Gravity''}
\item \href{https://arxiv.org/abs/1504.05561}{''Phases of holographic superconductors with broken translational symmetry''}
\item \href{https://arxiv.org/abs/1510.06363}{	
''Under The Dome: Doped holographic superconductors with broken translational symmetry''}
\item \href{https://arxiv.org/abs/1510.09089}{''Solid Holography and Massive Gravity''}
\item \href{https://arxiv.org/abs/1601.03384}{''Viscosity bound violation in holographic solids and the viscoelastic response''}
\item \href{https://arxiv.org/abs/1601.07897}{''On holographic disorder-driven metal-insulator transitions''}
\end{enumerate}

\chapter{Condensed Matter}
\label{CM}
\localtableofcontents
\noindent{\color{myblue}\rule{\textwidth}{1.5pt}}\par\medskip
\begin{figure}[h]
\includegraphics[scale=0.5]{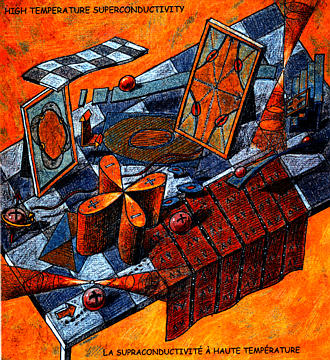}
\end{figure}
\epigraph{Bad times have a scientific value. These are occasions a good learner would not miss.}{\textit{Ralph Waldo Emerson}}
\textbf{Disclaimer:}
This represents an original and short recompilation by the author of very well known material regarding the condensed matter background necessary to introduce and motivate the original contributions present in part. \ref{results}.
This is not meant to be a detailed and complete solid state physics essay for which we refer to the well known condensed matter textbooks \cite{AshcroftMerminBook,AbrikosovBook,KittelBook} on which this chapter is based. Several inspirations are also taken from the nice reviews \cite{MITarticle,Sachdev:2011cs}.\\
\section{Solid state physics}
\begin{figure}[h]
\centering
\includegraphics[scale=0.6]{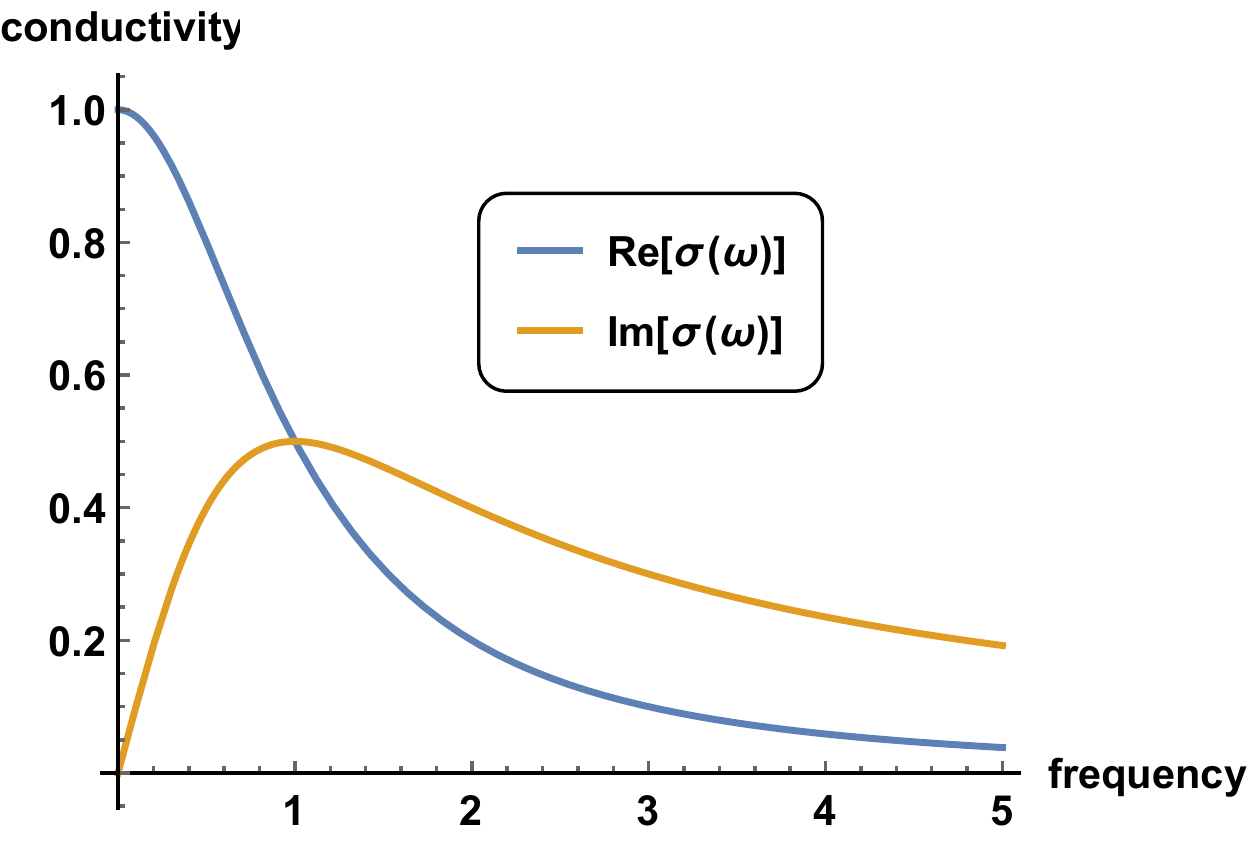}
\caption{Drude model. Optical conductivity $\sigma(\omega)$ with $\sigma_0=\tau=1$.}
\label{DrudePic}
\end{figure}

\begin{figure}[h]
\centering
\includegraphics[scale=0.4]{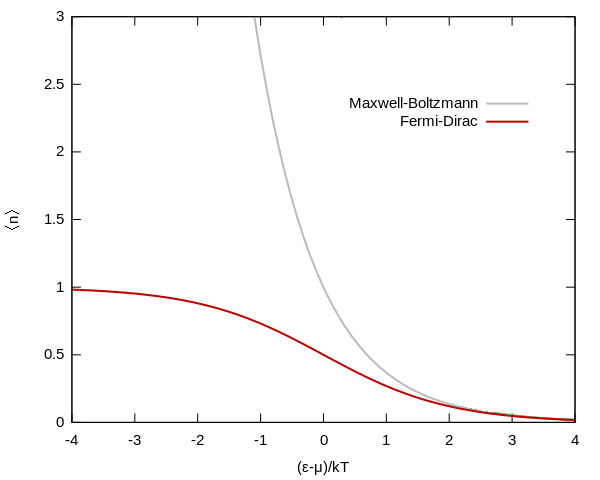}%
\qquad 
\includegraphics[scale=0.58]{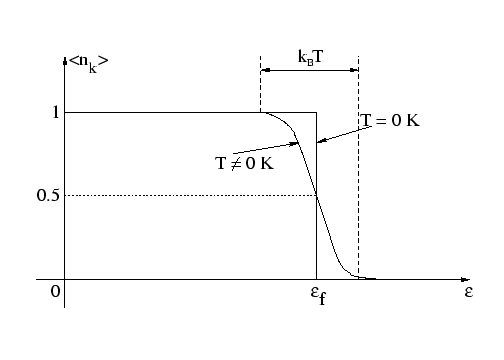}
\caption{\textbf{Left: }Comparison between the Maxwell-Boltzmann and the Fermi-Dirac distributions. \textbf{Right: }Fermi-Dirac distribution at zero and finite temperatures.}
\label{DistrPic}
\end{figure}

\begin{figure}[h]
\centering
\includegraphics[scale=0.6]{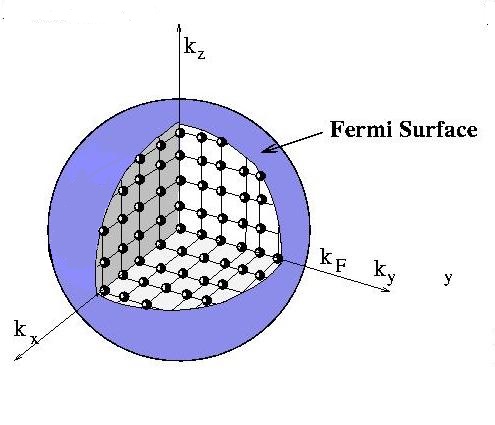}%
\caption{Fermi Sphere and Fermi Surface.}
\label{FermiSpic}
\end{figure}

The definition and the classification of materials according to their transport properties is one of the most important target of solid state physics. In particular, electric conduction appears as one of the first question to takle in order to separate metals, \textit{i.e.} materials which conduct electricity, and insulators, \textit{i.e.} materials who dont. To this regard, the most relevant quantity one has to introduced is the so-called DC conductivity $\sigma_{DC}$, which is the static conductivity defined at zero frequency $\omega=0$.
From the latter point of view one can distinguish metals and insulator through their DC conductivity temperature dependence as the following:
\begin{equation}
d\sigma_{DC}/dT\,<\,0\,\longrightarrow\,\text{METAL}\,,\qquad\,d\sigma_{DC}/dT\,>\,0\,\longrightarrow\,\text{INSULATOR}\,.
\label{DefInsulator}
\end{equation}
or more appropriately from the value of their DC conductivity at zero temperature:
\begin{equation}
\sigma_{DC}(T=0)\,=\,\text{finite}\,\longrightarrow\,\text{METAL}\,,\qquad\sigma_{DC}(T=0)\,=\,0\,\longrightarrow\,\text{INSULATOR}\,,
\end{equation}
\textbf{The Drude model}\\[0.2cm]
The first attempt of describing the experimentally observed features of a metal goes back to the beginning of the previous century, just three years after the discovery of the electron by J.J.Thomson. The model, proposed in 1900 by P.Drude and inspired by Kinetik theory, takes indeed the name of \textit{Drude model} and despite the several approximations and simplifications its success has been considerably high and the model still represents a practical and quick way to form a sketchy picture of what is really happening in an ordinary metal.\\
To be more precise, we consider the metal as a classical and diluite collection of free electrons which move, as balls in a pinball, within the lattice of the material represented by the massive and immobile ions. The electrons scatter in an instantaneous way on the non dynamical ions with a characteristic \textit{relaxation time} $\tau$, which measures indeed the average time between two consecutive collisions. The collisions result in a change of the electron velocity and they are the only ''interactions'' considered in this simple picture. In absence of any external perturbation, the electrons undergo a zig-zag random motion; their average velocity remains null and no correspondent average current is produced. On the contrary, when an external electric field $\vec{E}$ is switched on a non null average velocity for the electrons cloud arises as:
\begin{equation}
<v>\,=\,\underbrace{<v_0>}_{\,=\,0}\,-\,\frac{e\,E\,<t>}{m}\,=\,-\frac{e\,E\,\tau}{m}
\end{equation}
where $\tau=<t>$ is precisely the average time of collision and m the effective mass of the charge carriers. We can then express the correspondent net current as:
\begin{equation}
J\,=\,-\,n\,e\,\vec{v}\,=\,\left(\frac{n\,e^2\,\tau}{m}\right)\,E
\end{equation}
where n is the density of electrons which in an ordinary metal is order $n\sim 10^{28}/\text{cm}^3$.\\
From the previous expression one can directly derive the static (DC) electric conductivity at zero frequency:
\begin{equation}
\boxed{\sigma_{DC}\,=\,\frac{n\,e^2\,\tau}{m}}
\label{DCdrude}
\end{equation}
This formula represents the most important achievement of Drude theory and for ordinary metals it gives numbers which are surprisingly close to the experimental data.\\
The same model is more powerful than what we just explained and we can consider also time dependent situations where the external fource is not static anymore:
\begin{equation}
\frac{d\,\vec{p}(t)}{dt}\,=\,-\frac{\vec{p}(t)}{\tau}\,+\,\vec{f}(t)
\label{ACdrude}
\end{equation}
with $f(t)=-e\,E(t)$ in our case.\\
The damping term proportional $\sim 1/\tau$, produced by the collisions, incorporates all the effects which in real materials are provoked by the lattice, the impurities and disorder. Technically speaking, it explicitely represents the breaking of translational invariance and it has a crucial role for the computation of the conductivity. As one can see from \eqref{DCdrude}, if no collision are introduced in the model, and the average time $\tau\rightarrow \infty$, the correspondent static conductivity becomes infinite and the charge carriers can be infinitely accelerated by the static electric field. This is clearly not what happens in a real material.\\
That said, writing down equation \eqref{ACdrude} in Fourier space, one suddenly realizes that:
\begin{equation}
J(\omega)\,=\,-\frac{n\,e\,p(\omega)}{m}\,=\,\frac{(n\,e^2\,/m)}{1/\tau\,-\,\imath\,\omega}\,E(\omega)\,=\,\frac{\sigma_{DC}}{1\,-\,\imath\,\omega\,\tau}\,E(\omega)
\end{equation} 
which is the results for the optical AC conductivity within the Drude model (see fig.\ref{DrudePic}).
Although its considerable success in explaining and predicting the measured conductivity in ordinary metals, the model has several shortcomings. It is clear that, from the point of view of kinetik theory from which the Drude model is inspired, increasing the temperature T would increase the collision rate $\sim 1/\tau$ and therefore decreasing the conductivity, leading to just the description of a metallic behaviour \eqref{DefInsulator} (unless the density of mobile electrons is null). In a way, only a class of trivial insulators, the ones characterized by no available charge carriers, can be achieved within this framework. There are several other problems which can't be resolved with the Drude model, see the table \ref{tableDrude}.
The biggest mistery that the Drude Model leaves us is the answer to the question:\\[0.1cm]
\centerline{\textit{Why are some materials metals and other insulators?}}\\[0.1cm]
\begin{table}[ht]
\centering 
\begin{tabular}{c} 
\textbf{Drude Model successes}\\ [0.5ex] 
\hline \hline \\ [0.2ex]
First theoretical proof of Ohm's law   \\
Predicts the Hall effect  \\
Predicts the presence of a Plasma frequency   \\
Predicts electric and thermal conductivities to a very good accuracy\\
Weidemann-Franz law  \\ [1.5ex] 
\textbf{Drude Model failures}\\ [0.5ex]
\hline \hline \\ [0.2ex]
\textbf{Presence of materials which are insulators and semiconductors (i.e. not metals)}\\
Temperature dependence of the electric conductivity\\
Temperature dependence of the thermal conductivity\\
Temperature dependence of the specific heat\\
Overestimating the electronic heat capacity\\
Too long mean free path $l\sim v_f \tau$\\ [0.5ex]
\hline
\end{tabular}
\label{tableDrude} 
\caption{Not all the points of the present table are analyzed in detail through the main text. For an exhaustive discussion see the textbooks \cite{AshcroftMerminBook,AbrikosovBook,KittelBook}.}
\end{table}\\
The answer to this question and the origin of all the other failures have to be recasted in the several approximations which have been done:
\begin{enumerate}
\item The electrons are treated as classical and individual particle.
\item The electrons are considered as free and all the interactions are taken as negligible in the system.
\item The ions are not dynamical because designed as immobile and infinitely massive.
\end{enumerate}
\textbf{The Sommerfeld model and band theory}\\[0.2cm]
The first step towards a more complete description is the promotion of classical mechanics to its quantum version, which will lead us to the introduction of the so-called \textbf{Sommerfeld model}.
Treating the electrons like quantum particles (i.e. fermions) rather than molecules of a classical gas represents the first main improvement to the Drude model. Pauli exclusion principle replaces the Maxwell-Boltzmann distribution with the Fermi-Dirac one and at the temperatures of interest ($T<10^3$ K) those two can be amazingly different (see fig.\ref{DistrPic}).
In the quantum description, we need to consider the Schrodinger equation for the electrons:
\begin{equation}
-\frac{\hbar}{2\,m}\,\nabla^2\,\Psi(r)\,=\,\epsilon\,\Psi(r)
\end{equation}
whose solution takes the plane wave form:
\begin{equation}
\Psi_k(r)\,=\,\frac{1}{\sqrt{V}}\,e^{\imath\,k\,r}
\end{equation}
which implies the following energetic spectrum:
\begin{equation}
\epsilon(k)\,=\,\frac{\hbar^2\,k^2}{2\,m}\,,\qquad p\,=\,\hbar\,k\,,\qquad v\,=\,\frac{\hbar\,k}{m}\,.
\end{equation}
Because of the volume restrictions given basically by the Pauli esclusion principle, with the appropriate boundary conditions, the momenta quantizes into:
\begin{equation}
k_i\,=\,\frac{2\,\pi\,n_i}{L}\,,\qquad i\,:\,x\,,\,y\,,\,z
\end{equation}
and in a k-space region of volume $\Omega$ there therefore exist only $\frac{\Omega\,V}{8\,\pi^3}$ allowed values for k. Now we can build the ground state ($T=0$) of N electron states placing the electrons in the one-electron levels we just found. The correspondent volume occupied by piling up the electrons will have the topology of a Sphere (\textit{i.e.} the Fermi Sphere) with radius $k_F$ and whose surface takes the name of the \textbf{Fermi Surface} (see fig.\ref{FermiSpic}).
This is a direct application and consequence of the Pauli Exclusion principle and the density distribution $f_k$ of the electrons in the case of $T=0$ (\textit{i.e.} the ground state) takes just the form of a step function centered at $\epsilon_F$:
\begin{equation}
\begin{cases}
    f_k\,=\,1      & \quad \text{if } \epsilon_k<\epsilon_f\\
    f_k\,=\,0      & \quad \text{if } \epsilon_k>\epsilon_f\\
\end{cases}
\end{equation}
Through some standard computations is easy to show that the total number of electrons and the relative electron density are function of the Fermi momentum $k_F$ as following:
\begin{equation}
N\,=\,\frac{k_F^3}{3\,\pi^2}\,V\,\,\,\,\,\rightarrow\,\,\,\,n\,=\,\frac{k_F^3}{3\,\pi^2}\,.
\end{equation}
One can also define a Fermi velocity $v_F$:
\begin{equation}
v_F\,=\,\frac{\left(3\,\pi^2\,n\right)^{1/3}}{m}
\end{equation}
which results to be not zero also at $T=0$ because generated entirely by the Pauli esclusion principle. In conclusion the total energy of the N electrons ground state is given by summing the single state energies up to the fermi momentum $k_F$ (and taking into account the spin degeneracy) as:
\begin{equation}
E\,=\,2\,\sum_{k<k_F}\,\frac{\hbar^2\,k^2}{2\,m}
\end{equation}
Introducing some temperature T and a chemical potential $\mu$, the distribution of the states gets smoothed out and takes the form of the well established Fermi Dirac distribution:
\begin{equation}
f_i\,=\,\frac{1}{e^{(\epsilon_i-\mu)/k_B T}+1}
\end{equation}
which defines the mean number of electrons in the i level $i:\{k,s\}$, labelled by momentum k and spin s, whose energy takes the value $\epsilon(k)=\frac{\hbar^2k^2}{2m}$. The total number of electrons is just given by summing up what just found as $N=\sum_i f_i$. In the $T=0$ limit we recover the situation we discussed before.\\
Despite the several improvements given by the Sommerfeld theory, the theoretical description is still lacking of an answer for the simple and fundamental question:\\
\textit{Why some materials are insulators and semiconductors (and not metals) ?}\\
All the models considered so far assume the electron density n being an effective parameter, parametrizing the microscopic physics, and they can't account for such a difference. In other words, within that effective fashion, one can just distinguish metals, \textit{i.e.} materials with a finite density of mobile electrons n, from ''trivial'' insulators, \textit{i.e.} materials with zero density of mobile electrons n.\\ \color{black}
Technically speaking, it is intellectually unsatisfying to completely disregard the interactions between the electrons and the ionic cores, except as a source of instantaneous collisions. To get rid of the failures of the Sommerfeld model, and account also for insulating states, we must add interactions between these two sectors; in other words, we have to take into account the periodic potential due to the lattice.\\
The free electrons assumption accounts for a wide range of metallic features but has to improved to reach more efficient descriptions for solids.
Despite all the previous oversimplifications must be abandoned to achieve an accurate model for solids, a remarkable amount of progress can be made by first just abandoning the free electrons approximation (without modifying the single electron approximation or the relaxation time approximation).\\
Once aware of this fact, waving the free electrons assumptions proceeds in different stages:
\begin{itemize}
\item The electrons do not move in empty space but inside a static potential created by the ionic structure of the metal.
\item The ionic cores are not immobile anymore and the dynamics of the ionic position (phonons modes) has to be taken into account. In other words electron-phonon interactions might be not negligible.
\item Eventually Coulomb self-interactions among electrons have to be considered.\color{black}
\end{itemize}
\begin{figure}[h]
\centering
\includegraphics[scale=0.48]{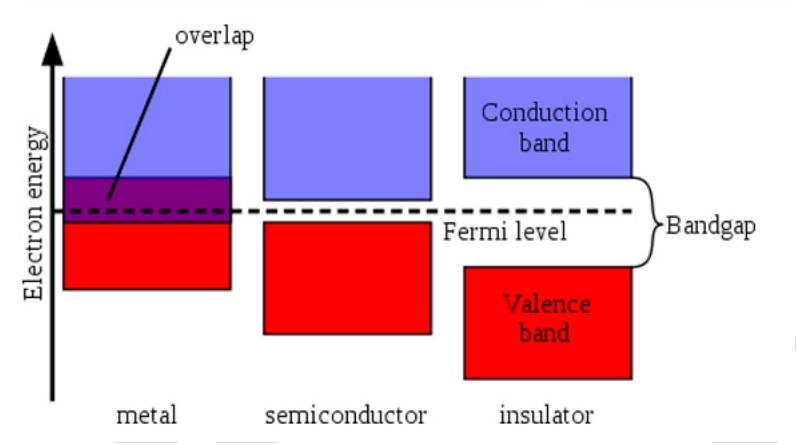}%
\caption{A sketchy explanation of how Band Theory works and distinguishes metals from semiconductors and insulators.}
\label{BandPic}
\end{figure}
For the moment we stick to the first stage, which will be already able to provide the wanted distinction between a metal and an insulator.
Therefore the main point is to include in the single electrons dynamics a potential term due to the ionic cores $V(r)$ which modifies the Schrodinger equation into:
\begin{equation}
\left(-\frac{\hbar^2\,\nabla^2}{2\,m}\,+\,V(r)\right)\,\Psi(r)\,=\,\epsilon\,\Psi(r)
\end{equation}
where the potential V is defined to be periodic $V(r+R)=V(r)$ with periodicity R, defined as the \textit{Bravais lattice vector}.\\
Skipping all the following details, the important thing is that one can prove (using also the famous Bloch theorem) that the energy levels $\epsilon_{n,k}$, for given k, now vary continuosly as k varies. The descriptions of electrons in a periodic potential is therefore given in term of the continuous functions $\epsilon_n(k)$. The information contained in those function is referred to the \textbf{Band structure} and the electrons level specified by $\epsilon_n(k)$ is called an \textit{energy band}.\\
If the free electrons approximation predicted a discrete set of allowed energies, now with the introduction of a periodic potential the available energy states form bands which are somehow the results of the overlap of atomic orbitals. Moreover, the concept of \textit{Fermi Surface} is still the same as before with the only difference that now the single electron states are labelled by two quantum numbers n and k, where n is the level of the band.\\
Now the crucial point for conduction is the position of the Fermi Surface within this electronic band structure. Two possible situations can arise:
\begin{itemize}
\item A certain number of bands are completely filled with all the others remaining empty. The difference in energy between the highest occupied one and the lowest unoccupied one defines the \textit{band gap} $\epsilon_{GAP}$. If $\epsilon_{GAP}\gg k_B T$ then we are in presence of an insulator, whereas if $\epsilon_{GAP}\approx k_B T$ we are speaking of a semiconductor. In the second case the gap is not big and thermal or other fluctuations can bridge it.
\item A specific band is partially filled and the Fermi energy $\epsilon_F$ lies within the energy range of that band. In this case we have a metal.
\end{itemize}
Let's rephrase this concept in a different way. We can define a delocalized band of energy levels in a crystalline solid which is partly filled with electrons as a \textit{conduction band}. The electrons present in the conduction band are vacant, they have great mobility and are responsible for electrical conductivity. On the other way the highest range of electron energies in which electrons are normally present at absolute zero temperature is called \textit{valence band}. The position of the fermi level respect to the conduction band is a crucial fact in determining the electric transport properties of a material. If the Fermi level lies on top of the conduction band, which overlaps with the valence one, then the material is a metal\footnote{If the overlap is small we are in presence of a \textit{semimetal} with very peculiar features. We do not discuss this case here.}; if, on the contrary, there is a big gap between the two bands and the fermi level turns out to be just on top of this gap, the correspondent material will be an insulator (see fig.\ref{BandPic}).\\
At this stage, we are finally able to distinguish the materials accordingly to their conductivity properties into metals and insulators and to provide a simple but quite often accurate description of the observed physics. We will see in the next section that this is sometimes not enough and that the idea of a static lattice and eventually the single electron approximation have to be relaxed as well.\\
An important final remark we need to stress is that in all the models considered the momentum relaxation time $\tau$ has been treated as an effective parameter encoding the breaking of translational symmetry. In an effective field theory fashion, such a parameter encodes several microscopic effects present in real materials which can lead to momentum dissipation via diverse mechanisms. This idea will be mantained also in the holographic models, described in the following, which will represent a strongly coupled version of such effective field theory for momentum dissipation.\\[0.2cm]
\textbf{Phonons}\\[0.2cm]
So far we have considered the ions as a fixed, immobile and rigid array. This is of course an approximation since the ions are not infinitely massive. In a classical theory this is true just at $T=0$; in a quantum theory even at $T=0$ this statement is false because of the indetermination principle $\Delta x \Delta p \geq \hbar$. This oversimplified assumption resulted to be impressively succesful whenever the physical property considered is dominated by the conduction electrons. To understand in a complete fashion the features of the metals (for example the temperature dependence of the DC transport coefficients) and especially to achieve a more accurate description that a rudimentary theory of insulators we must go beyond. One point which is already clear is that under the assumption that the lattice is a static object, in insulators, where the electrons are quiescent, there are no degrees of freedom left to account for their varied features. There is a list of other important problems which needs the presence of the phonons to be solved like:
\begin{itemize}
\item The temperature scaling of the specific heat.
\item The explanation of the Wiedemann-Franz law at intermediate temperature
\item Sound propagation
\item The BCS superconducting instability
\end{itemize}
Once the lattice is not static anymore we can consider the normal modes of vibrations of the crystal as a whole and the dynamics of the small displacements around the equilibrium configuration. The correspondent standing waves, if longitudinally polarized, are called \textit{sound waves} and the quanta of the lattice vibrational field are referred to as \textbf{phonons} \cite{ZimanBook}. The presence of such light modes is ensured by the spontaneous symmetry breaking (SSB) of translational symmetry with the phonons representing indeed the \textit{goldstone} modes associated to such a SSB \cite{LVEFT2}.\color{black} The easiest possible picture is given by replacing the lattice by a volume formed by a gas of phonons carrying energy and momentum and considering the relative normal modes in the so-called harmonic approximation\footnote{Of course there could be and there are anharmonic terms resulting in interactions.}. We will not enter in details the full quantum description of phonons theory which can be find in any CM textbook; we restrict ourselves just in collecting the major results and conclusions.\\
The theory of phonons gives rise in its continuum description to the elastic property of materials and it is much wider than what discussed here. This allows for example to distinguish clearly solids and fluids by the fact that fluids support just longitudinal waves and their rigidity is null. Additionally we can explore the thermodynamical properties of phonons considering them as a gas and applying the Bose-Einstein statistic:
\begin{equation}
n(\omega_{k,s})\,=\,\frac{1}{e^{\hbar\, \omega /k_B T}-1}
\end{equation}
and constructing the so-called Debye theory which for example predicts that the phonon gas energy U takes the form of:
\begin{equation}
U\,=\,3\,k_B\,T\,N\left(\frac{k_B\,T}{\hbar \Omega_D}\right)\,\int_0^{\beta\,\hbar\,\Omega_D}\,\frac{x^3}{e^x-1}\,dx
\end{equation}
which turns out to be very successful in explaining the thermodynamical features of metallic and insulating materials.\\
In conclusion, starting from the classical Drude Model, inserting the effects of quantum mechanics, relaxing the free electron approximation and finally introducing the dynamics of the ionic cores we reached a good description of lots of phenomena which real solids show off.\\[0.5cm]
We end here our quick journey through the basics of solid state physics. This is of course not meant to be a complete, precise and detailed discussion of the topics followed but just a small appetizer for the reader. We end up with a successful, even if simple and approximated, description of several features of metallic and insulating states. A considerable percentual of realistic materials are well enough described by the frameworks we presented and just in recent years we had to face new challenges linked with novel exotic phases. These new situations, which do not fit in what already explained, are direct consequence of \textbf{strong coupling} and \textbf{strong correlation} and will force us to take a new perspective and rely on new tools.
\section{Metal-insulator transitions and disorder}
\begin{figure}[h]
\centering
\includegraphics[scale=0.85]{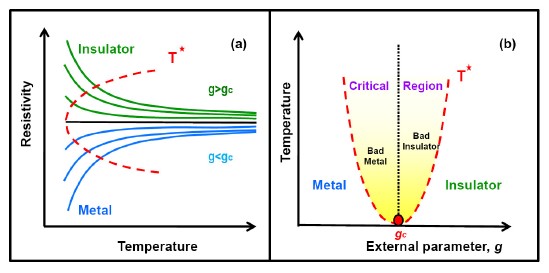}
\caption{Metal insulator transitions. Picture taken from \cite{MITarticle}.}
\label{MITpic}
\end{figure}

The surprisingly simple models described in the previous section appeared to be extremely successful in the description of weakly interacting electronic systems. The argument motivated by the filling of the electronic bands turned out to be very efficient in distinguishing good metals like Gold from insulating materials such as Silicon. Although its long-lived success, in 1937, a material has been found to have an half-filled band but nevertheless to not be able of conducting electricity. In other words, such a material was showing insulating properties even if according to band theory it should have been an ordinary metal. 
This discovery raised the following questions:\\[0.2cm]
\centerline{\textit{How materials with partially filled bands could be insulators?}}\\[0.1cm]
\centerline{\textit{How could an insulator transform into a metal as a continuous external parameter is varied?}}\\[0.2cm]
What we are after, is what is known as \textit{Metal-Insulator transition}, which represents one of the most striking example of quantum phase transition (QPT) \cite{MITbelitz,MITkravchenko,MITimadaetal,DobrajevskiBook, GebhardBook,MITarticle}.\\
QPTs are phase transitions where the ground state undergoes a dramatic change of phase at strictly zero temperature $T=0$. Quantum effects, rather than thermal fluctuations, are the responsable for such a phase transition; close to an QPT, an MIT for example, the properties of the system are dramatically modified upon dialing an external not thermal parameter such as pressure, charge carrier concentration, etc. In the case of an MIT, the electric conductivity of a material can change over a huge range of values upon modifying a parameter (see \ref{MITpic}).\\
There are several reasons why attacking the MIT problem is very challenging; metals and insulators appear to be very distinct, stable and robust systems. Moreover their excitations have completely opposite nature (fermionic quasiparticles VS bosonic d.o.f.) which one can hardly imagine to connect together in a continuous way varying a physical parameter. Not only that, but additionally, MIT transitions do not exhibit any symmetry breaking pattern and there is no clear order parameter to use within a continuous description \'a la Landau.\\[0.1cm]
\centerline{\textit{How can band theory fail ?}}\\[0.1cm]
Band theory relies on the fact that the kinetik energy of the charge carriers is parametrically larger that the interactions scales present in the system. As a result, a good description of the system can be realized through a diluite collection of quasiparticles which are weakly coupled as defined in the Fermi Liquid theory. This might be not the case if interactions of several types become comparable with the Fermi energy of the material. Whenever that happens, band theory is not a reliable scheme anymore and its predictions often fail. There are nowadays several known examples of this type such as the famous V$_2$O$_3$ and La$_{2-x}$Sr$_x$CuO$_4$.\\
In the scenario when the potential energy becomes competitive or even predominant over the Fermi energy scale, the electrons mobility is dramatically stopped and the charge carriers localized without being able to transport anymore.\\
Historically, we can divide the MIT into two classes:
\begin{enumerate}
\item MIT triggered by electronic correlations (or electron-electron interactions): \textbf{Mott transitions}
\item MIT triggered by disorder: \textbf{Anderson transitions}
\end{enumerate} 
Although the increasing interest and progress in understanding Mott transitions and the effects of the electron-electron self interactions through, the rest of this work we will focus our attention just to the second case where the responsable for localization is disorder \cite{DisorderAbrahams,DisorderLee}.
The problem have been analyzed for the first time by Anderson in 1958 \cite{AndersonOriginal} as the study of the diffusion of electrons in a random potential. The cardinal idea is that quantum interference produced by disorder can stop completely the classical expected diffusion. The mean free path becomes very short and the electronic wavefunction becomes a collection of exponentially localized states:
\begin{equation}
\Psi(r)\,=\,e^{-|r|/\xi_{loc}}
\end{equation}
where $\xi_{loc}$ is the so called \textit{localization length}.
Once the electronic states localize into $\delta$-function peaks, their overlap becomes very low and as a consequence the conductivity strongly decreases.\\
Diffusion in the material becomes null and the correspondent conductivity:
\begin{equation}
\sigma\,=\,e^2\,D\,v\,,\qquad v\,=\,\frac{dn}{d\mu}
\end{equation}
falls down to zero.\\
Despite the large progress in the field and the various resolutive proposals (phenomenological $\beta$ function, scaling theories, random matrix models, DMFT) Anderson Localization still remains an open and intriguing question. Indeed the effects of disorder and electron-electron interactions are usually comparable in realistic situations and the number of particle playing in the game is usually very high. These two factors make the non-interacting single particle reasoning made by Anderson too naive and open the door towards:
\begin{itemize}
\item Anderson-Mott transitions, where both the effects of disorder and electrons self-interactions have to be both taken into account;
\item Many-Body Localization (MBL) where the single particle wavefunction is not a reliable tool anymore.
\end{itemize}
\centerline{\textit{What is the fate of Anderson Localization when the constituent particles interact between themselves?}}\vspace{0.3cm}
\centerline{\textit{What happens to Anderson Localization in a many-body problem}}
\centerline{\textit{where the single electron approximation is not valid anymore?}}\vspace{0.5cm}
\textbf{Strongly correlated systems}, which cannot be effectively described in terms of independent and non-interacting entities, still constitute one of the most intriguing and misterious research field in modern solid state physics. The absence of a single particle approximation and a perturbative regime makes the theoretical description of such a systems very hard and call for new innovative tools. \textbf{Gauge gravity duality} could possibly be one of them.
\section{High-Tc superconductivity and quantum criticality}
\begin{figure}[h]
\centering
\includegraphics[scale=0.7]{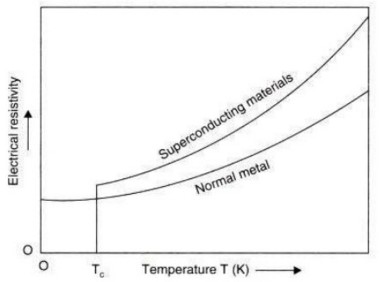}%
\qquad
\includegraphics[scale=0.7]{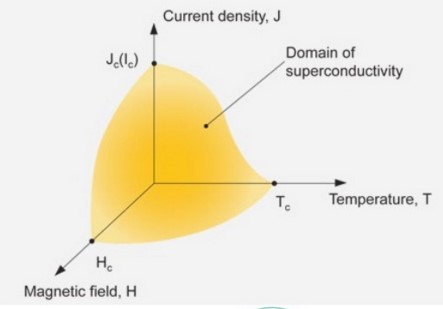}
\caption{Superconducting materials. \textbf{Left: }Comparison of the electric resistivity in the normal metallic phase and in the SC one. \textbf{Right: } Sketch of the phase diagram for the SC state.}
\label{SCpic1}
\end{figure}

\begin{figure}[h]
\centering
\includegraphics[scale=0.9]{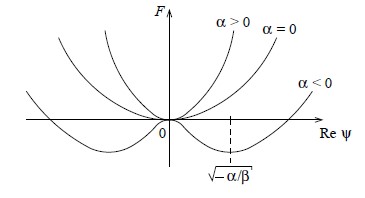}%
\qquad
\includegraphics[scale=0.9]{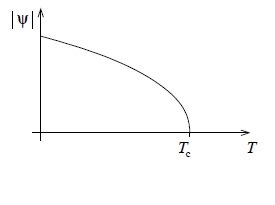}
\caption{Landau theory for super-(fluids/conductors). \textbf{Left: } Free energy. \textbf{Right: } Order parameter.}
\label{SCpic2}
\end{figure}

\begin{figure}[h]
\centering
\includegraphics[scale=1]{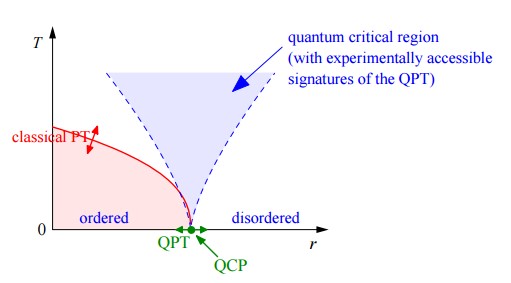}
\caption{Generic phase diagram close to a continuous quantum phase transition.}
\label{pappa}
\end{figure}

\textbf{Conventional Superconductivity}\\[0.5cm]
Superconductivity is a state of matter characterized by a vanishing static electrical resistivity and an expulsion of the magnetic field from the interior of the sample \cite{SchriefferBook,TinkhamBook}.\\
After H.K. Onnes had managed to liquify Helium, it became for the first time possible to reach temperatures low enough to achieve superconductivity in some chemical elements. In 1911, he found that the static resistivity of mercury abruptly fell to zero at a critical temperature $T_c$ of about 4.1 K. In a normal metal, the resistivity decreases with decreasing temperature but saturates at a finite value for $T \rightarrow 0$. That was not the case and he immediately realized that he was standing in front of a new state of solid matter. Under a certain temperature, defined as the \textit{critical temperature}, the system undergoes a phase transition into this novel phase where the resistivity drops down to 0 (see fig.\ref{SCpic1}) which takes the name of \textbf{Superconducting state}. He also realized that a certain amount of magnetic field (\textit{critical magnetic field}) $H_c(T)$ and a critical current $J_C(T)$ would destroy that state of matter and restore the usual metallic normal phase (see fig.\ref{SCpic1}).
A second striking feature is the so called \textit{Meissner Effect}, namely the strong repulsion of the magnetic field from the SC sample. This somehow qualifies a superconductor as a perfect diamagnetic material with zero magnetic permeability $\mu=0$\footnote{This can be better formalized using the so called \textit{London equations} $\nabla^2 B\,=\,\frac{1}{\lambda_c}B$ and $J\sim A$ where $\lambda_c$ is the penetration length and A the gauge field. We refer to CM textbooks for such details.}.\\
Further experiments indicated that the critical temperature, at which the SC transition appears, $T_c \approx \Omega$ where $\Omega$ is the typical oscillation frequency of the ions in the materials. This constituted a strong indication that the SC mechanism is somehow linked to the oscillations of the ionic lattice, \textit{i.e.} the phonons. Conventional superconductivity is indeed the physics of the \textbf{Cooper Pairs}, bound states of two electrons glued together by electron-phonon interactions. \textbf{BCS} (Bardeen, Cooper and Schrieffer) theory predicts that at sufficiently low temperatures, electrons near the Fermi surface become unstable against the formation of Cooper pairs. Cooper showed that such binding will occur in the presence of an attractive potential, no matter how weak. In conventional superconductors, an attraction is generally attributed to an electron-phonon interaction. The BCS theory, however, requires only the potential to be attractive, regardless of its origin. \\
Naively we can imagine the following picture: let us take an electron $e_1$ with defined momentum, energy and spin $e_1=(k,\epsilon_k,\uparrow)$ and another one with same energy but opposite momentum and spin $e_2=(-k,\epsilon_k,\downarrow)$. The Coulomb interaction between the first electron $e_1$ and the ions provokes a displacement in the ionic structure which takes the name of \textit{polarization}; as a consequence the region around $e_1$ is now more positive charged than its equilibrium configuration. This account for an attractive potential U for the second electron $e_2$ which is now forced to follow and form a bond to the first one, creating indeed the so called \textit{Cooper pair}. In conventional SC this is driven by electron-phonon interactions and can be explicitely computed in a diagrammatic fashion. As a result the correspondent critical temperature $T_c$ is directly proportional to the coupling of the electron-phonon interactions:
\begin{equation}
\boxed{T_c\,\sim\,g_{e-ph}}
\end{equation}
and because of this reason BCS theory predicts a maximum critical temperature of order $T_c \sim 30 K$\footnote{The highest BCS superconductor turns out to be $Nb_3Ge$ with $T_c\approx 23K$.}. To increase the critical temperature  the electron-phonon interactions should be stronger and this fact would lead to a strongly couple regime avoiding any perturbative approach such as the BCS theory itself.\color{black}\\
Once the \textit{Cooper pairs} are formed the electrons are not obliged anymore to follow the Fermi-Dirac statistic and the pairs themselves, now bosonic objects, can undergo \textit{Bose-Einstein} condensation and create a macroscopic ground state which is energetically favoured and whose electric resistivity becomes null. In this regard, superconductivity can be strictly related to superfluidity and analyzed in the optic of \textit{Landau Theory}.\\
The main idea is to identify an \textit{order parameter}: a thermodynamical variable which is 0 on one side of the transition and not null on the other one. Let us assume that this order parameter $\zeta$ is constant in space and time and let us follow the so called \textit{mean field theory}. In analogy with superfluidity we can build up the free energy F as a function of the temperature T and the order parameter $\zeta$ and we can expand it as:
\begin{equation}
F\,=\,\alpha\,\zeta^2\,+\frac{\beta}{2}\,\zeta^4
\end{equation}
In a superfluid:
\begin{equation}
\int d^3r \Big|\Psi_s(r)\Big|^2\,=\,n_s\,V
\end{equation}
where $n_s$ is the superfluid density and the wavefunction module $\Big|\Psi_s\Big|^2$ can indeed take the place of the order parameter such that $F=\alpha \Big|\Psi\Big|^2+\frac{\beta}{2}\Big|\Psi\Big|^4$. The reasoning can follow, with some caveats, also for the SC scenario. Now if $\alpha>0$ there is only a single minimum (see fig.\ref{SCpic2}) at $\Psi=0$ where the superfluid/superconductor density is 0. On the contrary for $\alpha<0$ there is another minimum at $\Psi=\sqrt{-\frac{\alpha}{\beta}}$ where the density $n_s$ is finite. If one then defines:
\begin{equation}
\alpha\,=\,\alpha'\,(T\,-\,T_c)\,,\qquad\beta\,=\,const
\end{equation}
the phase transition appears indeed at a critical temperature $T_c$ and the order parameter scales like:
\begin{equation}
\Psi\,\sim\,\sqrt{T\,-\,T_c}
\end{equation}
which is a characteristic result of mean field theory.
BCS theory predicts that the correlations between the electrons, mediated by phonons, can be broken with a certain amount of energy $\Delta_{gap}$ and their ''binding energy'' can indeed defined as $=2\Delta_{gap}$. This quantity takes the name of the Superconducting gap. It represents the energy gain of the SC state and it is normally a function of the temperature (and eventually of momentum\footnote{We will restrict ourselves to isotropic situations, namely \textit{S-Wave SC} where the gap is constant and can be define as $\Delta(k=0)$.}). A SC material can therefore uniquely be defined by two parameters:
\begin{equation}
\boxed{\textbf{SC}\,\longrightarrow\,\{T_c\,,\,\Delta_0\}}
\end{equation}
BCS theory fixes in a universal way these two quantities to satisfy:
\begin{equation}
\frac{2\,\Delta_0}{k_B\,T_c}\,=\,3.52
\end{equation}
The physics of conventional SC materials is more intricated, complicated and wider than what we just described, for length constrictions, here. We refer to standard CM textbooks for a detailed analysis.\\[0.5cm]
\textbf{Beyond BCS theory}\\[0.5cm]
The BCS framework turned out to be very succesful and led Bardeen, Cooper and Schrieffer to the Nobel Prize in 1972 ''for their jointly developed theory of superconductivity''. Years later, in 1986, two IBM researchers G.Bednorz and K.A. Muller\footnote{Who were awarded the 1987 Nobel Prize in Physics ''for their important break-through in the discovery of superconductivity in ceramic materials'' too.} found out that a particular material, whose electronic structure reads La$_{2-x}$Ba$_x$CuO$_4$, can undergo a superconducting transition at $T_c \sim 35 K$ \cite{BedMullHighTc}.
\begin{figure}[h]
\centering
\includegraphics[scale=0.4]{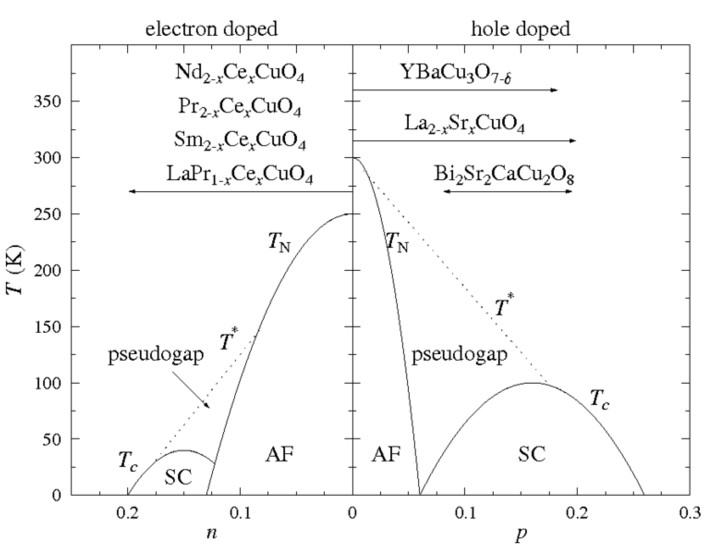}%
\qquad
\includegraphics[scale=0.4]{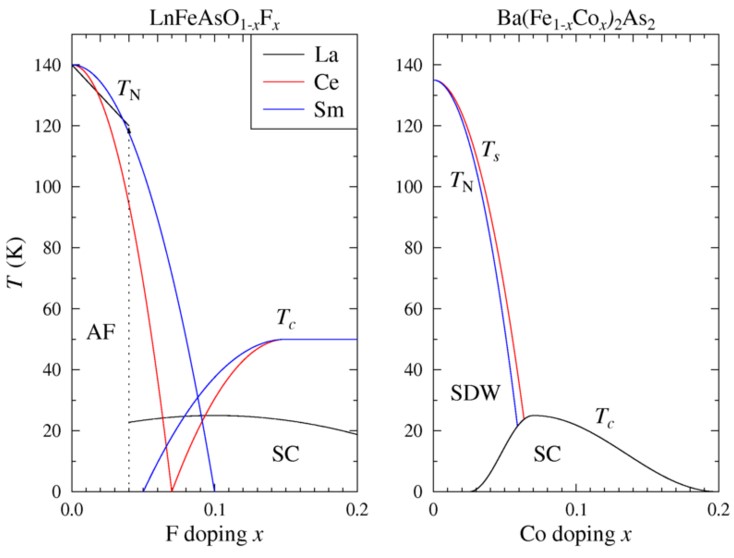}
\caption{Phase diagrams for various high-Tc SC compounds. Picture takne from Wikipedia.}
\label{SCpic3}
\end{figure}
That represented a shocking result and opened the scenario for a large class of new materials, called \textbf{High-Tc Superconductors}, whose critical temperature is unusually high and in contrast with the conventional BCS predictions \cite{GinzburgBook,LeggettHighTc}. Until 2008, only specific compounds of Copper and Oxygen, called ''Cuprates'', were thought to possess this unexplained feature but later on several other materials have been found such as the Iron-based compounds (''Pnictides'') \cite{IronSC}. Nowadays, the highest known critical temperature is about $T_c\sim 203 K$ and it referes to sulfur hydride H$_2$S at extremly high pressure \cite{HighTcrecord}.\\
High-Tc superconductors provide extremely challenging questions and unexplained features:
\begin{itemize}
\item The extremely high critical temperature can't be explained within BCS theory by electrons pairing through phonon interactions. If naively one pushes this further, realizes that such a high $T_c$ will require an interaction with a very strong coupling which would make the full framework not perturbative.
\item The normal phase of such High-Tc superconductors are not Fermi Liquids. They indeed exhibit a peculiar linear in T resistivity $\rho\sim T$ which is in constrast with the Fermi Liquid prediction\footnote{Fermi Liquid theory predicts the resistivity to be quadratic in temperature $\rho \sim T^2$; this result comes just from the T dependence of electron-eletron scatterings.}. In these novel phases there is no clear Fermi Surface  and therefore BCS fails just from the beginning. A fermi liquid instability requires a Fermi Surface! How do we get a SC from a non Fermi liquid?
\item The coherence length (which approximately measures the spatial ''extension'' of the Cooper pairs) in the high-Tc superconductors is much smaller ($\sim$ angstroms) than in normal SC described by BCS theory where it is usually around 100 nm.\color{black}
\item Within the phase diagram of these materials (see fig.\ref{SCpic3}) there are several open questions (Antiferromagnetic ordering, Pseudogap phases, etc\dots) and the interplay between superconductivity and magnetism appears to be very relevant.
\end{itemize}
'' After two decades of intense experimental and theoretical research, with over 100000 published papers on the subject, several common features in the properties of high-temperature superconductors have been identified. As of 2016, no widely accepted theory explains their properties.''\footnote{From Wikipedia.}\\[0.2cm]
\textbf{Quantum Criticality}\\[0.2cm]
Mysterious phenomena such the MITs and High-Tc superconductors,\textit{i.e.} dealing with quantum phase transitions, can be put on a common ground using the concept of quantum criticality \cite{Sachdev:2011cs}. QPT are occuring at zero temperature $T=0$ where the driving parameter is not a thermal one but rather than an external parameter, like pressure, magnetic field or carrier density. As every respected phase transition, they feature what is known as universality, namely an insensitivity with respect to the microscopic details of the system. Such universality can be understood directly from the divergenge of the correlation length around the phase transition which forces the system to be fluctuationg at all scales making its physical observables dependent just on the macroscopic or IR physics. Such observables follow, close to the transition, power law scalings which can be experimentally detected and which represent a clear signature of the QPT. Because they are characterized just at null temperature, these situations seem to be a purely academic and abstract exercise. Nevertheless, as far as the thermal excitations are small compared to the quantum one, in the so called quantum critical region, the physical observable are still determined by the property of the quantum phase transition despite the temperature is not strictly speaking zero. This has been observed in the actual phase diagram of High-Tc superconductors \ref{pappa} and it is an important subject of investigation and a possible solution to the problem.
The idea of universality along with the concept of scale invariance call for a description of such quantum phenomena through \textbf{conformal field theories}. This qualifies the class of issues described in this section as suitable for being tackled by the so called \textbf{AdS-CFT} correspondence, which will be indeed the main tool presented and exploited along this thesis.\\
The achievement of a unified understanding about thermal phase transitions and the concept criticality was a triumph of the last century. The interpretation and description of quantum phase transitions, a necessary ingredient to control phenomena like high temperature superconductivity or MITS will rely on the development of a new theory of QPT. In this sense Gauge Gravity duality is a promising direction to pursue.

\clearpage
\chapter{AdS-CFT correspondence}
\localtableofcontents
\noindent{\color{myblue}\rule{\textwidth}{1.5pt}}\par\medskip
\label{ads}
\begin{figure}[h]
\includegraphics[scale=0.65]{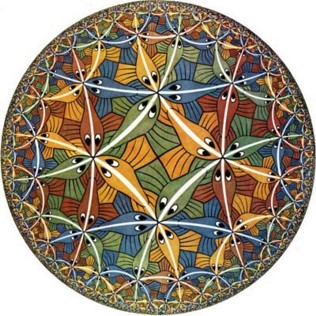}
\end{figure}
\epigraph{The distance between insanity and genius is measured only by success.}{\textit{Bruce Feirstein}}
\centerline{\textit{What electrons moving in a strongly correlated material and}}
\centerline{\textit{strings moving in a 11-dimensional spacetime have in common?}}\vspace{0.35cm}
\centerline{\textit{What Black Holes and Quantum Gravity can tell us}}
\centerline{\textit{about High-Tc Superconductivity and Disordered systems?}}\vspace{0.35cm}
\centerline{\textit{How can a String Theorist and a Solid state physicist eat now at the same table?}}\vspace{0.5cm}
These are few of the questions we will adress in this chapter.\\
An unpredictable and astonishing connection between completely different branches of physics is out in the market providing a revolutionary point of view for lots of the long standing problems of modern physics.\\
The ''magic stick'' goes under the name of \textbf{AdS-CFT correspondence} and it is one of the most important result of theoretical physics in the last decades. It is a powerful duality between a Quantum Field Theory in $d$ dimensions (without dynamical gravity) and a theory of Quantum Gravity in $d+1$ dimensions with the following suprising characteristics:
\begin{itemize}
\item The number of dimensions of the two sides does not correspond! This is why the theory is denominated \textit{holographic}.
\item One side contains dynamical (and quantum) gravity while the other one is defined on a fixed background and it is defined by completely different degrees of freedom.
\item When one side is strongly coupled the other one is weakly coupled and viceversa. For this reason AdS-CFT lies in the class of \textit{Strong-Weak} dualities.
\end{itemize}
The number of new directions, perspectives and connections that this discovery has introduced in the field of physics (and not only, \textit{i.e.} maths) is unbelievable and represented by the incredible amount of efforts and published papers in the last 20 years.\\[0.5cm]
\textbf{Disclaimer: }\\[0.3cm]
In this chapter I will just attempt to present in a very coincise and compact (therefore incomplete) way the main features of the duality using my own understanding of the subject.\\
There are several very good reviews about the Gauge Gravity duality nowadays ; we list here just some of the main ones we will be following with particular attention to the bottom-up setup and the applicative side \cite{adscftRamallo,adscftMcGreevy,adscftZaffaroni,adscftMateos}.\\
I will not enter any discussion nor derivation of the duality via String Theory, as it was originally discovered (\cite{MaldacenaOriginal,Wittenadscft,adscft1,mateosreview}) nor any technical details about String Theory \cite{string1,string2,string3,string4,string5,string6,stringhistory}, the large N limit \cite{largeNQFT,thooftN}, the holographic principle \cite{holprinrev,holprin1,holprin2}, the CFT-ology \cite{CFT1,CFT2,CFT3} and the AdS spacetime \cite{ADS1}.\\
All the material will be presented in a bottom-up fashion and doest not qualify as original contributions by the author himself.




















\section{What is AdS-CFT and its motivations}
\begin{figure}
\centering
\includegraphics[width=10cm]{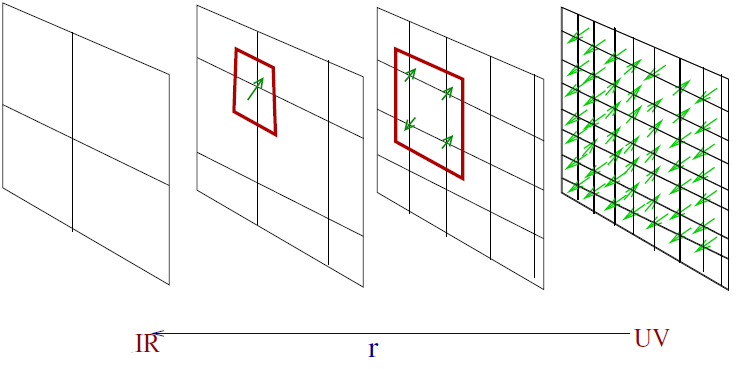}
\caption{Sketchy representation of the Wilsonian RG flow idea. Figure taken from J.McGreevy lectures \cite{McGreevy:2009xe}.}
\label{RGpic1}
\end{figure}

\begin{figure}
\centering
\includegraphics[width=10cm]{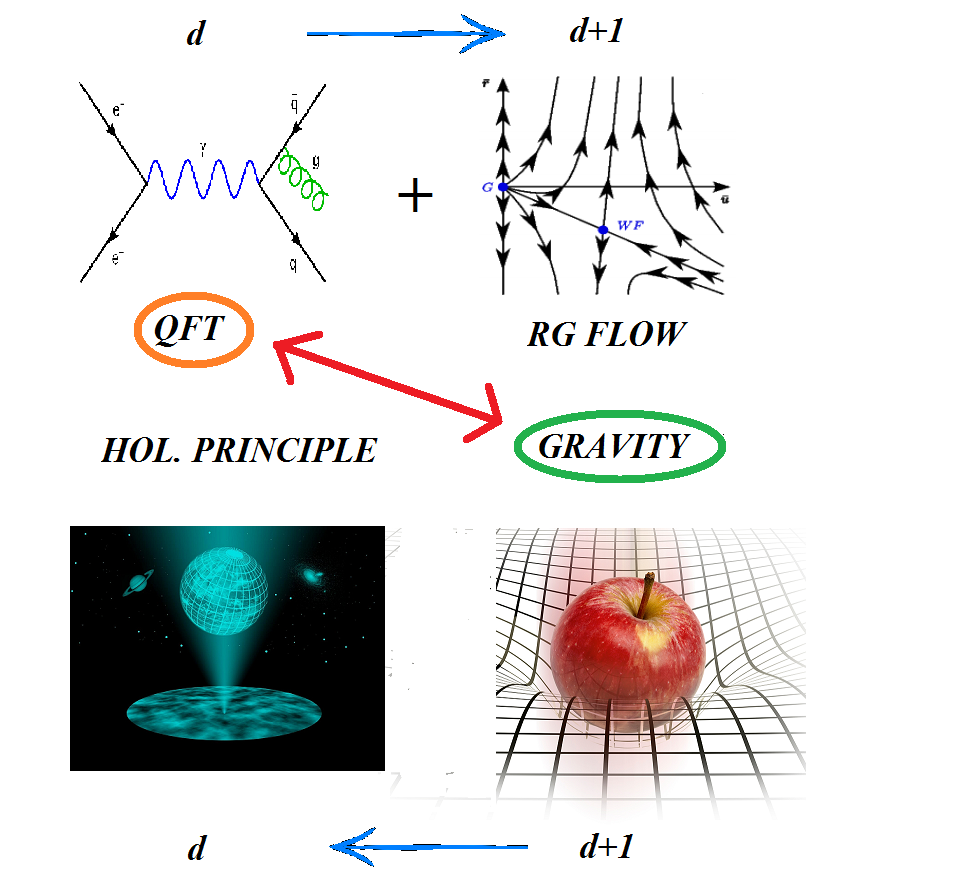}
\caption{Sketchy relations between QFTs in d dimensions and theories of gravity in $d+1$ dimensions hinting at the existence of a duality between the two pictures.}
\label{dd1}
\end{figure}

\begin{figure}
\centering
\includegraphics[width=6.5cm]{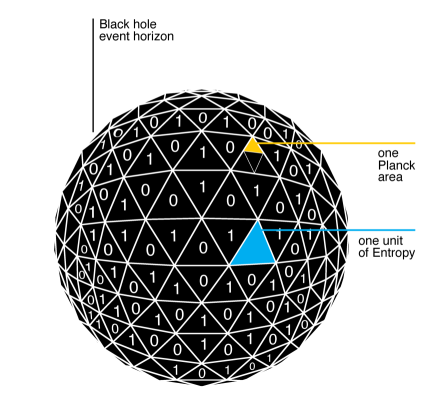}
\caption{Black Hole entropy and Bekeinstein-Hawking idea.}
\label{BHs}
\end{figure}

From a generic point of view the AdS-CFT correspondence represents a duality between a quantum field theory QFT defined in d-dimensions and a gravitational theory living in $d+1$-dimensions. In its original formulation, provided in 1998 by Maldacena \cite{MaldacenaOriginal,Wittenadscft,adscft1}, it is presented as the equivalence between the two following theories $[N=4]\,SYM\,\leftrightarrow[AdS_5\times S_5]$ and it is described within the framework of String Theory.\\
Nevertheless, considering the Planck length $l_p$, the string length $l_s$ and the typical spacetime length scale L set by the curvature there exists a sensible limit within the gravitational picture:
\begin{equation}
\frac{L}{l_p}\,\gg 1\,\qquad \frac{L}{l_s}\,\gg 1
\end{equation}
where quantum effects can be neglected and the fundamental degrees of freedom can still be parametrized by pointlike particles.\\
This weaker version of the duality takes the name of \textbf{Bottom-Up} and we will be using it for the rest of this thesis. The idea is to forget about Strings and Branes and just consider and study classical theory of gravity which reduces to:
\begin{equation}
\boxed{\text{General Relativity}\,+\,\text{bunch of fields on curved spacetime}}
\end{equation}
The point is that this limit is sensible and especially interesting and useful because it corresponds to that regime of the QFT side which is less known and less tractable with standard and perturbative techniques.\\
Indeed we can rewrite the previous inequalities as:
\begin{equation}
N\,g^2\,=\,\lambda\,\gg\,1\,,\qquad N\,\gg\,1\,.
\end{equation}
where $N$ is the rank of the gauge group of the QFT and $g$ its coupling.\\
This directly implies that the simplified version of the gravitational description refers to the regime of the QFT where the coupling is strong and therefore the computations not accessible via standard diagrammatic methods.\\
Despite nowadays we have several hints and we know several examples, beyond  the previous original case, there is no non perturbative proof of the conjecture available yet. In full generality we now refer to the \textbf{Gauge-Gravity duality} as a generic duality between a specific theory of gravity and a universal strongly coupled sector of a specific QFT. The question of searching such ''duals'' and the requirement of both sides in order to have a ''dual'' is still an open and active question we will not adress in this work.\\
The most relevant features of this conjectured equivalence are the following:
\begin{itemize}
\item The duality is \textit{Holographic} in the sense that it relates theories with different number of spacetime dimensions. In particular the QFT lives always in one dimension less with respect to the gravitational one. We will enter into the motivations and ideas behind it.
\item We can mark the correspondence as a \textit{Weak-Strong duality}. The two equivalent descriptions are connected in a way that when one is weakly coupled (and therefore simple) the other is strongly coupled (and therefore interesting) and viceversa. The latter feature stands as the most valuable and powerful characteristic of the duality and it is the origin of all the interest in applying such a tool to realistic situations. In the same way, this aspect of the duality represents also one of the biggest obstacle in order to prove it.
\item The degrees of freedom on which the two equivalent description live in completely different frameworks and are deeply unsimilar. The QFT, often a CFT, is described in terms of operators $\mathcal{O}$ of dimension $\Delta$, while the gravitational theory is written down in terms of bulk fields $\phi$ of mass $m^2$. The relation between the objects is higlhy non trivial and non local and a spoiler of the full map is presented in table \ref{tabledic}.
\end{itemize}
\begin{table}
\centering
\begin{tabular}{| a | b |}
\hline 
\rowcolor{LightCyan}\multicolumn{2}{ |c| }{\textbf{The Dictionary}}\\
\hline
\textbf{AdS$_{d+1}$} & \textbf{CFT}$_d$ \\
$d+1$ dimensions & $d$ dimensions \\
radial dimension r & energy scale $\mu$ \\
fields $\phi_I(r,x)$ & operators $\mathcal{O}_I(x)$\\
spin $\mathcal{J}$ & spin $\mathcal{J}$ \\
mass $m^2$ & conformal dimensions $\Delta$ \\
gauged symmetries & global symmetries \\
gauge invariance & currents conservation \\
confining geometry $r_0 \sim 1/m_{gap}$ & mass gap $m_{gap}$ \\
Hawking temperature T & QFT temperature T \\
metric $g_{\mu\nu}$ & stress tensor $T^{\mu\nu}$ \\
gauge field $A_\mu$ & current $J^{\mu}$ - charge density $\rho$ \\
diffeomorphism invariance & stress tensor conservation \\
black hole instabilites & QFT phase transitions \\
\hline
\end{tabular}
\label{tabledic}
\caption{Sketch of the AdS/CFT dictionary.}
\end{table}
Despite the suspicious feeling with regard to the existence of such an equivalence there are several hints which can motivate it and which can in particular give credibility to the holographic picture.\\
A quantum field theory with extensive degrees of freedom can be characterised at different scales/energies and it is an interesting question to see how such a QFT changes under measuring the system with coarser and coarser rulers (see fig.\ref{RGpic1}).\\
This was the original idea behind the \textbf{RG flow} as Wilson formulated it in the 70's.\\
From a technical point of view one can describe the behaviour of the physical observables and the couplings $g$ of the theory under changing the energy scale $\mu$ using the so-called $\beta$-function:
\begin{equation}
\mu\,\partial_\mu\,g(\mu)\,=\,\beta_g(g(\mu))\,.
\label{RGeqs}
\end{equation}
where indeed $\beta_g$ is defined as the \textit{Beta function} of the coupling g.\\
In few words the idea is that a QFT has to be ideally thought as sliced by the scale $\mu$ as a family of trajectories of the RG flow which are governed by the previous equation. A fundamental characteristic of the latter is that the system is explicitely local in the energy scale $\mu$. Therefore we can intepret that energy scale as an additional coordinate for the QFT, which can be imagined to be living in a dimension more ($d+1$) of the ''usual'' spacetime coordinates d. This constitutes a strong hint that a QFT in d dimensions can somehow be described by a different theory which lives in $d+1$ dimensions and that it incorporates in its dynamics also the RG evolution of the QFT itself. Holography can be indeed thought as the \textit{geometrization} of the QFT RG flow and the AdS spacetime seems to be the exact ingredient in that direction.\\
So far, we focused on the QFT side showing that perhaps the idea of describing it with a theory with an additional extra dimension is not that surprising. Now we jump to the other side, showing on the contrary that a gravitational theory can be described by a theory of degrees of freedom living on a spacetime with one less dimension. With both these ingredients it will be natural to conjecture a correspondence between a QFT in d dimensions and a theory of gravity in $d+1$ dimensions (see fig.\ref{dd1} which will be formalized in details in the formulation of the AdS/CFT duality.
The idea that Einstein's equations of General Relativity contain singular solutions was realized immediately after its formulation, in 1916, by K.Schwarzschild \cite{SBH}. These solutions, named \textbf{Black Holes}, are singular spacetime configurations provided by an \textit{event horizon}  outside which nothing can escape.\\
It was soon after realized that such BH objects are not static and quiescent objects but on the contrary they turn out to be characterized by thermodynamical quantities like entropy and temperature and moreover they emit thermal radiation via production of pairs at their horizon \cite{SBH,BHradiation,BekBH}. Characterizing these properties has been one of the most famous succes of high energy physics of the last decades and it left us the following prescriptions:
\begin{equation}
S_{BH}\,=\,\frac{\mathcal{A}}{4\,l_p^2}\,,\qquad T\,=\,\frac{\kappa}{2\,\pi}
\end{equation}
where the Boltzmann constant is fixed to $k_B=1$.\\
The temperature of a black hole is proportional to its surface gravity $\kappa$, the gravitational acceleration experienced at its horizon; its entropy, even more surpsingly, is directly proportional to the area $\mathcal{A}$ of such horizon. But that's not all! The black holes, as proper thermodynamical objects, fully satisfies all the fundamental thermodynamical law, such as for example the famous 2nd law:
\begin{equation}
\Delta\,S\,\geq\,0
\end{equation}
and they contains information \ref{BHs}.
The generalization of the 2nd law for BH objects (Bekenstein 1973 \cite{BekBH}), leads us directly to the main point. 
In absence of gravity, the number of d.o.f. $N_S$ of an ordinary system is extensive and it is related to the volume as $N_S\sim e^V$. This basically means that the maximum entropy $S\sim ln N_S$ results to be proportional to the volume $V\sim L^d$ of the system itself.\\
For gravity the story is different! For gravitational theories indeed the \textit{Holographic principle} states that: ''the maximum entropy of a region with volume V is the entropy of the biggest BH that fits''.\\
This means that:
\begin{equation}
S_{max}\,=\,S_{BH}\,=\,\frac{1}{4\,\pi\,G_N}\,\times\,\text{horizon area}
\end{equation}
The reason why the number of the degrees of freedom in theory with gravity scales like an area and not like a volume can be derived from the generalized 2nd law of thermodynamics we just mentioned.\\
The punchline is that the number of d.o.f for a gravitational theory in $d+1$ dimensions scales exactly in the same way of a QFT in fewer (d) dimensions!\\
This last remark, joined with the previous consideration about the nature of QFT and its RG flow, is sketched in figure \ref{dd1} and it represents an handwaving hint towards the formulation of the AdS/CFT correspondence.\\
As a final aside, let's note that an equivalence between QFT at large N and theories of gravity (in the specific String Theory) was already suggested long time ago \cite{largeNQFT,thooftN} and was one of the most important motivations giving rise to the discovery of the AdS-CFT correspondence.
\section{CFT and AdS}
\begin{figure}
\centering
\includegraphics[width=10cm]{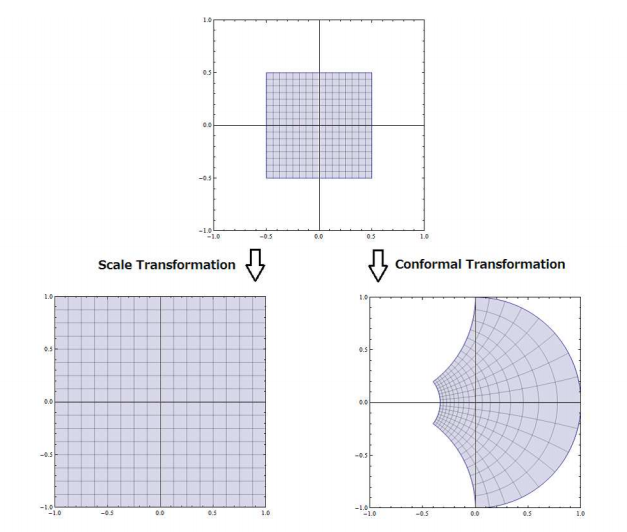}
\caption{Scale transformations versus Conformal transformations. Figure taken from \cite{sVSc}.}
\label{CFTpic}
\end{figure}

Starting from the name itself of the correspondence we can identify immediately two fundamental ingredients which have to be discussed:
\begin{equation}
\boxed{AdS}\qquad \qquad \boxed{CFT}
\end{equation}
Scale invariant theories are characterized by having no dimensionful parameter or scale in the system. We define scale invariant models the ones which do not change under the scale transformation:
\begin{equation}
\vec{x}\,\rightarrow\,\lambda\,\vec{x}\,,\qquad \phi(x)\,\rightarrow\,\lambda^\Delta\,\phi(\lambda\,x)\,.
\end{equation}
and their physics does not depend on the scale itself. The parameter $\Delta$ is defined as the scale dimension of the field $\phi$ and it describes its behaviour under the previous transformation.\\
A very easy example of a scale invariant theory is provided by the massless scalar field where clearly the introduction of a mass $m^2$ would break such a property.\\
Very often, a theory which enjoys scale invariance enjoys \textbf{conformal invariance} as well. It is ''folk theorem'' that scale invariance + Poincar\'e symmetry implies conformal invariance. This is not always true and there are various caveats (see \cite{sVSc} for details) and also some easy counterexamples such as electrodynamics in $d\neq 4$ \cite{scalexample2}. We will avoid this discussion here. \\
We can imagine a conformal transformation as a sort of spacetime dependent dilatation:
\begin{align}
&\text{scale:}\qquad x_\mu\,\rightarrow\,\lambda\,\x_\mu\,\qquad ds^2\,\rightarrow\,\lambda^2\,ds^2\nonumber\\
& \text{conformal:}\qquad x_\mu\,\rightarrow\,x'_\mu\,\qquad ds^2\,\rightarrow\,ds'^2\,=\,\Omega^2(x)\,ds^2
\end{align}
where in the limit $\Omega(x)=\lambda$ we recover the usual scale transformation. A conformal transformation rescales the lengths but it preserves the angles between vectors (see fig.\ref{CFTpic}).
In dimensions $d>2$ the conformal group is composed by the following generators:
\begin{itemize}
\item Translations: $\delta x_\mu\,=\,a_\mu$ whose generator $P_\mu$ is defined as $P_\mu\,=\,\partial_\mu$ ;
\item Lorentz transformations: $\delta x_\mu\,=\,\omega_{\mu\nu}\,x^\nu$ with $\omega_{\mu\nu}\,=\,-\omega_{\nu\mu}$; the generator $J_{\mu\nu}$ is defined by $J_{\mu\nu}\,=\,\frac{1}{2}\,\left(x_\mu \partial_\nu\,-\,x_\nu \partial_\mu\right)$ ;
\item Dilatations: $\delta x_\mu\,=\,\lambda\,x_\mu$ whose generator D is a scalar defined by $D\,=\,x^\mu\,\partial_\mu$ ;
\item Special conformal transformations: $\delta x_\mu\,=\,b_\mu\,x^2\,-2\,x_\mu (b\,x)$ where the generator $K_\mu$ is defined by $K_\mu\,=\,x^2\,\partial_\mu\,-\,2\,x_\mu\,x^\nu\,\partial_\nu$. The correspondent finite transformations are:
\begin{equation}
x_\mu\,\rightarrow\,\frac{x_\mu\,+\,c_\mu\,x^2}{1\,+\,2\,cx\,+\,(cx)^2}
\end{equation}
\end{itemize}
Altogether we have:
\begin{equation}
d\,+\,\frac{d(d-1)}{2}\,+1\,+d\,=\,\frac{(d+1)\,(d+2)}{2}
\end{equation}
generators.\\
Through some technical steps we will skip, in fact one can prove that the conformal group is isomorphic to SO(2,d), which will match exactly with the isometries for the AdS spacetime. This represents a strong hint about the duality of these two objects.\\
CFTs are very peculiar theories and they can be encountered in two distinct ways:
\begin{itemize}
\item $\beta(g*)=0$: we say that at $g=g*$ the theory has a fixed point of its renormalization group. At that particular point of the phase space, scale, and under some mild assumptions conformal, invariance emerges and we are in presence of a CFT.
\item $\beta(g)\,=\,0$: the theory is fully conformal also at the quantum level; there is no RG flow. This is for example the case for the well known $N=4$ SYM theory.
\end{itemize}
That said, conformal invariance represents a very strong and constraining symmetry of the system which leads to several implications:
\begin{itemize}
\item The stress tensor operator of the theory has to be traceless:
\begin{equation}
Tr\left(T_{\mu\nu}\right)\,=\,0\,.
\end{equation}
\item The correlation functions of the theory are very constrained. 1-point functions have to vanish. The 2-point functions for an operator $\mathcal{O}$ are forced to have the following structure:
\begin{equation}
\langle\,O_i(x)\,,\,O_j(y)\,\rangle\,=\,\frac{A\,\delta_{ij}}{\big|x-y\big|^{2\,\Delta_i}}
\end{equation}
where $\Delta$ is the conformal dimension of the correspondent operator.
\item CFTs have to be unitary and such a requirement restricts consistently the possible dimensions of the CFT operators.
\end{itemize}
Checking these constrained results constitutes a first good test of the AdS/CFT correspondence.\\
From the point of view of classifying the spectrum of the possible operators of the theory, CFTs differ consistently from standard QFTs. In particular the $P_\mu P^\mu$ operator is not a good Casimir anymore and the mass of an operator has not a definite meaning. One can always apply a dilatation to that field and such a mass would change accodingly to the algebra. In other words in a CFT we can just distinguish between massless and massive field. In this case a good quantum number to associate with the spin to classify the spectrum of the theory can be identified in the conformal dimension $\Delta$ of the operator which can be defined as:
\begin{equation}
\left[D\,,\phi(x)\right]\,=\,i\,\left(\Delta\,+\,x_\mu\,\partial^\mu \right)\,\phi(x)
\end{equation}
where D is the quantum generator for dilatations.\\
Although CFT they are very specific theories, naively implying that the AdS-CFT correspondence is not a so generic tool, it is nowadays clear that is not the case and that the duality can be generalized in order to accomodate also theories lacking of conformal invariance.\\
The other character of the duality is the so-called AdS spacetime whose metric takes the form:
\begin{equation}
ds^2\,=\,\frac{L^2}{z^2}\,\left(-\,dt^2\,+\,d\vec{x}^2\,+\,dz^2\right)
\end{equation}
where L is the length of the AdS$_{d+1}$ spacetime.\\
Anti de Sitter spacetime represents a maximal symmetric solution of General Relativity provided by a negative cosmological constant $\Lambda$:
\begin{equation}
\mathcal{S}\,=\,\frac{1}{16\,\pi\,G_N}\,\int\,d^{d+1}x\,\sqrt{-g}\,\left(R\,-\Lambda\right)
\end{equation}
This AdS solution enjoys the maximal number of independent possible \textit{killing vectors}, \textit{i.e.} isometries generators and its Ricci scalar curvature is constant and equal to:
\begin{equation}
\mathcal{R}\,=\,\frac{d+1}{d-1}\,\Lambda
\end{equation}
The $z=0$ position is defined as the ''boundary'' of AdS. To be mathematically precise, it represents a \textit{conformal boundary} while the $z=\infty$ position is the AdS horizon where the vector $\partial_t$ gets null.\\
In a particular coordinate system one can check that AdS contains Minkowski slices at finite value of the radial direction, and this is crucial in most of our applications of the AdS/CFT
correspondence.\\
The most important point about AdS spacetime is the fact that its isometry group $S0(2,d-1)$ corresponds exactly with the conformal group in d dimensions. This represents a strong argument which somehow AdS$_{d+1}$ can be dual to a CFT living in d dimensions, which we take as the actual statement of the AdS/CFT correspondence.\\
Furthermore, this is much more general and it corresponds to the statement that gauged symmetries in the bulk (such as the AdS isometries) are ''dual'' to global symmetries in the QFT side.\\
As a last remark it is important to stress that the radial dimension of the bulk geometry directly encodes the energy scales of the dual QFT:
\begin{equation}
z\sim\,\frac{1}{E}
\end{equation}
Therefore, excitations with energy E will be localized in the bulk at the correspondent radial position z defined by the latter. This is again a manifestation of the fact that AdS spacetime and the gravitational picture represents a geometrization of the RG flow dynamics of the correspondent QFT.
\section{Field, Operators and Correlation Functions}
We can finally introduce the map between the operators $\mathcal{O}_i$ of the conformal field theory and the bulk fields $\phi_i$ of the gravitational side and show how to extract informations from one side to the other in order to make statements about physical quantities like correlation functions.\\
The gravitational theory, in $d+1$, is defined via a bulk action:
\begin{equation}
\mathcal{S}_{bulk}\,\left(g_{\mu\nu}\,,\,A_{\mu}\,,\,\phi\,,\dots\right)
\end{equation}
including fields of different spins.\\
Looking at the other side, the CFT, we are left with a set of \textit{operators}:
\begin{equation}
\mathcal{O}_i
\end{equation}
with their correspondent spins and conformal dimensions $\Delta_i$\\
The main idea is that a field $\phi$ living in the bulk relates to an operator of the CFT with same quantum numbers and that their coupling shows up through a boundary term.\\
From the CFT perspective the latter represents a deformation, due to the source $\phi_0$, given by:
\begin{equation}
\mathcal{L}_{CFT}\,+\,\int\,d^dx\,\phi_0\,\mathcal{O}
\end{equation}
In standard QFT language we can define in this way the functional generator of the correlation functions $W(\phi_0)$ as:
\begin{equation}
e^{W(\phi_0)}\,=\,\langle\,e^{\int\,\phi_0\,\mathcal{O}}\,\rangle_{QFT}
\end{equation}
Following this prescription, we can derive the correlations function of the operator $\mathcal{O}$ inserting functional derivatives on the previous object with respect to the source $\phi_0$:
\begin{equation}
\langle\,\underbrace{\mathcal{O}\,\dots\,\mathcal{O}}_{1\,,\,\dots\,,\,n}\,\rangle\,=\,\frac{\delta^n\,W}{\delta\,\phi_0^n}|_{\phi_0=0}
\end{equation}
In the gravitational picture the source $\phi_0$ for the operator $\mathcal{O}$ will be represented by the boundary value of the $d+1$ dimensional bulk field $\phi(z,x)$.\\
More explicitely a bulk scalar field, for example, will have an asymptotic expansion close to the boundary of AdS given by:
\begin{equation}
\phi(x,z)\,\sim\,A(x)\,z^{d\,-\,\Delta}\,+\,B(x)\,z^\Delta
\label{UVscalarexp}
\end{equation}
where we have identified:
\begin{equation}
\Delta\,=\,\frac{d}{2}\,+\,\nu\,,\qquad
\nu\,=\,\sqrt{\frac{d^2}{4}\,+\,m^2\,L^2}
\end{equation}
with $m^2$ its bulk mass.\\
The parameter $\Delta$ is indeed the conformal dimension of the operator $\mathcal{O}$ which can result to be relevant, irrelevant or marginal depending if $\Delta$ is smaller, bigger or equal to the spacetime dimensions $d$. From the previous expression it is clear that the relevance of the QFT operator is dialed by the mass of the bulk field.\\
In the standard quantization scheme we can spot the source $\phi_0$ as the non-renormalizable part of such a boundary expansion:
\begin{equation}
\phi_0(x)\,=\,A(x)\,=\,\lim_{z\rightarrow 0}\,z^{\Delta\,-\,d}\,\phi(z,x)
\label{PPP}
\end{equation}
That said we are already in the position to write down the deepest equation for the AdS-CFT correspondence, known as the GPKW (Gubser, Polyakov,
Klebanov, Witten) master rule:
\cite{Witten1,GPKW1}
\begin{equation}
\boxed{e^{W(\phi_0(x)}\,=\,\langle\,e^{\int\,\phi_0(x)\,\mathcal{O}}\,\rangle_{QFT}\,=\,e^{\mathcal{S}_{AdS}[\phi(x,r)]}\,=\,\mathcal{Z}_{gravity}\left[\phi(x,r)_{boundary}\,=\,\phi_0(x)\,\right]}
\end{equation}
In simple words, the on-shell gravitational action, provided the correct boundary conditions, gives us the generating functional of the QFT and therefore all the informations about the correlation functions of its operators.\\
To give a glimpse, taken the field $\phi$ defined in \eqref{PPP} its two point function in Fourier space is going to take the form:
\begin{equation}
\langle\,\mathcal{O}(k)\,\rangle_{\phi_0}\,\sim\,B(k)
\end{equation}
which is indeed considered to be the vacuum expectation value (VEV) for the operator $\mathcal{O}$.\\
The correspondent 2-point function, \textit{i.e.} Green Function, is on the contrary related to the ratio of the normalizable mode $B$ over the non-renormalizable (source) $A$ as:
\begin{equation}
\mathcal{G}(k)\,=\,\langle\,\mathcal{O}(k)\,\mathcal{O}(k)\,\rangle_{\phi_0}\,\sim\,\frac{B(k)}{A(k)}
\end{equation}
It is an important check of the AdS-CFT tool to compute the Green function for a scalar operator in AdS and end up with the following result:
\begin{equation}
\langle\,\mathcal{O}(x)\,\mathcal{O}(0)\,\rangle\,=\,\frac{2\,\nu\,\eta\,L^{d-1}}{\pi^{d/2}}\,\frac{\Gamma\left(\frac{d}{2}\,+\,\nu\right)}{\Gamma(-\nu)}\,\frac{1}{|x|^{2\,\Delta}}
\end{equation}
which is indeed what we expect in a conformal field theory for a primary operator of dimension $\Delta$!\\
In conclusion, this is a very powerful tool which allows us to get control on the physical observables of the strongly coupled QFT just solving the classical Einstein equations in the bulk of the gravitational dual.\\
So far we have not explicitely told how to identify the dual couple $\{\phi_0,\mathcal{O}\}$ ! The roubst way to find such couples is fundamentally given by symmetries and by the requirement that both the bulk field and the CFT operator are labelled the same quantum numbers with respect to the $O(2,d-1)$ group.\\
As a concrete example, we can write down:
\begin{equation}
\mathcal{L}_{CFT}\,+\,\int\,d^dx\,\sqrt{g}\,\left(g_{\mu\nu}\,T^{\mu\nu}\,+\,A_\mu\,J^\mu\,+\,\phi\,\mathcal{O}\,\right)
\end{equation}
which already shows us part of the map:
\begin{align}
&graviton\qquad g_{\mu\nu}\qquad\longrightarrow\qquad stress\,tensor\qquad T^{\mu\nu}\nonumber\\
&gauge\,field\qquad A_{\mu}\qquad\longrightarrow\qquad current\qquad J^{\mu}\nonumber\\
& scalar\,field\qquad \phi\qquad\longrightarrow\qquad scalar\,operator\qquad \mathcal{O}
\end{align}
where for example in a gauge theory $\mathcal{O}=F_{\mu\nu}F^{\mu\nu}$.\\
Generically we can have different collection of gauge symmetries in the bulk associated to the various fields, for example:
\begin{align}
&graviton\qquad g_{\mu\nu}\qquad\longrightarrow\qquad diffeomorphisms\nonumber\\
&gauge\,field\qquad A_{\mu}\qquad\longrightarrow\qquad U(1)\nonumber\\
\end{align}
Gauge invariance relates to the conservation of the correspondent currents in the CFT and it fixes their conformal dimensions to the one of conserved quantities. A mass term in the bulk would generically break it and would modify the conformal dimension of the correspondent operator $\mathcal{O}$ which would aquire an anomalous part signaling its non conservation.\\[0.5cm]
There are several details and caveats about this mapping and the extraction of the correlation functions within the AdS-CFT framework which we will not analyze here. The interested reader can find them within the excellent material available in the literature and present in the actual bibliography of this work to which we refer.
\section{Introducing temperature and charge}
So far we have focused our attention to the original formulation of the AdS/CFT correspondence which relates a conformal field theory to a pure AdS bulk geometry. As we already explained, conformal field theories are very particular ''beasts'' dealing with quantum critical points or very fine tuned QFTs. This is not for example the case for a generic condensed matter system which usually lives at finite temperature T and/or finite charge density $\rho$. In such a way we of course introduce a scale into the problem breaking the original conformal invariance of the full theory. Such a deformations (if relevant) make the theory to clearly depart from the original UV conformal fixed point and to undergo an RG flow towards another infrared fixed point. The AdS/CFT correspondence can be generalized easily to describe also these situations such that its name can be mutated into the more generic one of \textbf{Gauge-gravity duality}.\\
From the bulk point of view the departure from conformal invariance renders the spacetime geometry different from the pure AdS case, which is recovered just asymptotically in the UV. The bulk spacetime encodes directly the RG flow due to such a deformation of the now non-conformal QFT.\\
The easiest and most important example we are going to analyze in this section is the so-called \textbf{Reissner Nordstrom black hole} which is the dual gravitational description of a QFT at finite temperature T and finite charge density $\rho$. This example is the first application we consider of the AdS/CMT correspondence and it has been subject of a huge amount of research under lots of directions. For generic discussions about its role among the applications to condensed matter we refer to \cite{HartnollCMreview,plumbers}.\\
The first step we have to make is to modify the geometry and the gravitational solution to account for a finite temperature. As already explained before, there is a definite gravitational object which has this feature, the Black Hole. We will consider generical BH solutions embedded in AdS spacetime, whose metrics are of the form:
\begin{equation}
ds^2\,=\,\frac{L^2}{u^2}\,\left(-f(u)\,dt^2\,+\,\frac{du^2}{f(u)}\,+\,dx^2\,+dy^2\right)
\end{equation}
in $3+1$ bulk dimensions.\\
The function $f(u)$ is known as the emblackening factor and it has a zero at the position of the so-called \textit{horizon} $u=u_h$:
\begin{equation}
f(u_h)\,=\,0
\end{equation}
The form of $f(u)$ is strongly dependent on the details of the theory and will be irrelevant for the following discussion.\\
The correspondent temperature associate to such a gravitational object can be easily deduced from the generic formula:
\begin{equation}
T\,=\,\frac{2\,\pi}{\kappa}
\end{equation}
where $\kappa$ is the surface gravity of the BH.\\
Within our conventions the temperature reads:
\begin{equation}
T\,=\,-\,\frac{f'(u_h)}{4\,\pi}
\end{equation}
and represents a tunable parameter of the theory.\\
\begin{figure}
\centering
\includegraphics[width=13cm]{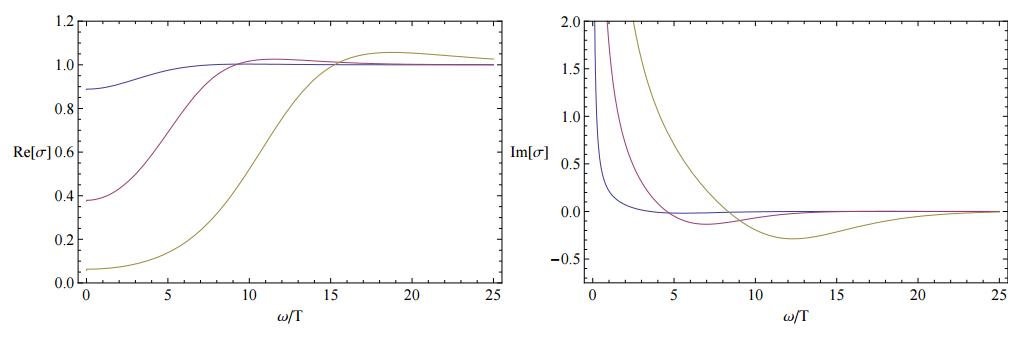}
\caption{The holographic results for the  real (left) and imaginary (right) parts of the electrical conductivity in the RN background. The various curves represent different values for the chemical potential $\mu$. Figure is taken from \cite{HartnollCMreview}}
\label{sigmafig}
\end{figure}
That is indeed the temperature of the BH detectable from an observer at infinity and it appears natural to identify it as the correspondent temperature T of the QFT.\\
With the scope of adding also the second ingredient, namely the charge density $\rho$ of our QFT, we need to introduce an additional ingredient in the game. We must consider a U(1) vector field $A_\mu$ in the bulk such that the original action gets modified into:
\begin{equation}
\mathcal{S}\,=\,\int\,d^4x\,\sqrt{g}\,\left(R\,-\,2\,\Lambda\,-\frac{1}{4}\,F_{\mu\nu}F^{\mu\nu}\right)
\end{equation}
which takes the name of Einstein-Maxwell model.\\
The charge density of the system results to be encoded in the temporal component of the gauge field which aquires a non trivial bulk profile $A_t(u)$. In particular the Maxwell equation in curved spacetime forces the solution to be of the form:
\begin{equation}
A_t(u)\,=\,\mu\,-\,\rho\,u
\end{equation}
where $\mu$ is the chemical potential and $\rho$ the charge density of the system. These two quantities are not indipendent but are constrained by the requirement of regularity for the gauge field at the horizon $A_t(u_h)=0$ which fixes $\mu=\rho\,u_h$.\\
The deformed CFT now contains a new dimensionless tunable parameter given by the ratio $T/\mu$. Note that both the charge density $\rho$ and the temperature T represent low energy IR deformation of the CFT and this is reflected by the fact that asymptotically the geometry does not cease to be AdS, \textit{i.e.} conformal.\\
The full gravitational BH solution of the Einstein-Maxwell system takes the name of Reissner-Nordstrom (RN) black hole and it is the main character of the AdS-CMT program.\\
Within this framework we can identify the Grand Potential $\Omega$ of the QFT at finite temperature and charge density with the on shell euclidean action as:
\begin{equation}
-T\,S^E_{on-shell}\,=\Omega\,=\,E\,-\,T\,S\,-\,\mu\,\rho
\end{equation}
From such a quantity, using standard QFT and statistical mechanics techniques, we can extract all the thermodynamical quantities of interest; we will not enter into these details.
\section{Some applications}
\begin{figure}
\centering
\includegraphics[width=14cm]{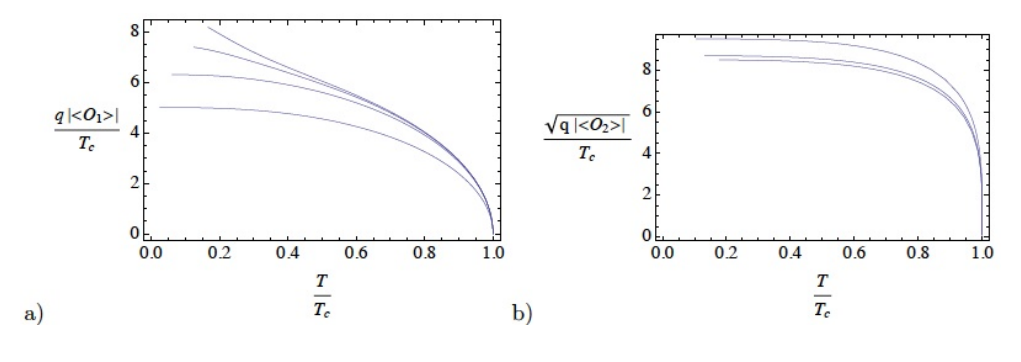}
\caption{Example of the development of a condensate in holographic theories. Figure taken from \cite{holSC2}.}
\label{condpic}
\end{figure}

\begin{figure}
\centering
\includegraphics[width=7cm]{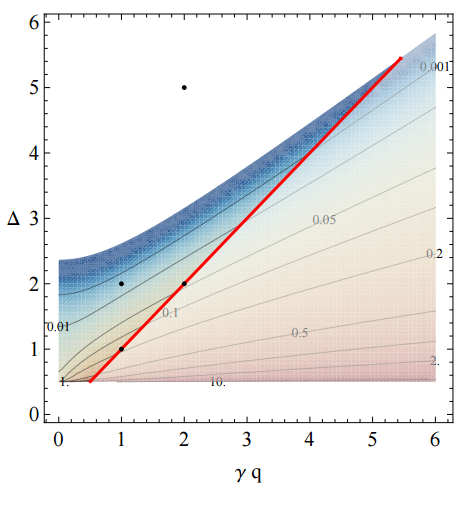}
\caption{The phase diagram for the RN case as a function of the conformal dimension of the scalar operator $\Delta$ and its charge q. The colour part indicates the unstable region where the SC phase appears. The top boundary, separating the blank and coloured regions, identificate the $T=0$ critical line obtainable with the BF argument. Figure taken from \cite{DenefSC}.}
\label{SCscanfig}
\end{figure}

Given the RN solution found in the previous section we are now ready to extract physical observable from such a framework and check our expectations. The electric conductivity can be computed holographically through linear response techniques; a good reference is given by \cite{tongsigma}.\\
Given an external oscillating electric field $E(\omega)$ sourcing a correspondent current $J(\omega)$ the electric conductivity is defined as the ratio:
\begin{equation}
\sigma(\omega)\,=\,\frac{J_x(\omega)}{E_x(\omega)}
\end{equation}
In the language of linear response theory, the conductivity could be compute through \textit{Kubo formulas} as the 2-points function of the $J_x$ operator:
\begin{equation}
\sigma(\omega)\,=\,-\,\frac{\,i}{\omega}\,G^R_{J_x\,J_x}\,(\omega)
\end{equation}
where R indicates the retarded Green function.\\
From the gravitational point of view we introduce an electric field in the x direction turning on a source $A_x=\frac{E}{i\,\omega}\,e^{i\,\omega\,t}$ on the boundary.\\
The radial profile of the gauge field perturbation takes the form:
\begin{equation}
A_x(u)\,=\,\frac{E}{i\,\omega}\,e^{i\,\omega\,t}\,+\,\langle J_x\rangle\,u\,+\,\dots
\end{equation}
where the subleading term $\langle J_x\rangle$ in the expansion can be derived by solving the equations of motion coming from the Einstein-Maxwell action.\\
In a more specific way, given the previous expansion, the electric conductivity can be obtained (numerically) via:
\begin{equation}
\sigma(\omega)\,=\,\frac{A_x'}{i\,\omega\,A_x}|_{boundary}
\end{equation}
after fixing the appropriate ingoing boundary conditions at the horizon.\\
The results for RN black hole are shown in fig.\ref{sigmafig}. At very large energy/frequency $\omega/T \gg 1$ the conductivity asymptotes a constant value given by the normalization of the Maxwell term in the gravitational action.\\
On the contrary at low frequency it gets depleted under a certain energy scale fixed by $\mu$. The present gap appears deeper for larger chemical potential.\\
A last very important feature is the presence of an $\omega=0$ pole in the imaginary part of the conductivity, which relates, via Kramers-Kronig relations, to a delta function $\delta(\omega)$ in its real part. The divergence appearing in the conductivity at zero frequency is related to momentum conservation and it will be one of the most important point of the present work. In other words, we are in presence of a system characterized by a background charge density and preserving translational invariance. If you force the system with a constant, $\omega=0$, electric field then the charge charge will necessarily accelerate. But, because of the presence of translational invariance, momentum is conserved. This implies that there is no way for the charges to dissipate their momentum, thus the current will persist forever. This represents the origin of the delta-function!\\
In real materials, the appearance of impurities, lattice and disorder breaks the conservation of momentum. These effects directly relax momentum with a timescale $\tau$ and the conductivity at law frequency acquires the well-known Drude structure:
\begin{equation}
\sigma(\omega)\,=\,\sigma_0\,+\,\frac{\rho^2}{\epsilon\,+\,P}\,\frac{1}{1/\tau\,-\,i\,\omega}
\end{equation}
which clearly reproduces the AdS/CFT results in the translational invariant limit $\tau \rightarrow \infty$.\\
Introducing the effects of impurities and momentum dissipation into the framework of holography will represent the main topic of this thesis and will permit to reconcile with the experimental expectations.\\[0.2cm]
Another interesting topic and application of the RN solution is related to its superconducting instability. From a theoretical perspective superconductivity can be associated to the spontaneous symmetry breaking (SSB) of the U(1) symmetry associated to charge conservation. We thus need an operator charged under the U(1) symmetry which aquires a non null vacuum expectation value (VEV). The minimal setup, known as \textit{S-wave} superconductivity, refers to the identification of such a field with a scalar operator of spin 0 and it will constitute our benchmark playground.\\
What we consider is the introduction of a complex massive scalar field charged under the U(1) global symmetry such that the related action takes the form of:
\begin{equation}
\mathcal{L}\,=\,\left(\,R\,-\,2\,\Lambda\,\right)\,-\,\frac{1}{4}\,F^2\,-\,|\,\nabla \phi\,-\,i\,q\,A\,\phi\,|^2,\,-\,m^2\,|\phi|^2)\,.
\label{Genaction}
\end{equation}
The scalar $\phi$ constitutes the dual bulk field of the charged operator which will condense, \textit{i.e.} getting a finite expectation value (VEV) $\langle \mathcal{O}\rangle$, and break the global $U(1)$ symmetry of the CFT producing the SC state.\\
Following the previous discussion we know that such a scalar bulk field will behave close to the boundary as:
\begin{equation}
\phi\,\approx\,\phi_0\,\left(\frac{u}{L}\right)^{d-\Delta}\,+\,\phi_1\,\left(\frac{u}{L}\right)^{\Delta}
\end{equation}
where the coefficient of the first term would result the dominant contribution at the boundary and therefore identified as the source for the operator $\mathcal{O}$ while the second term, the subleading one, would define the desired VEV of such operator:
\begin{equation}
\phi_1\,\equiv\,\langle \mathcal{O} \rangle
\end{equation}
The spontaneous breaking of the U(1) symmetry would correspond to a normalisable solution ($\phi_0=0$, \textit{i.e.} no source for $\mathcal{O}$) which develops dynamically a non trivial VEV:
\begin{equation}
\langle \mathcal{O}\rangle\,\neq\,0
\end{equation}
This would correspond to an instability of the RN black hole driven by the bulk scalar $\phi$ and would represent our SC phase.\\
Such a new solution, different from the normal phase with trivial profile $\phi=0$, will appear just under a critical temperature $T<T_c$ which will represent the temperature of the continuous phase transition. The corresponding VEV can be indeed numerically computed and it takes a non zero value just below $T_c$ in a mean-field fashion:
\begin{equation}
\langle \mathcal{O}\rangle\,\sim\,\left(T\,-\,T_c\right)^{1/2}
\end{equation}
The results are shown in fig.\ref{condpic}.
Note that such a solution has to be searched imposing infalling boundary condition at the horizon such that the instability would be represented in the quasinormal mode spectrum of the BH as a pole of the Green function appearing in the upper complex plane \cite{hydroSC}, \textit{i.e.} a mode growing exponentially in time.\\
The appearance of the instability at $T=0$ can be furthermore checked analytically and it corresponds to the so-called violation of the BF bound in the bulk. To be more specific, the scalar $\phi$ aquires at the extremal horizon geometry an effective mass:
\begin{equation}
M_{eff}^2\,L^2\,=\,m^2\,L^2\,+\,q^2\,g^{tt}\,A_t^2\,L^2
\end{equation}
which contains an additional contribution from the gauge field configuration. Because of the presence of $g^{tt}$ this new term appears to be negative such that at the extremal horizon the scalar can violate the BF bound:
\begin{equation}
M_{eff}^2\,L^2\,<\,-\,\frac{1}{4}
\end{equation}
and produce an instability, namely an exploding excitation in the bulk. This is of course the signal that the RN solution considered is not stable anymore and it does not represent the true ground state of the system. The latter is indeed represented by the Hairy BH solution which constitutes our  SC phase, whose details are skipped for the sake of simplicity.\\
Joining the numerical techniques coming from solving the Einstein-Maxwell system and reading the VEV for the operator $\mathcal{O}$ holographically and the analytical argument at $T=0$ one can produce the phase diagram for the dual QFT in function of the dimension $\Delta$ of the scalar field and its charge $q$. An example taken from \cite{DenefSC} is given in fig.\ref{SCscanfig}.
Additionally one can compute the conductivity across the phase transition and in particular in the SC state. As expected it presents an energy gap appearing for the presence of the condensate and an infinite contribution at zero frequency signaling the existence of a SC medium. Strictly speaking, this last feature appears to be very tricky and confusing since such an infinite is already present in the normal phase of the system because of translational invariance.\\
It is indeed hard to disentangle the two infinities and read of the superfluid density from the coefficient of the $\delta$ function.\\
The resolution of this issue will be one of the main concerns of this thesis. Introducing momentum dissipation in the original holographic superconducting models will provide a more realistic description and will open the room for extracting more physical quantities to be compared with experiments. We will come back on this issue in the section devoted to the original results of this thesis.
\chapter{Massive Gravity}
\label{MGchapter}
\localtableofcontents
\noindent{\color{myblue}\rule{\textwidth}{1.5pt}}\par\medskip
\begin{figure}[h]
\includegraphics[scale=0.5]{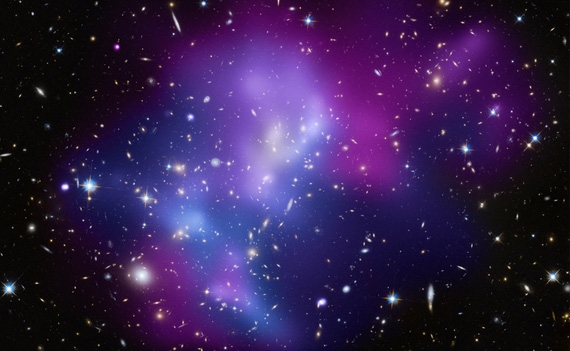}
\end{figure}
\epigraph{For everybody in their busy lives, you need to invest in sharpening your tools, and you need to invest in longevity.}{\textit{Ryan Holmes}}
General Relativity (GR) is one of the most successful, elegant and shining scientific result of the last century \cite{100GR}. Its agreement with experiments is incredibly high; from the confirmation of the deflection of light led by Eddington in 1919 to the most recent gravitational waves detection announced by LIGO collaboration just this year \cite{LIGO}. Nevertheless, some of the large distance properties of our universe, such as the origin of its late time acceleration, still remain not explained and consistently incorporated in the framework of General Relativity. As a consequence, soon after its discovery, an increasing industry of scientific efforts arose with the target of studying its possible modifications and the corresponding effects on physical observables. One of such attempts, \textit{i.e.} \textbf{massive gravity}, refers to the proposal of modifying the large distance GR dynamics giving the graviton a mass and breaking the full diffeomorphism invariance. Such a modification, mainly motivated by cosmology, provides for new degrees of freedom but also possible pathologies one should take care of.\\[0.3cm]
\textbf{Disclaimer:}\\[0.3cm]
Massive gravity is a subject with old history and a enormous amount of features. Due to the presence of several excellent reviews \cite{MG1,MG2,MG3} and books about the subject, in this section I will just sketch the ideas, the problems and the features connected to MG in a very simple and not detailed fashion.\\
All the material present in this section is not part of the original contributions by the author but just a recompilation of known material.
\section{What is massive gravity}
General relativity (GR) describes the full non-linear dynamics of spacetime in a remarkably elegant way based on the \textit{Equivalence Principle} and the idea of \textit{General Covariance}. Such requirements, associated to the existence of a force mediated by a spin-2 object, imply directly just one answer: GR! Although coordinates invariance and/or the equivalence principle are historically the main motivations and the pillars of GR they are not the real underlying principles of the theory. In modern language, we know that GR can be uniquely defined by the feature of being the only consistent theory for an interacting particle with helicity $h=2$ and zero mass. This has been proven in \cite{DeserGR} and all the rest follows from this statement and not vice versa.\\
Einsten connected in an elegant way the effects of gravity with the geometrical structure of spacetime, making it a dynamical object in rather elegant mathematical formalism. The framework allows to reconcile classical Newton's law with the requirements of Special Relativity and represents the current description of classical gravitation in modern physics.
GR is of course not a UV complete (and renormalizable) theory; it represents an effective field theory with a UV cutoff set at the so called Planck scale $M_p$. The only known UV completion which recovers GR at low energy is String theory. This is very fortunate since string theory is also at the roots of the AdS-CFT correspondence.\\
What we consider in this section is the regime of GR where the theory can be still be treated as classical but being highly non linear in its structure. Despite an impressing and universal consensus on the accuracy and the efficiency of general relativity, in the recent years increasing interest to modify it and test its modifications appeared. The main motivation comes from Supernova data \cite{SUPERNOVA1,SUPERNOVA2}: the acceleration of the Universe and its origin are still unknown and they can't be easily predicted within the framework of General Relativity.\\
If we firmly believe that GR is the ultimate theory, a ridiculously small density of \textit{dark energy} $\rho_{dark}$ has to be present in our Universe. The only way GR can explain such a density is via the introduction of a non zero Cosmological constant $\Lambda$, which relates to the latter as:
\begin{equation}
\rho_{dark}\,\sim\,M_p^2\,\Lambda
\end{equation}
Following the computations we realize that an incredibly small value of such a constant $\Lambda/M_p^2 \sim 10^{-65}$ is necessary and this leads to the known \textit{cosmological constant problem}. This is very analogous of the hierarchy problem present in the Standard Model in relation to the Higgs mass. Such a small value is not protected by any simmetry and it is therefore technically not natural unless one invokes some sort of Anthropic principles \cite{AnthropicP}.\\
Being GR the unique consistent model for an interacting massless particle with spin 2, in order to modify it we need to break some of the assumptions. One possibility is to make the force mediator to be massive in analogy to what happens to the gauge bosons in the Electroweak theory. Giving a mass to the graviton accounts for additional, and often dangerous, degrees of freedom besides the 2 usual polarizations of the massless one. Once the graviton becomes massive, the corresponding gravitational force takes the Yukawa form $\sim \frac{1}{r}e^{-m r}$ and at distances $r\geq \frac{1}{m}$ drops off in comparison to the GR expectation. If the mass is order the Hubble constant $r\sim H$ then one could try to explain the acceleration of the universe using the massive gravity idea. Now the small value of the cosmological constant translates into the ratio $m/M_p$ being very small and here it comes the novelty. Since the $m=0$ case provides for an enhancement of the symmetries of the system, namely diffeomorphism invariance, such as small value  is consequently protected by such a symmetry and no longer unnatural. Building at linear level a theory of non-interacting massive graviton is a pretty simple task and was already achieved in 1939 by Fierz and Pauli \cite{FIERZPAULI}. Promoting such a construction to a full non linear and interacting theory is a much more challenging task which has been pursued for decades and just in recent years has encountered some positive answers.
\section{MG in more detail}
At linearized level there are a priori two possible mass terms one can think of, $h_{\mu\nu}^2=h_{\mu\nu}h^{\mu\nu}$ and $h=Tr(h_{\mu\nu})$, such that a generic mass deformation of the linearized Einstein-Hilbert action takes the form\footnote{The linearized Einstein-Hilbert action reads:
\begin{equation}
\mathcal{L}_{EH}\,=\,-\frac{1}{2}\,\partial_\lambda h_{\mu\nu}\partial^\lambda h^{\mu\nu}+\partial_\mu h_{\nu\lambda}\partial^\nu h^{\mu\lambda}-\partial_{\mu}h^{\mu\nu}\partial_\nu h+\frac{1}{2}\partial_\lambda h \partial^\lambda h
\end{equation}}:
\begin{equation}
\mathcal{L}\,=\,\mathcal{L}_{EH}\,+\,\mathcal{L_{MASS}},\qquad \text{with}\qquad\mathcal{L_{MASS}}\,\sim\,m^2\,\left(h_{\mu\nu}^2\,-\,A\,h^2\right)
\end{equation}
The introduction of such a mass term clearly breaks the original diffeomorphism invariance of the theory:
\begin{equation}
h_{\mu\nu}\,\,\rightarrow\,h_{\mu\nu}\,+\,\partial_{(\mu}\zeta_{\nu)}
\end{equation}
and introduces new degrees of freedom.\\
Indeed performing the counting analysis we end up with 6 of them. In particular there exists an additional scalar mode compared to the only expected for a massive spin 2 field\footnote{which is expected to have (in $d=4$) 5 degrees of freedom}. The latter is the dangerous character and it represents a ghosty excitation unless the A parameter is fixed to be $A=1$ (Fierz Pauli Tuning).
The final obtained result is the so called \textbf{Fierz-Pauli} action, which takes the form:
\begin{equation}
\mathcal{L_{FP}}\,\sim\,m^2\,\left(h_{\mu\nu}^2\,-\,h^2\right)
\end{equation}
and it represents the only one consistent modification of the Einstein-Hilbert action producing a massive graviton.\\
We underline that what discussed here so far is valid on Minkowski background, for Lorentz invariant systems and at linear level. This is anyway a good exercise which already suggests that the most severe problems of MG are connected with the scalar sector. Indeed the helicity 0 mode is the responsable of most of the consistency issues and the phenomenology features of MG theories.\\
In order to analyze the features and the possible issues of MG theories is very convenient to use the so called \textit{St\"uckelberg trick}. Because it will be of fundamental importance for all the rest of the thesis, but a bit complicated in the non-linear context of GR, we prefer to first consider a simpler case, the one of a massive gauge field $A_\mu$ with spin 1. The same argument could be lifted without any caveat to the spin 2 case.\\
Let's start with the action for a massive U(1) vector field of the form:
\begin{equation}
\mathcal{S}\,=\,\int\,d^dx\,\left(-\frac{1}{4}\,F_{\mu\nu}F^{\mu\nu}-\frac{1}{2}\,m^2\,A_\mu A^\mu+A_\mu J^\mu\right)
\end{equation}
whose mass term spoils explicitly the gauge symmetry:
\begin{equation}
\delta A_\mu\,=\,\partial_\mu\,\Lambda
\end{equation}
and propagates (in $d=4$) 3 degrees of freedom.\\
It is straightforward to notice that the massless limit $m\rightarrow 0$ is not smooth, in the sense that it does not conserve the number of d.o.f. ( a massless gauge field in $d=4$ has 2).\\
The St\"uckelberg trick consists in introducing an additional scalar degree of freedom $\phi$ such that the original gauge symmetry is restored. The dynamics of the theory is equivalent to the original one and no d.o.f gets lost while driving the mass to zero. This can be achieved by the following:
\begin{equation}
A_\mu\,\rightarrow \,A_\mu\,+\,\partial_\mu\,\phi
\end{equation}
Note that: i) it does not represent  a change of field variables; ii) we are not decomposing the vector field $A_\mu$ in its longitudinal and transverse parts; iii) we are not in the presence of a gauge transformation.\\
Rescaling $\phi\rightarrow \frac{1}{m}\phi$ the new action takes the form:
\begin{equation}
\mathcal{S}\,=\,\int\,d^dx\,\left(-\frac{1}{4}\,F_{\mu\nu}F^{\mu\nu}-\frac{1}{2}\,m^2\,A_\mu A^\mu+A_\mu J^\mu\,-\,m\,A_\mu\,\partial^\mu \phi-\frac{1}{2}\,\partial_\mu \phi \,\partial^\mu \phi-\frac{1}{m}\,\phi\, \partial_\mu J^\mu\right)
\end{equation}
and it enjoys the gauge symmetry:
\begin{equation}
\delta A_\mu\,=\,\partial_\mu \Lambda\,,\qquad \delta \phi\,=\,-\Lambda\,.
\end{equation}
If we now go back to $\phi=0$, using the so called \textit{unitary gauge}, we exactly recover the previous theory for a massive U(1) field. Note that this is only possible if the current is conserved $\partial_\mu J^\mu=0$ otherwise such a limit does not exist.\\
We can now schematically reproduce the same trick for the spin 2 FP case:
\begin{equation}
\mathcal{L}_{m=0}\,-\,\frac{1}{2}\,m^2\,\left(h_{\mu\nu}h^{\mu\nu}-h^2\right)\,+\kappa\,h_{\mu\nu}T^{\mu\nu}
\end{equation}
but this time we have to introduce two additional fields: a vector $A_\mu$ and a scalar $\phi$ as the following:
\begin{equation}
h_{\mu\nu}\rightarrow h_{\mu\nu}+\partial_\mu A_\nu+\partial_\nu A_\mu\,,\qquad A_\mu\rightarrow A_\mu +\partial_\mu \phi
\end{equation}
In this way the gauge symmetry is restored again and takes the form of:
\begin{align}
& \delta h_{\mu\nu}=\partial_\mu \zeta_\nu +\partial_\nu \zeta_\mu,\,\qquad\delta A_\mu\,=\,-\zeta_\mu\,,\nonumber\\
& \delta A_\mu\,=\,\partial_\mu \Lambda\,,\qquad \delta \phi\,=\,-\Lambda\,.
\end{align}
After some unphysical rescaling the original action becomes:
\begin{align}
\mathcal{S}\,-\,\mathcal{S}_{m=0}\,=\,&\int\,d^dx\,\Big[-\frac{1}{2}\,m^2\,\big(h_{\mu\nu}h^{\mu\nu}-h^2\big)-\frac{1}{2}\,F_{\mu\nu}F^{\mu\nu}-2\,m\,\big(h_{\mu\nu}\,\partial^\mu A^\nu-h\,\partial_\mu A^\mu\big)\nonumber\\&-2\big(h_{\mu\nu}\,\partial^\mu \partial^\nu \phi-h\,\partial^2 \phi\big)+\kappa\,h_{\mu\nu}\,T^{\mu\nu}-\frac{2}{m}\,\kappa\,A_\mu \,\partial_\nu\,T^{\mu\nu}+\frac{2}{m^2}\,\kappa\,\phi\,\partial_\mu \partial_\nu\,T^{\mu\nu}\Big]
\end{align}
Assuming now that the stress tensor T is conserved and redefining the spin 2 field in an appropriate way  we can go ahead with the massless limit of the theory obtaining:
\begin{equation}
\mathcal{L}\,-\,\mathcal{L}_{m=0}\,=\,-\frac{1}{2}\,F_{\mu\nu}F^{\mu\nu}-\,2\,\frac{d-1}{d-2}\,\partial_\mu\phi \partial^\mu \phi\,+\,\kappa\,h_{\mu\nu}'\,T^{\mu\nu}\,+\,\frac{2}{d-2}\,\kappa\,\phi\,T
\end{equation}
Now the theory propagates 5 d.o.f. but the scalar mode is still coupled to the tensor one through the trace of the stress tensor $T=Tr[T_{\mu\nu}]$ and this coupling survives in the $m\rightarrow 0$ limit ! This means that in the massless limit we do not recover GR plus a set of decoupled fields.\\
This is the origin of the so called \textbf{zDVZ discontinuity} \cite{vdvz1,vdvz2} which refers to the fact that generically:
\begin{equation}
MG(m=0)\,\neq\,\text{massless gravity}
\end{equation}
and that the observable predictions of massive gravity in the $m\rightarrow 0$ limit are different to the ones expected by General Relativity. Because of this remaining coupling the scalar field $\phi$ does not change the light bending effect but it does affect for example the definition of the Newtonian potential with all its consequences.\\
As a side note, with the same trick one can also prove that a violation of the FP tuning leads to a ghosty excitation. In that case indeed the scalar mode would aquire a four derivatives term $\sim \left(\Box \phi \right)^2$, signaling the presence of 2 propagating d.o.f., one of which ghostlike \cite{ghost1,ghost2}. To this extent, Fierz Pauli can be thought as the exact combination that cancels such a higher derivative term.\\
The vDVZ problem is present just if we consider the MG theory in Minkowski space; once we promote the background to be curved such a issue disappears \cite{noV1,noV2,noV3}. In other words in curved spacetime the $m\rightarrow 0$ limit is again smoothly recovered and the counting of degrees of freedom works well.\\
This is due to the fact that in curved spacetime the Stuckelberg gauge field $A_\mu$ aquires already a mass given by the curvature:
\begin{equation}
\mathcal{S}\,=\,\int d^d_x\,\mathcal{L}_{m=0}\,+\,\sqrt{-g}\,\big(-\frac{1}{2}F_{\mu\nu}F^{\mu\nu}+\frac{2\,R}{d}\,A_\mu A^\mu+\kappa \,h_{\mu\nu}T^{\mu\nu}\big)
\end{equation}
and no additional and dangerous scalar Stuckelberg field has to be added.\\
Everything seems now consistent and resolved but this is not the end of the story. As we know GR is a full non linear theory for an interacting graviton and the FP modification we have considered refers just to its linear structure.\\
What we want to build is a non linear version of MG. With that we mean a non linear theory whose expansion around a fixed geometrical background results to be of the Fierz Pauli form. In this case there is no symmetry protecting the non linear structure of the theory, as diffeomorphims invariance does for GR, and the action could be very generic.\\
In particular we can write down the most generic non linear theory of massive gravity using an additional fixed metric $f_{\mu\nu}$, known as the reference metric, which will be the responsable for the breaking of diffeomorphism invariance. Along with the original metric $g_{\mu\nu}$ we can define the generic action of the form:
\begin{equation}
\mathcal{S}\,=\,\frac{1}{2\,\kappa^2}\,\int\,d^dx\,\left[\sqrt{-g}\,R\,-\,\sqrt{-g}\,\frac{m^2}{4}\,V\left(g,f\right)\right]
\end{equation}
The potential reproduces at the lowest order, the quadratic one $V_2$, the Fierz Pauli structure while all the terms represent non linear corrections. As we can see the form of the theory is completely undetermined and very general.\\
Using ADM techniques \cite{ADM1,ADM2} one can get through the counting of degrees of freedom dealing with the various constraints appearing in the theory. In that way we can generically test the presence of ghosts and other instabilities. Indeed that is the case: the instability appears via a ghosty excitation which takes the name of \textbf{Boulware-Deser Ghost} (BD) \cite{Hmg}. Its mass will be infinite, and therefore its presence not dangerous, in flat space but on the contrary it would result finite, and as a consequence problematic, on a non trivial background.\\
The BD ghost was introduced in \cite{BDUN} and killed all the hope about a possible consistent and non linear theory of MG, but that was too quick! Non trivial and additional interactions could possibly eliminate the presence of such a ghost \cite{NOBD}.
\section{dRGT and lorentz violating MG}
In 2010 de Rham, Gabadadze and Tolley succeed in constructing a full non linear theory of MG where the unwanted BD ghost was not present \cite{originalDRGT}, named after that dRGT MG. The idea was basically the one of tuning the coefficients of the non linear potential to cancel order by order the higher derivative producing the BD ghost. They managed to repack everything in a non linear structure which has been proven to be healthy beyond the decoupling limit \cite{dRGT1,dRGT2}.\\
We start again from the generic action for non linear MG given by:
\begin{equation}
\mathcal{S}_{dRGT}\,=\,\frac{M_P^2}{2}\,\int\,d^4x\,\sqrt{-g}\,\left(\,R\,+\,\frac{m^2}{2}\,\sum_{n=0}^{4}\,\alpha_n\,\mathcal{L}_n\,\left[\,\mathcal{K}\,[g,f]\,\right]\right)\,.
\end{equation}
where $\alpha_n$ are arbitrary coefficients and we have defined the matrix object:
\begin{equation}
\mathcal{K}^\mu_{\,\,\nu}[g,f]\,=\,\delta^\mu_{\,\,\nu}\,-\,\left(\sqrt{g^{-1}\,f}\,\right)^\mu_{\,\,\nu}
\end{equation}
In this way the corresponding Langragians $\mathcal{L}_n$, associated to the dRGT choice, take the form:
\begin{align}
&\mathcal{L}_0\,[\mathcal{K}]\,=\,4!\,,\nonumber\\
&\mathcal{L}_1\,[\mathcal{K}]\,=\,3!\,[\mathcal{K}]\,,\nonumber\\
&\mathcal{L}_2\,[\mathcal{K}]\,=\,2!\,\left(\,[\mathcal{K}]^2\,-\,[\mathcal{K}^2]\,\right)\,,\nonumber\\
&\mathcal{L}_3\,[\mathcal{K}]\,=\,\left(\,[\mathcal{K}]^3\,-\,3\,[\mathcal{K}]\,[\mathcal{K}^2]\,+\,2\,[\mathcal{K}^3]\,\right)\,,\nonumber\\
&\mathcal{L}_4\,[\mathcal{K}]\,=\,\left(\,[\mathcal{K}]^4\,-\,6\,[\mathcal{K}^2]\,[\mathcal{K}]^2\,+\,3\,[\mathcal{K}^2]^2\,+\,8\,[\mathcal{K}]\,[\mathcal{K}^3]\,-6\,[\mathcal{K}^4]\,\right)\,.
\end{align}
The zero order part $\mathcal{L}_0$ is the usual cosmological constant term and the second order term $\mathcal{L}_2$ reduces to the known FP structure. All the higher order terms constitute non trivial interactions which make the theory fully non linear and whose tuned coefficients ensure that the resulting theory is \textbf{ghost-free}.\\
The discovery of dRGT massive gravity has introduced a renewed interest into the field. A theoretical consistent non linear theory of massive gravity is possible and its applications to the reality have just to be tested.\\[0.1cm]
Through all the previous discussion we have strongly assumed Lorentz invariance as a fundamental symmetry of our system showing explicitly all their possible problems connected with the existence of an extra ghosty scalar mode, \textit{i.e.} the BD ghost. If one stick to Lorentz invariant situations the only viable theory is the famous dRGT massive gravity, but as soon as this assumption gets relaxed a plethora of new possibilities appear. They go under the name of \textit{Lorentz violating massive gravity theories} and they are nicely reviewed and described in \cite{infrubakov,phasesMG}.\\
Naively one would expect these models to be less pathologic than their Lorentz invariant version because they admit the possibility of preserving certain subgroups of the diffeomorphism one and indeed that is the case. We do not enter the discussion of the phenomenological viability of breaking Lorentz simmetry for which we refer to \cite{LV} but we will just accept it as a theoretical possibility.\\
The most generic LV massive gravity theory in $d=4$ which preserves the rotation group in the spatial subspace can be written in the following form \cite{LVMG1}:
\begin{equation}
\mathcal{L}_m\,=\,\frac{M_P^2}{4}\,\left(m_0^2\,h_{00}\,h_{00}\,+\,2\,m_1^2\,h_{0i}\,h_{0i}\,-\,m_2^2\,h_{ij}\,h_{ij}\,+\,m_3^2\,h_{ii}\,h_{jj}\,-\,2\,m_4^2\,h_{00}\,h_{ii}\right)
\label{LVmassterms}
\end{equation}
According to these notations the FP theory can be defined by:
\begin{equation}
\text{FP:}\qquad m_0\,=\,0\,,\,\,\,\,m_1\,=\,m_2\,=\,m_3\,=\,m_4\,=\,m_G\,.
\end{equation}
The model now contains a massive tensor sector:
\begin{equation}
\mathcal{L}_m^{TT}\,=\,-\,\frac{m_2^2}{4}\,h_{ij}^{TT}\,h_{ij}^{TT}
\end{equation}
where $m_2$ qualifies indeed as the mass for the tensor gravitons and in order to avoid tachyonic excitations its square would better be positive:
\begin{equation}
m_2^2\geq 0
\end{equation}
In the vector sector both $m_1$ and $m_2$ appears whereas all the other masses are present in the scalar action. There are rigid constraints on the masses to avoid ghosts, tachyons and other instability issues. One can consider all the possible combinations and classify the healthy, and numerous, phases of lorentz violating massive gravity which are theoretically consistent \cite{phasesMG}. The resulting zoology can be efficiently classified by looking at the residual gauge symmetries \cite{phasesMG}. Despite the recent efforts in discussing the possible UV completions of such theories (see for example \cite{UVLVMG}) we just rely on them as low energy effective field theory (EFT) with a certain cutoff and we completely avoid such a topic.\\
A convenient way of describing these theories is by introducing St\"uckelberg fields. We can introduce a set of four scalars ($\phi^0, \phi^a$), \textit{i.e.} the Goldstones, with $a=1,2,3$. The \textit{spontaneous} breaking of LI appears when the latter aquire background expectation values which depend on the space-time coordinates. More precisely the background fields have non trivial vevs fixed by:
\begin{align}
&\bar{\phi}^0\,=\,a\,t\,,\nonumber\\
&\bar{\phi}^a\,=\,b\,x^a\,.
\end{align}
The previous expressions are of course a solution of the system if those fields enter in the action just with derivative terms. The latter property automatically implies that the corresponding action results invariant under a shift symmetry $\phi^a\rightarrow\phi^a+\lambda^a$ with constant $\lambda^a$. Likewise in order to preserve isotropy of the spatial 3-dimensional subspace we need to ask the action to be invariant upon the transformation $\phi^i\rightarrow \Lambda^i_{\,\,j}\phi^j$ as well.\\
All in all the most generic action for the gravity + scalars system is constrained to have the following form:
\begin{equation}
\mathcal{S}\,=\,\mathcal{S}_{EH}\,+\,\int\,d^4x\,\sqrt{-g}\,\Lambda^2\,F(Y,V^i,X^{ij})\,.
\end{equation}
with:
\begin{align}
&Y\,=\,\frac{1}{\Lambda^4}\,g^{\mu\nu}\,\partial_\mu \phi^0\,\partial_\nu \phi^0\,,\nonumber\\
&V^i\,=\,\frac{1}{\Lambda^4}\,g^{\mu\nu}\,\partial_\mu \phi^0\,\partial_\nu \phi^i,\nonumber\\
&X^{ij}\,=\,\frac{1}{\Lambda^4}\,g^{\mu\nu}\,\partial_\mu \phi^i\,\partial_\nu \phi^j\,.
\end{align}
For reasons which will be clearer in the following in this work we will only consider theories with non trivial scalar profiles just in two spatial directions $x,y$. In this language this corresponds to fix $a=0$ and therefore making the field $\phi^0$ disappear from the game. With just two spatial directions the symmetries fix the action to be just a function of two scalar object constructed from the $X^{ij}$ matrix, its trace and its determinant:
\begin{equation}
X\,=\,Tr[X^{ij}]\,,\qquad Z\,=\,det[X^{ij}]\,.
\end{equation}
In conclusion the most generic (within the assumptions we made) Lorentz violating theory of massive gravitons takes the form:
\begin{equation}
\mathcal{S}\,=\,\mathcal{S}_{EH}\,+\,\int\,d^4x\,\sqrt{-g}\,\Lambda^2\,V(X,Z)\,.
\label{LVMGgeneral}
\end{equation}
where V is a generic potential function.\\
There is a very strong analogy between the LV theories of massive gravity and the EFT for spontaneous Lorentz symmetry breaking (\textit{i.e.} the EFT for fluids and solids) \cite{LVEFT1,LVEFT2,LVEFT3,LVEFT4}. The latter are defined in flat Minkowski space but their construction and action is exactly the same. Recasting LV MG theories in the language of General Relavity + a scalar sector is not only very convenient from the point of view of the consistency and healthiness checks but from the phenomenological perspective too. One can indeed classify again the various phases using the internal symmetries of the scalar sector. As a simple example, we can distinguish \textit{solid} from \textit{fluids} in this way; whereas solids enjoy just internal translational symmetry, fluids are invariant also under internal volume-preserving diffeomorphism:
\begin{align*}
&\text{SOLIDS:}\qquad\{\,\phi^i\rightarrow\phi^i\,+\,c^i\,\}\,,\\
&\text{FLUIDS:}\qquad\{\,\phi^i\rightarrow\phi^i\,+\,c^i\,\}\,\,+\,\,\{\,\phi^i\rightarrow \xi^i(\phi^j)\,\}\qquad\text{with}\qquad det(\partial \xi^i/\partial \phi^j)\,=\,1\,.\\
\end{align*}
This fact has a strong influence on the allowed action. The larger symmetry , which fluids enjoy, forces the action to be a function of the only determinant Z:
\begin{equation}
\mathcal{L}_{solid}\,\sim\,V(X,Z]\,,\qquad\mathcal{L}_{fluids}\,\sim\,V(Z]\,.
\end{equation}
and in the language of \eqref{LVmassterms} constraints the mass of the traceless transverse part of the graviton to vanish, \textit{i.e.} $m_2=0$. This account to say that, despite the solid case, there are no propagating transvers phonons in a fluid.\\
Note that this formulation in terms of scalars with non vanishing V.E.V.s allows to construct the most generic massive gravity theories and is able to reproduce also the Lorentz Invariant case\footnote{One has just to fix the vevs of the scalars to be all the same such that Lorentz invariance is restored.} and for example the dRGT scenario. An important point to make is that a theory defined as in \eqref{LVMGgeneral} is \underline{much more general} than the dRGT case and that this is consistent and allowed thanks to the breaking of Lorentz simmetry.\\
We will make use of these theories defined on an Anti de Sitter background in the context of the Gauge-Gravity duality in order to mimick particular Condensed Matter situations.











\section{A jump into AdS-CMT}
\begin{figure}
\centering
\includegraphics[width=8cm]{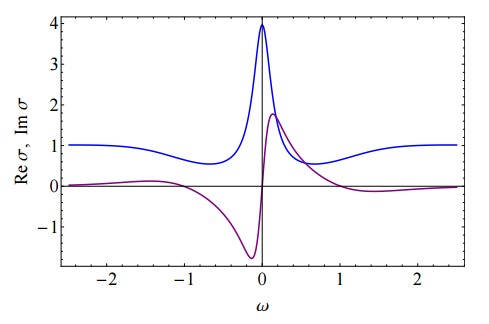}
\caption{AC electric conductivity $\sigma(\omega)$ for the dRGT holographic model. The plot is taken from \cite{Veghoriginal}.}
\label{MGsigma}
\end{figure}

\begin{figure}
\centering
\includegraphics[width=10cm]{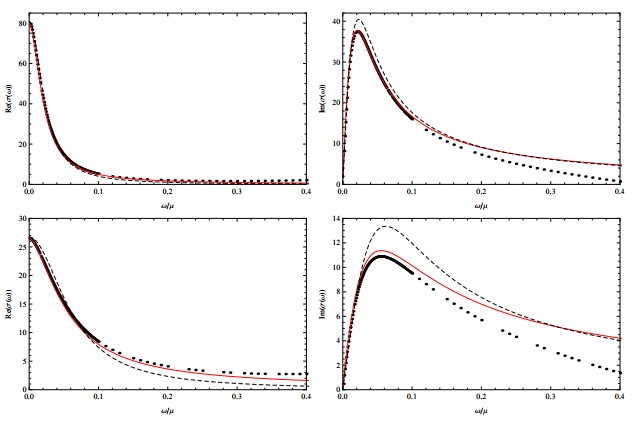}
\caption{An example of AC electric conductivity $\sigma(\omega)$ results in presence of a massive gravity term taken from \cite{DavisonHydro}.}
\label{MGvsDrude}
\end{figure}

\begin{figure}
\centering
\includegraphics[width=10cm]{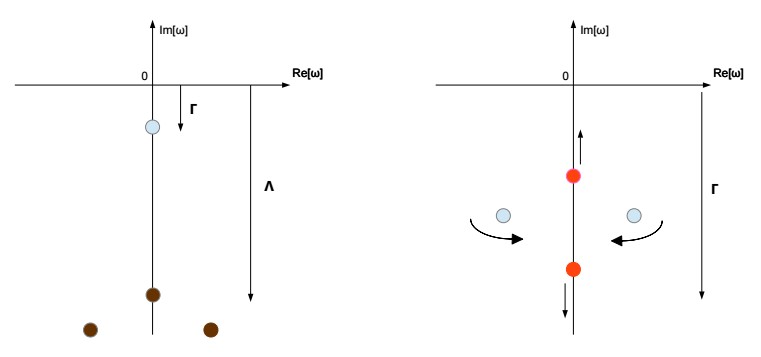}
\caption{Schematic representation of the quasinormal modes locations in the massive gravity system. For details see \cite{DavisonGouteraux}.}
\label{INCpoles}
\end{figure}

\begin{figure}
\centering
\includegraphics[width=12cm]{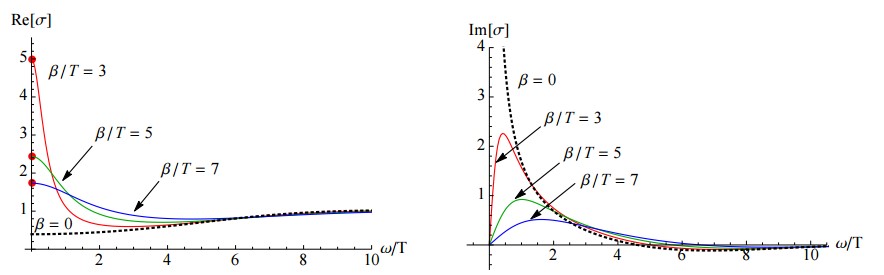}\\[0.3cm]
\includegraphics[width=12cm]{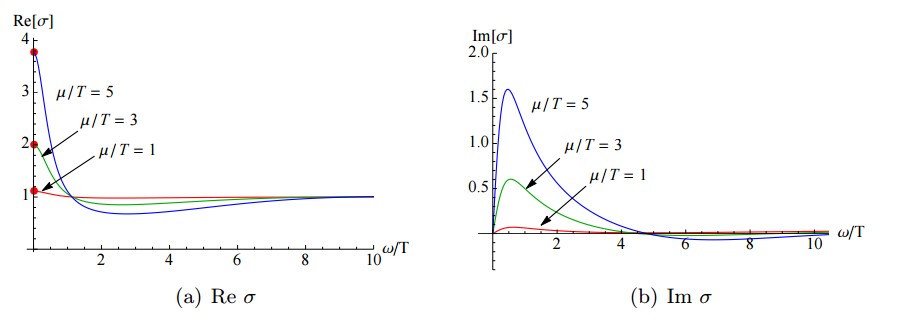}
\caption{AC electric conductivity $\sigma(\omega)$ results in presence of a massive gravity term $\beta$ coming from the scalar model with profile $\psi^I=\beta x^I$. Plots are taken from \cite{KimIncoherent}.}
\label{Axionssigma}
\end{figure}

\textbf{dRGT-CMT}\\[0.5cm]
Systems with perfect translational simmetry cannot dissipate momentum. As a consequence whenever a finite density of charge carrier is present in such a system the correspondent electric DC conductivity $\sigma_{DC}=\sigma(\omega=0)$ is infinite. In a weakly coupled fashion this can be easily seen from the DC formula given by the Drude Model:
\begin{equation}
\sigma_{DC}\,=\,\frac{n\,e^2\,\tau}{m}
\label{DrudeDC1}
\end{equation}
where $\tau$ is the already mentioned collision/relaxation time coming from the equation which controls the dynamic of the momentum $\vec{p}$:
\begin{equation}
\frac{d \vec{p}}{dt}\,=\,e\,\vec{E}\,-\,\frac{\vec{p}}{\tau}
\end{equation}
Note again that the $\tau$ parameter is a priori an effective parameter of the theory which is not determined by any microscopic physics; this will happen also in holographic theories.\color{black}
Whenever translational symmetry is preserved, momentum cannot be relaxed meaning that the relaxation time $\tau=\infty$. It follows directly from \eqref{DrudeDC1} that the electric DC conductivity is infinite\footnote{Note that this infinite is significantly different from the one encountered in a Superconducting medium:
\begin{equation}
\sigma\,\sim\,\frac{\rho_S\,i}{\omega}
\end{equation}
where the delta function is due to a Bose-Einstein condensation mechanism.}. Hydrodynamics arguments give that in the presence of a conserved momentum operator the low frequency conductivity reads:
\begin{equation}
\sigma(\omega)\,=\,s\,T\,\left(\delta(\omega)\,+\,\frac{i}{\omega}\right)
\end{equation}
and it is characterized indeed by a $\delta$ function at zero frequency which in the presence of momentum dissipation gets smoothed out into the so-called \textit{Drude Peak}:
\begin{equation}
\sigma(\omega)\,=\,\frac{\sigma_{DC}}{1\,-\,i\,\omega\,\tau}
\end{equation}
revealing a pole shifted in the lower half of the imaginary axes.\\
The same phenomenon can be re-expressed in modern language \cite{Hartnollsus} stating that whenever the current operator $\vec{J}$ has a finite overlap with the momentum operator $\vec{P}$ (meaning the susceptibility $\chi_{\vec{J}\vec{P}}\neq0$) and momentum is a conserved quantity than the conductivity coming from the $\vec{J}\vec{J}$ correlator through the Kubo formula:
\begin{equation}
\sigma\,=\,\lim_{\omega\rightarrow0}\,\frac{Im\, G^R_{\vec{J}\vec{J}}(\omega)}{\omega}
\end{equation}
has an infinite DC value at zero frequency.\\
Holography does not evade such a generic prescription and holographic systems with translational symmetry, like the Reissner-Nordstrom benchmark model described in the previous section, indeed show an infinite DC conductivity\footnote{Strictly speaking the numerical procedure does not show the $\delta$ function in the real part of the conductivity but just a pole $1/\omega$ in the imaginary part. Through Kramers-Kronig relations one can then argue the presence of the $\delta$ function in the real part.}.\\
In the recent years various ways of avoiding the infinite DC conductivity have been introduced by treating the charge carriers in the probe limit \cite{probebraneKarch,probebraneHartnoll} (\textit{i.e.} as a small part in a larger system of neutral fields where they can dump momentum), or by introducing spatial inhomogeneities thereby breaking translational invariance explicitly \cite{holattice1,holattice2}. The first scenario consists in freezing the fluctuations of the metric. Strictly speaking, the dual field theory does not have a proper energy momentum tensor and there is no overlap between the current and the momentum operators. In the second case there is an explicitly inhomogeneous background which involves hard numerical efforts in order to solve complicated systems of PDEs.\\
Motivated by the main goal of building a framework for translational symmetry breaking and momentum dissipation in holography without the
need for complex numerical computations Massive gravity was introduced in the context of holography in \cite{Veghoriginal}.\\
The AdS-CFT dictionary tell us that the metric field $g_{\mu\nu}$ in the bulk is dual to the Stress Tensor operator $T_{\mu\nu}$ of the correspondent dual boundary CFT. Momentum (density) operator is defined as $T^{0i}$ and it is part of such a object whose conservation reads:
\begin{equation}
\nabla_\mu\,T^{\mu\nu}\,=\,0
\end{equation}
In more details, translational symmetry (in the spatial coordinates) implies that momentum is a conserved quantity. From the point of view of the AdS-CFT correspondence the conservation of the stress tensor is encoded in the gauge symmetry of the metric field, \textit{i.e.} diffeomorphism invariance. It is therefore clear that one method to make momentum be not conserved consists in breaking (at least the spatial part) diffeomorphims invariance in the bulk.\\
On the other hand the temporal part of the Stress Tensor $T^{00}$ encodes the energy density of the system and it is kept to be conserved. This forces univoquely the symmetry breaking pattern we are interested in. Working in $d=3+1$ dimensions $\{t,r,x,y\}$, we break translational symmetry in the spatial directions $\{x,y\}$ expressed in the linear diffeomorphism transformation:
\begin{equation}
x^i\,\rightarrow\,x^i\,+\,\zeta^i\,,\qquad i\,=\,x\,,\,y\,.
\end{equation}
but we preserve the temporal part of such a transformation.\\
In the charged black brane background described by the RN solution, Ward identities for translational invariance in the x direction imply a shift symmetry in the $g_{tx}$ field. The simplest option to break such a symmetry is to add a mass term for the graviton:
\begin{equation}
\mathcal{L}_m\,\sim\,\sqrt{-g}\,m^2\,g^{tx}\,g_{tx}
\end{equation}
The idea is to replace the usual Einstein-Maxwell theory in the bulk with dRGT-Maxwell generalization, namely:
\begin{equation}
\mathcal{S}\,=\,\frac{1}{2\,\kappa^2}\,\int\,d^4x\,\sqrt{-g}\,\left[\underbrace{\,R\,+\,\Lambda\,-\,\frac{L^2}{4}\,F^2\,}_{Einstein-Maxwell}+\,m^2\,\sum_{i=1}^4\,c_i\,\mathcal{U}_i(g,f)\right]
\end{equation}
where $f$ is the usual fixed reference metric, $c_i$ are constants and $\mathcal{U}_i$ are polynomials of the eigenvalues of the matrix $\mathcal{K}=\sqrt{g^{\mu\alpha}\,f_{\nu\alpha}}$ defined as\footnote{The square root in $\mathcal{K}$ is understood to denote the matrix square root  and the rectangular brakets the matrix trace.}
\begin{align}
&\mathcal{U}_1\,=\,[\mathcal{K}]\,,\nonumber\\
&\mathcal{U}_2\,=\,[\mathcal{K}^2]\,-\,[\mathcal{K}]^2\,,\nonumber\\
&\mathcal{U}_3\,=\,[\mathcal{K}^3]\,-\,3\,[\mathcal{K}^2]\, [\mathcal{K}]\,+\,2\, [\mathcal{K}^3]\,,\nonumber\\
&\mathcal{U}_4\,=\,[\mathcal{K}^4]\,-\,6\,[\mathcal{K}^2]\, [\mathcal{K}]^2\,+\,8\,[\mathcal{K}^3]\, [\mathcal{K}]\,+\,3\,[\mathcal{K}^2]\,[\mathcal{K}]^2\,-\,6\,[\mathcal{K}^4]\,.
\end{align}
This massive gravity construction is built to avoid the already discussed BD ghost and in the limit $m\rightarrow 0$ it boils down to the usual translational invariant Einstein-Maxwell setup. To implement the wanted symmetry breaking pattern we choose the refence metric to be (in the basis $(t,r,x,y)$) :
\begin{equation}
f^{SP}_{\mu\nu}\,=\,diag(0,0,1,1)
\end{equation}
where $SP$ stands for spatial\footnote{In the St\"uckelberg language this corresponds to switch on just the two spatial St\"uckelberg fields $\phi^x,\phi^y$. We will come back on this point later.}. In this way the mass term $\sim m^2 \mathcal{U}(g,f^{SP})$ preserves general covariance in the t,u, coordinates but breaks it in the two spatial dimensions x,y. This is exactly what we need and it corresponds to allow momentum (but not energy) density to dissipate in the dual picture.\\
Because of this choice and the number of dimensions we are working on only $\mathcal{U}_{1,2}$ are independent objects and the generic mass term considered takes the form of:
\begin{equation}
\sim m^2 \left[\,\alpha \,[\mathcal{K}]\,+\,\beta\,\left([\mathcal{K}]^2\,-\,[\mathcal{K}^2]\right)\right]
\end{equation}
With these assumptions the charged BH solution reads:
\begin{align}
&ds^2\,=\,L^2\,\frac{1}{r^2}\left(\,\frac{dr^2}{f(r)}\,-\,f(r)\,dt^2\,+\,dx^3\,+\,dy^2\right)\,,\nonumber\\
&A(r)\,=\,A_t(r)\,dt\,=\,\mu\,\left(1\,-\,\frac{r}{r_h}\right)\,dt\,.
\end{align}
where the emblackening factor is:
\begin{equation}
f(r)\,=\,1\,+\,\alpha\,L\,\frac{m^2}{2}\,r\,+\,\beta\,m^2\,r^2\,-\,M\,r^3\,+\,\frac{\mu^2}{4\,r_h^2}\,r^4
\end{equation}
The mass of the BH object M is fixed in such a way that $f(r_h)=0$ and $r_h$ is the proper event horizon of such a BH.\\
In the limit $m=0$ we recover the usual RN solution; otherwise we have two new parameters in the system $\alpha,\beta$.\\
The temperature of such a background is given by:
\begin{equation}
T\,=\,\frac{1}{4\,\pi\,r_h}\,\left(3\,-\,\left(\frac{\mu\,r_h}{2}\right)\,+\,m^2\,r_h\,(\alpha\,L\,+\,\beta\,r_h)\right)
\end{equation}
and the geometry represents a finite density state with definite entropy $s$, energy density $\epsilon$ and charge density $\rho$ which satisfy the usual first law of thermodynamics:
\begin{equation}
d\epsilon\,=\,T\,ds\,+\,\mu\,d\rho
\end{equation}
The corresponding spin 2 perturbation around the background, \textit{i.e.} the graviton, will aquire a radial dependent mass term given by:
\begin{equation}
m^2(r)\,=\,-\,2\,\beta\,-\,\frac{\alpha\,L}{r}
\end{equation}
where r is the radial coordinate and the UV is fixed at $r=0$.\\
It is immediately clear that one requisite of stability is dictated by imposing that such a mass is positive and real. That would correspond to require that the momentum relaxation time $\tau$ is positive.\\
The solution asymptotes from the AdS$_4$ boundary to an infrared AdS$_2\times$ R$_2$ geometry.\\
On top of this background we can run the machinery to compute holographically the conductivity, namely perturbing the solution as the following:
\begin{align}
&ds^2\,\rightarrow\,ds^2\,+\,g_{tx}(r)\,e^{i\,\omega\,t}\,+\,g_{rx}(r)\,e^{i\,\omega\,t}\,,\nonumber\\
&A(r)\,\rightarrow\,A(r)\,+\,a_x(r)\,e^{i\,\omega\,t}\,dx
\end{align}
The Maxwell equation becomes:
\begin{equation}
\left(f\,a_x'\right)'\,+\,\frac{\omega^2}{f}\,a_x\,=\,-\,\frac{A_t'\,r^2}{L^2}\,\left(g_{tx}'\,+\,\frac{2}{r}\,g_{tx}\,-\,i\,\omega\,g_{rx}\right)
\end{equation}
At the same time the remaining Einstein equations read:
\begin{align}
&\left(g_{tx}'\,+\,\frac{2}{r}\,g_{tx}\,-\,i\,\omega\,g_{rx}\,+\,A_t'\,L^2\,a_x\right)'\,=\,\frac{m^2(r)}{f}\,g_{tx}\,,\nonumber\\
&\left(g_{tx}'\,+\,\frac{2}{r}\,g_{tx}\,-\,i\,\omega\,g_{rx}\,+\,A_t'\,L^2\,a_x\right)\,=\,-\,\frac{i\,f\,m^2(r)}{\omega}\,g_{rx}\,,
\end{align}
These two equations are no longer equivalent, like it happens in the usual translational symmetric case, and therefore we are obliged to turn on also the $g_{rx}$ component which is usually set consistently to 0.\\
Nevertheless the 2 equations imply the constraint:
\begin{equation}
\frac{i\,\omega\,m^2(r)}{f}\,g_{tx}\,=\,\left(\,m^2(r)\,f\,g_{rx}\right)'
\end{equation}
which can used to eliminate $g_{tx}$ as in the usual RN case.\\
All in all, after redefining $\tilde{g}_{rx}=f g_{rx}$ we are left with the two differential equations:
\begin{align}
&\left(f\,a_x'\right)'\,+\,\frac{\omega^2}{f}\,a_x\,=\,\rho^2\,r^2\,a_x\,+\,\frac{\rho\,m^2\,r^2}{i\,\omega\,L^2}\,\tilde{g}_{rx}\,,\nonumber\\
&\frac{1}{r^2}\,\left(\frac{r^2\,f}{m^2}\,(m^2\,\tilde{g}_{rx})'\,\right)'\,+\,\frac{\omega^2}{f}\,\tilde{g}_{rx}\,=\,\rho\,L^2\,i\,\omega\,a_x\,+\,m^2\,\tilde{g}_{rx}\,.
\label{drgtEQ}
\end{align}
where $\rho=\mu/r_h$ is the charge density of the system.\\
These equations can be used to extract numerically the electric conductivity through the usual prescription:
\begin{equation}
\sigma(\omega)\,=\,\frac{a_x'}{i\,\omega\,a_x}|_{UV}
\end{equation}
There has been extensive effort in studying numerically the electric (and not only) conductivity in the context of massive gravity theories \cite{Veghoriginal,DavisonHydro,KimIncoherent,DavisonGouteraux, genova1}.\\
The, not surprising result, is that indeed the DC conductivity gets finite and the $\delta$ function coming from momentum conservation gets broaden up. Benchmark results are shown in fig.\ref{MGsigma} from ref.\cite{Veghoriginal}. It is also possible, through methods which we will explain in details in the next sections, to extract the DC value of the electric conductivity for this setup \cite{BlakeTongDC}:
\begin{equation}
\sigma_{DC}\,=\,\frac{1}{e^2}\,\left(1\,+\,\frac{\rho^2\,r_h^2}{m^2(r_h)}\right)
\label{DC1}
\end{equation}
with $m^2(r_h)\,=\,-\,2\,\beta\,-\,\frac{\alpha\,L}{r_h}$.\\
The DC value contains two different terms with their own physical meaning; for the moment we just notice how the second term looks like very similar to the Drude formula we have already encountered many times during our journey. We will give more details about this formula in the following.\\
In order to understand better the dynamics of the momentum dissipation introduced by the presence of a graviton mass a convenient way is to rely on an hydrodynamical low energy description \cite{DavisonHydro}. The low energy description of the theory now enjoys a modified conservation law for the energy-momentum $T^{\mu\nu}$ which takes the form:
\begin{equation}
\partial_a\,T^{at}\,=\,0\,,\qquad
\partial_a\,T^{ai}\,=\,-\,(\epsilon\,+\,P)\,\tau^{-1}\,u^i\,=\,-\,\frac{T^{ti}}{\tau}\,.
\end{equation}
where where $\epsilon$, P and $u^i$ are the energy density, pressure and velocity of the fluid state, and the $\tau$ constant represents the characteristic timescale for momentum relaxation.\\
Such a relaxation time scales turns out to be inversely proportional (up to thermodynamical quantities) to the graviton mass:
\begin{equation}
\tau\,\sim\,\frac{1}{m^2}
\end{equation}
We can therefore understand the physical interpretation of the instability related to the $m^2\geq 0$ constraint:  whenever $\tau<0$ the system will absorb momentum rather than dissipating it, and thus once a small perturbation is introduced it will grow
exponentially in time producing the instability itself.\\
In the limit of small graviton mass $m^2/\mu^2 \,\ll\,\omega/\mu\,\ll\,1$, where momentum is almost conserved, the conductivity shows the typical Drude form:
\begin{equation}
\sigma(\omega)\,=\,\frac{\sigma_{DC}}{1\,-\,i\,\omega\,\tau}
\end{equation}
However the full expression, which can be derived perturbatively in $m^2$, deviates from the Drude formula and it contains corrections to the latter (see fig.\ref{MGvsDrude} taken from \cite{DavisonHydro}). The introduction of these corrections produces spectral weight transfer from the Drude peak towards an incoherent state. It is anyway pretty amazing that massive gravity, at small graviton mass, namely at weak momentum dissipation, provides an strongly coupled analogue of the Drude model.\\
Note how the identification of the graviton mass with the inverse of the relaxation time $\tau$ is in perfect agreement with the formula for the DC conductivity, whose second term $\sim \frac{\rho^2}{m^2}\sim \rho^2 \tau$ takes indeed the common Drude form.\\
\\
It has been later analyzed through a quasinormal modes analysis \cite{DavisonGouteraux} that whenever momentum is slowly dissipated a well defined Drude pole dominates the response, making it appear as ''coherent''. On the contrary, when momentum conservation is brutally violated, there is no such a excitation and what is left is diffusion which leads to an \textit{incoherent} behaviour. This confirms the previous expectations that for small graviton mass the conductivity takes a Drude-like form with a purely imaginary pole $\omega=-i\Gamma=-i \tau^{-1}$ dominating the conductivity. The current is carried by a long-lived collective excitation whose decay rate $\Gamma$ is parametrically larger the the others. In fig.\ref{INCpoles} from \cite{DavisonGouteraux} we provide a sketch of the two different situations.\\[0.3cm]
Finite DC conductivity was already found in the context of holography working with the so-called holographic lattices \cite{holattice1,holattice2,holattice3,holattice4,holattice5,holattice6,holattice7}.
These models consist of Einstein-Maxwell action plus a neutral scalar field:
\begin{equation}
\mathcal{S}\,=\,\int\,d^4x\,\sqrt{-g}\,\left[\frac{1}{2\,\kappa^2}\,\left(R\,+\,\frac{6}{L^2}\right)\,-\,\frac{1}{4\,e^2}\,F_{\mu\nu}F^{\mu\nu}\,-\,\frac{1}{2}\,g^{\mu\nu}\,\partial_\mu \phi\,\partial_\nu \phi\,-\,\frac{m^2}{2}\,\phi^2\,\right]\,.
\end{equation}
We would like to break translational invariance via introducing a modulated source for the scalar operator $\mathcal{O}$ dual to the bulk field $\phi$.\\
For static solutions the near-boundary expansion of the scalar $\phi$ reads:
\begin{equation}
\phi_0(r,x,y)\,\sim\,\phi_{-}(x,y)\,\left(\frac{r}{L}\right)^{\Delta_{-}}\,+\,\phi_{+}(x,y)\,\left(\frac{r}{L}\right)^{\Delta_{+}}
\end{equation}
where $\Delta_{\pm}\,=\,\frac{3}{2}\,\pm\,\sqrt{\frac{9}{4}\,+\,m^2L^2}$.
Assuming standard quantization, meaning fixing the value of $\phi_{-}$ and identifying it as the source for the $\mathcal{O}$ operator, we work with the profile:
\begin{equation}
\phi_{-}\,=\,\epsilon\,\cos(k_L\,x)
\end{equation}
where $\epsilon$ is a small parameter which will allow us to treat the lattice perturbatively. The lattice radial profile gets determined by the equations of motion themselves. The solution in the bulk takes the form $\phi(r,x,y)=\epsilon\,\phi_0(r)\cos(k_L \,x)$ with the additional constraint:
\begin{equation}
\frac{d}{dr}\,\left(\frac{f}{r^2}\,\frac{d\phi_0}{dr}\right)\,-\,\frac{k_L^2}{r^2}\,\phi_0\,-\,\frac{m^2\,L^2}{r^4}\,\phi_0\,=\,0\,.
\end{equation}
with $f(r)$ the emlackening factor of the charged black brane.\\
We will choose $m^2<0$ such that the operator $\mathcal{O}$ will be relevant with the profile $\phi_0(r)$ growing in the infra-red.
The method consists in treating such a periodic scalar deformations as a small perturbations and solve the system at leading order in $\epsilon$.\\
The main result is that at leading order the graviton aquires a mass:
\begin{equation}
\mathcal{S}_{eff}\,=\,\frac{1}{2}\,\int\,d^4x\,\sqrt{-g}\,M^2(r)\,g_{tx}\,g^{tx}
\end{equation}
with a radial dependent mass given by:
\begin{equation}
M^2(r)\,=\,\frac{1}{2}\,\epsilon^2\,k_L^2\,\,\phi_0(r)^2
\end{equation}
It has the same form as the mass terms arising in massive gravity model, albeit with a different radial profile.\\
This suggests a deep connection between holographic models for lattices and/or explicit disorder (in this case the scalar profile is not periodic but a more complicated and random function) and massive gravity theories. At least at leading order those models are ''equivalent'' (we will come back to this topic with more details) to massive gravity, which seems to realize an effective and efficient description of momentum dissipation in the context of holography. Despite the first attempts were focused on the dRGT choice it should be now clear that such a theory is not the most general massive gravity theory we can construct and that in absence of Lorentz invariance there are many more possibilities available. We will return to this aspect.\\[1cm]
\textbf{Dissipating momentum via St\"uckelberg fields}\\[0.3cm]
The holographic stress tensor obeys (\cite{AndradeWithers}) a conservation equation which reads:
\begin{equation}
\nabla_i\,\langle T^{ij}\rangle \,=\,\nabla^j\,\psi^{(0)}\,\langle O \rangle\,+\,F^{(0)ij}\,\langle J_i\rangle\,.
\end{equation}
where i,j label the boundary spacetime directions. This Ward identity suggests a route to holographic momentum relaxation ($\nabla_i\,\langle T^{ij}\rangle \neq 0$) by turning on spatially dependent source terms for the scalar $\psi$, meaning $\nabla^j\,\psi^{(0)}\neq 0$.\\
This is not surprising since spatially dependent sources have been utilised to construct holographic lattices which exhibit finite DC conductivity \cite{holattice1,holattice2,holattice3,holattice4,holattice5,holattice6,holattice7}.
In all those scenarios the stress tensor becomes dependent on the spatial coordinates $x^i$ and the correspondent Einstein equations turn out to be PDEs whose numerical integration is not trivial. Anyway the scalar(s) $\psi$ enter into the stress tensor just via first derivatives:
\begin{equation}
T^{ij}\,\sim\,\nabla^i\,\psi\,\nabla^j\,\psi
\end{equation}
showing the presence of a scalar field shift symmetry. Therefore the idea is to exploit such a symmetry noticing that if we turn on sources for the scalars which are linear in the boundary coordinates:
\begin{equation}
\psi^{(0)i}\,\sim\,\beta_i\,x^i
\end{equation}
the stress tensor is blind to the boundary coordinates and one can find homogenous bulk solutions for the system. In general though, such a configuration will not be isotropic. To render it isotropic we need to introduce a totals of $\tilde{d}$ scalar fields $\psi^I$ where $\tilde{d}$ is the number of spatial dimensions of the boundary. We can then arrange their sources such that the bulk solution is also isotropic.\\
In particular let's consider the following action:
\begin{equation}
\mathcal{S}\,=\,\int\,d^{d+1}x\,\sqrt{-g}\,\left[R\,-\,2\,\Lambda\,-\,\frac{1}{2}\,\sum_{I=1}^{\tilde{d}}\,(\partial \psi^I)^2\,-\,\frac{1}{4}\,F^2\right]
\end{equation}
where $\Lambda=-d(d-1)/(2L^2)$ is the d-dimensional cosmological constant.\\
The model admit a bulk solution which reads:
\begin{align}
&ds^2\,=\,-\,f(r)\,dt^2\,+\,\frac{dr^2}{f(r)}\,+\,r^2\,\delta^a_b\,dx^a\,dx^b\,,\qquad A\,=\,A_t(r)\,dt\,,\qquad \psi_I\,=\,\beta_I\,x^I\,,\nonumber\\
&f(r)\,=\,r^2\,-\,\frac{\beta^2}{2\,(d-2)}\,-\,\frac{m_0}{r^{d-2}}\,+\,\frac{(d-2)\,\mu^2}{2\,(d-1)}\,\frac{r_h^{2(d-2)}}{r^{2(d-2)}}\,,\nonumber\\
&A_t(r)\,=\,\mu\left(1\,-\,\frac{r_h^{d-2}}{r^{d-2}}\right)\,.
\end{align}
where for simplicity we have fixed $\beta_I=\beta$ in order to retain isotropy and the BH mass can be identified with the usual condition $f(r_h)=0$. These solutions have been first investigated in \cite{scalI} and in the anisotropic case in \cite{scalANI}.\\
We can again perturb this background in order to compute the conductivity and, after some easy manipulations (see \cite{AndradeWithers}), we are left with the following two equations:
\begin{align}
&r^{3-d}\,(r^{d-3}\,f\,a_x')'\,+\,\frac{\omega^2}{f}\,a_x\,=\,(d-2)^2\,\mu^2\,\frac{r_h^{2(d-2)}}{r^{2(d-1)}}\,a_x\,+\,i\,(d-2)\,\mu\,\frac{r_h^{d-2}}{r^{2(d-1)}}\,\phi\,,\nonumber\\
&r^{d-1}\,\left(r^{1-d}\,f\,\phi'\right)'\,+\,\frac{\omega^2}{f}\,\phi\,=\,-\,i\,(d-2)\,\beta^2\,\mu\,\frac{r_h^{d-2}}{r^2}\,a_x\,+\,\frac{\beta^2}{r^2}\,\phi\,.
\end{align}
where $a_x$ is the perturbation of the vector field $A_\mu$ and $\phi$ is the metric perturbation. The results are shown for various value of the mass $\beta$ in figures \ref{Axionssigma} taken from \cite{KimIncoherent}.\\ Also in this case, with appropriate methods, we can derive the zero frequency value of the conductivity, which reads:
\begin{equation}
\sigma_{DC}\,=\,r_h^{d-3}\,\left(1\,+\,(d-2)^2\,\frac{\mu^2}{\beta^2}\right)\,.
\label{axionsDC}
\end{equation}
where $d$ is the number of  boundary spacetime dimensions ($d=3$ in the previous examples). From this formula, comparing with the massive gravity scenario, it is clear that such scalars provide a mass for the graviton:
\begin{equation}
m^2\sim\,\beta^2
\end{equation}
which is proportional to their vev.\\
Despite at the beginning this model was supposed to be independent of the massive gravity results, and even incompatible \cite{AndradeWithers}
, it has been realized later on that there is a clear link between these two models (see next chapter) which are indeed ''equivalent''. These scalars represent nothing else than the additional degrees of freedom which breaking diffeomorphism invariance produces; in other words they are the St\"uckelberg fields themselves.
\\
Being the simplest model available to introduce momentum dissipation into the holographic scenario, it surely contains some shortcomings like the apparent insensitivity of the DC conductivity (at fixed chemical potential $\mu$) to the temperature T. This represents a big obstacle to try to classify the dual CFTs in terms of metals and insulators and to face the phenomenology of strongly correlated materials.\\
One way to overcome this issue is to complicate the scalar sector introducing new terms allowed by shift symmetry. Inspired by the DBI action  which arises in the tensionless limit of extended objects such as thin branes, in \cite{TaylorScalar} new terms proportional to the square root of the scalar kinetik term have been introduced:
\begin{equation}
\mathcal{L}\,\sim\,-c_{1/2}\,\sum_I\,\sqrt{(\partial\psi_I)^2}
\end{equation}
The bulk solution can still be retained to be homogeneous and isotropic but the transport and thermodynamical properties of the system get modified. In particular the DC conductivity now becomes:
\begin{equation}
\sigma_{DC}\,=\,r_h^{d-3}\,\left(1\,+\,\frac{(d-2)^2\,\mu^2}{\tilde{\beta}\,+\,\tilde{\alpha}/r_h}\right)
\end{equation}
where the $\tilde{\beta}$ parameter is due to the linear term for the scalar $(\partial \psi^I)^2$ as in \cite{AndradeWithers} while the new $\alpha$ term is entirely due to the introduction of the new square root terms $\sqrt{(\partial\psi_I)^2}$ and it is proportional indeed to its coefficient $\tilde{\alpha}\sim c_{1/2}$.\\
The first thing to notice is that now in $d=3$ the DC conductivity, thanks to the $\tilde{\alpha}$ term, becomes temperature dependent and can be suitable to phenomenological discussions. It was indeed claimed in \cite{TaylorScalar} that the square root term is exactly the one associated to a linear growth of the resistivity $\rho_{DC}=1/\sigma_{DC}$ at small temperatures:
\begin{equation}
\rho_{DC}\sim T
\end{equation}
which hints to what observed experimentally in the so-called \textit{strange metals} where the resistivity does not scale quadratically with the temperature $\rho\sim T^2$, as imposed by the \textit{Fermi Liquid theory}, but it does linearly.\\
Moreover, adding this new term, the model defined through the scalars seems to give a DC formula which, upon the correct identifications, corresponds exactly to the one extracted by dRGT theory \ref{DC1}. It has been showed in \cite{TaylorScalar} that the two theories are indeed equivalent at linear level, giving the same DC transport coefficients, but they still differ when finite frequency and momentum are considered, namely at non linear level. We will get back to the interpetation and the explanation of this issue.\\[0.3cm]
\textbf{DC quantities}\\[0.3cm]
The particulary simple setup provided by the massive gravity effective description allows a strong analytical control as well. The DC, \textit{i.e.} zero frequency, part of the conductivity can be indeed derived analytically using various techniques inspired by the \textit{membrane paradigm} \cite{IqbalMembrane}. It has been first derived in \cite{BlakeTongDC} and then refined with more efficient methods in \cite{DonosDC,AmorettiDC}. Since the DC conductivity will represent one of the main character of the next chapter, containing the original results of this thesis, we will enter in the details of both the available methods following their historical order. We will focus only on the electric conductivity although the same methods are suitable to extract the full thermo-electric conductivities and their scalings also in the presence of a magnetic background field and more complicated situations \cite{DonosDC,AmorettiDC,DonosStokesDC,BanksDC,DonosDCmore,Davisondissecting
,BlakeFluid1,LucasHydro,BlakeMagneto,GouterauxDC,BlakeB,AmorettiMagneto}.\\
Let's go back to the fluctuations equations for the dRGT model \ref{drgtEQ}; we can rewrite them as:
\begin{equation}
\left( \begin{array}{cc}
L_1 & 0  \\
0 & L_2  \end{array} \right)\,\left( \begin{array}{c}
a_x  \\
\tilde{g}_{rx} \end{array} \right)\,+\,\frac{\omega^2}{f}\,\left( \begin{array}{c}
a_x  \\
\tilde{g}_{rx} \end{array} \right)\,=\,\mathcal{M}\,\left( \begin{array}{c}
a_x  \\
\tilde{g}_{rx} \end{array} \right)
\end{equation}
where $L_{1,2}$ are linear differential operators and $\mathcal{M}$ is the ''mass matrix'':
\begin{equation}
\mathcal{M}\,=\,\left( \begin{array}{cc}
 \rho^2 \,r^2 &  m^2\,r^2/\,i\,\omega \\
 \rho\,i\,\omega& m^2  \end{array} \right)
\end{equation}
where $m^2$ is importantly a function of the radial coordinate r.\\
The main point is that we have $det(\mathcal{M})=0$ ! This means that even at finite charge density $\rho \neq 0$ there is a particular combinations of the fields which is not evolving radially in the limit of $\omega \rightarrow 0$.\\
The eigenmodes read:
\begin{align}
&\lambda_1\,=\,\left(1\,+\,\frac{\rho^2\,r^2}{m^2}\right)^{-1}\,\left[a_x\,-\,\frac{\rho\,r^2}{i\,\omega}\,\tilde{g}_{rx}\right]\,,\nonumber\\
&\lambda_2\,=\,\left(1\,+\,\frac{\rho^2\,r^2}{m^2}\right)^{-1}\,\left[\frac{\rho}{m^2}\,a_x\,+\,\frac{\tilde{g}_{rx}}{i\,\omega}\right]\,.
\end{align}
where $\lambda_1$ corresponds to the vanishing eigenvalue of the matrix $\mathcal{M}$.\\
Since $\lambda_1$ corresponds in the UV, at $r=0$, to the vector field perturbation $a_x$, we can therefore compute the conductivity as:
\begin{equation}
\sigma(\omega)\,=\,\frac{\lambda_1'}{i\,\omega\,\lambda_1}|_{UV}
\end{equation}
The equation for such a mode reads\footnote{Since the $\lambda$ fields do not diagonalize the derivative terms, the correspondent equations are not decoupled.}:
\begin{equation}
\left[f\,\left(1\,+\,\frac{\rho^2\,r^2}{m^2}\right)\,\lambda_1'\,-\,\frac{\rho\,f\,r^4}{m^2}\,\left(\frac{m^2}{r^2}\right)'\,\lambda_2\right]'\,+\,\frac{\omega^2}{f}\,\left(1\,+\,\frac{\rho^2\,r^2}{m^2}\right)\,\lambda_1\,=\,0\,
\end{equation}
which makes evident that in the DC limit $\omega \rightarrow 0$ there is a conserved quantity $\Pi$:
\begin{equation}
\Pi\,=\,f\,\left(1\,+\,\frac{\rho^2\,r^2}{m^2}\right)\,\lambda_1'\,-\,\frac{\rho\,f\,r^4}{m^2}\,\left(\frac{m^2}{r^2}\right)'\,\lambda_2
\label{Pidef}
\end{equation}
We can therefore define a DC membrane conductivity associated to each radial slice r:
\begin{equation}
\sigma(r)\,=\,\lim_{\omega\rightarrow 0}\,\frac{\Pi}{i\,\omega\,\lambda_1}|_r
\end{equation}
At $r=0$ this expression coincides with the electric conductivity and because this quantity does not evolve radially\footnote{There are some caveats to actually prove it in details that such statement is true; see\cite{BlakeTongDC}.} we can compute it at whichever radial position and in particular at the horizon $r=r_h$. This can be easily achieved remembering that both fields obey ingoing boundary conditions $a_x\sim\tilde{g}_{rx}\sim f(r)^{-i \omega/4\pi T}$ at the horizon. With this in mind, we find out that, at the horizon position, the $\lambda_2$ term in \ref{Pidef} vanishes while the $\lambda_1'$ term survives. All in all we end up with an analytic expression for the DC conductivity:
\begin{equation}
\sigma_{DC}\,=\,1\,+\,\frac{\rho^2\,r_h^2}{m^2(r_h)}
\end{equation}
which has already introduced before in \ref{DC1}.\\
Whenever the second term is the dominant one we expects the standard Drude form to appear; in contrast, in the other case we have an incoherent metal.\\
We can now generalize such a model and consider a more general setup where we can still compute with analytical methods the DC conductivity following the work of \cite{DonosDC}. We now consider the more generic action given by:
\begin{equation}
\mathcal{S}\,=\,\int\,d^4x\,\sqrt{-g}\,\left\{R\,-\,\frac{1}{2}\,\left[(\partial \phi)^2\,+\,\Phi_1(\phi)\,(\partial \chi_1)^2\,+\,\Phi_2(\phi)\,(\partial \chi_2)^2\right]\,-\,V(\phi)\,-\,\frac{Z(\phi)}{4}\,F^2\right\}
\end{equation}
where $\chi_I$ are the scalars field we were considering before (we allow them to be different) and $\phi$ is an additional scalar, \textit{i.e.} the dilaton. We assume that the model admits a unit radius AdS$_4$ solution with $\phi=0$ (and $V(0)=-6$ along with $Z(0)=1$).\\
The solutions that we shall consider all lie within the ansatz:
\begin{align}
&ds^2\,=\,-\,U\,dt^2\,+\,U^{-1}\,dr^2\,+\,e^{2\,V_1}\,dx^2\,+\,e^{2\,V_2}\,dy^2\,,\nonumber\\
&A\,=\,a\,dt\,,\qquad \chi_1\,=\,k_1\,x\,,\qquad\chi_2\,=\,k_2\,y\,.
\end{align}
where $U,V_I,a$ and $\phi$ are only functions of r. In general the solutions are anisotropic, $V_1\neq V_2$ but we can enforce isotropy choosing $k_1=k_2$ and $\Phi_1=\Phi_2$.\\
We will furthermore assume that there exists an event horizon at $r=r_h$ where the functions have the following behaviour:
\begin{align}
&U\,\sim\,4\,\pi\,T\,(r\,-\,r_h)\,+\,\dots \qquad V_I\,=\,V_{h}\,+\,\dots\nonumber\\
&a\,\sim\,a_h\,(r\,-\,r_h)\,+\dots\qquad\phi\,\sim\,\phi_h\,+\,\dots
\end{align}
where T is the temperature of the BH background.\\
On the other side at the UV AdS$_4$ position $r\rightarrow\infty$ we assume that:
\begin{align}
&U\,\sim\,r^2\,+\,\dots \qquad e^{V_I}\,=\,r^2\,+\,\dots\nonumber\\
&a\,\sim\,\mu\,-\,\frac{\rho}{r}\,+\dots\qquad\phi\,\sim\,\lambda\,r^{\Delta-3}\,+\,\dots
\end{align}
with $\Delta<3$.\\
The current density $J^a=(J^t,J^x,J^y)$ in the dual field theory has the form:
\begin{equation}
J^a\,=\,\sqrt{-g}\,Z(\phi)\,F^{ar}
\end{equation}
evaluated at the boundary $r\rightarrow\infty$.\\
The left non trivial Maxwell equation within our ansatz reads:
\begin{equation}
\text{maxwell equation: }\,\,\sqrt{-g}\,\nabla_r\left(\sqrt{-g}\,Z(\phi)\,F^{r t}\right)\,=\,0\,.
\end{equation}
which looks like indeed as the conservation of the charge density:
\begin{equation}
\rho\,=\,e^{V_1\,+\,V_2}\,Z(\phi)\,a'
\end{equation}
In order to compute the electric conductivity we have to introduce an external electric field through the small perturbations:
\begin{align}
&a_x\,=\,-E\,t\,+\,\delta a_x(r)\,,\nonumber\\
&g_{tx}\,=\,\delta g_{tx}(r)\,,\nonumber\\
&g_{rx}\,=\,e^{2\,V_1}\,\delta g_{rx}(r)\,,\nonumber\\
&\chi_1\,=\,k_1\,x\,+\,\delta \chi_1(r)\,.
\end{align}
Again the only non trivial component of the Maxwell equation at linear order in these fluctuations is $\nabla_r\left(\sqrt{-g}\,Z(\phi)\,F^{r x}\right)$; therefore we can deduce that the following quantity:
\begin{equation}
J\,=\,-\,e^{V_2\,-\,V_1}\,Z(\phi)\,U\,\delta a_x'\,-\,\rho\,e^{-\,2\,V_1}\,\delta g_{tx}
\end{equation}
is radially conserved and it corresponds at the boundary with the electric current $J_x$ in response to the electric field E.\\
We then consider the linearized Einstein equations. There is just one of them relevant for the present discussion\footnote{Nevertheless of course the full system of equations is consistent.} and it can be algebrically solved giving the constraint:
\begin{equation}
\delta g_{rx}\,=\,\frac{E\,\rho\,e^{-\,V_1,-\,V_2}}{k_1^2\,\Phi_1(\phi)\,U}\,+\,\frac{\delta \chi_1'}{k_1}
\end{equation}
Under some assumptions (see \cite{DonosDC}) we can derive the regular behaviour of the various fields at the horizon getting:
\begin{align}
&\delta a_x\,\sim\,-\,\frac{E}{4\,\pi\,T}\,\ln(r\,-\,r_h)\,+\,\dots\nonumber\\
&\delta g_{tx}\,\sim\,-\,\frac{E\,\rho\,e^{V_1\,-\,V_2}}{k_1^2\,\Phi_1(\phi)}\,+\,\dots
\end{align}
At this stage we can already define the electric conductivity as $\sigma=J/E$ and computing the following quantity at the horizon we are left with the generic DC formula:
\begin{equation}
\sigma_{DC}\,=\,\left[\frac{Z(\phi)\,s}{4\,\pi\,e^{2\,V_1}}\,+\,\frac{4\,\pi\,\rho^2}{k_1^2\,\Phi_1(\phi)\,s}\right]_{r=r_h}\,.
\end{equation}
where $s=4 \pi e^{V_1+V_2}$ is the entropy density of the unperturbed black hole.\\
Note that the scalar model of \cite{AndradeWithers} we consider before can be re-obtained fixing:
\begin{equation}
\phi\,=\,0\,,\qquad \Phi_i\,=\,1\,,\qquad Z\,=\,1\,,\qquad V\,=\,-6\,.
\end{equation}
along with $k_1=k_2=k$ obtaing the well known:
\begin{equation}
\sigma\,=\,1\,+\,\frac{\mu^2}{k^2}
\end{equation}\\
Nowadays, since the first paper \cite{Veghoriginal} in 2013, there has been a huge amount of activity, efforts and results in the present field which of course we are not able to cover in this small room. We refer to the bilbiography for more details.\\
We will enter more advanced and hot topics in the next chapter where we will summarize the original results presented in the published papers by the author and collaborators.
\cleardoublepage
\part{Results}
\label{results}
\begin{figure}
\includegraphics[scale=0.43]{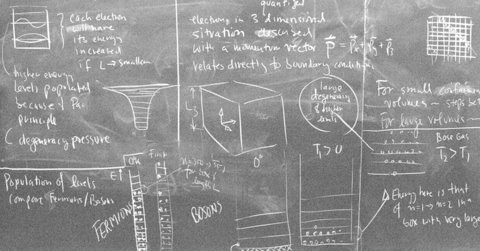}
\end{figure}
\epigraph{However beautiful the strategy, you should occasionally look at the results.}{\textit{Winston Churchill}}
In this section we summarize the major original results generated in the papers constituiting the main body of this Ph.D thesis. We will be rather minimal and schematic. For details on the computations and the procedures we refer directly to the papers themselves.\\[0.3cm]
After the realization that Massive Gravity could represent an effective tool to introduce momentum dissipation in the framework of holography \cite{Veghoriginal} lots of efforts and works have been done in the direction of understanding and improving such a tool under various aspects:
\begin{itemize}
\item Understanding the tool \cite{DavisonHydro,MGlattice,HPtrans,DavisonGouteraux,Davisondissecting,BlakeFluid1,Banks1,Panfluid};
\item Enlarging the tool with additional ingredients (anisotropies, dilaton) or more generic models \cite{DavZaanen,Gouteraux1,DCan,KiritsisRen};
\item Exploiting the Stuckelberg mechanism to simplify the model to GR + scalar sector \cite{AndradeWithers,TaylorScalar,AndradeLif};
\item Identifying efficient methods to extract the thermoelectric transport coefficients both numerically and analytically \cite{BlakeTongDC,genova1,genova2,KimIncoherent,ChengDC,LucasMemory,AmorettiMagneto
,BlakeMagneto,KimMagneto,ZhouDC,DonosDC,AndradeDrude};
\item Studying the stability and the consistency requirements for such a MG theories living in AdS spacetime \cite{LasmaAndrei1,CaiMG,Lasma2};
\item Applying MG to various phenomenological applications \cite{MGSC1,AndradeSC,KimSC1,Cai2,KiritsisPena,KiritsisLi,ZengEE,AndradeKrikun,HuMG,KimHomes};
\item Establishing and studying the possible existence of universal bounds on physical quantities in these frameworks \cite{genovabounds,Grozdanov1,Hartnollincoherent,Hartnolleta,nicketa,Ikedabound,Gouterauxbound,BBound,Liubound,Lingshear}.
\end{itemize}
In the following we will review the major contributions of our work to the subject. For the sake of simplicity, we will skip all the technical parts and the details of the computations. The interesting reader can find them in the original published papers:
\begin{enumerate}
\item \href{https://arxiv.org/abs/1411.1003}{''Electron-Phonon Interactions, Metal-Insulator Transitions, and Holographic Massive Gravity''}
\item \href{https://arxiv.org/abs/1504.05561}{''Phases of holographic superconductors with broken translational symmetry''}
\item \href{https://arxiv.org/abs/1510.06363}{	
''Under The Dome: Doped holographic superconductors with broken translational symmetry''}
\item \href{https://arxiv.org/abs/1510.09089}{''Solid Holography and Massive Gravity''}
\item \href{https://arxiv.org/abs/1601.03384}{''Viscosity bound violation in holographic solids and the viscoelastic response''}
\item \href{https://arxiv.org/abs/1601.07897}{''On holographic disorder-driven metal-insulator transitions''}
\end{enumerate}
\section{Momentum dissipation \& Massive Gravity: towards a more generic framework}
\begin{figure}
\centering
\includegraphics[width=7.3cm]{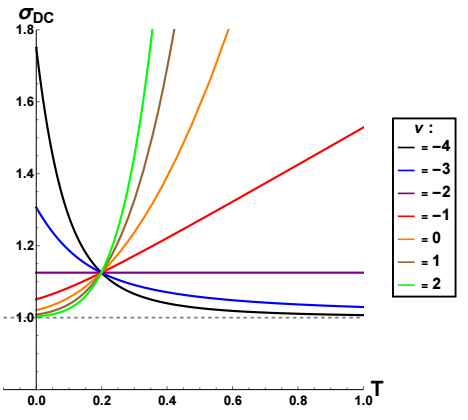}%
\qquad
\includegraphics[width=6.5cm]{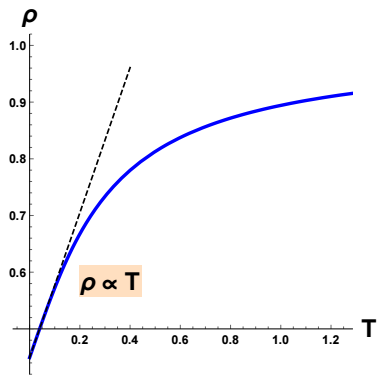}
\caption{DC conductivity for a generic function $M^2(r)\sim r^\nu$. \textbf{Left: }DC conductivity for various values of $\nu$ showing the transition between a metallic to an insulating phase. \textbf{Right:} Linear in T resistivity (at low T) for the choice $\nu=-3$.}
\label{DCDC}
\end{figure}

\begin{figure}
\centering
\includegraphics[width=14.5cm]{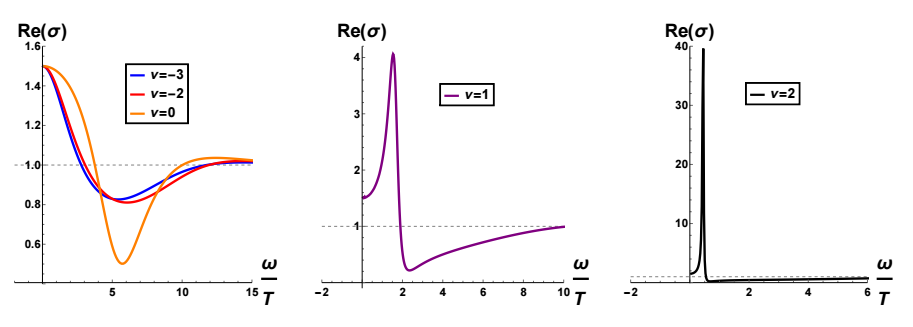}
\caption{Real part of the optical conductivity for different choices of  $\nu$.}
\label{ACAC}
\end{figure}

The idea of exploring more generic Massive Gravity models in the context of the AdS/CMT applications has been initiated in \cite{BaggioliPRL} and done later in a more sistematic way in \cite{BaggioliSolid}. We are going to present the papers with a historical perspective of how ideas came out at that time.\\[0.3cm]
\textbf{Towards more generic MG models and their dual}\\[0.3cm]
Massive gravity model can be thought as a family of infra-red deformations of General Relativity described generically by the introduction of a new ''mass term'' of the form:
\begin{equation}
\sim\,m^2\,V\left(P_{\mu\nu}\,g^{\mu\nu}\right)
\label{GenMG}
\end{equation}
where $P_{\mu\nu}$ is an external fixed metric of convenience.\\
In order to break just the translational invariance of the dual CFT the projector P has to be choosen non null only in the spatial coordinates $P_{ij}=\delta_{ij}$ providing a Lorentz violating version of massive gravity (LVMG).\\
The new term \eqref{GenMG} introduces extra degrees of freedom into the spectrum of the theory which it turns out convenient to explicitate via covariantizing the theory and restoring the original gauge symmetry. The minimal amount of fields which has to be added is a collection of scalar fields $\phi^I$ transforming under an internal Euclidean group of translations and rotations in field-space.\\
All in all, the most generic action\footnote{This statement is not totally true. We will see soon that an additional term can be introduced \cite{BaggioliSolid} and can assume an important role for various aspects.} for the MG theory (in $3+1$ dimensions) can be written down as:
\begin{equation}
\mathcal{S}_{MG}\,=\,M_P^2\,\int\,d^4x,\,\sqrt{-g}\,\left[\,\frac{R}{2}\,+\,\frac{3}{L^2}\,-\,m^2\,V\left(X\right)\right]
\label{PRLaction}
\end{equation}
where $X=\frac{1}{2}\partial_\mu\phi^I\partial^\mu \phi^I$.\\
This action admits a simple solution where the scalar field get a VEV linear in the spatial coordinates:
\begin{equation}
\phi^I\,=\,\alpha\,\delta^I_i\,x^i
\end{equation}
With this choice the original $P_{\mu\nu}$ projects just on the spatial coordinates and assumes the form $\sim \partial_\mu \phi^I \partial_\nu \phi^I$.\\
The crucial point is that most of the literature to date unnecessarily restrict to a very narrow families of potential V. In particular most of the work has been focused on dRGT MG \cite{Veghoriginal} and on the choices $V(X)\sim X,\sqrt{X}$ \cite{AndradeWithers, TaylorScalar}. Nevertheless, in the Lorentz violating case the set of consistent choices is definitely bigger and unexplored.\\
We therefore consider action \eqref{PRLaction} and we study the 
consistency of the model and the phenomenological implications which a completely generic potential $V(X)$ can produce in the transport properties of the dual field theory.\\
In order to avoid ghosty excitations and other pathological instabilities the potential V has to satisfy some requirements which can be derived performing an analysis of the fluctuations in the decoupling limit. Perturbing the Goldstone fields $\phi^I=\bar{\phi}^I+\delta \phi^I$ and expanding the corresponding action up to 2nd order we get:
\begin{equation}
V(\bar{X})\,\partial_\mu \delta \phi^i \partial^\mu \delta \phi^i\,+\,\bar{X}\,V''(\bar{X})\,\left(\partial_i \delta \phi^i\right)^2
\end{equation}
Absence of ghosts, then, leads to monotonic potentials:
\begin{equation}
V'(\bar{X})\,>\,0
\end{equation}
The local (sound) speed of longitudinal phonons is:
\begin{equation}
c_S^2\,=\,1\,+\,\frac{\bar{X}\,V''(\bar{X})}{V'(\bar{X})}
\end{equation}
which is safest to require it to be everywhere positive to
guarantee the absence of gradient instabilities\footnote{This represents a local speed of sound in the bulk which does not coincide, at least a priori, with the speed of sound of any CFT excitations.}.\\
No further constraints arise from the vector and tensor sectors nor at nonlinear level in the scalars, which is not surprising since a substantial advantage of the Lorentz non-invariant mass terms is that they can be free from the Boulware-Deser ghost. The requirement of having an asymptotical AdS spacetime implies a further constraint on the potential which has to vanish at the boundary $u=0$. This gets translated into the condition:
\begin{equation}
\lim_{X\rightarrow 0}\,V(X)\,=\,0
\end{equation}
All in all, an economic and safe way to satisfy all constraints is to assume that $V(X)$ is a positive definite polynomial function of X. We will therefore consider the benchmark example:
\begin{equation}
V(X)\,=\,X\,+\,\beta\,X^N\,,\qquad \beta\,>0\,,\,\,\,\,\,\,\,\,N\,>\,1
\label{BenchmarkPRL}
\end{equation}
which is rich enough to give rise to new and interesting phenomenological results. It is somehow very generic because it can be thought as the Taylor expansion of any function $V(X)$ satisfying at non linear level the requirements above.\\
In any asymptotically AdS solution $\bar{X}\,=\,u^2\,\alpha^2$ and close to the AdS boundary ($u=0$) the Goldstone gradient $\bar{X}$ vanishes. This implies that the mass term for metric modes is
also vanishing. So this is a weaker form of massive gravity than is usually discussed in cosmology - the Compton wavelength is at most of order the curvature radius of the spacetime. In the CFT interpretation, the stress tensor $T_{\mu\nu}$ does not develop an anomalous dimension. Still this is enough to lead to momentum relaxation in the CFT.\\
Additionally, noticing that X is getting smaller at the boundary, it is clear that at large temperature T (u small) the physics will be dominated by the smallest power of X appearing in the potential V. On the contrary at small temperature T (u large) the higher powers, \textit{i.e.} non linear new corrections, get important and can abruptly affect the physics of the system.\\
In CFT language, the scalars $\phi^I$ correspond multiplet of operators $\mathcal{O}^I$ with
internal shift symmetries and which are somehow related with phonons and impurities. A consistent interpretation seems to be that the strength $\alpha$ of the linear vevs $\langle \mathcal{O}^I \rangle=\alpha \delta^I_i x^i$ is the density of homogeneously-distributed impurities. The fluctuations $\delta \mathcal{O}^I$ around this distribution are CFT operators that create ''phonon'' excitations\footnote{This point is quite subtle because from the point of view of the CFT side the breaking of translational symmetry is definitely explicit and not spontaneous. There is anyway a limit in which these excitations can be thought as ''pseudo-goldstone'' \'a la pions. We avoid a deep discussion on this point at this stage.}.\\
In order to study the transport properties of the dual CFT, we need to add one ingredient, the
charge carriers. So we assume that the CFT also contains a conserved current operator $J_\mu$. This is implemented in the gravity dual by adding to the model a Maxwell field,
\begin{equation}
\mathcal{S}\,=\,\mathcal{S}_{MG}\,-\,\frac{M_p^2}{4}\,\int\,d^4x\,\sqrt{-g}\,F_{\mu\nu}F^{\mu\nu}
\end{equation}
The dual of the CFT ground state at finite charge density $\rho$, temperature T and impurity strength $\alpha$ is a planar black brane (BB) with:
\begin{align}
&ds^2\,=\,\frac{L^2}{u^2}\,\left(\frac{du^2}{f(u)^2}\,-\,f(u)\,dt^2\,+\,dx^2\,+\,dy^2\right)\nonumber\\
&f(u)\,=\,u^3\,\int_{u}^{u_h}\,dy\,\left[\frac{3}{y^4}\,-\,\frac{\rho^2}{2\,L^2}\,-\,\frac{m^2\,L^2}{y^4}\,V\left(\frac{\alpha^2\,y^2}{L^2}\right)\,\right]\nonumber\\
&A_t(u)\,=\,\mu\,-\,\rho\,u\,,\qquad \phi^I\,=\,\alpha\,\delta^I_i\,x^i\,.
\end{align}
where $u_h$ stands for the BB horizon.\\
The regularity condition for the gauge field implies $\mu=\rho\, u_h$ and the correspondent Hawking temperature can be found to be $T=-f'(u_h)/(4\pi)$.\\
We want now to switch on the vector perturbations around the BB above by setting:
\begin{equation}
A_\mu\,=\,\bar{A}_\mu\,+\,a_\mu\,,\qquad g_{\mu\nu}\,=\,\bar{g}_{\mu\nu}\,+\,h_{\mu\nu}\,,\qquad \phi^I\,=\,\bar{\phi}^I\,+\,\delta\,\phi^I
\end{equation}
with bars denoting the background solutions.\\
\begin{figure}
\centering
\includegraphics[width=6.3cm]{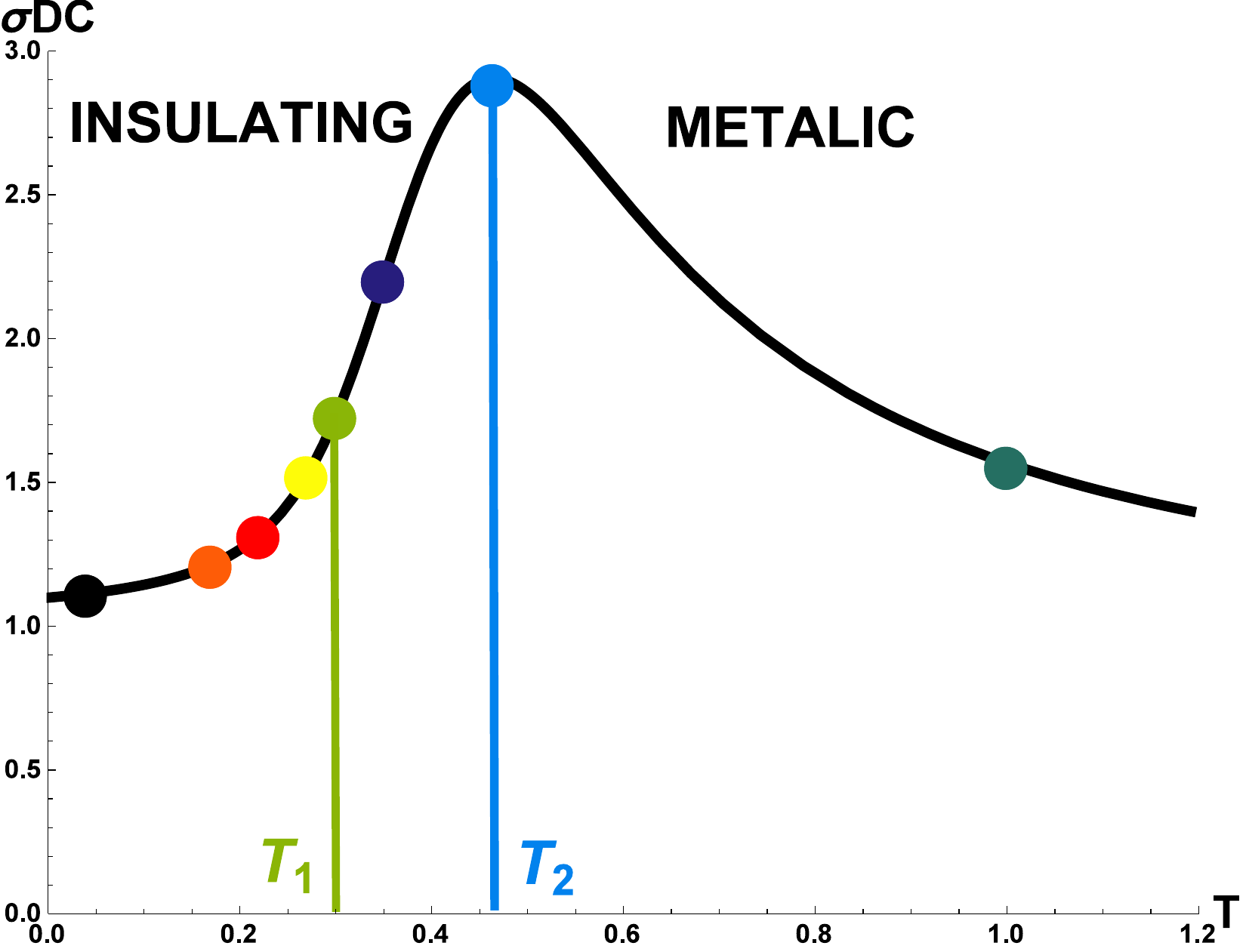}%
\qquad
\includegraphics[width=7.5cm]{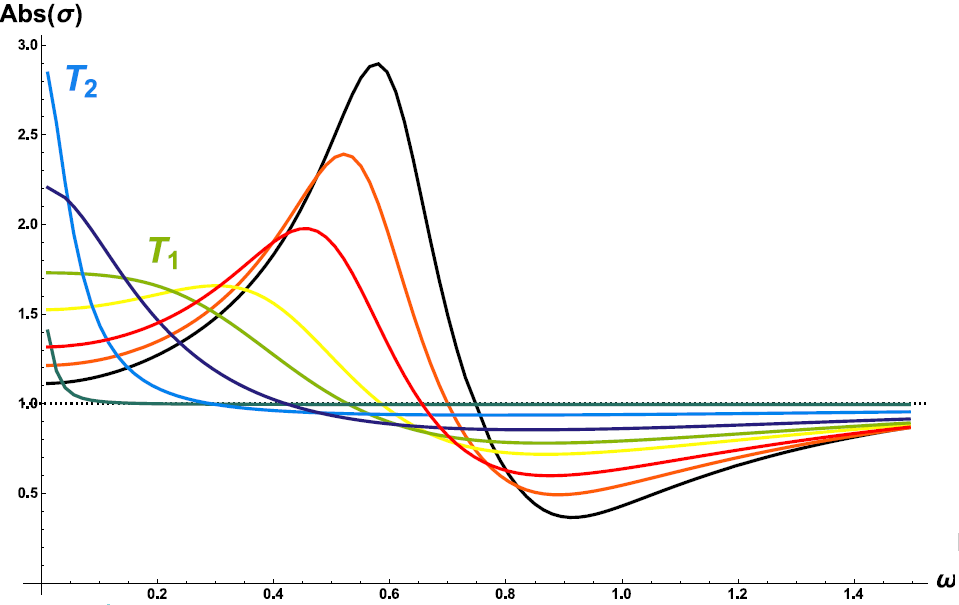}
\caption{Electric conductivity. \textbf{Left: }DC conductivity in function of temperature and MIT crossover. \textbf{Right: }Optical conductivity $\sigma(\omega)$ at various temperatures $T=1,\,0.46 (T_2),\, 0.35,\,0.3 (T_1),\,0.27,\,0.22,\,0.17,\,0.04$ and \textit{Polaron} formation.}
\label{DCopticalPRL}
\end{figure}
It is convenient to work with gauge invariant variables defined by:
\begin{equation}
T_i\,\equiv\,u^2\,h_{ti}\,-\,\frac{\partial_t \phi^i}{\alpha}\,,\qquad U_i\,=\,f(u)\,\left[h_{ui}\,-\,\frac{\partial_u \phi^i}{\alpha\,u^2}\right]\,,\qquad B_i\,=\,b_i\,-\,\frac{\phi_i}{\alpha}\,.
\end{equation}
with $h_{ij}\,=\,\frac{1}{u^2}\,\partial_{(i}b_{j)}$.\\
With this choice of variables the linearized equations can be written as:
\begin{align}
&\partial_u\,\left(f\,\partial_u\,a_i\right)\,+\,\left[\frac{\omega^2}{f}\,-\,k^2\,-2\,u^2\,\rho^2\right]\,a_i\,-\,\frac{i\,\rho\,u^2\,(2\,\bar{m}^2\,+\,k^2)}{\omega}\,U_i\,+\,\frac{i\,f\,\rho\,k^2}{\omega}\,\partial_u\,B_i\,=\,0\,,\nonumber\\
&\frac{1}{u^2}\,\partial_u\,\left[\frac{f\,u^2}{\bar{m}^2}\,\partial_u(\bar{m}^2\,U_i)\right]\,+\,\left[\frac{\omega^2}{f}\,-\,k^2\,-\,2\,\bar{m}^2\right]\,U_i\,+\,2\,i\,\rho\,\omega\,a_i\,-\,\frac{f'\,k^2}{u^2}\,B_i=\,0\nonumber\\
&k\,\left\{u^2\,\partial_u\,\left(\frac{f}{u^2}\partial_u B_i\right)\,+\,\left[\frac{\omega^2}{f}\,-\,k^2\,-\,2\,\bar{m}^2\right]\,B_i\,+\,2\,\frac{\bar{m}'}{\bar{m}}\,U_i\right\}\,=\,0\,.
\end{align}
where we introduced $\bar m^2(u) = \alpha^2 m^2 V'(\alpha^2 u^2)$, namely the radial dependent effective grativon mass.
\begin{figure}
\centering
\includegraphics[width=7.6cm]{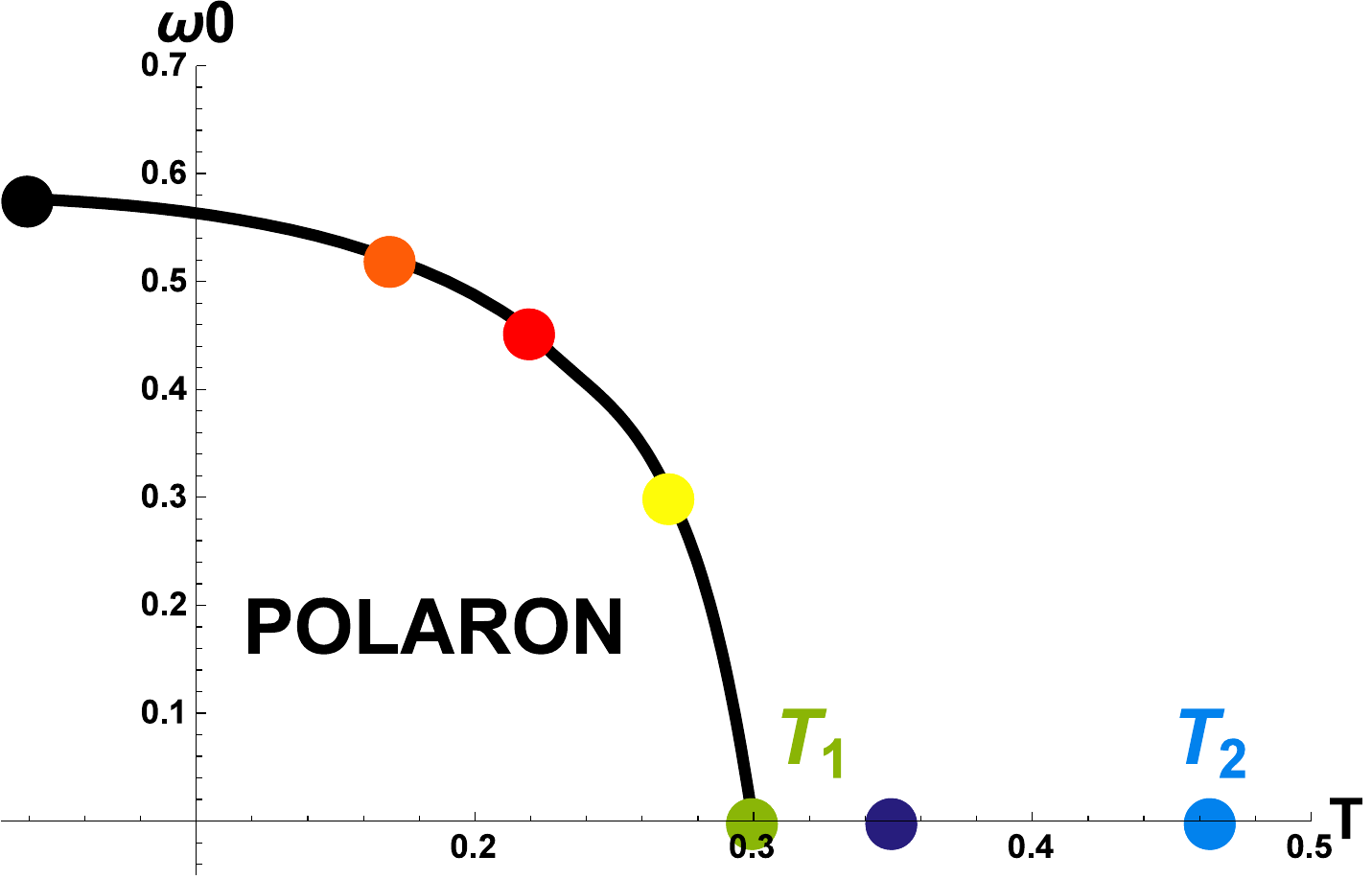}%
\qquad
\includegraphics[width=6.5cm]{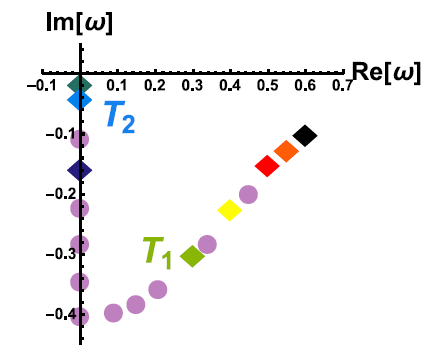}
\caption{Excitations of the system. \textbf{Left: }Polaron formation. \textbf{Right: }QNM spectrum of the model. At large $T$, the QNM separates from the real axis with decreasing $T$, until it collides with the next QNM (near $T_1$) and forms a pair of conjugated poles with positive and negative real parts -- the polaron particle/anti-particle poles. (Similar QNM collisions have been observed e.g. in \cite{DavisonGouteraux}. In our case, it is crucial that after collision the QNM approaches the real axis.)}
\label{polaronQNMPRL}
\end{figure}\\
From the numerical integration of these equations we can extract all the transport properties of the dual CFT and in particular the DC and AC electric conductivity and the correspondent QNM spectrum using the dictionary we discussed in the previous sections.\\
In the homogeneous limit $k\rightarrow 0$ we can compute the DC, $\omega=0$, value of the electric conductivity in terms of horizon data analytically and we get:
\begin{equation}
\sigma_{DC}\,=\,1\,+\,\frac{\rho^2\,u_h^2}{\bar{m}^2(u_h)}\,=\,1\,+\,\frac{\rho^2\,u_h^2}{\alpha^2\,m^2\,V'(\alpha^2\,u_h^2)}
\label{DCformulaPRL}
\end{equation}
which generalizes the previous results in the literature.\\
The previously considered models \cite{Veghoriginal,AndradeWithers, TaylorScalar} allowed just for metallic behaviour of the type $d\sigma_{DC}/dT<0$. Within this class of more generical potential we can not only provide a dual for insulating states\footnote{Strictly speaking the proper definition for an insulator would be $\sigma(T=0)\approx 0$. This is not the case in this kind of models because of the first unitary and temperature independent piece of the DC conductivity. We will therefore assume a milder definition $d\sigma_{DC}/dT>0$. See further discussions in the following.} of the type $d\sigma_{DC}/dT>0$ but also transitions between such states and metallic phases (MIT) as in figure \ref{DCopticalPRL}. From the analytical formula \ref{DCformulaPRL} we can extract the exact condition ($d\sigma/dT=0$) for such a crossover to happen which takes the form:
\begin{equation}
\bar{X}\,V''(\bar{X})\,=\,V'(\bar{X})
\end{equation}
\begin{figure}
\centering
\includegraphics[width=8cm]{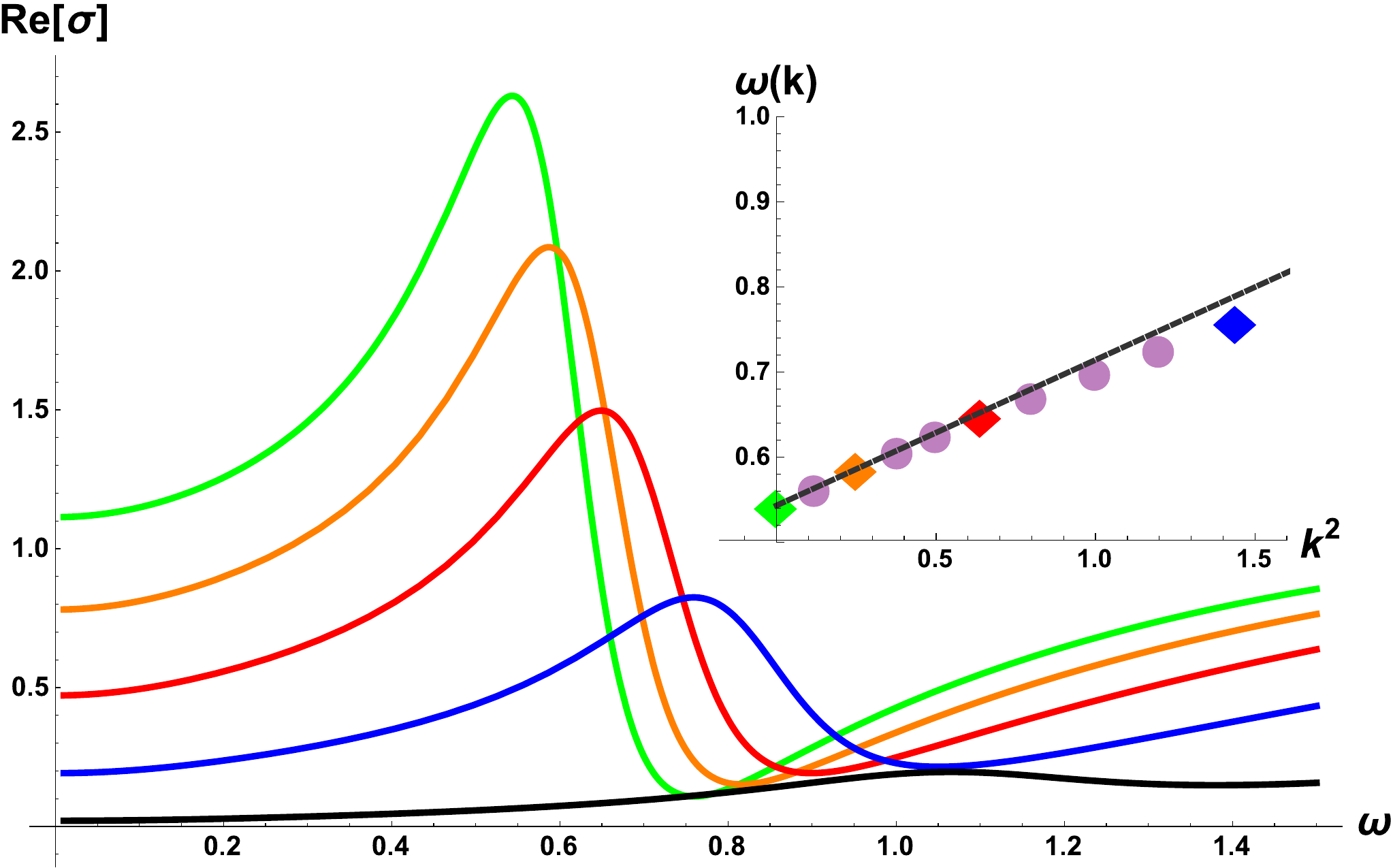}
\caption{Motion of the polaron peak with wavenumber. $T=0.04$, $k=0,\,0.5,\,0.8,1.2,\,2$. Inset: the extracted dispersion relation.}
\label{kPRL}
\end{figure}
It is easy to show that for the benchmark model \eqref{BenchmarkPRL} this condition translates in the definition of a critical temperature $T_2\sim (N-2)^\gamma$ showing that one needs an high exponent $N>2$ to have an MIT. Models with $N<2$ tend to give metals and incoherent metals. It is anyway clear and important that generic massive gravity models seem to be able to reproduce much more than the single metallic behaviour and can encorporate also strongly coupled insulating systems.\\
Along with the depletion of the DC conductivity at low temperatures further features appear in the optical part of the conductivity (see fig.\ref{DCopticalPRL}) with the appearance of a localized excitation in the mid-infrared region. This suggests the formation of a localized and propagating bound state which one is tempted to identify as a phonon-electron state known as \textit{polaron}. Such a bound state is created at a temperature $T_1$ which is generically different from the MIT temperature $T_2$ and always smaller. It is not clear therefore if there is a close correlation between the two effects but they turn out to be certainly related. Moreover, the formation of this ''peak'' is favoured by the amount of non linearities (which can be dialed increasing the \textit{impurity strength} $\alpha$ or the non-linear power in the potential N) and by lowering the temperature T. Note that despite the formation of the \textit{polaron} appears of the mean-field type (see fig.\ref{polaronQNMPRL}), usually related to 2nd order continous phase transitions, here we are not in presence of any spontaneous symmetry breaking mechanism. The phenomenon is more similar to a dynamical crossover which is generated by the non linearities of the massive gravity sector.\\
The features of the system can be efficiently collected into its QNM spectrum (see fig.\ref{polaronQNMPRL}). The model undergoes a transition between an overdamped regime with a clear and separated drude peak, appearing as a pole on the lower imaginary axes $\omega=-i \,\Gamma$ to an underdamped regime with a pole with both real and imaginary part $\omega=-i\,\Gamma+ \nu$. Eventually such a pole, lowering the temperature, can build up a bigger and bigger real part and approach the real axes, providing the peaked feature in the optical conductivity.\\
The emerging bound state represents a real propagating degree of freedom, whose real part of the frequency $\nu$ stands for its effective energy/mass and the imaginary part as its decay width. One can follow such a pole and picture its dispersion relation (see fig.\ref{kPRL}) which turns out to be $\omega\sim k^2$ at low momentum as expected for a non lorentz invariant propagating mode.\\
In summary, we have seen that a simple nonlinear extension of holographic massive gravity
captures interesting features of correlated materials such as polaron-localization and a (phonon-electron) interaction-driven Metal-Insulator transition (MIT). In other words, generalizing the massive gravity dual through a generic potential, introduces an effective graviton mass whose radial dependence can be arbitrary, within the consistency regime of the theory. Such a dependence gets translated in the field theory picture as the temperature dependence of the relaxation time such that the correspondent phenomenology becomes very rich and accounts also for insulating states and MIT transitions.\\[0.3cm]
\textbf{Phases of holographic massive gravity}\\[0.3cm]
In \cite{BaggioliSolid} we performed a systematic analysis of holographic massive gravity theories (HMG) which might store interesting condensed matter content. As already underlined, the set of consistent and healthy MG theories is much larger than the famous dRGT case firstly considered in the holographic literature. Many results are indeed not generic and this restricted view could somehow lead to misguided implications for the CM dual.\\
Realizing the existence of a broader class of consistent EFTs provides already an important distinction between \textit{solid-fluid} type massive gravity theories which can be separated by:
\begin{enumerate}
\item The nature of the unbroken spacetime symmetries.
\item The elastic response of the dual system.
\end{enumerate}
The first point has origins in the framework of EFT for fluids and solids which in modern language are written down through a set of phonon scalars $\phi^I$ (in $2+1$ dimensions, $I=1,2$) that enjoy internal shift and rotation symmetries for homogeneous and isotropic materials \cite{LVEFT1,LVEFT2,LVEFT3,LVEFT4}. The internal symmetry group for solids is the two-dimensional Euclidean group of translations and rotations. For fluids, the internal group is much bigger and includes also volume preserving diffeomorphisms (VPDiffs). The scalars acquire an expectation value 
$\langle\phi^I\rangle=\delta^I_i\,x^i$ and break the product of the (space transformations) $\times$ (internal transformations) to the diagonal subgroup. For fluids, the preserved symmetry includes a volume preserving diagonal subgroup.\\
The effective Lagrangian at the lowest order in derivatives in the two cases can be written as:
\begin{equation}
\mathcal{L}^{solids}\,=\,V_s(X,Z)\,,\qquad \mathcal{L}^{fluids}\,=\,V_f(Z)
\end{equation}
where $X=Tr[\mathcal{I}^{IJ}]$, $Z=Det[\mathcal{I}^{IJ}]$ and $\mathcal{I}^{IJ}=\partial_\mu \phi^I\partial^\mu \phi^J$.\\
The functions $V_{s,f}$ encode the linear and nonlinear properties of the solid and fluid, and they are free functions subject to mild consistency constraints. It is easy to realize that gauging these theories leads to graviton mass terms around the solution with non trivial vevs for the Goldstones $\phi^I$. The simplest way to see this is to replace $\eta_{\mu\nu}$ with $g_{\mu\nu}$ and to go to the unitary gauge where the scalar fields are fixed to be equal to their background configuration. The above solid/fluid Lagrangians then become nonlinear potential terms for the metric
\begin{equation}
V_s\left(Tr \left[g^{ij}\right],Det\left[ g^{ij}\right]\right)\,,\qquad V_f\left(Det \left[g^{ij}\right]\right)
\label{Pot}
\end{equation}
where $g^{ij}$ denotes the spatial part of the inverse metric.\\
The form of this action is dictated by requiring it to be invariant under certain subset
of the diffeomorphisms, that do not include the spatial diffeomorphisms $x^i=\tilde{x}^i(x^j,t)$.
The preserved diffeomorphisms are the ones enjoyed by the potential terms in \eqref{Pot}. Both
for solid and fluid MGs, these include the time-reparametrizations $t\rightarrow f(t)$ plus global
translations and rotations that force the potential not to depend explicitly on $x^i$ and to
contract the spatial indices with Kronecker delta $\delta_{ij}$ .For fluid MG the potential is also
invariant under the spatial VPDiffs, that forces it to be a function of $Det \left(g^{ij}\right)$ only. Importantly, as we shall see below, the VPDiff symmmetry protects the vanishing of the physical
mass parameter of the metric tensor modes.\\
The main idea and novel contribution of \cite{BaggioliSolid} is to consider fully generic HMG theories and to study in details the implications of such a separation in the dual CM picture. Such a description will lead to the definition of a new physical quantity encoded in the Green function of the tensor mode, which we will identify as the \textit{elasticity} of the system.\\
In the flat space language of \cite{phasesMG} an isotropic and homgeneous LV massive gravity theory can be realized as:
\begin{equation}
m_0^2\,h_00^2\,+\,2\,m_1^2\,h_{0i}^2\,-\,m_2^2\,h_{ij}^2\,+\,m_3^2\,h_{ii}^2\,-\,2\,m_4\,h_{00}\,h_{ii}
\end{equation}
and the distinction between solids and fluids, following from symmetry arguments, boils down to:
\begin{itemize}
\item Solids: $m_{1,2,3}\neq 0$.
\item Fluids: $m_{1,3}\neq 0$ and importantly $m_2=0$.
\end{itemize}
Let us insist that in both cases the spatial translations are broken and both cases lead to a finite DC conductivity in the electric response. The main differences between the two types of theories are i) that the solid phases exhibit propagating transverse phonons - the Goldstone modes of the broken space translations, inhomogenenous in spatial coordinates - whereas the fluids do not; and ii) that the tensor modes are massive/massless for solid/fluid phases respectively.\\
Once translated such a distinction into the holographic framework we will realize that the presence or not of the tensor mode mass $m_2$ is indeed linked to the elastic response and in particular to the \textit{shear modulus} of the dual CM system.\\
For holographic applications in condensed matter theory we are interested in massive
gravity theories that allow for asymptotically AdS charged black brane solutions. The
action that will be considered is the Einstein-Maxwell action with a negative
cosmological constant and a graviton mass term:
\begin{equation}
\mathcal{S}\,=\,\int \,d^4x\,\sqrt{-g}\,\left[\frac{1}{2}\,\left(R\,+\,\frac{6}{L^2}\right)\,-\,\frac{L^2}{4}\,F_{\mu\nu}F^{\mu\nu}\,+\,\mathcal{L}_{\phi}\right]
\end{equation}
where $\mathcal{L}_{\phi}$ stands for the massive gravity term written through the St\"uckelberg fields $\phi^I$.\\
A straightforward way of analyzing the stability and phenomenology of a general theory
of massive gravity is to start with the non-covariant form of the massive gravity action
with the mass term written it terms of the metric perturbation $h^{\mu\nu}$ and then restore the
diffeomorphism invariance by the St\"uckelberg trick. The most general quadratic mass term that preserves the rotations of the transverse coordinates in the following form:
\begin{align}\label{acth}
\mathcal L_\phi (h^{\mu\nu},r)&= \frac{1}{2}\left ( m_{0}^2(r) (h^{t t})^2+ 2m_{1}^2(r) h^{t i} h^{t i } - m_2^2(r) h^{i j} h^{i j}  \right. \notag \\
&\quad  + m_{3}^2(r) h^{i i} h^{j j} -2m_{4}^2(r) h^{t t} h^{i i} \notag\\
&\quad  + m_{5}^2(r) (h^{r r})^2 + m_{6}^2(r) h^{t t} h^{r r }   \notag \\
&\quad   +m_{7}^2(r) h^{r i} h^{r i } + m_{8}^2(r) h^{t i} h^{r i } +m_{9}^2(r) h^{r r} h^{i i} \notag\\
&\quad + m_{10}^2(r)h^{rt}h^{rt} +m_{11}^2(r)h^{rt}h^{ii}+m_{12}^2(r)h^{tt}h^{rt}+m_{13}^2(r)h^{rr}h^{rt}\Big) \;,
\end{align}
where all the masses $m_i^2$ are functions of the radial coordinate r.\\
Note that the mass parameters defined in \eqref{acth} can be classified with respect to the perturbations that they affect as:
\begin{enumerate}
\item scalar: $m_0,m_{2-6},m_{9-13}$
\item vector: $m_1,m_7,m_8$
\item tensor: $m_2$
\end{enumerate}
From the dual perspective the vector and tensor masses will be the respectively relevant for characterizing the electric and the visco-elastic responses.\\
The costrunction of the model follows exactly in the same way of the previous section and allow us to define the most general bulk action in AdS$_4$ for our HMG theories: 
\begin{equation}
\mathcal{S}_{\phi}\,=\,\int\,d^4x\,\sqrt{-g}\,\mathcal{L}_{\phi}\,=\,-\,\int\,d^4x\,\sqrt{-g}\,V(X,Z)
\end{equation}
which generalizes further what we analyzed in \cite{BaggioliPRL}.\\
The dRGT theory  considered in ref.\cite{Veghoriginal} is a particular case of the latter with the Lagrangian given by:
\begin{equation}\label{vdRGT}
V_{\text{dRGT}} = - \beta_1 \sqrt{ \frac12 \left({X + \sqrt{Z}} \right) } - \beta_2 \sqrt{Z} \;.
\end{equation}
Let's first concentrate to the electric response of the system which is determined by the only $m_{1,7,8}$ mass terms. Consistency conditions can be derived in details (see \cite{BaggioliSolid}) and force:
\begin{equation}
m_7^2\,+\,+2\,m_1^2\,=\,0\,,\qquad m_8^2\,=\,0\,.
\end{equation}
reducing the full problem to a single mass term whose radial dependence is anyway arbitrary and linked with the structure of the potential $V(X,Z)$.\\
Performing the usual vector perturbations in a gauge invariant formalism we are left with the system of differential equations described by:
\begin{align}\label{eom1a}
&\left(fa'\right)'+\omega^2\frac{a}{f}-\frac{2\mu^2r^2}{r_h^2}a+\frac{2r^2\mu}{r_h}\lambda=0 \; ,\\\label{eom2a}
&2m_1^2\frac{1}{fM^2}\omega^2\lambda-i\omega m_8^2 \frac{1}{M}\left(\frac{\lambda}{M}\right)'-m_7^2\left(\frac{f}{M^2}\lambda'\right)'+\frac{r^2\det\mathcal P}{2L^2}\left(\frac{\mu}{r_h}a-\lambda\right)=0 \; , 
\end{align}
with $\det\mathcal P=m_8^4-8m_1^2m_7^2\neq 0$ and $\lambda$ being a gauge invariant field whose structure is not relevant for the following.\\
While the optical conductivity (finite $\omega$) requires the numerical integration of such a system with infalling boundary conditions for the fields at the horizon, using standard techniques one can derive the DC conductivity for the generic model, which turns out to be:
\begin{equation}
\sigma_{DC}\,=\,1\,+\,\frac{\mu^2\,L^2}{m^2(r_h)}
\end{equation}
where $\mu$ is the chemical potential, $r_h$ the position of the horizon and:
\begin{equation}
m^2(r)\,=\,2\,m_1^2\,r^2\,M(r)^2
\end{equation}
the only left graviton mass. Note how all the freedom of the model, namely the choice of the potential $V(X,Z)$ is just incorporated in the function $M(r)$ which can be taken generically of the form:
\begin{equation}
M^2(r)\,=\,L^{-2}\,\left(\frac{r}{L}\right)^\nu
\end{equation}
Translating into the language of the potential $V(X,Z)$ such a function gets the form\footnote{The DC conductity becomes:
\begin{equation}
\sigma_{DC}\,=\,1\,+\,\frac{\mu^2}{V_X(r_h)\,+2\,r^2\,V_Z(r_h)}
\end{equation}} :
\begin{equation}
M^2(r)\,=\,\frac{1}{r^2}\,\left(V_X\,+2\,r_h^2\,V_Z\right)
\label{MassV}
\end{equation}
where $V_{X,Z}= \partial_{X,Z}\,V$.\\[0.2cm]
A phenomenologically interesting question to investigate is the different types of materials
that can be described within the framework of holographic massive gravity and, in
particular, their ability to conduct an electric current. The distinction between different
classes of materials is well captured by the temperature dependence of their DC electric
conductivity. In the models of massive holography proposed in this paper, it is controlled
by the radial dependence of the mass function $M^2(r)$ and in particular by the parameter $\nu$.
The latter, which determines the nature of the dual CFT, can be related in the case of the simple potential of the form $V(X)=X^n$ to the power n as follows:
\begin{equation}
n\,=\,\frac{4\,+\,\nu}{2}
\end{equation}
It is straightforward to realized that for the model to be consistent we need to impose $\nu>-4$ which corresponds to $n>0$.\\
We can define a distinction between a metallic behaviour ($d\sigma/dT<0$) and an insulating behaviour ($d\sigma/dT>0$). For $\nu<-2$ the behaviour is metallic while for the opposite case,
$\nu>-2$, we are in the presence of an insulator (see fig.\ref{DCDC}). Since the mass function \eqref{MassV} in the dRGT theory corresponds to the cases $\nu=\{-2,-3\}$, the dual materials exhibit a metallic
behaviour there (this happens for the linear and square root models of \cite{AndradeWithers,TaylorScalar} as well). The possibility to mimic metallic and insulating behaviour (and a transition between them) in the context of holographic massive gravity has been already
pointed out in \cite{BaggioliPRL}.
We further exploit the freedom offered by the generic power $\nu$ to investigate the option
of having a linear T resistivity $\rho=1/\sigma_{DC} \sim T$. This is a special feature of
strange metals which evades the usual scaling predicted by the Fermi liquid theory ($\sigma_{DC} \sim T^{-2}$). Within our class of models we observe a linear scaling in the resistivity at low temperature only for the power $\nu=-3$ as shown in fig. \ref{DCDC}.
As mentioned above, the $\nu=-3$ mass function coincides with the $\beta_1$ term of dRGT
massive gravity. The linear scaling of the DC resistivity for this particular case in the low
temperature regime has been already observed earlier in \cite{TaylorScalar}.\\
The AC conductivity can be found by numerically and details are presented in \cite{BaggioliSolid}. We fixed a generic mass function $M^2(r)=r^\nu$ and we performed the computation using the standard holographic methods. As already shown in fig \ref{DCDC} the power $\nu$ determines whether the model is dual to a metallic or to an insulating state. The choice of $M^2(r)$ influences also the AC response of the dual CFT as shown in fig. \ref{ACAC}. The value of the real part of the conductivity at zero frequency coincides with the analytic value for the DC conductivity given before. For $\nu\geq0$ we observe a formation of a peak, which becomes sharply localized for higher values of $\nu$ and moves towards zero frequency. This suggests the presence of a localized excitation whose width decreases with $\nu$ very fast. This resonance shows up only in the insulating state and it is reminescent of what has been found in \cite{BaggioliPRL}. We associate the appearance of the peak at these values of $\nu$ with the fact that the mass functions that are more localized near the AdS boundary give rise to an explicit source of disorder that seems equivalent to the lattice disorder discussed in introduction. We find this a reasonable interpretation because such a disorder amounts to an additional source of an explicit breaking of translations that makes the otherwise exactly massless phonons become pseudo-Goldstone bosons. This is indeed what can be seen in the solid HMGs:
one can identify almost massless transverse phonon poles in cases where the breaking is
localized near the horizon, while as the profile of the mass function is moved towards
the AdS boundary the phonons become massive. Hence, the physical phenomenon related to the appearance of peaks in the electric conductivity for $\nu\geq 0$ is expected to be the small collective field excitations - the phonons - due to the spontaneous breaking of the translational symmetry. In the presence of charge density this can be interpreted as a polaron formation as first suggested and observed in \cite{BaggioliPRL}.\\[0.3cm]
In addition to the classification of the materials into solids and fluids according to the
residual symmetries that are preserved in the low energy EFT, there is another perhaps
more intuitive way to distinguish them, namely, according to the type of response they
exhibit under a shape deformation.\\
In standard mechanical response theory, the magnitude that encodes the material deformation is a displacement vector $u_i$, and the direct measure of the deformation is the linearized strain tensor,
\begin{equation}
u_{ij}\,=\,\frac{1}{2}\,\partial_{(i}u_{j)}
\end{equation}
A clear distinction between fluids and solids is that they respond very differently to a
constant applied shear stress (given by a traceless stress tensor $T_{ij}$ leaving the volume
of the material unchanged). For fluids, this leads to a constant shear velocity gradient
(traceless part of $\dot{u}_{ij}$ ) and the corresponding response parameter is the shear viscosity.
For solids, instead, a small constant applied shear stress leads to a constant shear strain
(traceless part of $u_{ij}$), and the response parameter is the elasticity. Thus, a practical distinction between solids and fluids is that the static shear elasticity (or rigidity) modulus
is nonzero for solids and vanishes for fluids - fluids do not offer resistance to a constant
shear deformation. This distinction between the solid and fluid phases is exactly reproduced
in HMG: it is encoded exclusively in the $m_2(r)$ mass parameter, which vanishes for
the fluid HMGs but not for the solid HMGs.\\
Indeed this mass term is just related to the derivative of the potential with respect to $X=Tr[\mathcal{I}^{ij}]$:
\begin{equation}
m_2(r)\,\sim V_X
\end{equation}
and for the fluids case, where $V(X,Z)=\mathcal{H}(Z)$, it is trivially zero.\\
Of course, the response of different materials (and also black branes) is more complex
than the simplified picture above. Materials can, for instance, exhibit both elastic and
viscous (i.e., viscoelastic) response. Given that the elastic properties of HMG black branes
have not yet been unveiled, we shall restrict here to the elastic response by considering
only static applied stress and static deformation or strain. The full viscoelastic response
can be studied by considering time dependent applied stresses.\\
In homogeneous and isotropic materials, the (static) elastic shear modulus or modulus
of rigidity can be defined as the stress/strain ratio:
\begin{equation}
T_{ij}^T\,=\,G\,u_{ij}^T
\end{equation}
where the superscript T stands for the traceless part. Equivalently, one can extract the
modulus of rigidity, G, from a Kubo-like formula that relates it to the Fourier transform
at zero frequency and wavenumber of the retarded correlator as:
\begin{equation}
G\,=\,\lim_{\omega \rightarrow 0}\,Re\,\langle T_{xy}^T\,T_{xy}^T \rangle^R
\end{equation}
Once we have G expressed in terms of the stress tensor two-point functions, it is easy
to apply the holographic prescription to extract it. The bulk field dual to $T_{ij}$ is the
traceless tensor mode of the metric perturbation $h_{ij}^T(r,t,x^k)$ whose equation of motion reads:
\be\label{spin2}
\left[f\,\partial_r^2 +\left(f'-2\frac{f}{r}\right)\partial_r + \left(\frac{\omega^2}{f} -\,4\,r^2\, m_2^2(r)\right)\right]h_{ij}^T=0 \; .
\ee
\begin{figure}
\centering
\includegraphics[width=13.5cm]{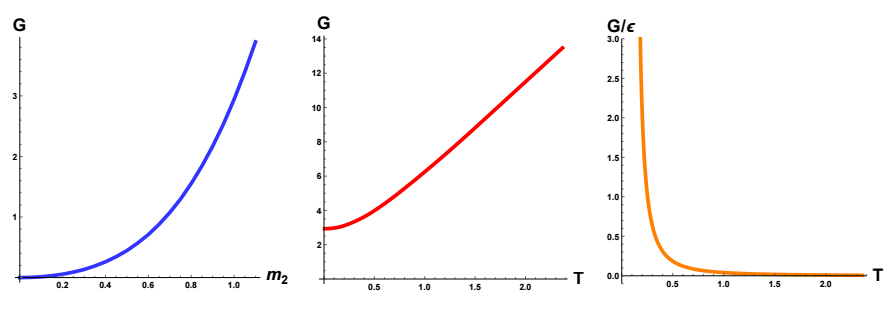}
\caption{Modulus of rigidity G for the choice $\nu=-2$. \textbf{Left: }mass dependence. \textbf{Center: }temperature dependence. \textbf{Right: } temperature dependence after renormalizing the modulus of rigidity G by the T dependent energy density $\epsilon$.}
\label{elasticfig}
\end{figure}
For the constant and homogeneous ($\omega=0$) response, this equation depends exclusively on the $m_2(r)$ mass parameter. The retarded Greens function is extracted as usual by solving \eqref{spin2} with ingoing boundary conditions and taking the ratio of  the subleading to the leading mode. The resulting response vanishes only for $m_2=0$, i.e. for the fluid HMGs. In fig.~\ref{elasticfig} we show the dependence on temperature of the shear modulus for some representative model choice. This shows that there is a well-defined sense in which the HMG black branes enjoy a nontrivial  elastic shear response and thus behave as solids. We observe that the rigidity modulus $G$ increases with temperature, whereas most ordinary solids display the opposite dependence and $G$ decreases with increasing $T$. However, in ordinary solids this behaviour occurs at roughly constant energy density while in the middle panel of fig.~\ref{elasticfig} the energy density is strongly increasing with $T$ (which relates to the fact that the CFT degrees of freedom are inevitably in a relativistic regime). The ratio of $G$ to the energy density $\epsilon$, instead, does display the more standard decreasing in $T$ form, so in this sense the result seems to be consistent with expectations from ordinary solids.\\[0.3cm]
We have not analyzed in details all the technical part dealing with the consistency and stability of the generical HMG we considered; the interested reader can find it in \cite{BaggioliSolid}. All in all we made some importants points which can be summarized here:
\begin{itemize}
\item We introduced the most generic holographic massive gravity framework, which goes beyond the dRGT case, and we showed explicitely its formulation in terms of the unitary gauge and via the St\"uckelberg field $\phi^I$.
\item We analzyed the all the possible pathological issues in details and we found out that the set of possible consistent HMG is pretty wide (and unexplored).
\item We can define two types of HMG: the solid type and the fluid one. They are distinguished by the symmetry breaking pattern and by a new physical observable which refers to \textit{elasticity}.
\item We performed the first computation of the \textit{shear modulus} showing that it is not null in solids while it turns out to be null in fluids as expected.
\item We analyzed in full generality the electric response of HMG theories, showing that they are able to encode not only metallic behaviours but insulating as well.
\end{itemize}
In the next section we will give more phenomenological details about the electric response of holographic EFT for condensed matter dealing with massive gravity.
\section{Electric response \& Metal Insulator transitions}
In the previous chapter we stated that HMG could mimick not only metallic phase ($d\sigma/dT<0$) but insulating ones ($d\sigma/dT>0$) as well.  Keeping a critical attitude, this is actually not totally correct because we never had:
\begin{equation}
\sigma_{DC}\left(T=0\right)\,\approx\,0
\end{equation}
which is the proper definition for a insulator.\\
This issue can be derived from the generic structure of the DC conductivity coming from that class of models:
\begin{equation}
\sigma_{DC}\,=\,\frac{1}{e^2}\,\left(\,1\,+\,\frac{\mu^2}{\mathbb{M}^2(T)}\,\right)
\label{genericDC}
\end{equation}
where $\mathbb{M}^2(T)$ is an effective graviton mass, which depends generically on the temperature T (through the position of the horizon $u_h$) and on the details of the specific model considered. This generic structure is quite robust and can be derived just through hydrodynamical arguments \cite{LucasHydro,LucasMemory}.\\
The issue just described, has been formalized in \cite{Grozdanov1} as the existence of a ''universal'' lower bound for the electric conductivity for HMG theories, which are thought to represent a \textit{mean-field} description of disorder:
\begin{equation}
\sigma_{DC}\geq \frac{1}{e^2}
\label{boundDC}
\end{equation}
The presence of a lower bound of course prevents the possibility of having a proper insulating state with a very small electric conductivity at zero temperature.\\
From a more phenomenological point of view, this is as to say that no disorder-driven metal-insulator transitions MIT appear in such a models. On the contrary, increasing the amount of disorder, \textit{i.e.} the graviton mass, the system passes from a good metal characterized by a clear Drude peak to an incoherent metal where no localized excitation is dominating the response. In nature increasing further disorder will turn the system into an insulating state which is the result of a \textit{localization} mechanism. In HMG theories this is not happening (yet)!\\
To be more precise, it has been claimed in \cite{Grozdanov1} that this is true for a large class of ''simple'' models, defined by the following benchmark structure:
\begin{equation}
\mathcal{S}_{bulk}\,=\,\int\,d^4x\,\sqrt{-g}\,\left({R\over2}-2\,\Lambda-\frac{F_{\mu\nu}F^{\mu\nu}}{4\,e^2}\,+\,\mathcal{V}(\phi^I)+\dots\right)
\label{ActPerm}
\end{equation}
where $\mathcal{V}(\phi^I)$ encodes the Lagrangian for a generic neutral translation-breaking (TB) sector.\\
The main point of this issue turns around the first term appearing in the generic DC formula \eqref{genericDC}: performing indeed a large disorder limit, $\mathbb{M}\rightarrow \infty$, the second term generically drops out but the first term remains and it is completely unaffected by the TB sector. In other words the first term is just dictated by the structure of the electromagnetic part of the action and without modifying the usual Maxwell term such a term will not vary neither with temperature nor disorder strength.\\
One possible way to avoid such a feature is to introduce a new degree of freedom into the model which couples directly to the Maxwell term. This is what happens for the strings inspired Einstein-Maxwell-Dilaton models whose action reads:
\begin{equation}
\mathcal{S}_{bulk}\,=\,\int\,d^4x\,\sqrt{-g}\,\left({R\over2}-2\,\Lambda-\frac{Z(\Phi)}{4\,e^2}\,F_{\mu\nu}F^{\mu\nu}\,+\,(\partial_\mu \Phi)^2\,-\,V(\Phi)\,+\,\mathcal{V}(\phi^I)+\dots\right)
\end{equation}
where the new neutral scalar field $\Phi$ is called the \textit{dilaton}.\\
This new ingredient represents from the dual perspective the running of the coupling of the dual field theories and can have a strong impact on the IR features of the system. From the point of view of the DC conductivity, it modifies the generic formula \eqref{genericDC} into:
\begin{equation}
\sigma_{DC}\,=\,\frac{1}{e^2}\,\left(\,Z(\Phi_h)\,+\,\frac{\mu^2}{\mathbb{M}^2(T)}\,\right)
\end{equation}
where $\Phi_h$ is the value of the dilaton at the horizon.\\
It is clear that as a conseguence of this modification the proposed bound \eqref{boundDC} is violated and depending on the dynamics of the dilaton sector, \textit{i.e.} the form of the function $Z(\Phi)$, possible insulating states could appear. This is indeed what happens \cite{KiritsisRen,GouterauxDC}.\\
In the present \cite{BaggioliDisorder}, we shall show instead that one can certainly avoid a bound
like \eqref{boundDC} in minimal and natural holographic models that contain the same dynamical
ingredients (operators) as well as the mutual (''electron-disorder'') interactions which
are still allowed by the symmetries\footnote{There are other effective ways of reaching this goal which have been recently proposed. One can introduce non-linearities in the EM sector of the form $K(F^2)$. This model has been investigated recently in \cite{BaggioliMott} and seems to give an effective holographic description of Mott Insulators. Eventually one can also couple the Ricci scalar directly to the St\"uckelberg sector $\sim R \mathcal{P}(\Phi^I)$ and get similar results explored in \cite{Garcia1}}. Indeed, using EFT-like reasoning, it is clear that
\eqref{ActPerm} is not the most general action allowed by the symmetries and the required field
content. Clearly, there are additional couplings between the charge and TB sectors
that can (and should) be included in the effective action. The crucial new ingredient
that will be relevant for the present discussion is a direct coupling between the charge
and TB sectors, which we can write schematically as:
\begin{equation}
Y[\phi^I]\,F_{\mu\nu}F^{\mu\nu}
\end{equation}
where $Y[\phi^I]$ stands for some function of the TB sector $\phi^I$ (or its derivatives) alone.\\
Physically, even before specifying how we shall implement the TB sector and choose
the Y function, it is clear that this effective interaction captures how much the TB
sector affects the charge sector. This coupling, then, encodes the charged disorder -
the effects from ionic impurities that directly couple to the mobile charge carriers.
From the point of view of an effective description it is all the more reasonable that
this kind of disorder is encoded in a direct coupling of this form.\\
Let us also emphasize one crucial difference between our proposal and some previous
models \cite{KiritsisRen,GouterauxDC} that use a running dilaton $\Phi$ that couples to the charge
sector through a bulk term like $Z(\Phi)F^2$. These models include a new dynamical
ingredient, a scalar CFT operator $\mathcal{O}$. The BB solutions are relevant deformations of
the CFT by the operator $\mathcal{O}$ that already in vacuum gives rise to confinement and
therefore an insulating-like behavior. In these cases, it is hard to argue that the
insulating behavior is driven by disorder. In our case, instead, there is no room for
doubt. There are no more dynamical ingredients in the CFT other than the TB and
the charge sector, so the BB solutions represent CFTs deformed by disorder (and
finite density). At this point we can also see that the new interaction will play a
role similar to the dilaton-Maxwell coupling in the sense that the physical magnitude
of the charge carried by a charge carrier gets renormalized along the RG flow - in
our case clearly due to disorder. Furthermore, we will show that dynamical consistency
of the model requires that the renormalization is such that the conductivity is
necessarily reduced at small temperatures (which is not necessarily the case for the
Maxwell-dilaton coupling).\\
\begin{figure}
\centering
\includegraphics[width=6cm]{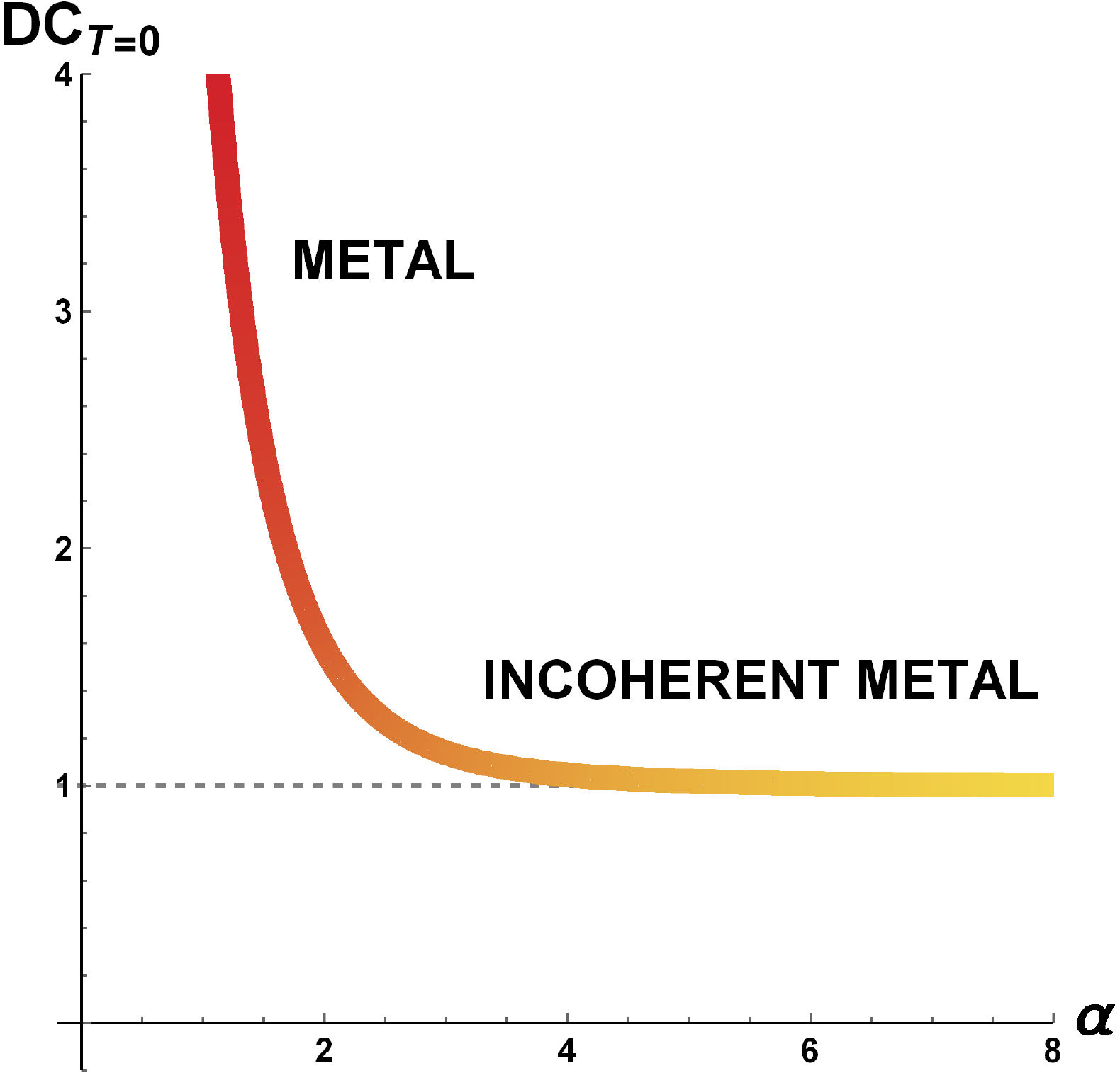}
\qquad \includegraphics[width=6cm]{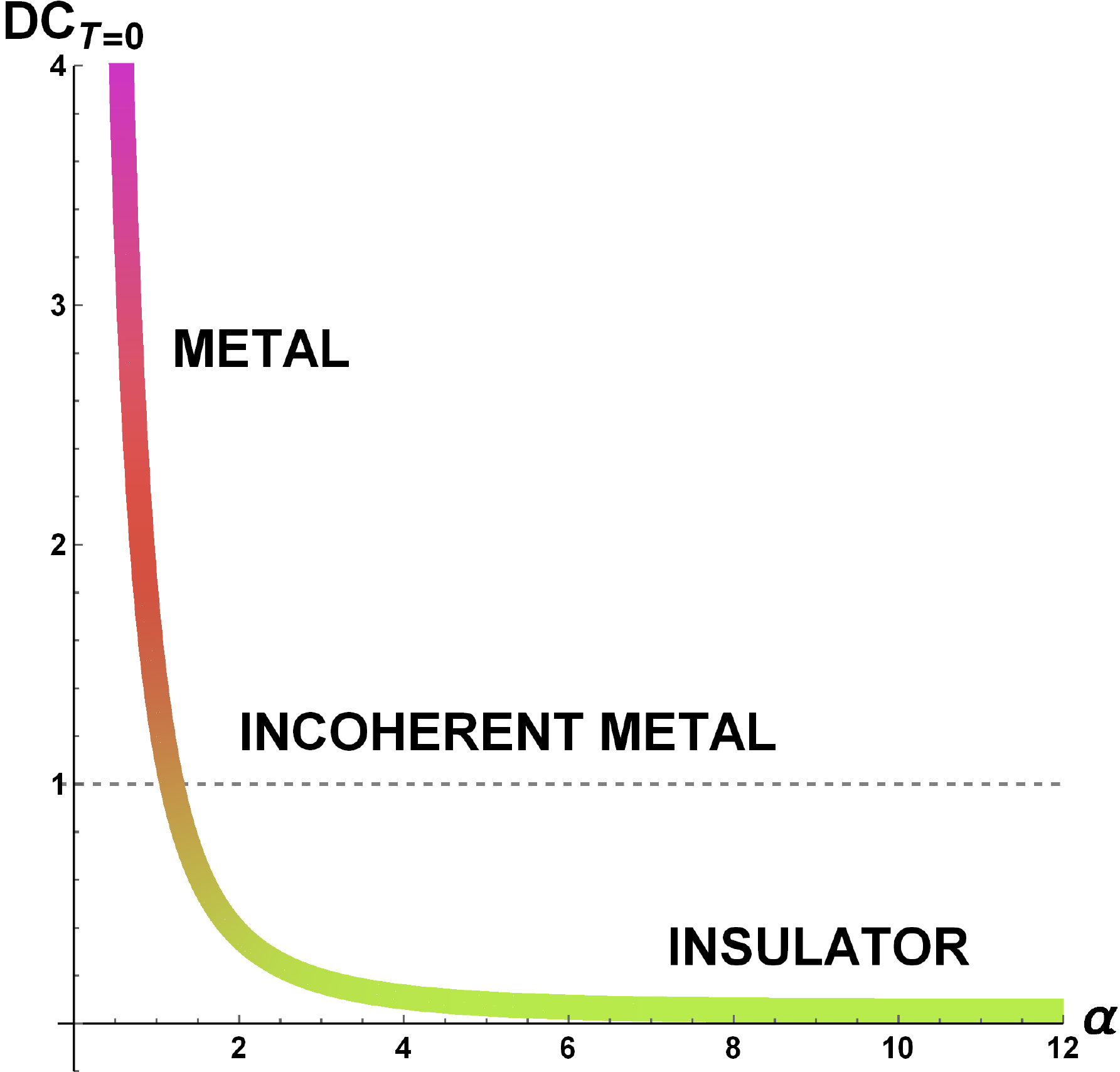}
\caption{Electric DC conductivity at zero temperature for the model \eqref{themodel} dialing the disorder strength $\alpha$ (i.e the graviton mass); \textbf{Left:} $\kappa=0$ (i.e. previous literature); \textbf{Right:} with the new coupling $\kappa=0.5$ (safe region) .}
\label{DC0plot}
\end{figure}
We therefore consider the minimal model in $3+1$ dimensions:
\begin{equation}
\mathcal{S}\,=\,\int d^4x\,\sqrt{-g}\,\left[{R\over2}-\Lambda-\frac{1}{4\,e^2}\,Y(X)\, F^2-\,m^2\, V(X)\right].\label{actionba}
\end{equation}
with $X=g^{\mu\nu}\p_\mu\phi^I\p_\nu\phi^I$ and $F^2=F_{\mu\nu}F^{\mu\nu}$.\\
It is quite clear from the structure of the action \eqref{actionba}, that the TB enters in
two distinct ways, encoded in the functions $Y(X),V(X)$. $V(X)$ represents a neutral
disorder - the disorder from neutral forms of impurities, that do not couple directly
to the charge carriers. Instead, $Y(X)$ captures the effects of disorder that are felt directly by the charged sector. In other words this model provides a generalization of the previous HMG models \cite{BaggioliPRL,BaggioliSolid}.\\
The model admits asymptotically AdS  charged black brane solutions with a planar horizon topology. For arbitrary choice of $V,Y$ they take the form:
\begin{align}
&ds^2 = \frac{1}{u^2}\left[-f(u)dt^2+\frac{1}{f(u)}\,du^2+dx^2+dy^2\right]\,,\nonumber\\
&f(u)\,=-\,u^3\,\int_{u}^{u_h} \left(\frac{\rho^2}{2\, Y\left(\alpha ^2 \xi^2\right)}+\frac{m^2\, V\left(\alpha ^2 \xi^2\right)}{\xi^4}+\frac{\Lambda }{\xi^4}\right)\,d\xi\,,\nonumber\\
&\phi^I=\alpha \,\delta^I_i x^i\,\,,\,\,\,\,\,I=\{x,y\}\,,\nonumber\\
&A_t(u)\,=\,\rho\,\int_{u}^{u_h}\frac{1}{Y(\xi^2\,\alpha^2)}\,d\xi\,,
\label{ansatz}
\end{align}
where $u_h$ denotes the horizon location.\\
The temperature of the background geometry reads:
\begin{equation}
T\,=\,-\frac{\rho^2\,u_h^3}{8\, \pi \, Y\left(\alpha ^2\,u_h^2\right)}-\frac{m^2\, V\left(\alpha ^2 u_h^2\right)}{4\, \pi  \,u_h}-\frac{\Lambda
   }{4\, \pi  \,u_h}
\end{equation}
An very important part of the present analysis concerns the conditions under which the models above are consistent -- they are free from  instabilities. Its main outcome is that the functions $V(X), Y(X)$ that appear in the Lagrangian are subject to the constraints:
\begin{equation}
V'(X)>0\,\,\,,\,\,\,Y(X)>0\,\,\,,\,\,\,Y'(X)<0
\label{resum}
\end{equation}
Crucially, the Maxwell-St\"uckelberg coupling $Y$ is allowed (and must be positive). Not only that, it must also be a decreasing function of $X$. Let us emphasize that the latter condition stems solely from the requirement that the transverse vector modes have a normal (non-ghosty) kinetic term (the actual condition is slightly less restrictive, but for simplicity we  \color{black}shall take $Y'<0$ \color{black} which is certainly sufficient and more robust). The fact \color{black}that $Y'<0$ \color{black} will have  a dramatic impact on the possibility to have a MIT.

We shall focus on a representative `benchmark' model,
\begin{equation}
Y(X)\,=\,e^{-\kappa\,X}\,,\,\,\,\,\,V(X)=X/(2\,m^2)\,\,.
\label{themodel}
\end{equation}
\begin{figure}
\centering
\includegraphics[width=6cm]{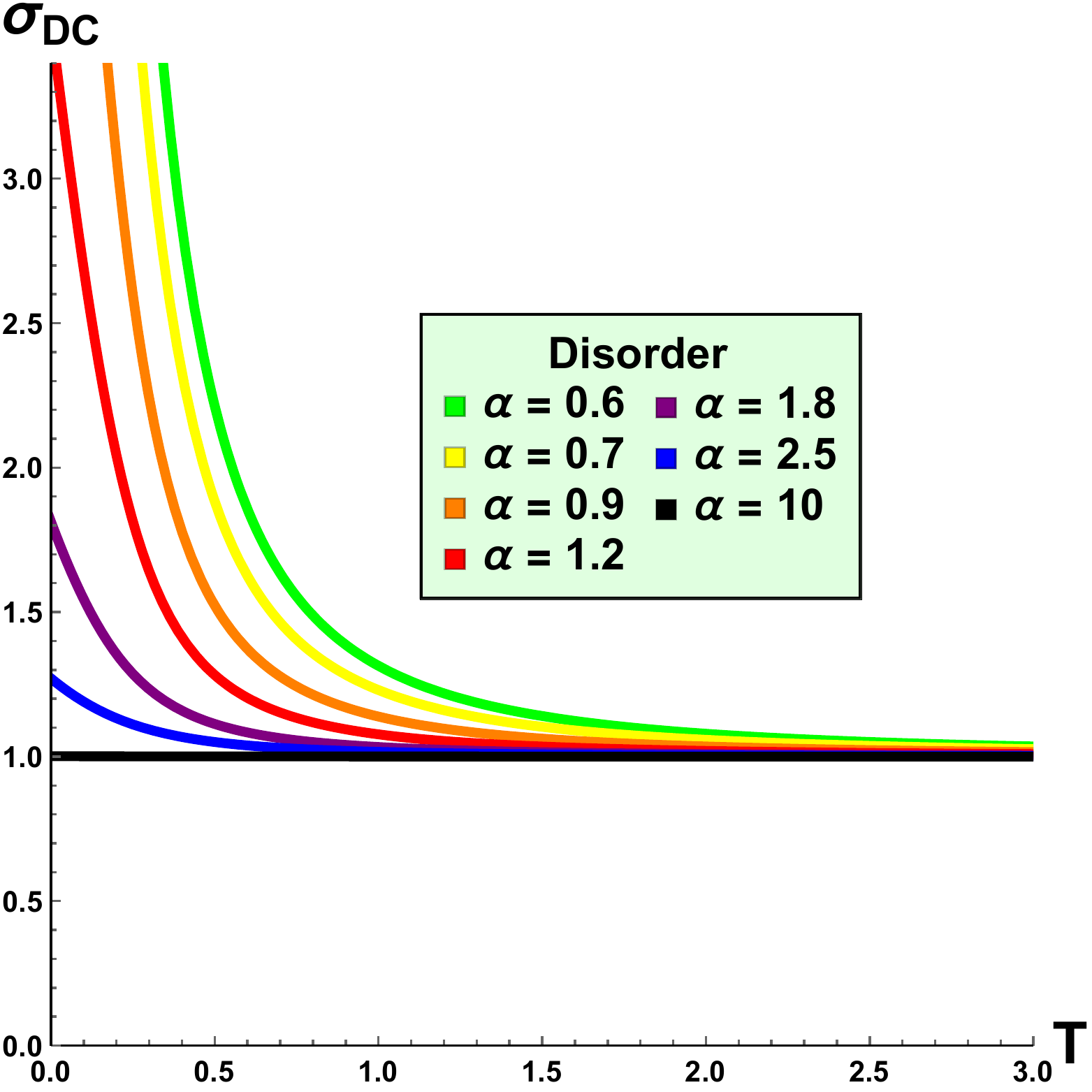}
\qquad
\includegraphics[width=6cm]{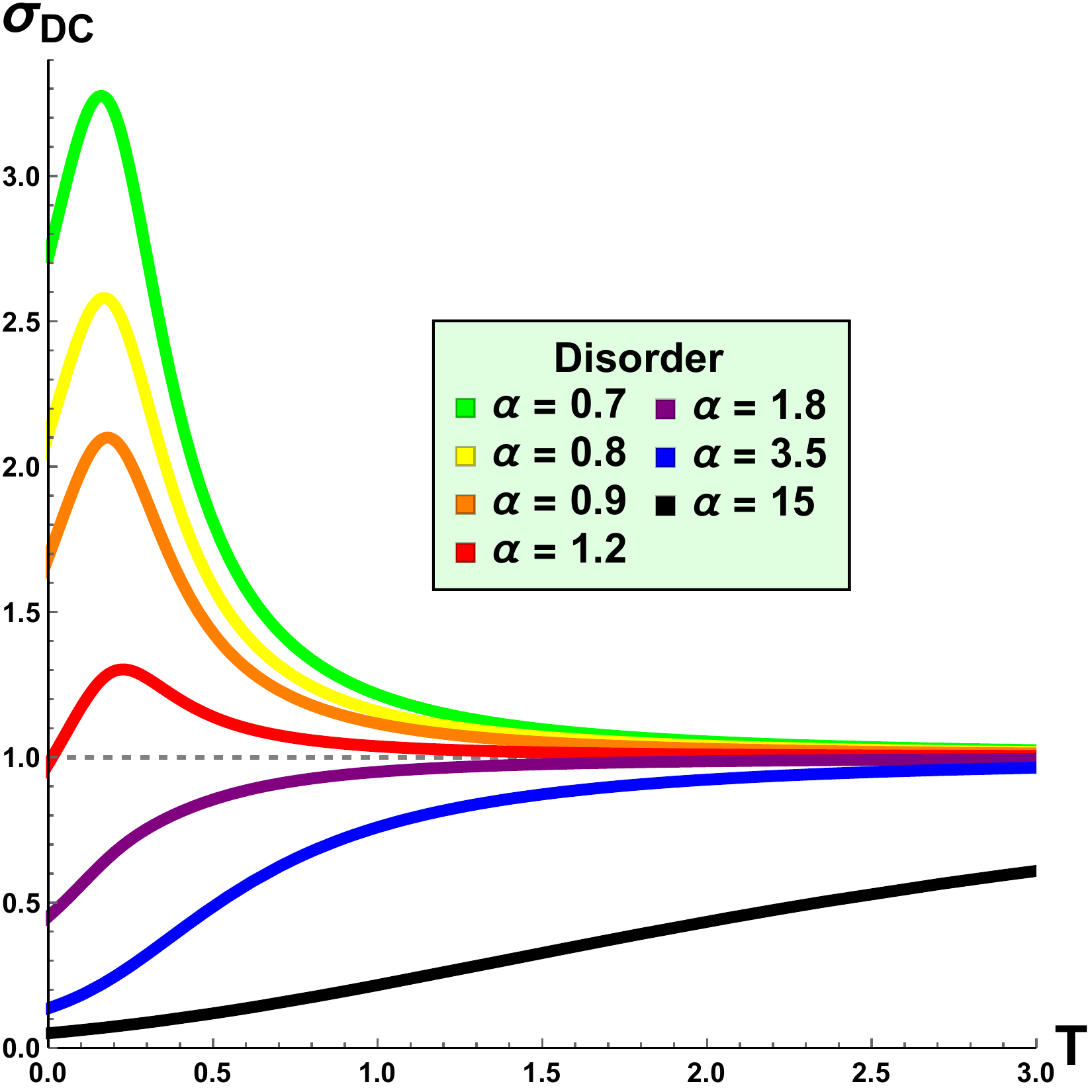}
\caption{Temperature features of the DC conductivity at different disorder strengths $\alpha$ (i.e. graviton mass) for the model considered in \eqref{themodel} with unitary charge density; \textbf{Left:} metal-incoherent metal transition for $\kappa=0$ (i.e. previous literature); \textbf{Right:} metal-insulator transition for $\kappa=0.5$ (safe region).}
\label{MITplot}
\end{figure}
This is by far not the most general model but it will suffice to illustrate the new features that can be modeled with this kind of coupling. 
Note that it suffices to take $\kappa>0$ to satisfy all the  consistency conditions. One can also anticipate that for order-one values of $\kappa$ the effects from this coupling can be rather important. \\
Proceeding with the vector perturbations on top of the background defined in \eqref{ansatz}, one can compute numerically the optical electric conductivity and analytically its DC value. We find the following analytic result for the  electric DC conductivity,
\begin{equation}
\sigma_{DC}\,=\left[\,Y(\bar{X})+\frac{\rho ^2\, u^2}{m_{\footnotesize eff}^2} \,\right]_{u_h}
\label{DCformula}
\end{equation}
with 
$$
m_{\footnotesize eff}^2\equiv \alpha ^2 \left(m^2\,  V'(\bar{X})-\rho ^2 \,u^4\, {Y'(\bar{X})\over{ 2\,Y^2(\bar{X}) }}\right)
\label{MMM}
$$
where $\bar{X}\,=\,u^2\,\alpha^2$ and all quantities have to be evaluated at the horizon, $u=u_h$.\\
The expression \eqref{DCformula} encodes all the interesting phenomenology which follows.\\
For the benchmark model \eqref{themodel}, the interesting quantities read:
\begin{align}
&T\,=\,-\frac{\rho ^2 \,u_h^3 \,e^{\kappa\,\alpha ^2 u_h^2}}{8\, \pi }-\frac{\alpha ^2 \,u_h}{8 \,\pi }+\frac{3}{4\, \pi \,u_h},\nonumber\\
&\sigma_{DC}\,=\,e^{-\kappa\,\alpha ^2 \,u_h^2}+\frac{2 \,\rho ^2 \,u_h^2}{\alpha ^2\, \left(\rho ^2 \,\kappa\,u_h^4\, e^{\kappa\,\alpha ^2 u_h^2}+1\right)}\,.
\end{align}
In this scenario we are left with only four parameters in our model: the temperature $T$, the charge density $\rho$, the neutral and charged disorder strengths $\alpha$ and $\kappa$.\\
\begin{figure}
\centering
\includegraphics[width=15cm]{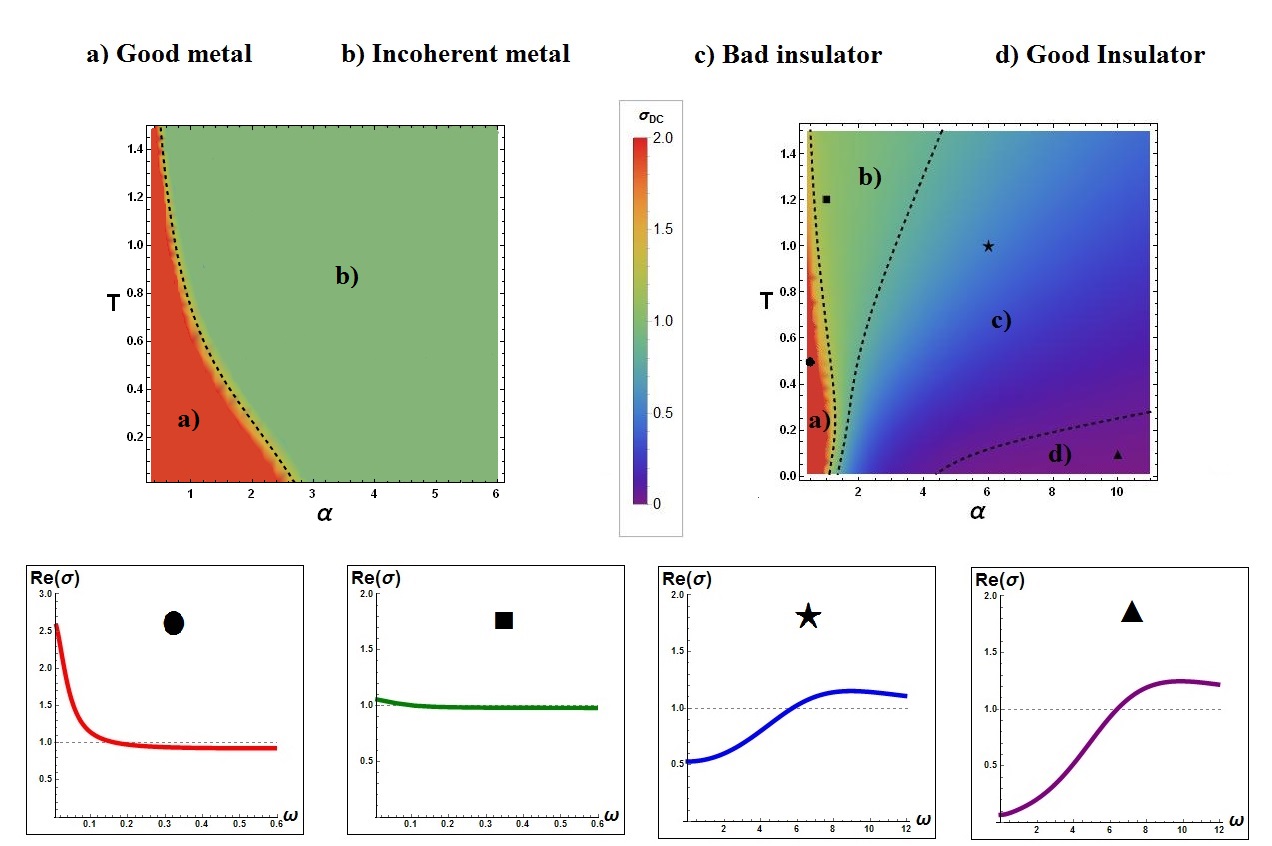}
\caption{Phase diagrams for the model \eqref{model} with unitary charge density $\rho=1$. Dashed lines correspond to $\sigma_{DC}=0.1,\,0.8\,,1.2$ and they divide the four regions: a) good metal, b) incoherent metal, c) bad insulator and d) good insulator . \textbf{Top Left:} Temperature-disorder plane with $\kappa=0$. \textbf{Top Right:} Temperature-disorder plane with graviton mass $\kappa=0.5$. \textbf{Bottom:} For every region (a,b,c,d) in the phase diagrams one representative example of $\Re\,(\sigma)$ is shown. The parameters for each one of the AC plots are pinpointed in the $T-\alpha$ phase diagram (top right) and they correspond to: $\left[\,\bullet:(\alpha=0.5,T=0.5),\,\blacksquare:(\alpha=1,T=1.2),\,\bigstar:(\alpha=6,T=1),\,\blacktriangle:(\alpha=10,T=0.2)\,\right]$ .}
\label{Phases}
\end{figure}
The analysis of the electric DC conductivity in function of disorder is our primary task. The first interesting and new feature of the model deals with the DC conductivity at zero temperature which characterizes the nature of our '\textit{material}' (i.e. metal/insulator). In the previous massive gravity models the system could be just in a metallic phase (with a sharp Drude peak) or in an incoherent metallic phase where there is no clear and dominant localized long lived excitation. The bound of \cite{Grozdanov1} can be indeed rephrased with the statement that such a models fall down in an extremely incoherent metallic state for very strong disorder without undergoing a metal-insulator transition. Note that this is a quite unnatural behaviour in real-life experiments where usually strong disorder produces clear insulating behaviours. In our case the scenario is more complex and increasing the disorder one can reach an insulating state where $\sigma_{DC}\approx 0$ at zero temperature. This is basically what we mean by disorder-driven MIT and to the best of our knowledge this model is the first holographic example of such a mechanism. It is fair to say that the link between massive gravity and disorder is still very blurred and that right now massive gravity is able to capture just some few features of it. The difference between our novel results and previous literature is summarized in fig.\ref{DC0plot}.\\

Most of the results of this short note are summarized in fig.\ref{MITplot} where the presence of a metal-insulator transition (in contrast with previous massive gravity models) is made evident . The picture again emphasizes how we can overcome the bound $\sigma_{DC}\geq 1/e^2$ proposed in \cite{Grozdanov1} (dashed line) increasing the disorder in the system and exploiting the new parameter $\kappa$.\\
This result also  suggests that there is no universal lower-bound for the electrical conductivity at least within the holographic models. Despite the absence of any lower bound in the electric conductivity provided by our simple generalization of the implicit assumptions in \cite{Grozdanov1}, it should be very clear that we are not giving in a derivation of \textit{localization}, or that localization is the phenomenon that lies behind the bad conductivity of these holographic materials. 
For this one would certainly need to abandon homogeneity and study more complicated models.
In conclusion, the main features of the model \eqref{action}  are summarized in Fig.~\ref{Phases} where we draw the phase diagrams in the temperature-disorder plane for $\kappa=0$ ({\em i.e.} previous literature) and $\kappa=0.5$ (which lies in the healthy region of the parameters space). For the known case $\kappa=0$ just metallic phases are accessible and only a crossover between metals and incoherent metals can be manifested (see {\em e.g.} \cite{DavisonGouteraux}). On the contrary, for the novel case, the phase diagram gets richer and incorporate several phases of matter depending on the parameters: good metal (a), bad or incoherent metal (b), bad insulator (c) and good insulator (d). Both the quantum phase transition (MIT) and the finite temperature crossover are present in the picture. For every phase of matter a representative example of optical conductivity is shown at the bottom of Fig.~\ref{Phases}.\\

The framework of effective holographic theories can be enriched and enlarged to account for several condensed matter wisdoms and it can represent an useful tool to reproduce a large set of unexplained phenomena. The study of insulating states in this context has been initiated bringing a collection of new questions and unexplored directions.\\
Whether holographic models account for the presence of universal bounds/values for certain physical observables, such as the electric conductivity we considered at this stage, is a very valuable question on which we will return in the next section and especially in the final remarks of this thesis.
\section{The $\eta/s$ bound in theories without translational symmetry}
It has been long known that black brane solutions can be characterized both by thermodynamic quantities like temperature and entropy as well as hydrodynamic entities like viscosity and diffusion. In gauge/gravity duality, the hydrodynamics of the black branes is mapped to the hydrodynamic properties in the dual field theory. One of the most prominent insights that the AdS/CFT correspondence have provided for the understanding of dynamics of strongly coupled condensed matter systems is that the shear viscosity to entropy density ratio takes on a universal value\footnote{We work in the units where $\hbar=k_B=8\pi G\equiv1$.} for all gauge theories with Einstein gravity duals \cite{viscosity1}:
\begin{equation}
\frac{\eta}{s}\,=\,\frac{1}{4\,\pi}
\end{equation}
This value was conjectured to set a fundamental lower bound on this ratio - the celebrated Kovtun-Son-Starinets (KSS) bound \cite{viscosity2}.
Amazingly enough, the bound seems to be satisfied for all known fluids where $\eta/s$ has been measured, including examples like superfluid helium \cite{Rupak} and the QCD quark gluon plasma  (see \emph{e.g.} \cite{Song}).\\
By now it is well established that the KSS bound is violated by higher curvature corrections to the Einstein theory. In particular, the violation of the bound was observed in Einstein gravity supplemented by the quadratic Gauss-Bonnet term \cite{Brigante1}. In terms of the Gauss-Bonnet coupling $\lambda_{GB}$ the viscosity to entropy density ratio was found to be
\be
\frac{\eta}{s}=\frac{1}{4\pi}\left[1-4\lambda_{GB}\right]\;.
\ee
For a positive coupling this would imply an arbitrary violation of the bound. However, the consistency requirements on the dual field theory impose constraints on the allowed values of the Gauss--Bonnet coupling constant. In particular, it was found that the field excitations in the dual field theory allow for superluminal propagation velocities for $\lambda_{GB}>9/100$, thus imposing a new lower bound on the viscosity to entropy ratio \cite{Brigante2}. In the light of these results it is at present not clear whether a universal fundamental bound on the shear viscosity to entropy ratio exists. For a review on the bound violation in higher derivative theories of gravity, see \cite{Cremonini1} and references therein.\\
Asking ourselves about the fate of the $\eta/s$ universal bound in the context of holographic theories with momentum dissipation is for sure a valuable and interesting question which we want to adress \cite{BaggioliViscosity}. We make use of the generic HMG theories described in \cite{BaggioliSolid} and we compute the viscosity of the system via the Kubo formula:
\be\label{viscosity}
\eta \equiv \lim_{\omega\to 0}\frac{1}{\omega}\text{Im }{\cal G}^R_{T_{ij}\,T_{ij}}\;
\ee
where ${\cal G}^R$ is the retarded Green's function of the stress tensor.\\
Note that such a generic theories possess a more complicated viscoelastic response, stress-tensor $T_{ij}$-displacement tensor $u_{ij}$ relation of the type:
\be\label{viscoel}
T_{ij}^{(T)}=G\,u_{ij}^{(T)}+\eta \,\dot u_{ij}^{(T)}\;.
\ee
where G is the so called modulus of rigidity, dealing with the elastic properties, which can be similarly computed through:
\be\label{elasticity}
G \equiv \lim_{\omega\to 0}\text{Re }{\cal G}^R_{T_{ij}\,T_{ij}}\;.
\ee
In terms of the two parameters defined in \eqref{viscosity} and \eqref{elasticity}, the static mechanical response of generic isotropic materials can be depicted in the $\{G,\eta\}$ plane. The $G=0$ axis corresponds to fluids. The $\eta=0$ axis to non-dissipative ({\emph{e.g.} at zero temperature)  solids. The rest of the two dimensional space is spanned by viscoelastic materials. As we shall see, solids dual to massive gravity black branes of \cite{BaggioliSolid} do lie inside this plane.\\
We consider, as in \cite{BaggioliSolid} holographic models defined by the generic $3+1$ dimensional gravity theory:
\be\label{action}
S=\int d^4x\,\sqrt{-g}\left[\frac{1}{2}\left(R+\frac{6}{L^2}\right)-\frac{L^2}{4}F_{\mu\nu}F^{\mu\nu}-\frac{m^2}{L^2}\,V(X,Z)\right]+\int_{r\to0}d^3x\sqrt{-\gamma}\,K\;,
\ee
where $L$ is the AdS radius, $m$ is a dimensionless mass parameter, and
\begin{align}
X  \equiv \half \tr[\I^{IJ}]\,,\qquad  Z \equiv \det [\I^{IJ}]\,,\qquad \mathcal I^{IJ}\equiv\d_\mu \phi^I \d^\mu \phi^J\,, 
\end{align}
and the indices $I,J=\{x,y\}$ are contracted with $\delta_{IJ}$. In \eqref{action}, we have included the Gibbons-Hawking boundary term where $\gamma$ is the induced metric on the AdS boundary, and $K=\gamma^{\mu\nu}\nabla_\mu n_\nu$ is the extrinsic curvature with $n^\mu$ - an outward pointing unit normal vector to the boundary. Around the scalar fields background $\hat\phi^I=\delta^I_ix^i$ the metric admits the black brane background solution 
\be
\label{solmetric}
ds^2 =   L^2\left(\frac{dr^2}{f(r)r^2}+\frac{-f(r)dt^2+dx^2+dy^2}{r^2}\right) \;,
\ee
with the emblackening factor given in terms of the background value of the mass potential: 
\begin{equation}\label{f2phi}
f(r) = 1 +  \frac{\mu^2 r^4}{2 r_h^2}+m^2\,r^3 \, \int^r d\tilde r\frac{1}{\tilde r^4} \hat V(\tilde r)\; ,
\end{equation} 
where $\hat V(r) \equiv V(\hat X, \hat Z)$. The solution for the Maxwell field is $\hat A_t = \mu\left(1-r/r_h\right)$.\\
The viscoelastic response of the boundary theory in the holographic description is encoded in the transverse traceless tensor mode of the metric perturbations which obeys:
\be\label{spin2}
\left[f\partial_r^2 +\left(f'-2\frac{f}{r}\right)\partial_r + \left(\frac{\omega^2}{f} -4m^2 M^2(r)\frac{r^2}{L^2}\right)\right]h_\omega=0 \; .
\ee
where we have defined a mass function 
\begin{equation} \label{m_ten}
M^2(r) \equiv \frac1{2r^2} \hat V_X(r)\;.
\end{equation}
It is very important to emphasize here that the mass of the tensor mode is only due to the $X$ dependence of the potential $V(X,Z)$. Hence, in the case when $V$ is only a function of $Z$ the graviton remains massless. In our previous work we have argued that in the case when $V=V(Z)$ the dual theory describes \emph{fluids}, whereas the presence of an $X$ dependence, \emph{i.e.} when $V=V(X,Z)$, indicates that the material is a \emph{solid} \cite{BaggioliSolid}. We have also shown that there is no elastic response in the case of fluids. Moreover, since for fluids the graviton mass is zero, the universality proof \cite{IqbalMembrane}  for the viscosity to entropy ratio based on the membrane paradigm is applicable and we expect no violation of the KSS bound. Without loss of generality we therefore only consider the theories describing solids with graviton mass terms of the form
\be\label{vx}
V(X)=X^n\;.
\ee
Here we are allowing for general values of $n$ in order to see what is the impact of this parameter on the elasticity and viscosity. As already discussed this choice corresponds to a mass function of the form:
\begin{equation}
M^2(r)\,=\,\frac{1}{2\,L^2}\,\left(\frac{r}{L}\right)^\nu
\end{equation}
with $n=\frac{4\,+\,\nu}{2}$.\\
From the equation of motion \eqref{spin2} it follows that in the near-boundary region the metric perturbations $h\equiv h_\omega$ behave as
\be
\lim_{r\to 0}\,h=h_0+\left(\frac{r}{L}\right)^3h_3+\dots
\ee
showing that the scaling dimension of $h$ is $\Delta = 3$ and is independent on the radial dependence of the graviton mass. The gauge/gravity duality prescription then allows one to find the retarded Green's function as the ratio of the subleading to leading mode of the graviton:
\be\label{sublead}
{\cal G}^R_{T_{ij}\,T_{ij}}=\frac{2\Delta-d}{2L}\,\frac{h_3}{h_0}
\ee
where $d=3$ is the number of spatial dimensions. We numerically solve the equation of motion for the graviton and extract the retarded Green's function by using the above expression. From the latter we are able to compute the complete viscoelastic response of the dual CFT.
\begin{figure}
\center
\includegraphics[width=7.5cm]{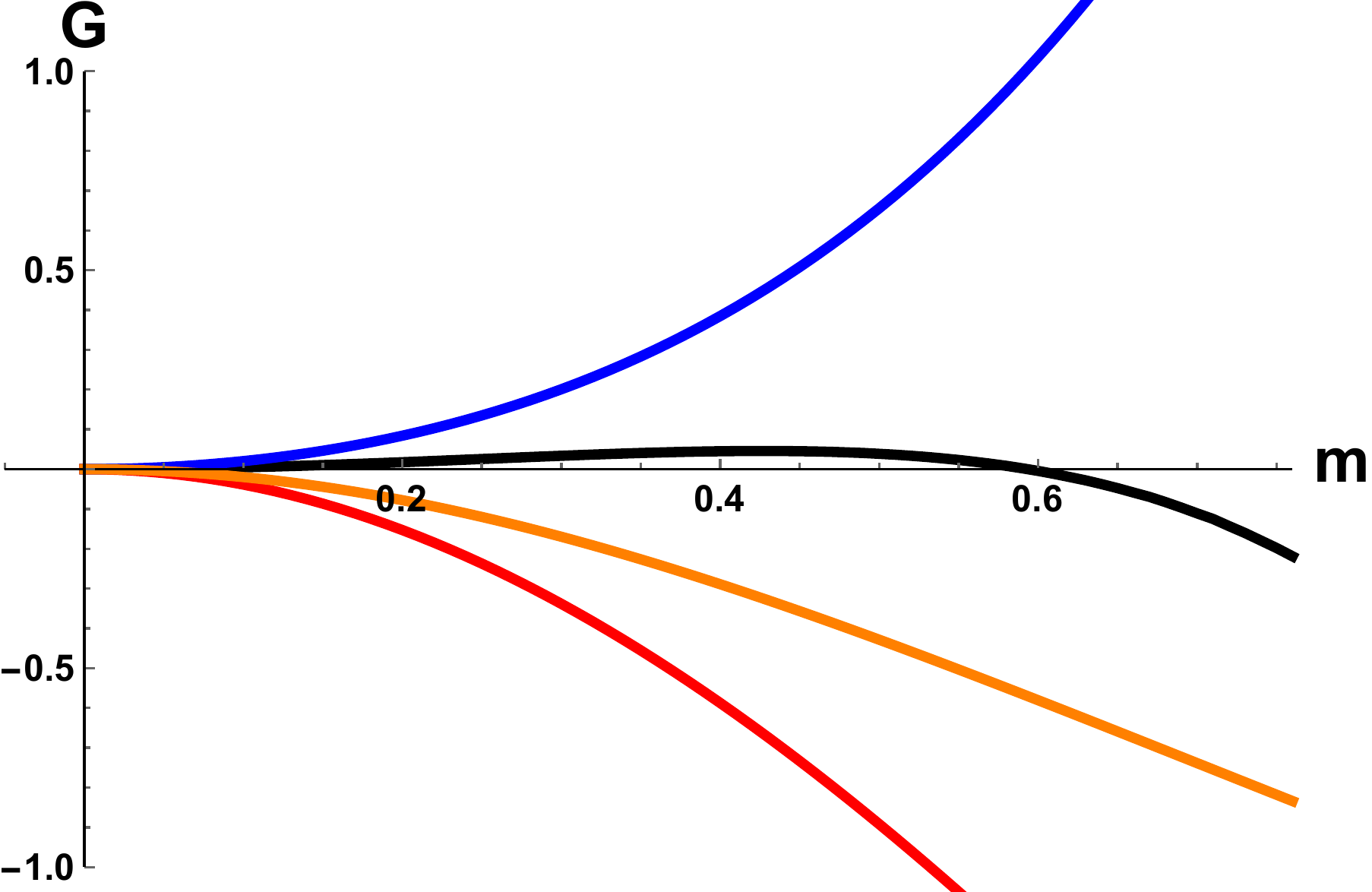}%
\qquad
\includegraphics[width=7.5cm]{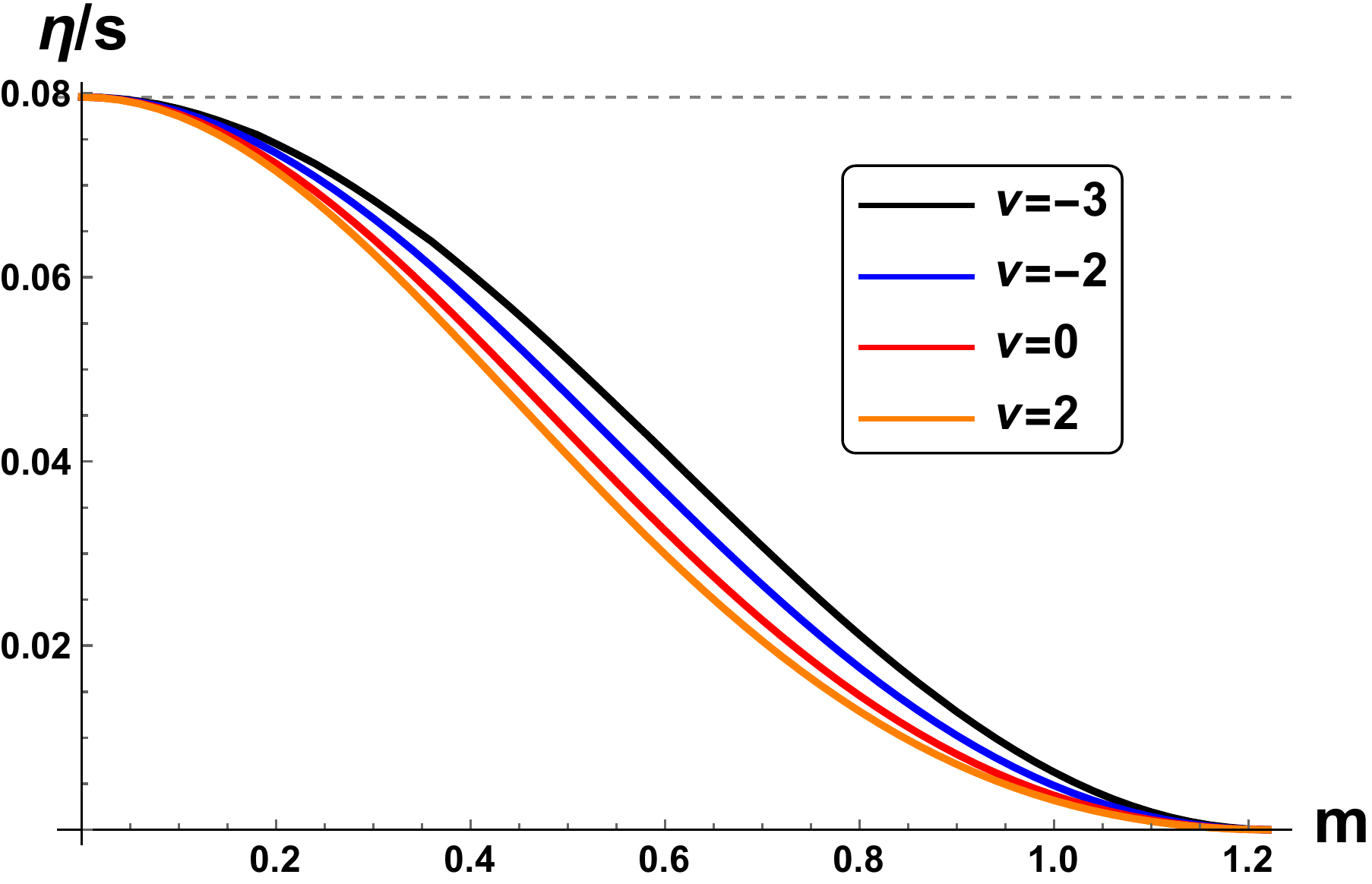}
\caption{Backreacted model $V(X)=X^n$ for different $n=\frac{4+\nu}{2}$, $\nu=-3,-2,0,2$. \textbf{Left:} Elasticity; \textbf{Right:} $\eta/s$ ratio, the horizontal dashed line shows the value $\eta/s=1/(4\pi)$. Plots from \cite{BaggioliViscosity}.}
\label{fig:main}
\end{figure}
In Fig. \ref{fig:main} we show the real part of the Green's function and the viscosity to entropy density ratio as a function of the graviton mass for different values of the exponent~$\nu$. We first observe that the $\eta/s$ ratio goes below the universal value $1/(4\pi)\approx 0.08$ for graviton mass parameter values $m>0$ and thus violates the KSS bound. As expected, in the fluid regime with $m=0$ we recover the standard universal value. The second observation that we make is that the real part of the Green's function becomes negative for all values of $\nu$ apart from $\nu=-2$. Although, negative modulus of elasticity can, in principle, be observed in nature it is always associated with instabilities. From the holographic perspective, the fact that there is an instability is not so surprising because the kinetic terms for the St\"uckelberg fields are non-canonical for $V(X)= X^n$ with $n>3/2$ and $n=1/2$ (corresponding to $\nu>-1$ and $\nu=-3$ respectively). Both the numerical and analytical results give a positive rigidity modulus for the canonical St\"uckelberg case,  $n=1$ ($\nu=-2$) with $V=X$, which can therefore be singled out as the most reasonable model from the phenomenological point of view.

We would like to point out that the fact that the KSS bound can be violated in theories with massive gravity duals was also noticed in \cite{DavisonGouteraux} for the case $\nu=-2$ corresponding to $V= X/2$ and $\mu=0$. It was then argued by the authors that this result is irrelevant for the physical viscosity due to the fact that for graviton masses of order $m/T\gtrsim 1$ the dual field theory does not admit a coherent hydrodynamic description. Instead a crossover from the coherent hydrodynamic phase of the system to an incoherent regime occurs for graviton mass that is comparable to the black brane temperature. In the results presented in this paper we see the violation of the KSS bound also at arbitrary small values of the graviton mass where the hydrodynamic description applies \footnote{See \cite{nicketa} for a detailed hydrodynamical analysis of the $\eta/s$ violation in holographic theories with momentum dissipation.}. We therefore believe that our findings are physically significant and suggest that the KSS bound can be violated in materials with non-zero elastic response. In general, however, we find that the question of whether or not the black branes are close to having a hydrodynamic description is not particularly relevant in the context of holographic solids. In these systems we do not expect the dynamics to be understood in terms of hydrodynamics while there does exist a well defined low energy effective field theory description of solids defined as an expansion at low frequencies and momenta.\\
As an additional remark, let's note (see fig.\ref{etapicT}) that in the extremal limit $T=0$ the viscosity-entropy bound is null. This was also observed and discussed in details in \cite{Hartnolleta} where the value of $\eta/s$ at $T=0$ has been related to the nature of the momentum dissipating sector in the deep IR. To be more precise, the graviton mass we consider does not vanish at the extremal horizon such that the momentum dissipation mechanism is still active in that limit leading to a vanishing KSS ratio. On the contrary at temperatures $T \gg m$ we expect the presence of a graviton mass to be completely irrelevant and indeed we recover the universal value $1/4\pi$ ((see fig.\ref{etapicT}).\\
Note also that the violation of the KSS bound in this framework is definitely stronger than the higher derivative case because no further lower bound appear at all: the $\eta/s$ ratio can go down to 0!
\begin{figure}[h!]
\center
\includegraphics[width=9cm]{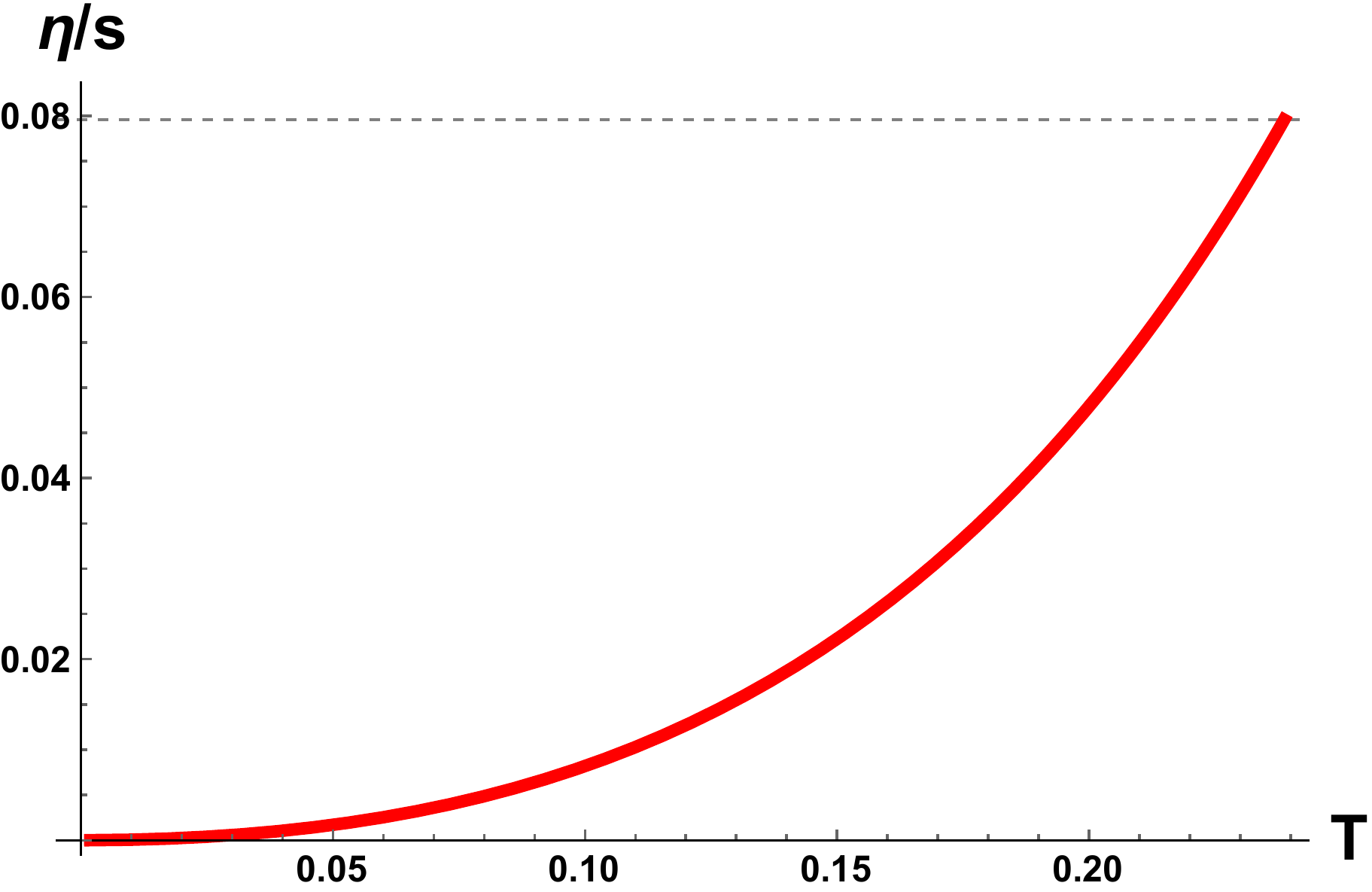}
\caption{$\eta/s$ in function of temperature T for the benchmark model $V(X)=X$. Note that is totally generically in our theory to have $\eta/s(T=0)=0$. Dashed line is the universal KKS value $1/4\pi$.}
\label{etapicT}
\end{figure}
We can analyze further the situation exploiting perturbative methods which allow us to compute at small graviton mass the value of G and $\eta$  analytically. What we find out (see \cite{BaggioliViscosity} for details) is that at low momentum dissipation those two quantities take the form:
\begin{align}
&G\,=\,\frac{L^2}{2\,r_h^3}\,c_\nu\,m^2\,+\,\dots\nonumber\\
& \frac{\eta}{s}\,=\,\frac{1}{4\,\pi}\left(1+\frac{2}{3}\,c_\nu\,m^2\,\mathcal H_{\frac{1}{3}(\nu+1)}\,+\,\dots\right)
\end{align}
with $c_\nu=-\frac{2}{\nu+1}\,\left(\frac{r_h}{L}\right)^{4+\nu}$\footnote{The case $\nu=-1$ is a particular one and it has to be treated separately.} and $\mathcal H_p$ the p-th harmonic number.
The numerical results for the real part of the Green's function and for the $\eta/s$ ratio are in good agreement with these analytic expressions for small values of the graviton mass parameter m.\\
In conclusion we have seen a violation of the KSS bound $\eta/s\geq 1/4\pi$ in holographic massive theory of the solid type. We suspect this violation to be correlated with the presence of a non zero shear elastic modulus G. Importantly the violation we found is not directly connected with the breaking of translational symmetry of the dual CFT because for fluid type HMG theories the bound is fully satisfied even if momentum is dissipating.\\
Given the putative connection between the elastic response and the violation of the KKS bound we provided it would be extremely interesting to perform experimental measures of the $\eta/s$ ration in viscoelastic materials.
\begin{figure}[h]
\center
\includegraphics[width=7.5cm]{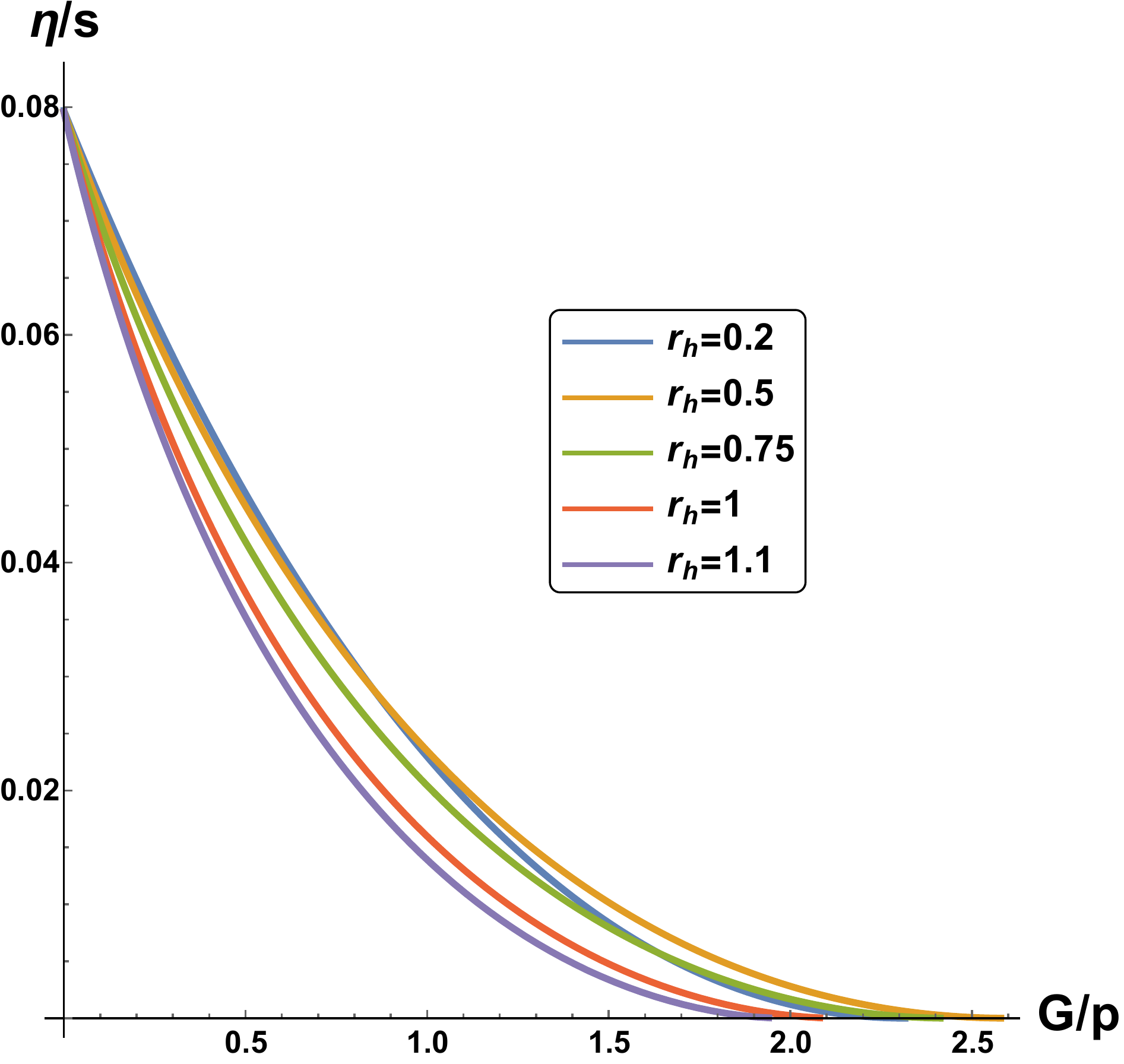}
\caption{Viscosity--elasticity diagram for the $V(X)=X$ model. The value of the graviton mass and the temperature are changing along the solid lines, with $m=0$ on the axes where $\eta/s=1/4\pi\approx 0.08$ and $T=0$ on the axes $\eta/s=0$ where the rigidity G takes its maximum value.}
\label{fig:diagram}
\end{figure} 
In addition, in the spirit of the KSS conjecture, one can also wonder whether or not there is any generalization of it that holds in solid systems. From dimensional analysis it is reasonable to expect that if there does exist a more general bound, it should involve the rigidity to pressure ratio, $G/p$, in addition to the $\eta /s$ ratio. 
In Fig.~\ref{fig:diagram} we plot $\eta/s$ against $G/p$ for the holographic solid with $\nu=-2$ and see a clear correlation. 
Keeping the KSS logic that the gravity solutions might represent the least dissipative materials, the Fig.~\ref{fig:diagram} then suggests that there might be a more general bound in (viscoelastic) solids. At relatively large temperatures  this would approximately take the form   
$$4\pi\frac{\eta}{s} + \mathcal{C}  \frac{G}{p} \lesssim 1$$ with  $\mathcal{C}$ being an order-one constant. Another possibility for a generalized KSS bound has been proposed in \cite{Hartnolleta} in connection with the rate of entropy production encoded by the viscosity $\eta$.
\section{Superconductors with momentum dissipation}
\textbf{Metal-SC phase transitions}\\[0.3cm]
The holographic superconductors model \cite{holSC1,holSC2} is one of the first and main result of the AdS/CMT program. Within its simplicity it describes a system which exists in two states: a superconducting state which has a nonvanishing charge condensate, and a normal state which is a perfect conductor. As a direct conseguence, already in the normal phase the static electric response, namely the DC conductivity ($\omega=0$), is infinite. This is a straightforward consequence of
the translational invariance of the boundary field theory, which leads to the fact that
the charge carriers do not dissipate their momentum, and accelerate freely under an
applied external electric field. This fact represents a shortcoming of the model which has to be cured in order to have a realistic metallic normal state with finite electric DC conductivity clearly distinguishable from the infinite one in the superconducting phase.\\
We \cite{BaggioliSC1} therefore introduce HMG theories into the original SC model \cite{holSC1,holSC2} to take care of such a lack  and to study the effects of disorder-driven momentum dissipation on the main features of the SC phase transitions, such as the critical temperature $T_c$ and the value of the charged condensate $\langle \mathcal{O} \rangle$.\\
The total action of our model is :
\beq
I=I_1+I_2+I_3\,,\label{Itotb}
\eeq
where we have denoted the Einstein-Maxwell terms $I_1$,
the neutral scalar terms $I_2$, and the charged scalar terms $I_3$;
\begin{align}
I_1&=\int d^{d+1}x\,\sqrt{-g}\,\left[R-2\Lambda-\frac{L^2}{4}F_{\mu\nu}F^{\mu\nu}\right]\,,\notag\\
I_2&=-2m^2\int d^{d+1}x\,\sqrt{-g}\,V\left(X\right)\,,\notag\\
I_3&=-\int d^{d+1}x\,\sqrt{-g}\,\left(|D\psi|^2+M^2|\psi|^2+\kappa\, H\left(X\right)\,|\psi|^2\right)\,.
\label{azione}
\end{align}
and we consider the following MG potentials as benchmark examples\footnote{\ref{model1} has been already studied in the context of holographic SC in \cite{AndradeSC,KimSC1}. We have inserted an additional coupling $m^2$ in front of the potential $V(X)$ which is going to be redundant for the monomial cases \ref{model1} and \ref{model3} where we decided in fact to reabsorb it into the definition of $V(X)$. In this way for those cases we are left with just one parameter $\alpha$ which is going to represent the disorder-strength in the system. In the case of the polinomial potential \ref{model2} $m^2$ is going to be an independent parameter in addition to $\alpha$.}
\begin{align}
{\bf model\;\; 1\;:}\qquad V(X)&=\frac{X}{2\,m^2}\,,\label{model1}\\
{\bf model\;\; 2\;:}\qquad V(X)&=X+X^5\label{model2}\\
{\bf model\;\; 3\;:}\qquad V(X)&=\frac{X^N}{2\,m^2}\,,\qquad N\neq 1\label{model3}
\end{align}
We have defined
\begin{equation}
X=\frac{1}{2}\,L^2\,g^{\mu\nu}\p_\mu\phi^I\p_\nu\phi^I\,.
\end{equation}
We denote $D_\mu\psi =(\p_\mu-i\,q\,A_\mu)\psi$ to be the standard covariant derivative of the scalar $\psi$
with the charge $q$.\\
The generic ansatz we consider is given by:
\begin{align}
&ds^2=L^2\left(-\frac{1}{u^2}f(u)e^{-\chi(u)}dt^2+\frac{1}{u^2}(dx^2+dy^2)+\frac{1}{u^2f(u)}du^2\right)\notag\\
&\phi^I=\alpha\, \delta^I_i\,x^i\,,\,\,\,\,\,\,\,\,\,\,\quad I,i=x,y\,.\label{ansb}\\
&A=A_t(u)du\,,\qquad \psi=\psi (u)\,.\notag
\end{align}
where $\psi$ can be taken to be real-valued.\\
Within this ansatz the EOMs coming from the generic action \eqref{azione} read:
\begin{align}
&\frac{q^2\,u\,e^\chi\, A_t^2\,\psi^2}{f^2}-\,\chi'+u\,\psi'^2=0\label{sEinst}\\
&\psi'^2-\frac{2\,f'}{u\,f}+\frac{e^\chi u^2 A_t'^2}{2 f}+\frac{M^2 L^2 \psi^2}{u^2 f}+\frac{\kappa\,L^2\, H \psi^2}{u^2 f}+\frac{e^\chi q^2 A_t^2 \psi^2}{f^2}\notag \\&+\frac{2 m^2 L^2 V}{u^2 f}+\frac{2\Lambda L^2}{u^2 f}+\frac{6}{u^2}\,=0\label{lEinst}\\
&\frac{2 q^2  A_t \psi^2}{u^2 f}-\frac{\chi'}{2}A_t'-A_t''=0\label{Maxb}\\
&\psi''+\left(-\frac{2}{u}+\frac{f'}{f}-\frac{\chi'}{2}\right)\psi'
+\left(\frac{e^\chi q^2 A_t^2}{f^2}-\frac{M^2 L^2}{u^2 f}-\frac{\kappa H\,L^2 }{u^2 f}\right)\psi\,=0\label{scfeq}
\end{align}
and the Hawking temperature of the black brane (\ref{ansb}) is given by:
\beq
T\,=\,-\,\frac{f'(u_h)}{4\pi}\,e^{-\frac{\chi(u_h)}{2}}\,.\label{Tdef}
\eeq
\begin{figure}
\centering
\includegraphics[width=.35\textwidth]{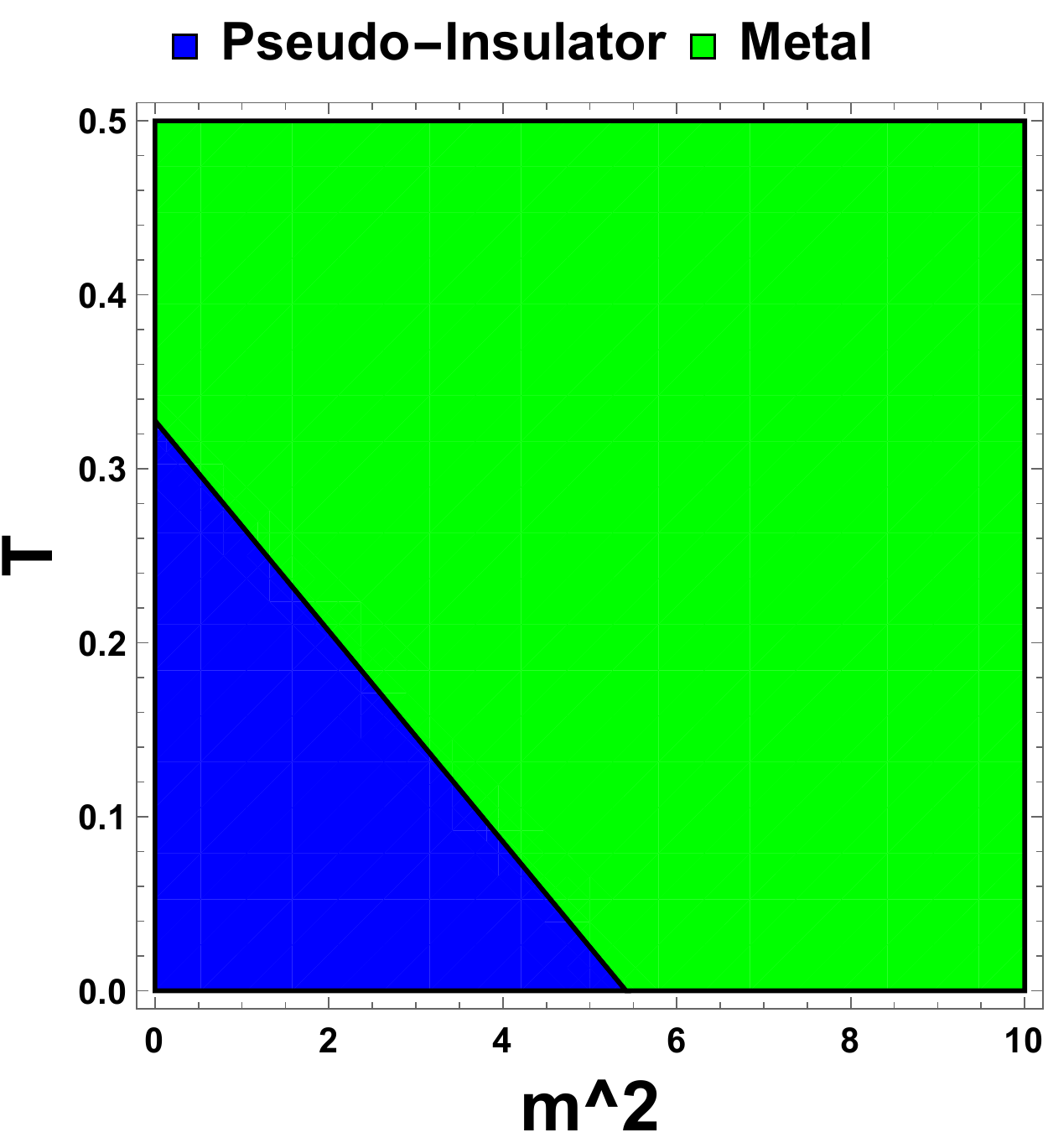}
\includegraphics[width=.35\textwidth]{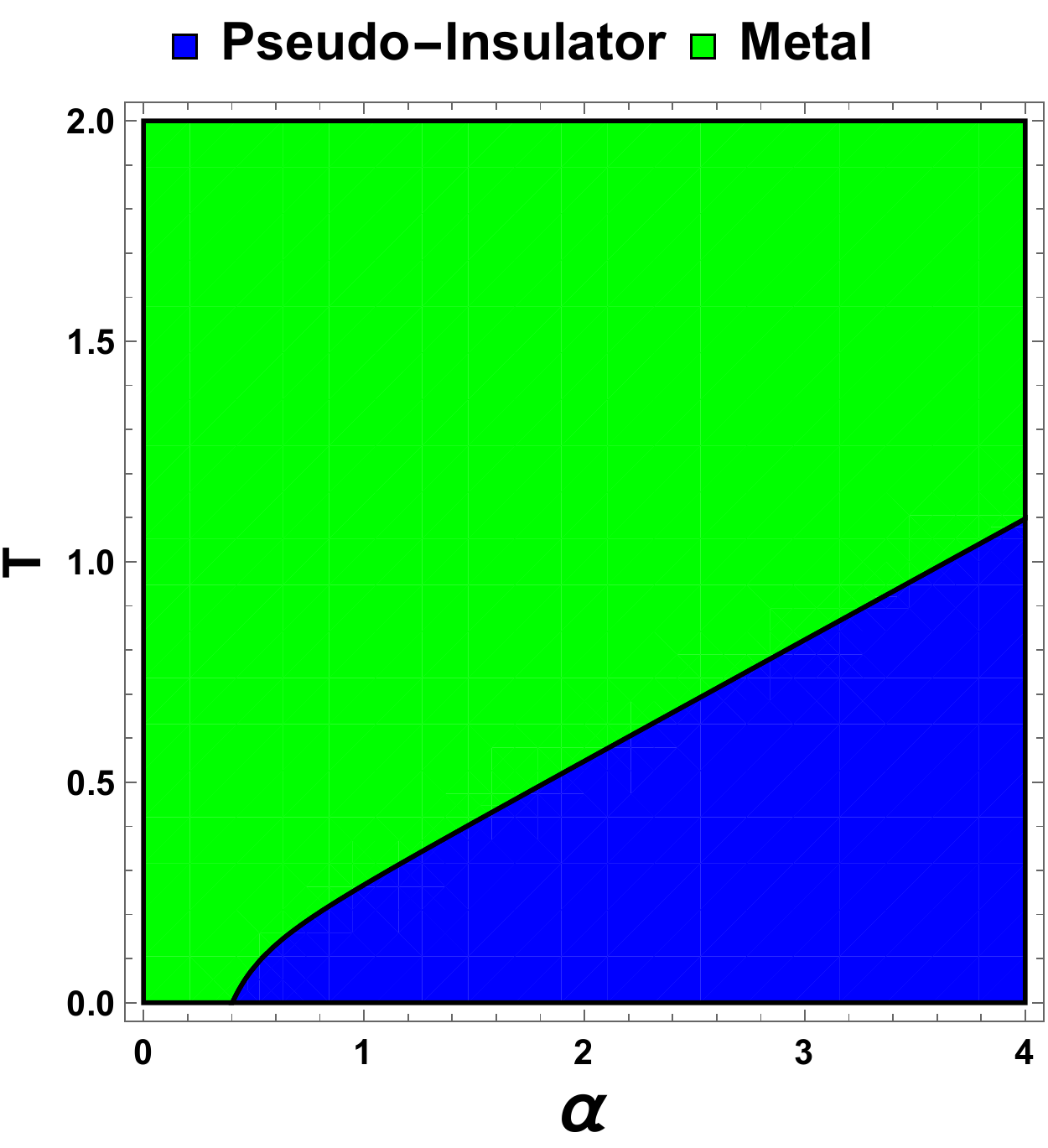}
\caption{Region plots for the model (\ref{model2}) in the normal phase. We choose
units where the density is $\rho=1$. Here we have fixed $\alpha=1$ (left plot) and $m=1$ (right plot).
The blue region is pseudo-insulating, $d\sigma_{DC}/dT>0$,
the green region is metallic, $d\sigma_{DC}/dT<0$.}
\label{NormalPhase}
\end{figure}
In the case of a non-trivial condensate $\psi(u)$ it is in general impossible to solve the
background equations of motion analytically. However, when $\psi(u)=0$, \textit{i.e.} the normal phase, the solution is known and it was given in \cite{BaggioliPRL}. The correspondent DC electric conductivity is finite and takes the value:
\begin{equation}
\sigma_{DC}=\frac{1}{e^2}\left(1+\frac{\rho^2\,u_h^2}{2\,m^2\,\alpha^2\,\dot V(u_h^2\,\alpha^2)}\right)\,.
\label{DCformula}
\end{equation}
With this formula at disposal we can already draw down the phase diagram of the normal phase accordingly to the following definitions:
\begin{itemize}
\item $d\sigma_{DC}/dT<0$ : metal
\item $d\sigma_{DC}/dT>0$ : pseudo-insulator\footnote{We use the ''pseudo'' prefix to make explicit the fact that we do not have $\sigma_{DC}(T=0)=0$ in this phase.}
\end{itemize}
The phase diagram of this normal phase is already rich and can give insights towards the interpretation about the various ingredients introduced into the model. In the case of the linear Lagrangian, which goes back to the original model \cite{AndradeSC}, the parameters $m$ and $\alpha$ are combined into $m\,\alpha$,
which can be interpreted as the strength of translational symmetry breaking. From the dual field theory point of view this is thought to be related to some sort of homogeneously distributed density of impurities,
representing the disorder-strength in the material.\\
\begin{figure}
\centering
\includegraphics[width=.35\textwidth]{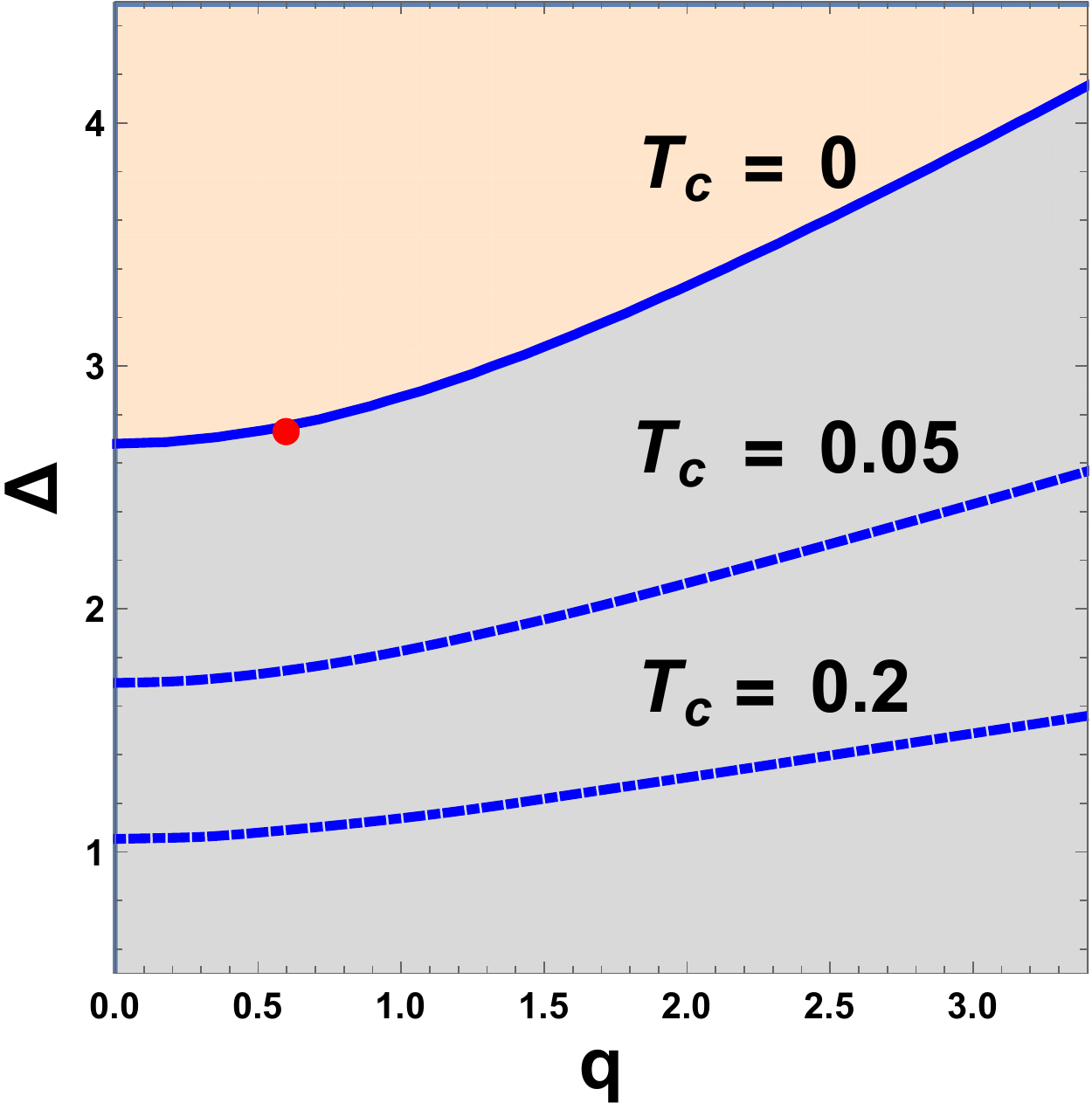}%
\qquad
\includegraphics[width=.35\textwidth]{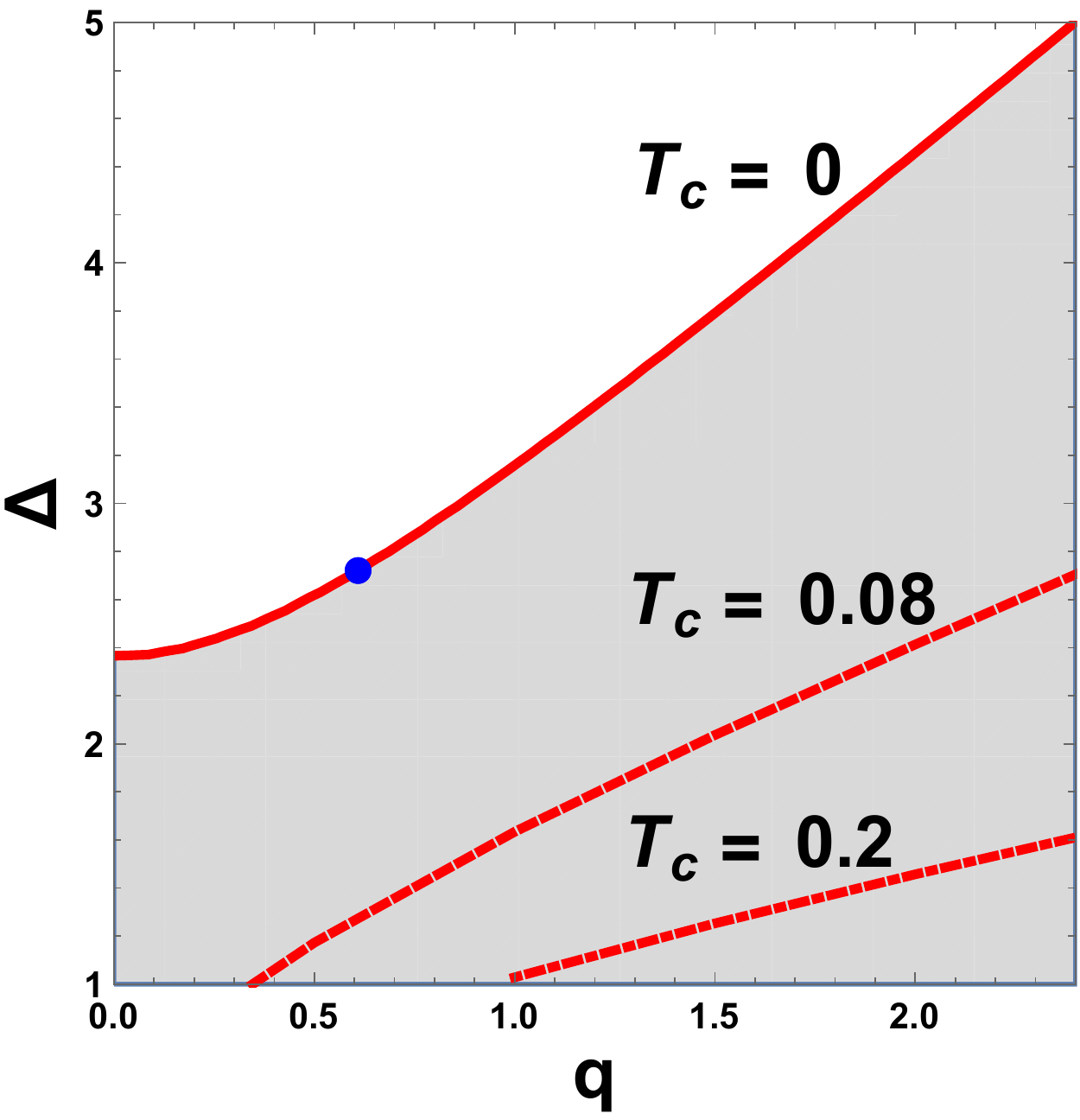}
\caption{\textbf{Left: }Region and contour plots for the model (\ref{model1}) with linear
potential for the neutral scalars.
The region of $\Delta$, $q$, satisfying the IR instability condition (\ref{domez})
is shaded in grey. The red dot is centered around
$(q_d,\Delta_d)=(0.6,\,2.74)$. These tuned $(q,\Delta)$
confine superconducting phase of the model (\ref{model1})
into a dome region. Notice the proximity
of the red dot to the boundary of the IR instability region, resulting 
in $T_c(q_d,\Delta_d)$ being very small. \textbf{Right: }Region and contour plots for the model (\ref{model2}) with non-linear potential for some representative parameters.}
\label{TcConsplot}
\end{figure}

In the case of a more general $V(X)$, the $m$ parameter keeps this kind of interpretation while the $\alpha$ one represents the strength of interactions of the neutral scalar sector. This reasoning is confirmed by the study of the phase diagrams of the system (figure \ref{NormalPhase}) which makes evident the difference between the two parameters. Indeed, while the $m$ parameter, which we are going to interpret as the disorder-strength of our High-Tc superconductor, enhances the metallic phase, the $\alpha$ one clearly reduces the mobility of the electronic sector driving the system towards the pseudo-insulating phase.\\[0.2cm]
The normal phase we just described is unstable towards the development of a non-trivial profile of the charged scalar field. This allows one to determine a line of the second order superconducting phase transition, $T_c(\alpha,m)$, in the boundary field theory, with broken translational symmetry.
We start by considering the system at zero temperature, which we are able to study analytically. 
Then we proceed to studying the normal phase at a finite temperature. Upon lowering the temperature, at a certain critical value $T=T_c$, the normal phase becomes unstable. This is the point of
a superconducting phase transition.
We construct numerically $T_c$ as a function of the parameters $\Delta$, $q$, $\alpha$ (or $m$),
for the models with various $V(X)$.

In the case of $T=0$ the normal phase geometry interpolates
between the $AdS_4$ in the ultra-violet and the $AdS_2\times \mathbb{R}^2$
in the infra-red. We can apply the known analytical calculation to study the stability
of the normal phase towards formation of a non-trivial profile of the scalar $\psi$.

Due to eq. (\ref{scfeq}), the effective mass $M_{eff}$ of the scalar $\psi$ is given by:
\beq
M_{eff}^2\,L^2 =\,M^2\,L^2+\kappa\,H\,L^2+q^2\,A_t^2\, g^{tt}\,L^2\,.\label{psiHmass}
\eeq
Notice that at the boundary the mass of the scalar is just $M^2$ but at the horizon it gets an additional contribution.
The normal phase is unstable towards formation of the scalar hair, if $M_{eff}$ violates 
the BF stability bound in the $AdS_2$, namely:
\beq
M_{eff}^2\,L^2_{2}< -\frac{1}{4}\,.\label{IRinst}
\eeq
In (\ref{IRinst}) we have denoted the $AdS_2$ radius as $L_2$.\\
All in all, the IR instability condition (\ref{IRinst}) finally reads:
\begin{equation}
D<0\,,\label{domez}
\end{equation}
where we have defined the function $D$ as:
\beq
\label{Ddef}
D=\frac{1}{4}+\frac{L^2 \left(\kappa  H+M^2\right) \left(L^2 m^2 \left(\alpha ^2 u_h^2 \dot V\left(\alpha ^2 u_h^2\right)-2 V\left(\alpha ^2 u_h^2\right)\right)+6\right)-q^2 \rho ^2 u_h^4}{\left(L^2 m^2 \left(\alpha ^2 u_h^2 \dot V\left(\alpha ^2 u_h^2\right)-2 V\left(\alpha ^2 u_h^2\right)\right)+6\right)^2}
\eeq
Consider the system at large temperature in a normal phase, which exists in a superconducting
phase at low temperatures. Therefore as we decrease the temperature, at certain critical value $T_c$ the superconducting phase transition occurs. If $T_c$ is non-vanishing, then for $T<T_c$ the system is in a superconducting phase, with a non-trivial scalar condensate $\psi(u)$. \\
Recall that near the boundary the scalar field with mass $M$:
\beq
M=\frac{1}{L}\sqrt{\Delta(\Delta-3)}\,,
\eeq
behaves asymptotically as:
\beq
\psi(u)=\frac{\psi_1}{L^{3-\Delta}}u^{3-\Delta}+\frac{\psi_2}{L^\Delta}u^\Delta\,,\label{nbexp}
\eeq
where $\psi_1$ is the leading term, identified as the source in the standard quantization.\\
Near the second order phase transition point $T=T_c$ the value of $\psi$ is small, and therefore one can neglect its backreaction on the geometry. 
The SC instability can be detected by looking at the motion of the QNMs of $\psi$ in the complex plane. To be more specific, it corresponds to a QNM going to the upper half of
a complex plane. Exactly at the critical temperature we have a static mode at the origin of the complex plane, $\omega=0$, and the source at the boundary vanishes, $\psi_1=0$.
We will solve numerically the equations for the whole background, and confirm this explicitly.\\
The scalar field is described by eq. (\ref{scfeq}), which in the normal phase becomes:
\beq
\psi''+\left(-\frac{2}{u}+\frac{f'}{f}\right)\psi'+\left(\frac{q^2\rho^2}{f^2}-\frac{M^2 L^2}{u^2 f}-\frac{\kappa H\,L^2 }{u^2 f}\right)\psi\,=0\,,\label{scfeqn}
\eeq
where $f(u)$ is the emblackening factor of the BH solution.\\
To determine the critical temperature $T_c$ we need to find the {\it highest} temperature,
at which there exists a solution to eq. (\ref{scfeqn}), satisfying the $\psi_1=0$ condition.
In this case for $T<T_c$ the system is in a superconducting state, with a non-vanishing condensate $\psi_2$.\\[0.2cm]
With these tools, we are able to describe completely the behaviour of the critical temperature $T_c$ in function of the various parameters of the system.\\
In figure \ref{TcConsplot} we plot the IR instability region on the $(\Delta,q)$ plane,
for the model~1, (\ref{model1}) and the model~2, (\ref{model2}) for some representative parameters.
The $T_c=0$ line is in perfect agreement with the analytical BF argument while the rest of the plane is built via the numerical routine described above. Two clear statements, which are generic for all the models we considered, can be extracted from those plots:
\begin{itemize}
\item Increasing the charge $q$ enhances the SC instability;
\item Increasing the conformal dimension $\Delta$ of the charged scalars on the contrary disfavoures it.
\end{itemize}
\begin{figure}
\centering
\includegraphics[width=.28\textwidth]{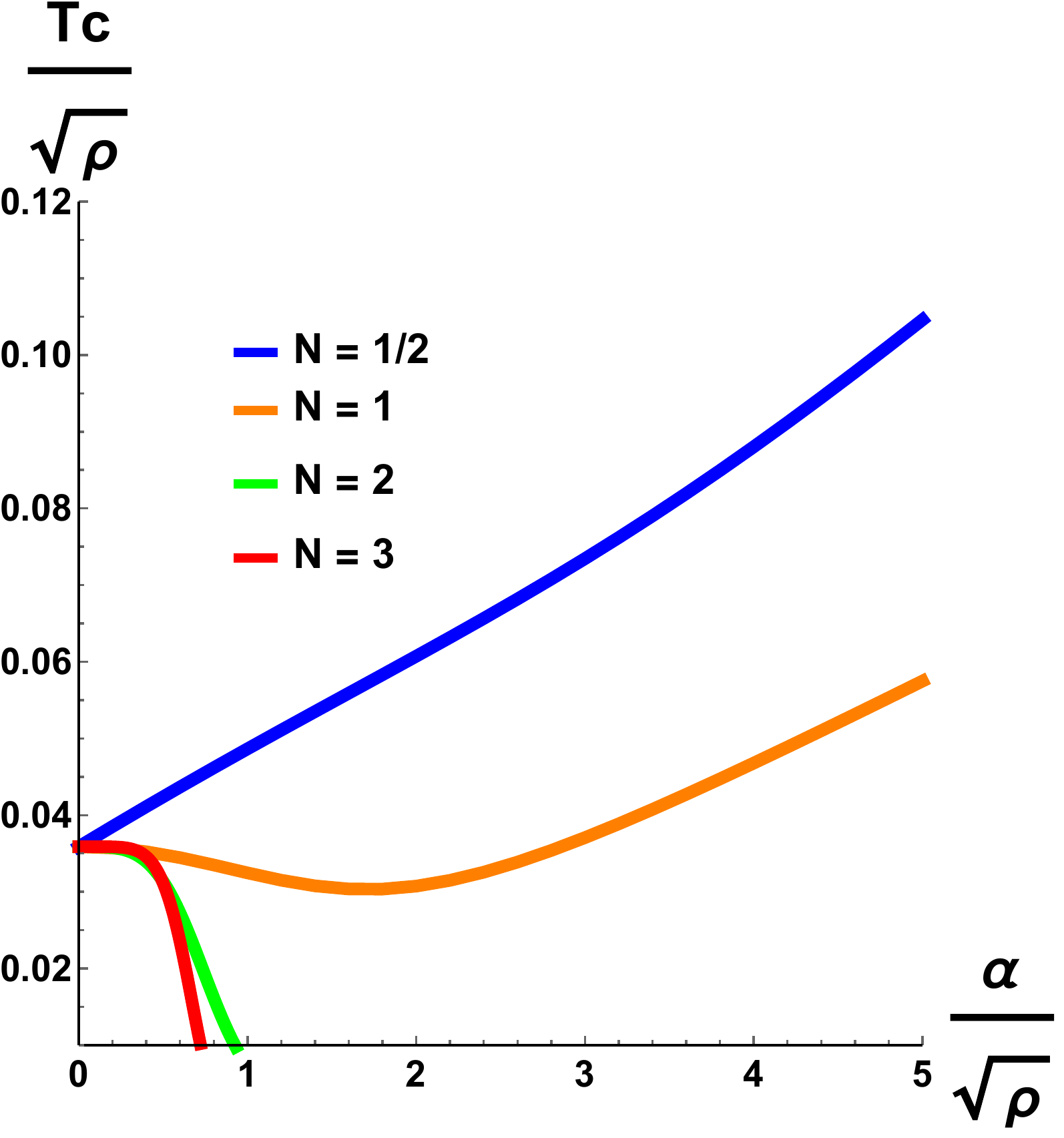}\qquad
\includegraphics[width=.28\textwidth]{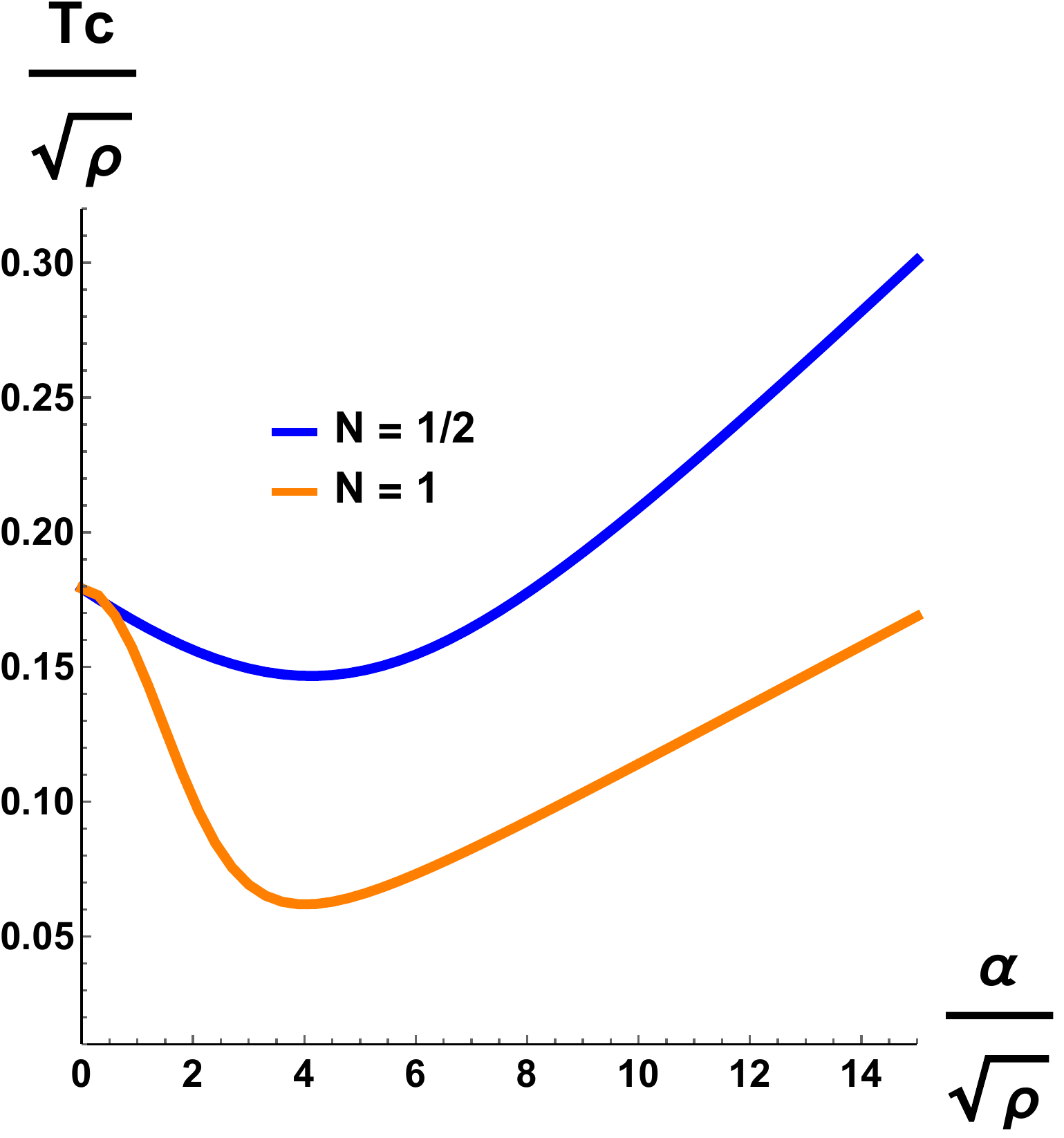}\qquad 
\includegraphics[width=.30\textwidth]{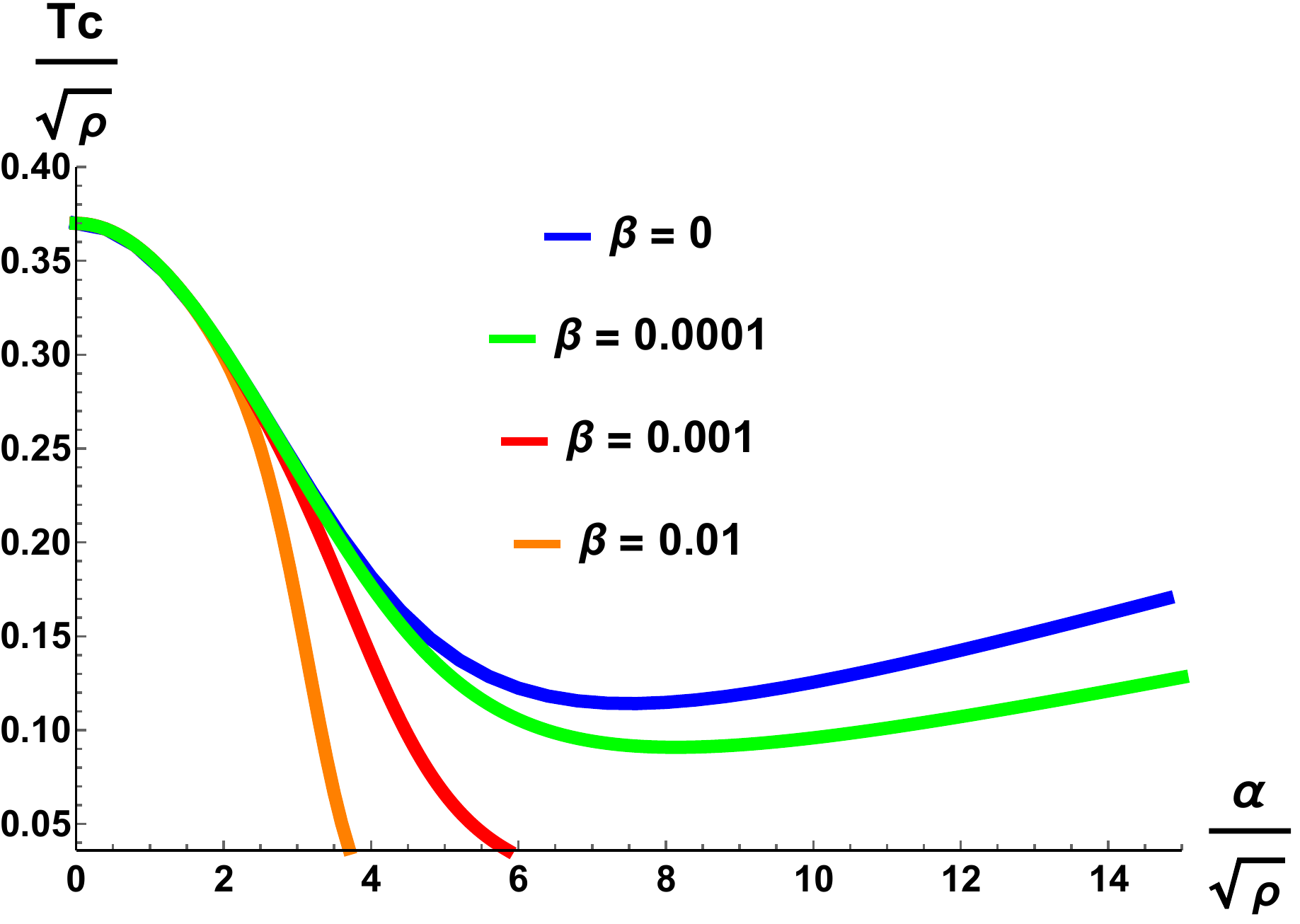}
\caption{Critical temperature as a function of $\alpha$ for:  \textbf{Left:} Model \ref{model3}
$q=1$, $\Delta=2$. \textbf{Centre:} Model \ref{model3} $q=3$, $\Delta=2$. \textbf{Right:}$V(X)= X/2m^2+\beta\,X^5/2m^2$ for different choices $\beta$. All the curves have a runaway behavior at $\alpha\rightarrow\infty$,
and only the shape depends on the value of $\beta$.}
\label{figTc}
\end{figure}
In addition, the behaviour of the critical temperature in function of the disorder strength, \textit{i.e.} graviton mass, is shown for the various models in fig.\ref{figTc}. The results are pretty curious because despite momentum dissipation disfavoures the SC instability, decreasing the critical temperature $T_c$, at its small values the behaviour then changes drastically showing that increasing further the graviton mass one can enhance the formation of the SC phase. We do not have a clear interpretation of this fact which is anyway observed also in other models with momentum dissipation such as \cite{AndradeSC}. One clear result is that non linearities in the MG potential do decrease in a generic way the critical temperature $T_c$ of the SC transition.\\[0.2cm]
We can then construct in a full fashion the complete SC solution and check via Free Energy analysis that the SC transition indeed appears and it does as a 2nd order phase transition. We skip the details of such procedure for which one can read \cite{BaggioliSC1}.\\
Additionally we can compute, using the standard holographic method, the electric optical conductivity of the system across the SC transition. The results are shown in fig. \ref{ACSC} for the non linear MG model \ref{model2}.
\begin{figure}
\centering
\includegraphics[width=.28\textwidth]{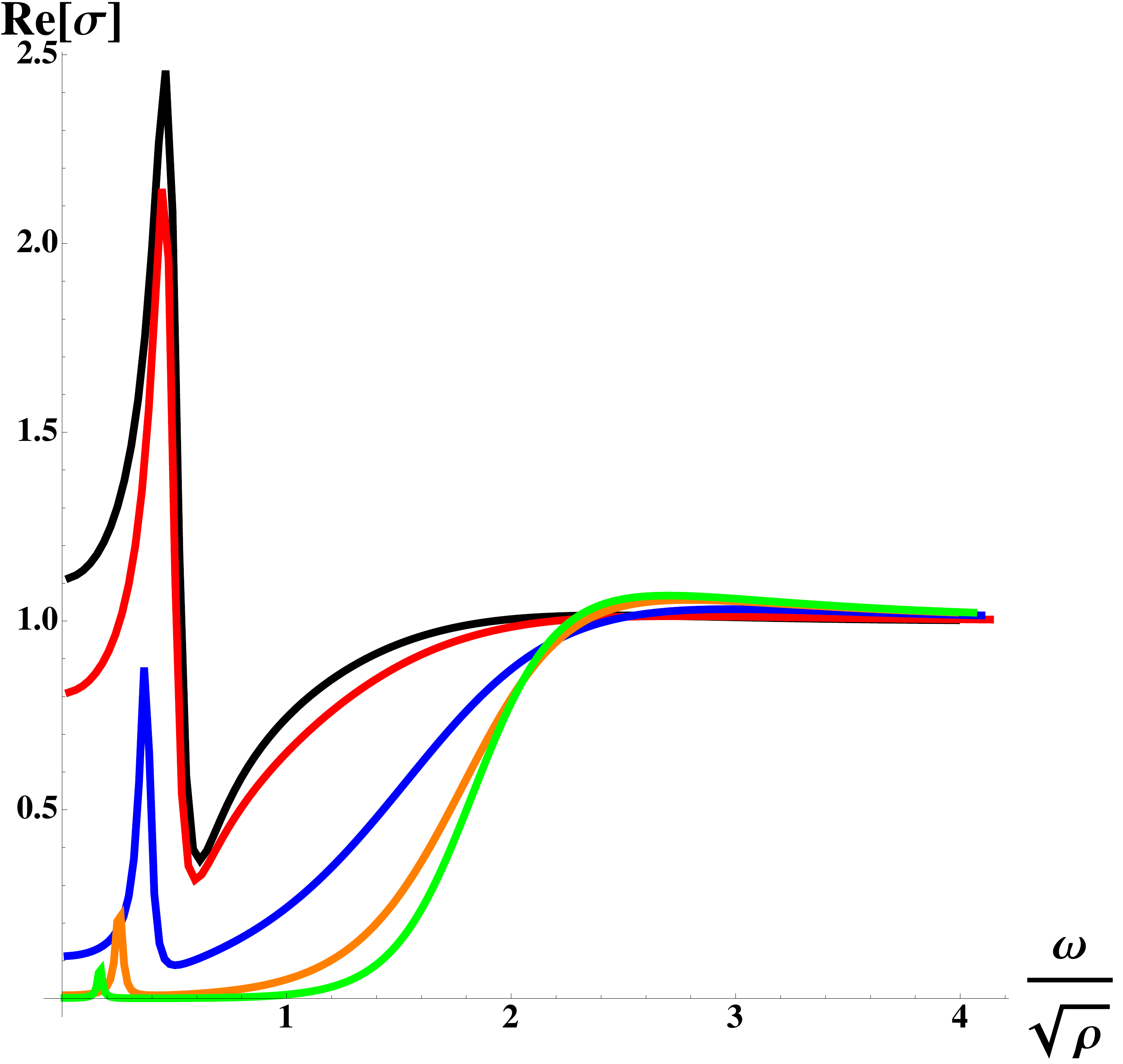}\qquad
\includegraphics[width=.28\textwidth]{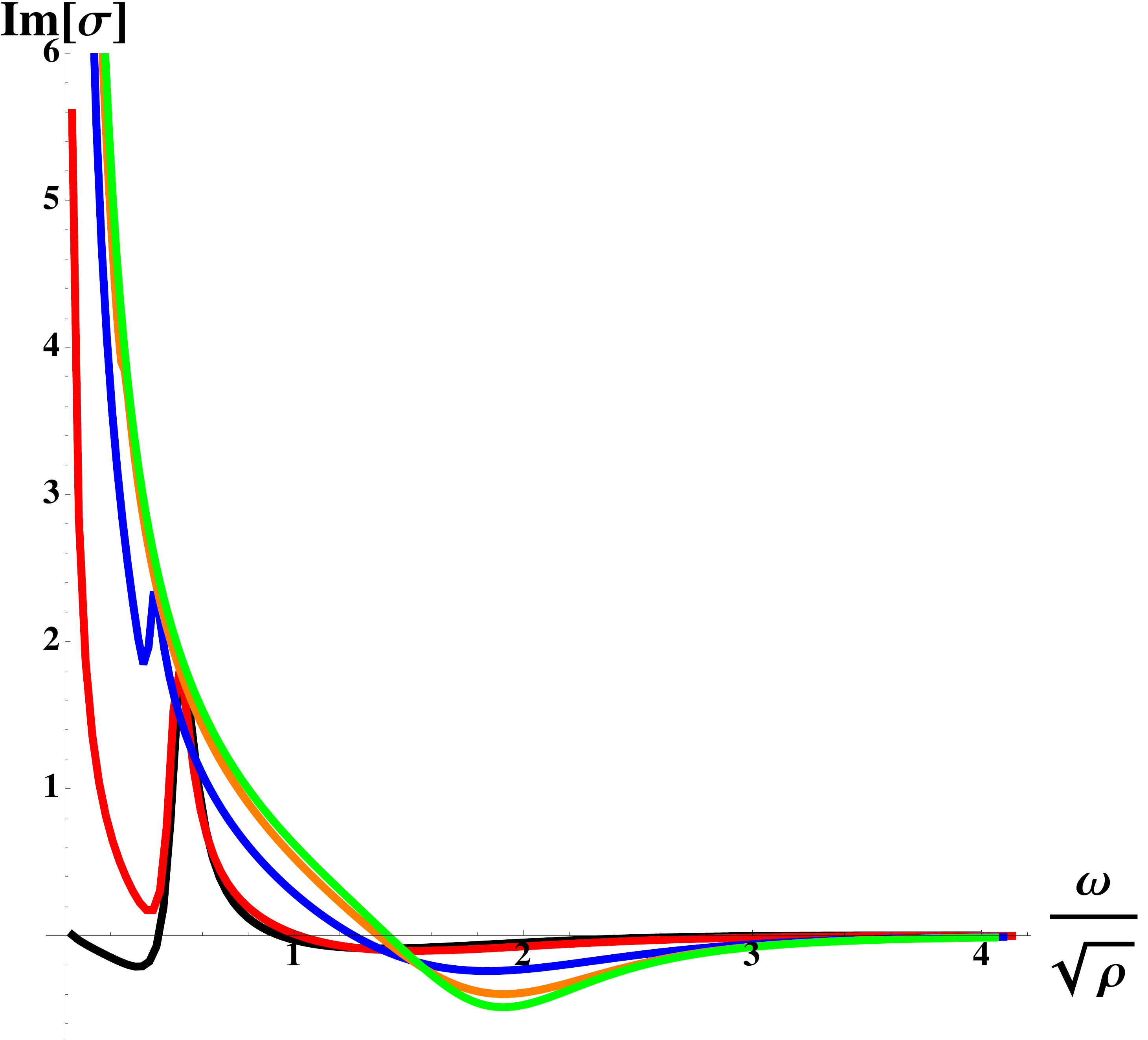}\qquad
\includegraphics[width=.28\textwidth]{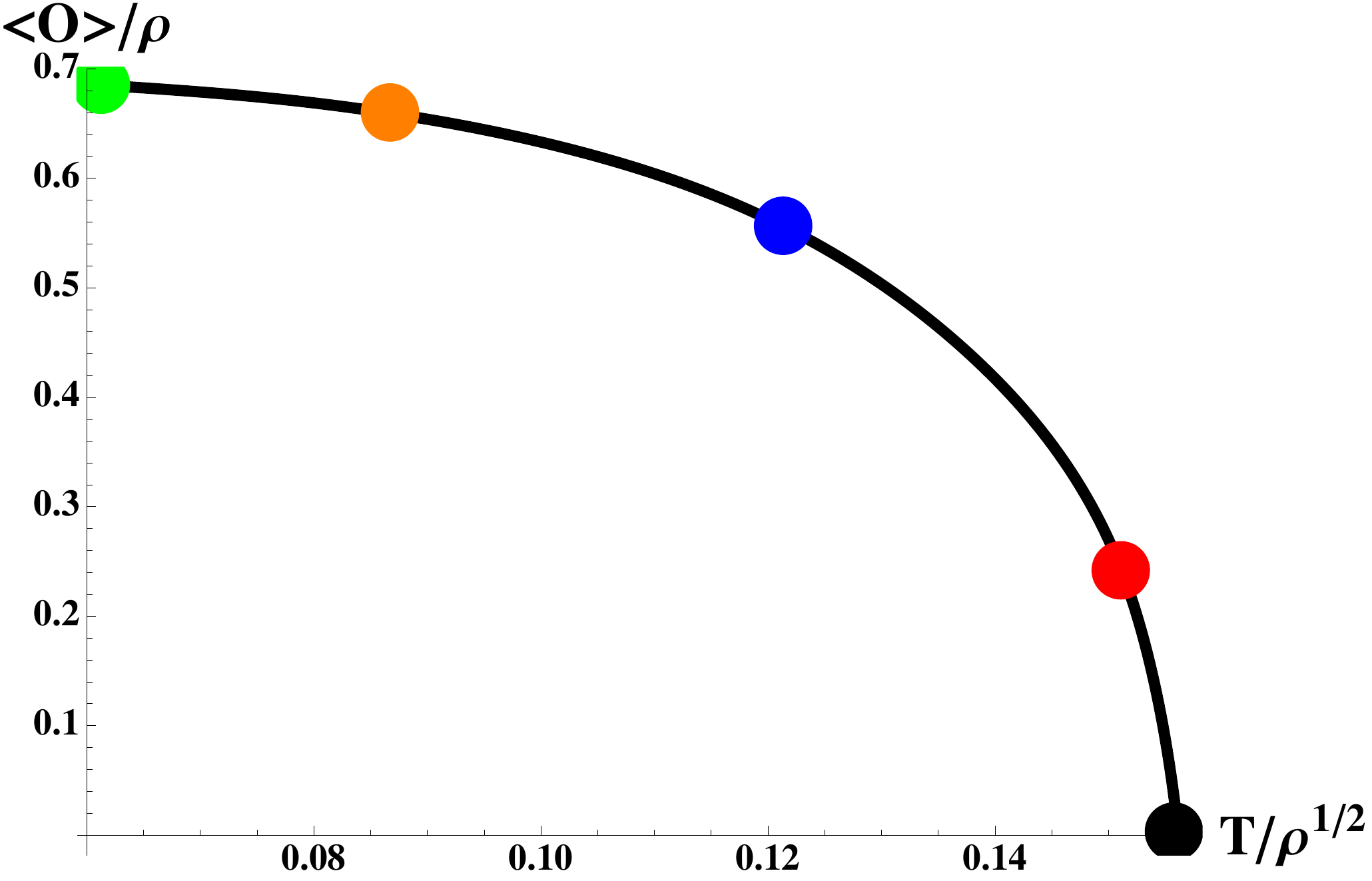}
\caption{The AC conductivity for the model (\ref{model2}) for some representative parameters.
Black line is at the temperature, slightly below the corresponding critical temperature $T_c/\rho^{1/2}$, and matches the result of the normal phase calculation at $T=T_c$.
Red, blue, orange and green lines are for $T/\rho^{1/2}=0.15, 0.12, 0.09, 0.06$, respectively. Notice that as we decrease the temperature, between blue and orange line, the peak in the imaginary part of the AC conductivity disappears. We call the corresponding critical temperature $T''/\rho^{1/2}$.
We also provide the condensate as a function of temperature and mark the points where we calculated the AC conductivity.}
\label{ACSC}
\end{figure} 
One can notice various fact. As we expected in the normal phase the DC electric conductivity is finite, it takes the value \ref{DCformula} and it is clearly distinct from the infinite DC conductivity appearing in the SC phase. In the normal phase the AC conductivity is characterized by a mid-infrared peak which was first observed in \cite{BaggioliPRL} and described in this thesis. Decreasing the temperature, and letting the condensate grow, this peak gets depleted and eventually disappear at a temperature $T''<T_c$. This suggests a possible competition between the superconducting mechanism and the momentum dissipating one. In particular it seems clear that a large superfluid density completely screens this collective excitation which in a sense
gets eaten by the large condensate.\\[0.1cm]
A final interesting question is studying the full phase diagram of the dual CFT which now can contain three different phases: the metallic one, the pseudo-insulating one and the SC one. The landscape of the possible outcomes is quite rich. We focus on three examples, shown in figure \ref{PDdue}.
\begin{figure}
\centering
\includegraphics[width=.28\textwidth]{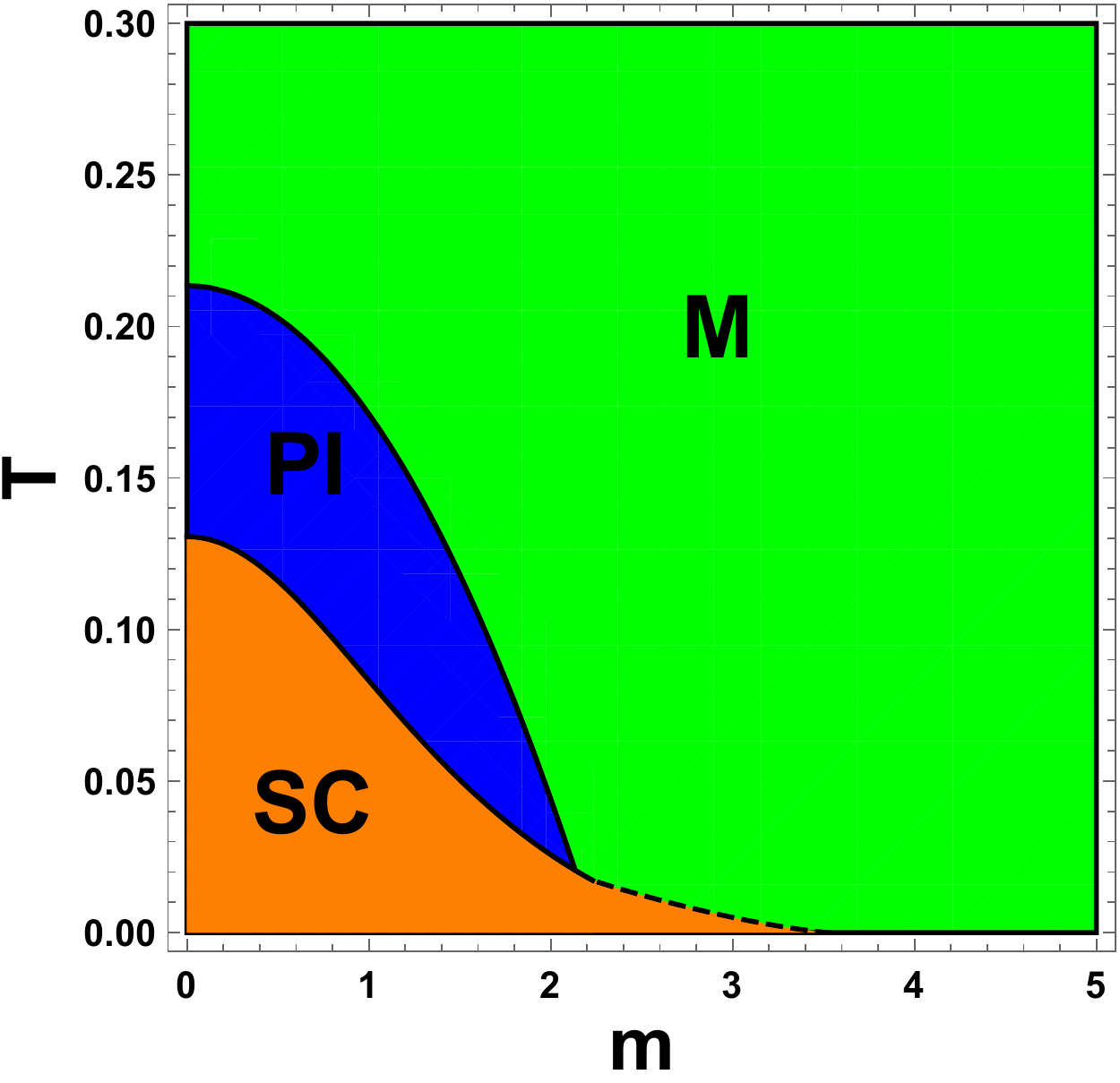}\qquad
\includegraphics[width=.28\textwidth]{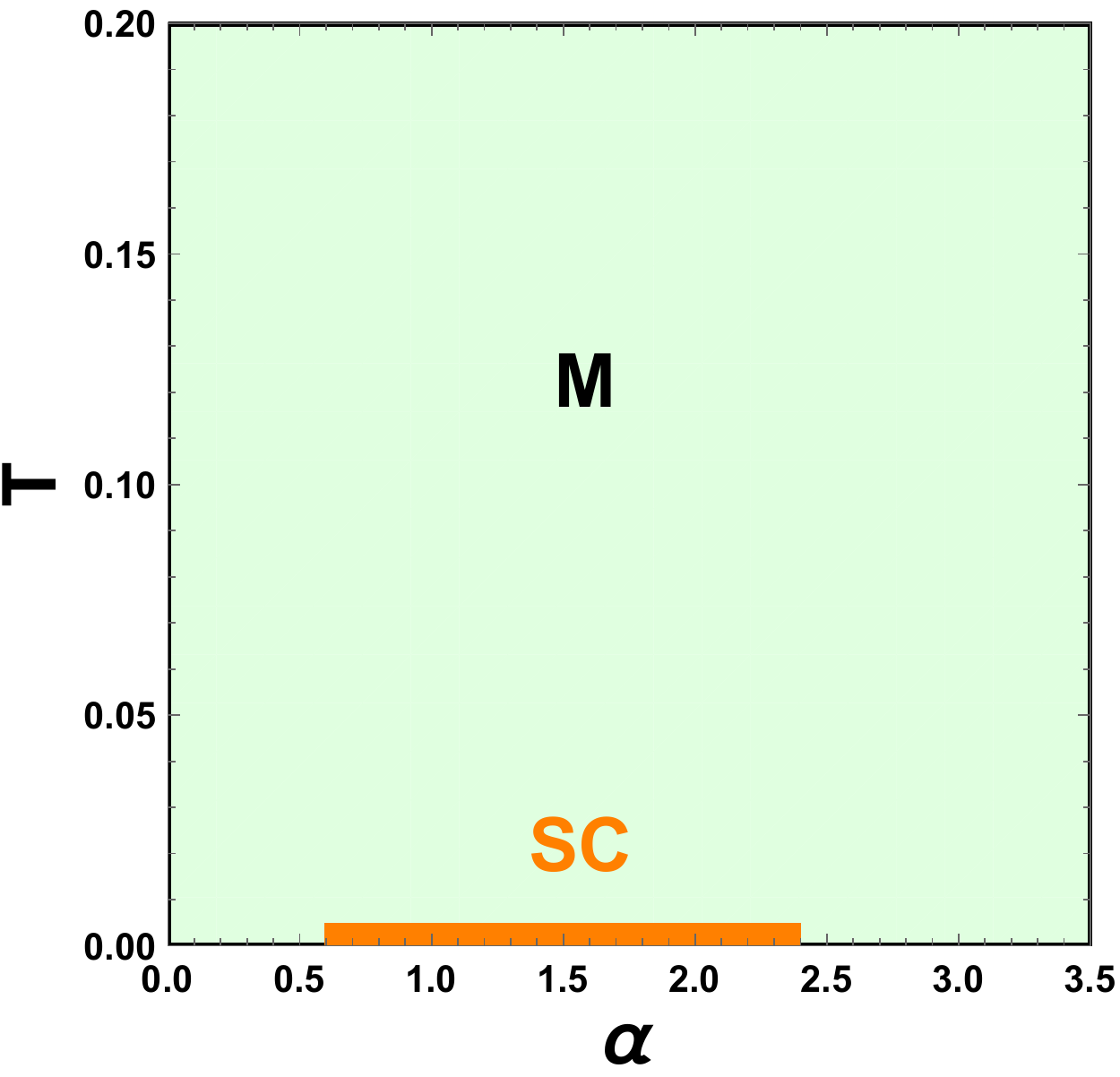}\qquad
\includegraphics[width=.28\textwidth]{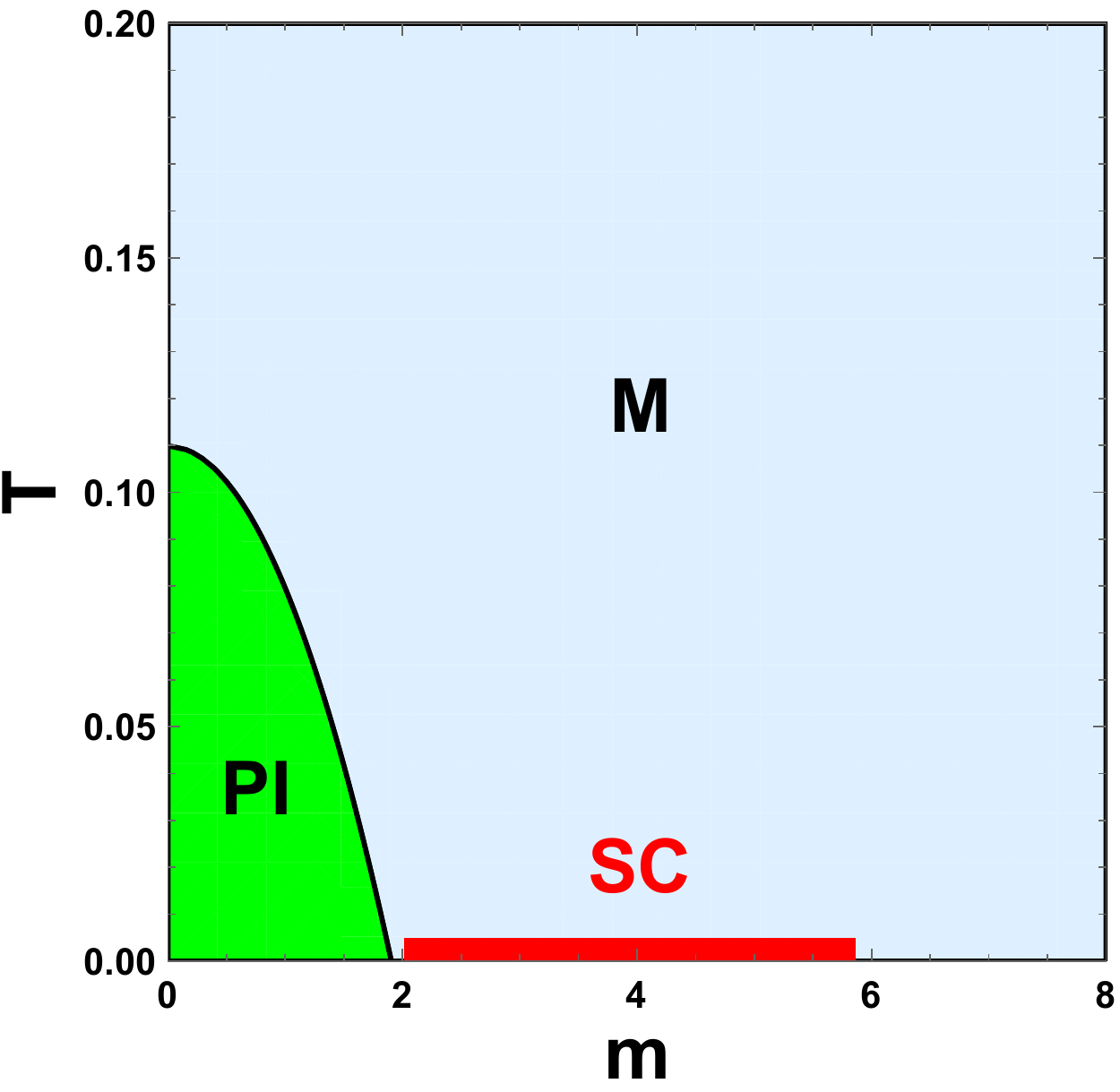}
\caption{Full phase diagram for three different cases. Green region is a normal pseudo-insulating phase, grey region is a normal metallic phase, red region is a superconducting phase.\textbf{Left: }\ref{model2} model for some representative parameters. \textbf{Center: }\ref{model2} model at the fine tuned ''dome-point''. \textbf{Right: }\ref{model1} model at the fine tuned ''dome-point''. Plots from \cite{BaggioliSC1}.}
\label{PDdue}
\end{figure} 
\begin{enumerate}
\item In the first example of fig. \ref{PDdue} a phase diagram for the non linear MG model it is shown. As expected one can provide the competition of three different phases with a quite rich phenomenology.
\item Close to the fine tuned point we discuss one can produce a SC dome shaped region in the middle of the phase diagram which:
\begin{itemize}
\item In the case of the linear potential \ref{model1} can be just surrounded by a metallic phase.
\item In the case of the non linear potential \ref{model2} can be embedded in a richer phase diagram which shows interesting features.
\end{itemize}
\end{enumerate}
The first important thing to notice is that we are considering the phase diagram on the temperature-disorder strength plane, which is very different from what is usually shown for the High Tc SC experiments. There the horizontal axes is doping, which should be related to our chemical potential $\mu$. Therefore the SC dome has no direct link with the famous CM results we are aiming to reproduce.\\
Additionally, because this behaviour, even if generic, is present just in a small fine tuned region in the plane $\{\Delta,q\}$ the critical temperature $T_c$ of its edge is very small and with the actual techniques we are not able to detect it nor describe it with accuracy.\\
In order to get closer to the actual High Tc phase diagram we need to introduce more ingredients into the holographic model in addition of the ones we already considered.\\[0.3cm]
\textbf{Towards the High-Tc phase diagram}\\[0.3cm]
The idea is to introduce novel fields in the bulk and a more generic action governing them and check if a phase diagram, which shares similarities with the experimental one characterizing High-Tc superconductors, is actually obtainable within the holographic framework. We take inspiration from the model described in \cite{KiritsisLi} adding to it an additional translational symmetry breaking sector, \textit{i.e.} massive gravity.\\
We consider the following bulk degrees of freedom: the metric $g_{\mu\nu}$,
two $U(1)$ gauge fields $A_\mu$, $B_\mu$, the complex scalar field
$\psi$, and two neutral scalars $\phi ^I$, $I=x,y$.
Here $x,y$ are spatial coordinates on the boundary. We will denote the radial bulk coordinate as $u$.
The boundary is located at $u=0$, the horizon is located
at $u=u_h$.

We want to describe a system of charge carriers,
coexisting with a media of impurities. The density of the charge
carriers is denoted by $\rho_A$ and is dual to the gauge field $A_\mu$ while the density of impurity  
$\rho_B$ is dual to the gauge field $B_\mu$.
The quantity
\begin{equation}
\label{dopingdefinition}
{\bf x}=\rho_B/\rho_A
\end{equation}
is called the doping parameter and represents the amount of charged impurities present in the system
\cite{KiritsisLi}.\\
The total action of the model is written as:
\begin{align}
\label{totalaction}
S=\frac{1}{16\pi}\int d^4x \sqrt{-g}\left(R+\frac{6}{L^2}+{\cal L}_c+{\cal L}_s\right)
\end{align}
where we fixed the cosmological constant $\Lambda=-3/L^2$, and denoted the Lagrangian densities for the charged sector \cite{KiritsisLi},
and the neutral scalar sector \cite{BaggioliPRL} as:
\begin{align}
{\cal L}_c&=-\frac{Z_A(\chi)}{4}A_{\mu\nu}A^{\mu\nu}
-\frac{Z_B(\chi)}{4}B_{\mu\nu}B^{\mu\nu}
-\frac{Z_{AB}(\chi)}{2}A_{\mu\nu}B^{\mu\nu}\label{chargesectorlagrangian}\\
&-\frac{1}{2}(\partial_\mu \chi)^2-H(\chi)(\partial_\mu \theta-q_AA_\mu-q_B B_\mu)^2
-V_{int}(\chi)\\
{\cal L}_s&=-2m^2 V(X)\,.
\end{align}
Here the $A_{\mu\nu}$ and $B_{\mu\nu}$ stand for the field strengths of the
gauge fields $A_\mu$ and $B_\mu$ respectively.
Following \cite{KiritsisLi} we decomposed the charge scalar as $\psi =\chi e^{i\theta}$.
We also defined:
\begin{equation}
X=\frac{1}{2}g^{\mu\nu}\p_\mu\phi^I\p_\nu\phi^I\,.
\end{equation}
The most general black-brane ansatz we consider is:
\begin{align}
\label{generalansatz}
ds^2&=\frac{L^2}{u^2}\left(-f(u)e^{-\tau(u)}dt^2+dx^2+dy^2+\frac{du^2}{f(u)}\right)\,,\\
A_t&=A_t(u)\,,\qquad B_t=B_t(u)\,,\\
\chi&=\chi (u)\,,\qquad \theta\equiv 0\,,\\
\phi^x&=\alpha\, x\,,\qquad \phi^y=\alpha\, y\,.
\end{align}
The corresponding equations of motion are provided in the original paper \cite{BaggioliSC2}.
The temperature of the black brane (\ref{generalansatz}) is given by:
\begin{equation}
T=-\frac{e^{-\frac{\tau(u_h)}{2}}f'(u_h)}{4\pi}\,.
\end{equation}
We will be considering:
\begin{align}
V_{int}(\chi)&=\frac{M^2\chi^2}{2}\,.
\end{align}
Solving the $\chi$ e.o.m. near the boundary $u=0$ one obtains $\chi(u)=C_- \, (u/L)^{3-\Delta}
+C_+\, (u/L)^\Delta $, where
$(ML)^2=\Delta (\Delta -3)$.
Here $C_-$ is the source term, which one demands to vanish, and
$C_+$ is the v.e.v. of the dual charge condensate operator, $C_+=\langle {\cal O}\rangle$.
The $\Delta$ is equal to the scaling dimension of the operator ${\cal O}$. Following \cite{KiritsisLi}we
fix the scaling dimension to be $\Delta =5/2$.\\
In the normal phase the charge condensate vanishes, and the charged scalar field is trivial, $\chi\equiv 0$. Solving the background equations of motion we obtain $\tau\equiv 0$, along with:
\begin{align}
\label{normalphase}
f(u)&=u^3\int _{u_h}^udy\,\frac{\rho_A^2(1+{\bf x}^2)\,y^4+4\,(mL)^2\, V(\alpha^2\, y^2)-12}{4y^4}\,,\\
\label{normalphase2}
A_t(u)&=\rho_A(u_h-u)\,,\hspace{0.5cm}B_t(u)=\rho_B(u_h-u)\,.
\end{align}
The temperature in the normal phase is given by:
\begin{equation}
\label{generaltemperature}
T=\frac{12-\rho_A^2(1+{\bf x}^2)\,u_h^4-4\,(mL)^2 \,V(\alpha^2\,u_h^2)}{16\pi u_h}\,.
\end{equation}
Using the membrane paradigm one can calculate analytically the DC conductivity in the normal phase. Its value for a general neutral scalars Lagrangian $V$ is given by:
\begin{equation}
\label{generalsigmaDC}
\sigma_{DC}=1+\frac{\rho_A^2u_h^2}{2\,m^2\,\alpha^2\,\dot V(u_h^2\alpha^2)}\,.
\end{equation}
The features of this normal phase have been already described in \cite{BaggioliPRL} and in the previous sections.\\
We can then perform the usual routine to check the instability of such a normal phase towards the spontaneous formation of a charged condensate encoded in the scalar profile $\chi(u)$. Following \cite{KiritsisLi} we define the following expansion of the couplings:
\begin{align}
\label{couplingexpansions}
&H(\chi)=\frac{n\,\chi^2}{2}\,,\hspace{0.3cm}
Z_A(\chi)=1+\frac{a\,\chi^2}{2}\,,\hspace{0.3cm}
Z_B(\chi)=1+\frac{b\, \chi^2}{2}\,,\hspace{0.3cm}
Z_{AB}(\chi)=\frac{c\, \chi^2}{2}\,.
\end{align}
and define the $U(1)_{A,B}$ charges to be $q_A=1\,,\,q_B=0$ .\\
A natural place to start searching for superconductor
is at zero temperature. The BF violation argument, within this setup, leads to the instability condition:
\begin{align}
(2\,M^2-u_0^4\,(a+2\,c\,x+b\,x^2))\,(6+m^2((\alpha\,u_0)^2\,\dot V-2V))-2\,n\,u_0^4\,(
q_A+q_B\,x)^2<0\,,\label{AdS2BFbound}
\end{align} 
where dot stands for derivative of $V$ w.r.t. its argument and $u_0$ for the radial position of the extremal horizon, $T(u_0)$=0.\\
To obtain a superconducting dome on the temperature-doping plane $(T,{\bf x})$, one needs to fix the parameters of the model in such a way that zero-temperature superconducting instability appears in an interval $[{\bf x}_1,{\bf x}_2]$, between two positive values ${\bf x}_{1,2}$ of the doping parameter.
\begin{figure}
\begin{center}
\includegraphics[width=.45\textwidth]{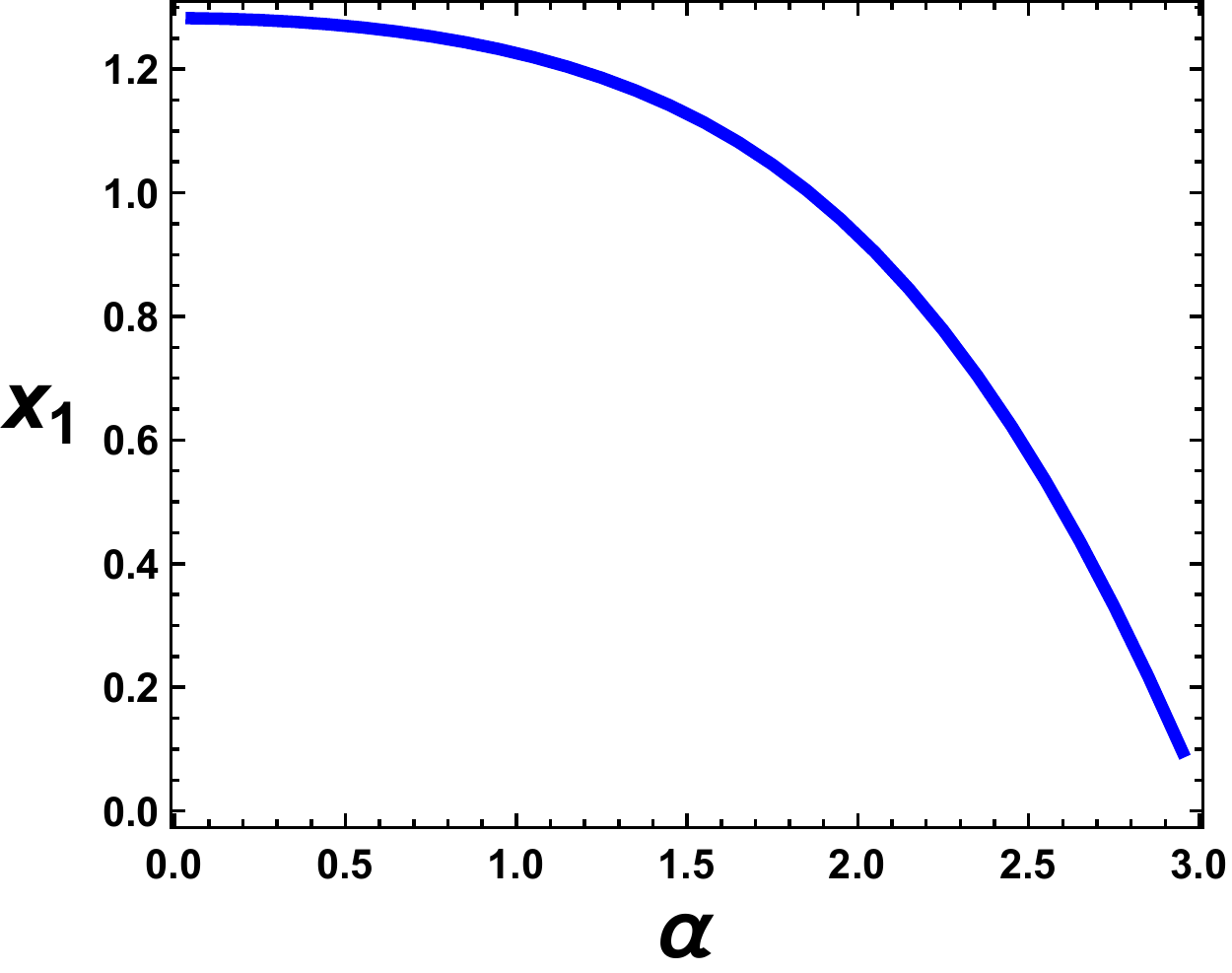}
\includegraphics[width=.45\textwidth]{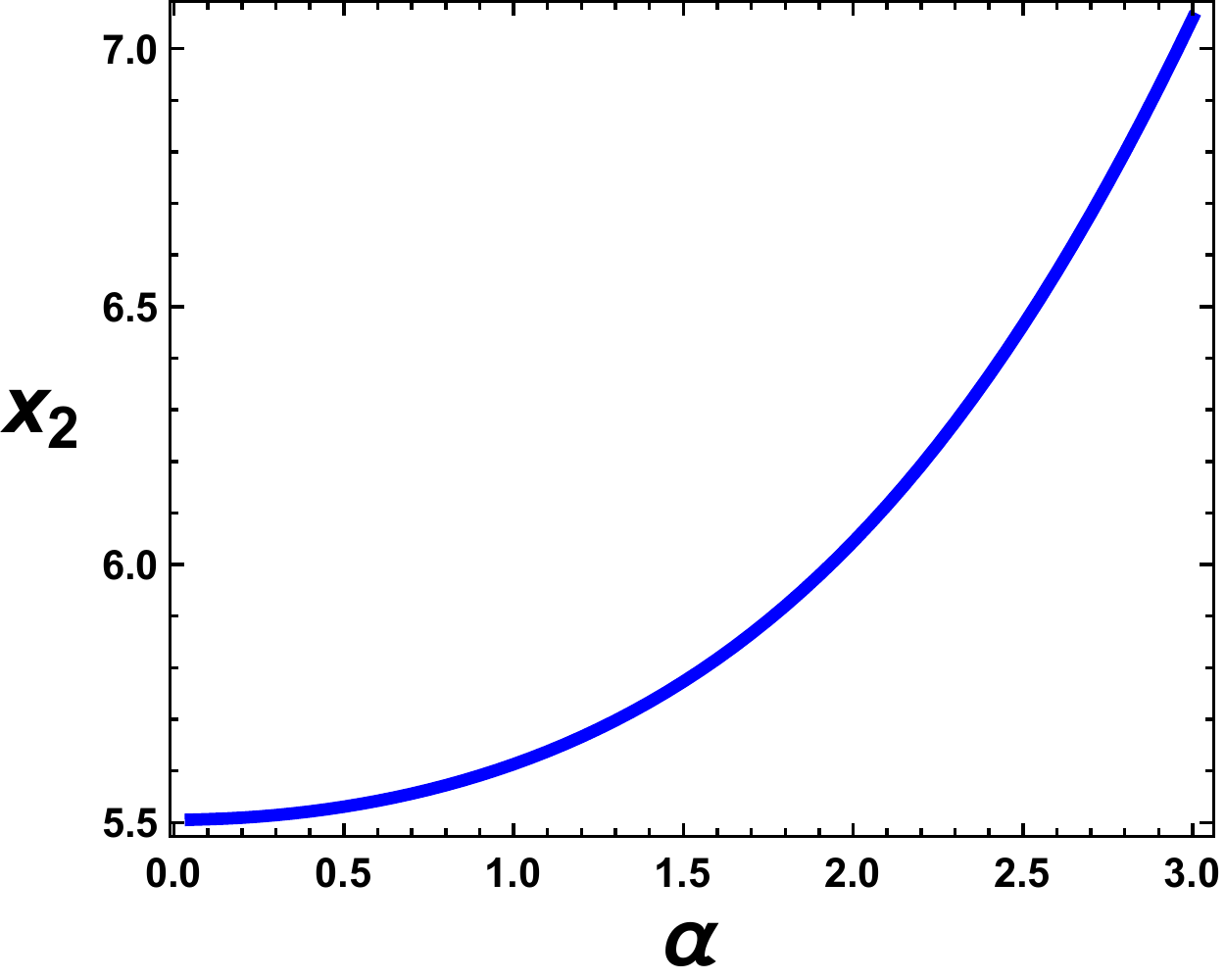}
\end{center}
\caption{The boundaries of the zero-temperature IR instability region 
on the doping line for the model (\ref{Kiritsisparameters}), with the
translational symmetry broken by the neutral scalars with the linear potential $V(X)\sim X$.}
\label{fig:x12ofalpha}
\end{figure}
 The specific model determined by the parameters:
\begin{equation}
\label{Kiritsisparameters}
a=-10\,,\quad b=-\frac{4}{3}\,,\quad c=\frac{14}{3}\,,\quad n=1\,.
\end{equation}
has been extensively studied, and it was pointed out that in the
interval ${\bf x}\in [{\bf x}_1,{\bf x}_2]$, ${\bf x}_1\simeq 1.28$, ${\bf x}_2\simeq 5.51$ at zero temperature the effective mass of the scalar field $\chi$ violates the $AdS_2$ BF bound in absence of momentum dissipation. Our main goal is to add the breaking of translational symmetry and study the consequences on such a SC dome found in \cite{KiritsisLi}.\\
We observe that for $\alpha\neq 0$ the instability persists, although
the `depth' of the $AdS_2$ BF violation becomes smaller, and therefore we expect the corresponding
critical temperature of the superconducting phase transition to be lower. At zero temperature the SC range of values of the doping parameter increases when the
translational symmetry breaking parameter $\alpha$ gets bigger.
We plot the $\alpha$-dependence of the boundary
points of the IR instability region, ${\bf x}_{1,2}(\alpha)$, in figure \ref{fig:x12ofalpha}.\\
\begin{figure}
\begin{center}
\includegraphics[width=.43\textwidth]{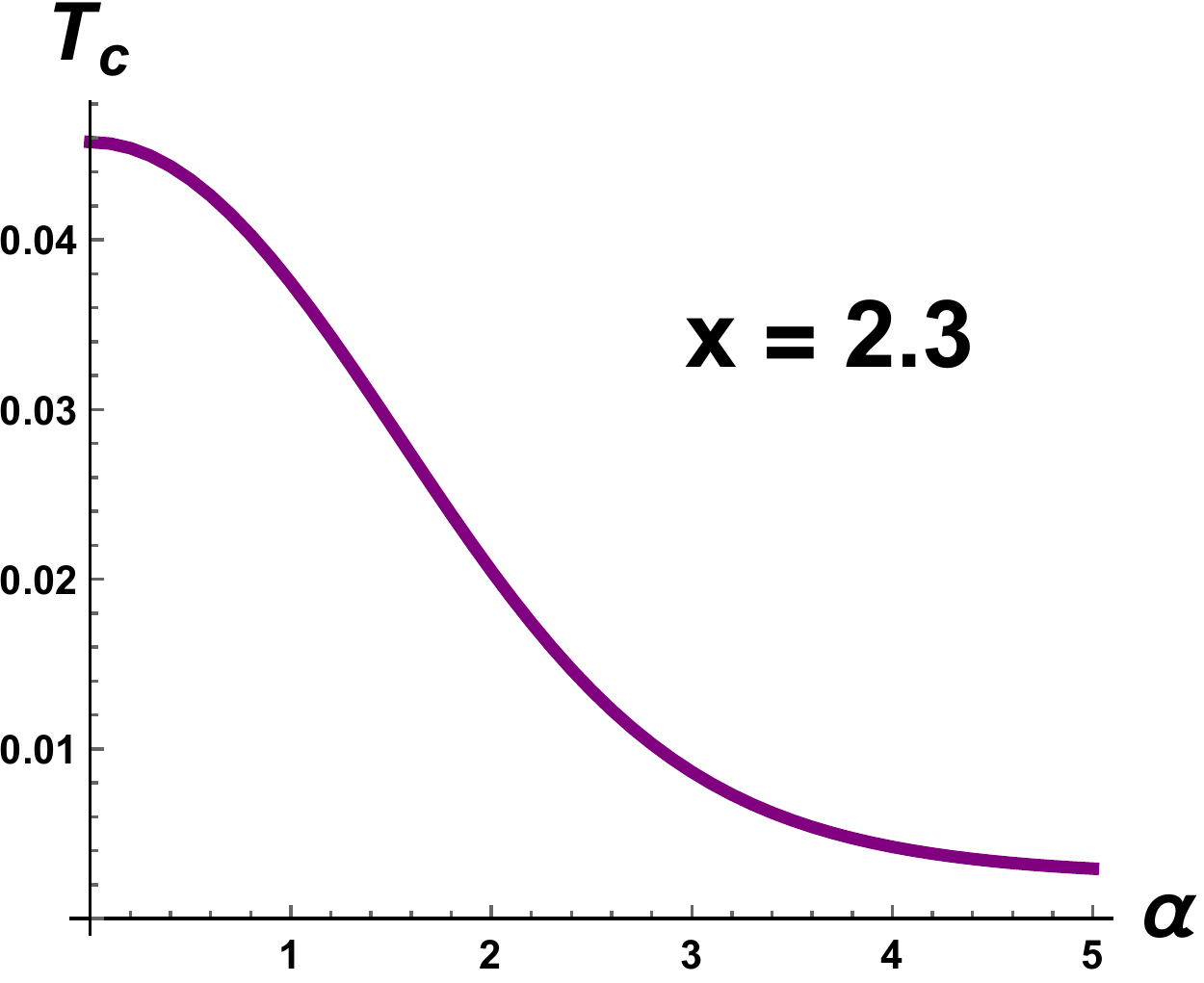}
\qquad
\includegraphics[width=.43\textwidth]{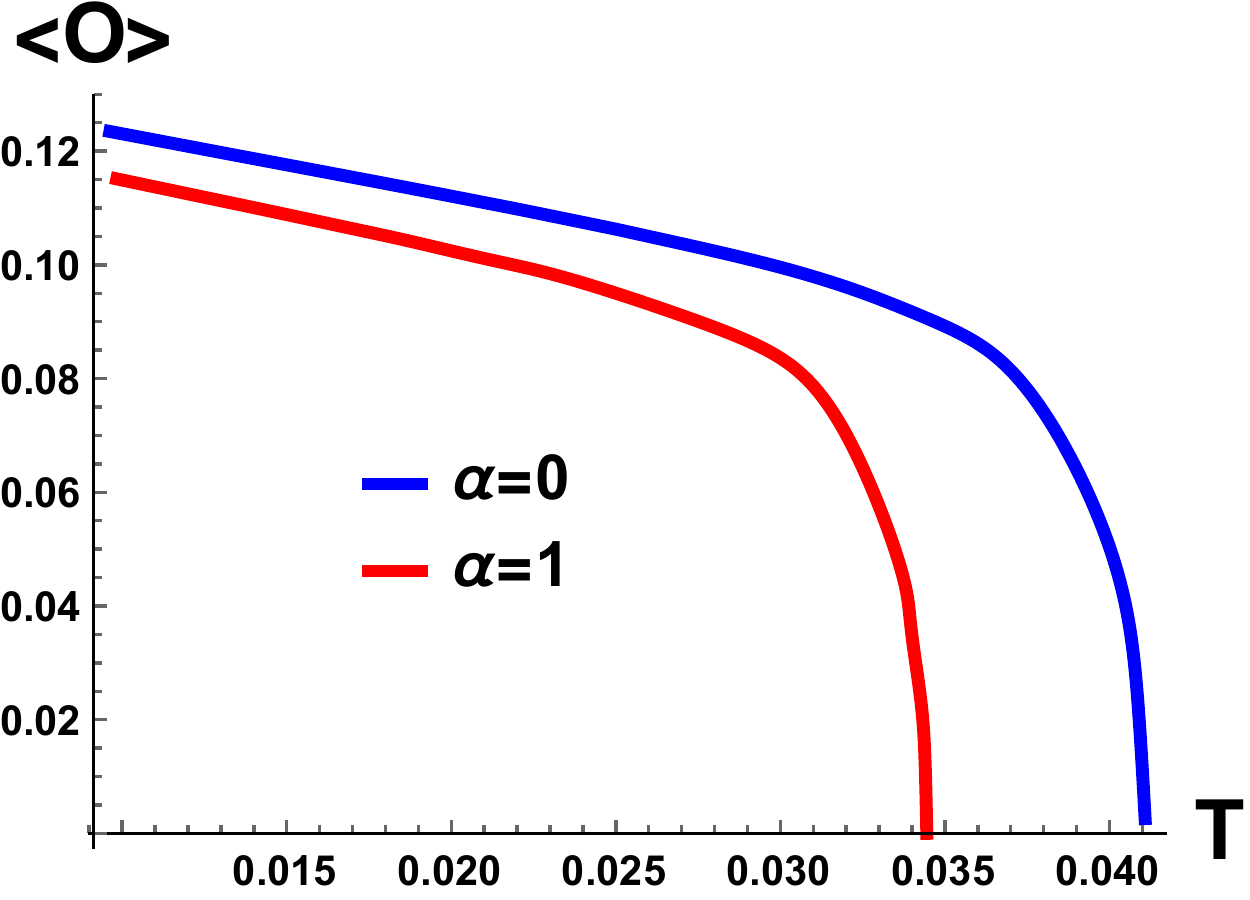}
\end{center}
\caption{\textbf{Left: }Critical temperature $T_c(\alpha)$ for the model (\ref{Kiritsisparameters}), with the
translational symmetry broken by the neutral scalars with the linear potential $V(X)\sim X$.
Here the doping is fixed to be ${\bf x}=2.3$. \textbf{Right: }Condensate for the model (\ref{Kiritsisparameters}), with the
doping fixed to ${\bf x}=2$, with broken translational symmetry.
{\bf Left:} The linear model linear potential $V(X)\sim X$ with $\alpha=1$,
plotted next to the translationally-symmetric system $\alpha=0$.}
\label{fig:TcMass}
\end{figure}

The critical temperature $T_c$ of a second-order phase transition can be determined by studying the dynamics of the scalar $\chi (u)$, considered as a probe in a finite-temperature normal phase background. In accordance with our expectations from the zero-temperature instability
analyses we observe a decrease of the critical temperature with $\alpha$,
as shown in figure \ref{fig:TcMass}. Moreover the presence of a non null disorder strength, \textit{i.e.} the graviton mass $\alpha$, depletes the value of the condensate as well as shown in figure \ref{fig:TcMass}.

\begin{figure}
\begin{center}
\includegraphics[width=.43\textwidth]{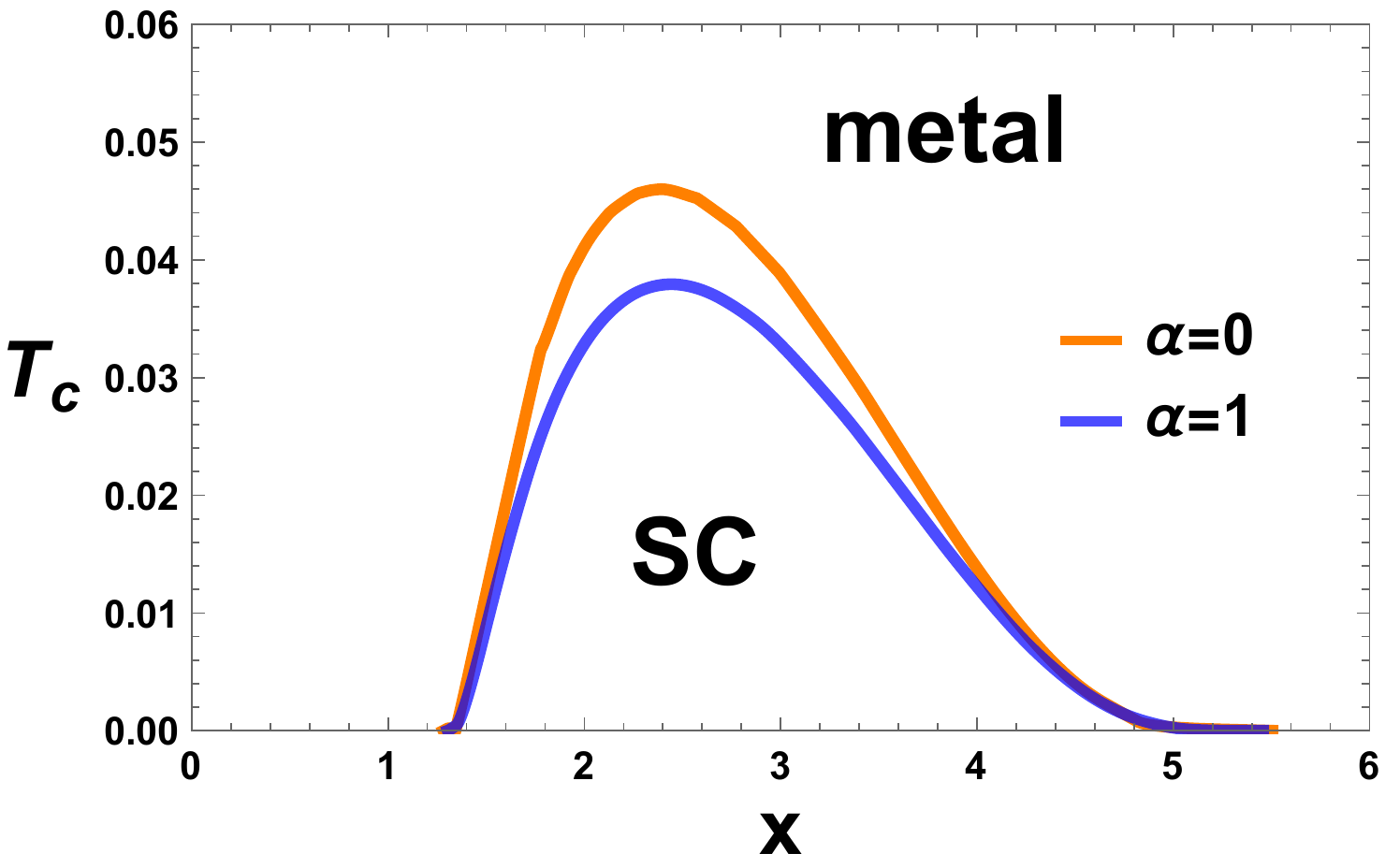}\qquad
\includegraphics[width=.43\textwidth]{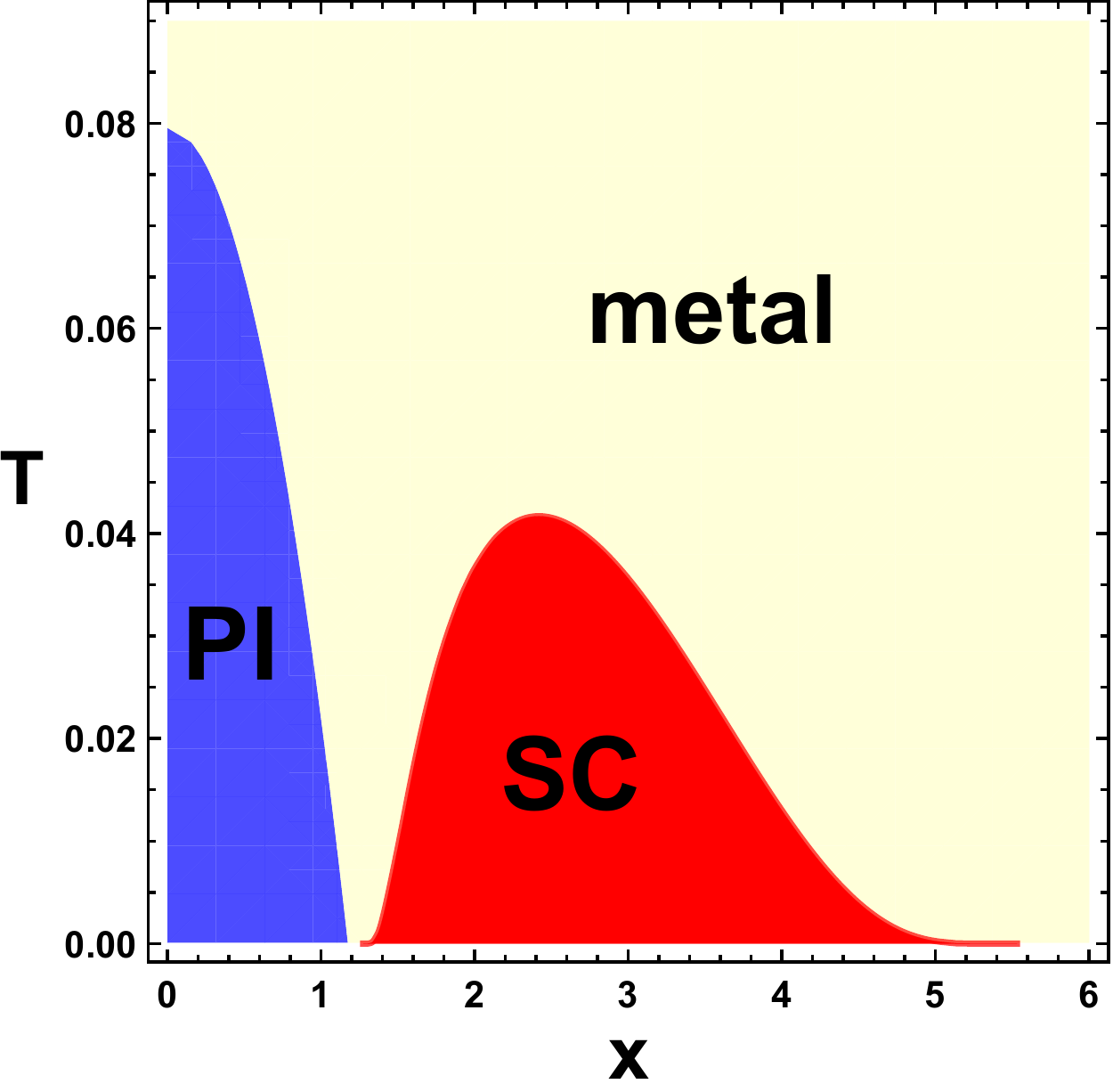}
\end{center}
\caption{\textbf{Left: }Phase diagram in the $(T,{\bf x})$ plane for the model (\ref{Kiritsisparameters}) coupled to
the neutral scalars with the linear potential $V(X)\sim X$). We compare the case $\alpha=0$ of \cite{Kiritsis:2015hoa}, and
the system with broken translational symmetry, at $\alpha=1$. \textbf{Right: }Phase diagram in the $(T,{\bf x})$ plane for the model (\ref{Kiritsisparameters}) coupled to
the neutral scalars with the non linear potential $V(X)\sim X+X^5$. We fixed $\alpha=0.5$ and $m=1$.}
\label{fig:phasediagram}
\end{figure}
 Now let us fix the value of $\alpha$ and plot the critical temperature
as a function of the doping parameter ${\bf x}$, see figure \ref{fig:phasediagram}.
The breaking of translation symmetry preserves the superconducting
dome structure exhibited by the model (\ref{Kiritsisparameters}), and merely diminishes a little
the critical temperature. Now let us consider the model with translational
symmetry broken by neutral scalars governed by the non-linear
non linear potential $V(X)\sim X+X^5$.
We fix $\alpha=0.5$, $m=1$ and determine the critical temperature $T_c({\bf x})$.
In figure \ref{fig:phasediagram} we combine this with the temperature $T_0({\bf x})$ of the
metal/pseudo-insulator phase transition (MIT),
and obtain the full phase diagram of the system
with the superconducting phase enclosed inside a dome.\\

This means that even if momentum dissipation unfavores the SC phase it is still possible to achieve a SC dome-shaped region as in actual High-Tc superconductors and having a normal phase with a finite DC conductivity. The main result is to show that the SC dome-shaped region built in \cite{KiritsisLi} can be completed with a simple momentum dissipation mechanism and embedded in a normal phase region featuring a finite DC conductivity. This represents a further step towards reproducing holographically the phase diagram for High-Tc superconductors \cite{Chen2}.

\cleardoublepage

\newpage\leavevmode\thispagestyle{empty}\newpage
\part{Final Remarks}
\label{part:Final Remarks}
\begin{figure}
\includegraphics[scale=0.6]{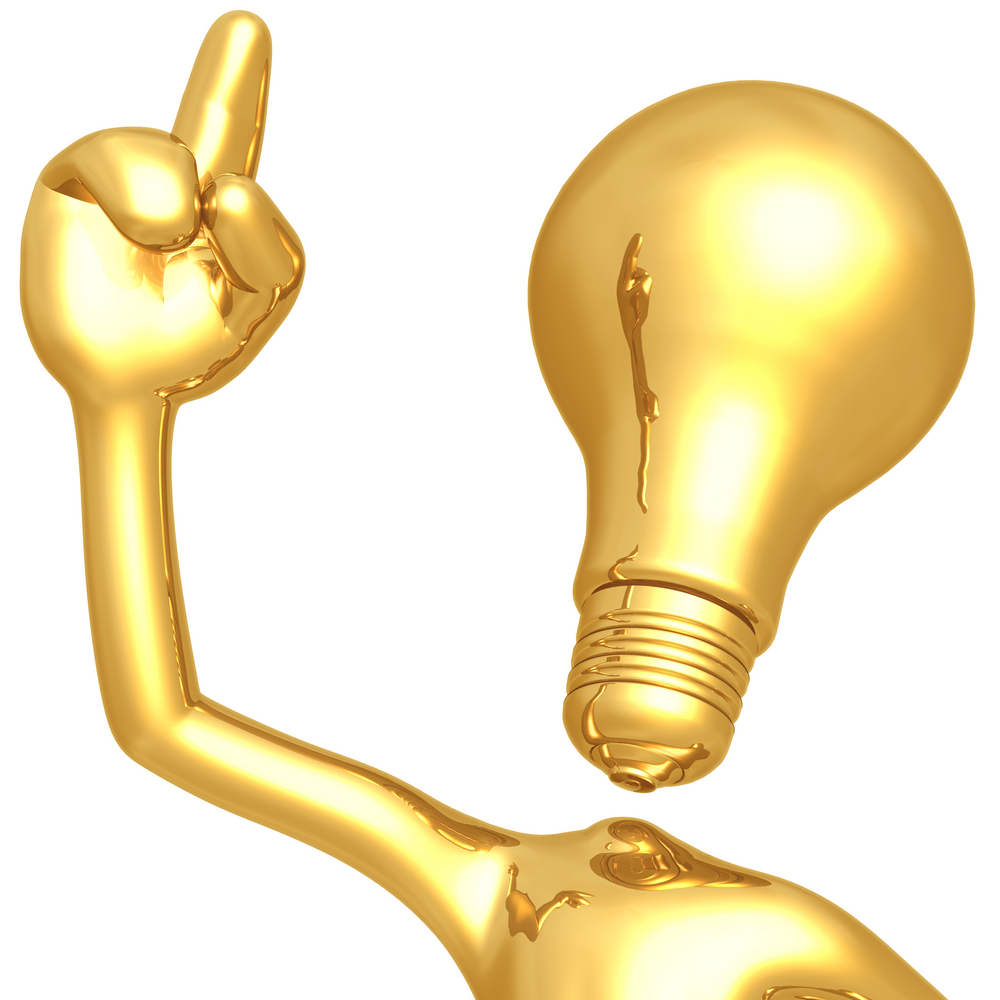}
\end{figure}
\epigraph{I think and think for months and years. Ninety-nine times, the conclusion is false. The hundredth time I am right.}{\textit{Albert Einstein}}
In this last section we aim to give a critical review of what AdS/CMT did, is doing and could do in the nearby future.\\[0.3cm]
\textbf{''What did gauge-gravity duality teach us about condensed matter physics ?''}\\[0.3cm]
\textbf{''What can gauge-gravity duality teach us (more) about condensed matter physics ?''}\\[0.3cm]
Quoting \cite{SachdevCrit}, these will be our main questions to review the actual status of the framework and its possible future achievements. A critical, and perhaps even provocative review, can be found in \cite{Khveshchenko}.
In the last years AdS-CMT started to become a novel effective theory approach able to capture several CM mechanisms and features opening unexpected new directions and techniques.\\
We will start summarizing our main contributions to the attempt of answering the first question. 
In what is left we will wandering about various condensed matter topics with a proposing attitude discussing the impact of gauge-gravity duality on their understanding, as an \textbf{holographic effective theory for condensed matter}.\\[0.3cm]
\section{What did gauge-gravity duality teach us about CM physics ?}
Since the awareness of the necessity of breaking translational symmetry in order to have finite holographic DC physical quantities such as the electric and heat conductivities, massive gravity theories acquired a special role in the AdS-CMT program. They qualify as simple and effective toy models able of realizing momentum dissipation without the need of introducing explicit spatial dependences nor advanced numerical techniques. In the last three years, since the appearance of the first paper \cite{Veghoriginal}, lots of progress has been made with the target of generalizing, understanding and exploiting such a framework. In this section we will summarize the main contributions we gave to this field which constitute the main scientific body of this thesis.\\[1cm]
\textbf{Generic massive gravity duals}\\[0.3cm]
The first papers on the subject \cite{Veghoriginal,AndradeWithers} focalized on  very peculiar models, namely dRGT massive gravity and the linear Stueckelberg model. We now understand that there is no particular reason to do so and that the landscape of healthy and interesting massive gravity theories, which could be embedded into the holographic machinery, is much wider than what discussed originally.\\
More in detail, despite dRGT being the only healthy and consistent non linear theory of massive gravity in Lorentz invariant setups, this turns out to be no longer true while the Poincar\'e group gets broken as it always happens in condensed matter scenarios. As a consequence, there is a plethora of MG bulk theories which one could consider and that could be conveniently written down via the Stueckelberg trick as General Relativity plus a sector of massless scalars (\textit{i.e} the Goldstones). This is exactly what has been suggested in \cite{AndradeWithers} where the most simple action has been considered.\\
The first important message present in \cite{BaggioliPRL} and later on expanded in \cite{BaggioliSolid} is that one can construct and embed in AdS spacetime a wide class of lorentz violating massive gravity theories, which would effectively provide the dissipation of momentum in the CFT duals. The stability and the consistency of this large class of models has been analyzed in detail in \cite{BaggioliSolid}. All these theories differ among themselves for the choice of a potential of two scalar quantities\footnote{This is true only if we consider two spatial dimensions.} which has to satisfy very mild constraining conditions to ensure no ghosty modes nor instabilities of any kind.\\[0.3cm]
\textbf{Electric response}\\[0.3cm]
From the bulk perspective all these different choices lead to different radial coordinate dependences of the graviton mass $m^2(r)$ which turns out to be a specific function of the potential cited before. On the contrary, from the phenomenological point of view, such a dependence of the graviton mass is reflected in the features of the DC electric conductivity and in particular in its temperature dependence $\sigma_{DC}(T)$\footnote{This can be understood in the following way. The DC quantities can be holographically determined just by IR horizon data. The graviton mass $m^2(r)$ gets (inversely) mapped into the momentum relaxation time $\tau$ which has to be computed at the horizon $r=r_h$ which determines the temperature of the CFT dual. As a consequence the graviton mass appears in the DC quantities computed at the horizon position and it determines directly the functional dependence $\tau(T)$ crucial for the phenomenological features of the DC conductivities.}. Playing and varying the potentials one can therefore dial the temperature dependence of such a quantity and passing from a metallic to an insulating behaviour:
\begin{equation}
d\sigma_{DC}/dT\,<\,0\,\longrightarrow\,\text{METAL}\,,\qquad\,d\sigma_{DC}/dT\,>\,0\,\longrightarrow\,\text{INSULATOR}\,.
\end{equation}
Not only that, one can also construct (see \cite{BaggioliPRL}) dynamical transitions between the two regimes which represent metal-insulator crossovers at finite temperature and can identify a specific shape of the potential such that at low temperature the resistivity scales linearly in temperature as in the most wanted realistic strange metals (see \cite{BaggioliSolid}).\\
It was afterward noticed in \cite{Grozdanov1} that despite all these nice features these MG models could not provide an insulator with $\sigma_{DC}(T=0)=0$ because their DC electric conductivity is bounded by below by the value:
\begin{equation}
\sigma_{DC}\,\geq\,\frac{1}{q^2}
\label{BABA}
\end{equation}
where q is the electromagnetic coupling appearing in the action in fron of the $F_{\mu\nu}F^{\mu\nu}$ kinetik term. In order to get rid of this shortcoming further improvements have to be introduced into the models we initially considered.\\[0.3cm]
\textbf{Holographic metal-insulator transitions}\\[0.3cm]
Taking an abstract point of view the theories considered so far all look like:
\begin{equation}
\underbrace{GR\,\,\,+\,\,\,STUECKELBERG\,\,SECTOR\,}_{MASSIVE\,\,GRAVITY}\,\,\,+\,\,\,EM\,\,SECTOR
\end{equation}
Anyway from an effective field theory perspective, as the one we would like to take, it is natural to include a direct coupling between the Stueckelberg sector and the electromagnetic one. If one dares to do that, one realizes that the previously conjectured and observed bound \ref{BABA} does not hold anymore and that the DC electric conductivity at $T=0$ can drop down to zero upon increasing the strength of momentum dissipation encoded in the graviton mass.\\
Within this new generalized models \cite{BaggioliDisorder} insulating CFTs with very low conductivity at zero temperature appear. In addition, upon dialing the graviton mass, infact the parameter encoding how fast momentum dissipates, one can build quantum ($T=0$) transitions between metallic and insulating states providing one of the first cases of holographic MIT.\\
Despite the interpretation of those MITs is not clear yet\footnote{It has been suggested that such MITs are driven just by charge screening, namely the renormalization of the electromagnetic coupling, and not by any interaction or disorder effect.} the present represents without any doubt a striking result which can open the room for the study of metal-insulator transitions via holographic techniques.\\[0.3cm]
\textbf{Solid vs Fluids and the viscoelastic response}\\[0.3cm]
The general character of the massive gravity theories introduced in \cite{BaggioliSolid} allows also to distinguish them between solid and fluid type. Technically speaking this distinction can be made at the level of the internal symmetries of the Stueckelberg sector but it exhibits strong consequences in some physical quantities which match with the theoretical expectations. In more detail, one can make a distinction between the two types of gravity dual looking at the viscoelastic response of the corresponding CFT. The latter, encoded in the Green function of the transverse and traceless spin 2 operator, is fully determined by the mass acquired by the helicity 2 mode of the graviton\footnote{Note that on the contrary the graviton mass appearing in the computation of the electric conductivity is the one dealing with the vectorial part of the graviton, which in general differs from the one discussed here.}. Surprisingly enough, such a mass, is non zero just for the solid massive gravity theories while is trivially null in the fluid case. This fact leads to two important consequences analyzed in \cite{BaggioliSolid,BaggioliViscosity} :
\begin{itemize}
\item The CFT duals to solid type MG theories exhibit a non zero rigidity modulus which grows with the ''tensorial'' graviton mass. In other words, the solid CFT duals show a non zero elastic response which is not present in the case of the fluids. This is of course what one would expect from a solid and not from a fluid where no transverse dynamical phonons are present and therefore no elastic response appears.
\item In the solid CFT duals the well-known $\eta/s$ KSS bound is brutally violated and drops down to zero at zero temperature. Again this is not the case for the fluid type where the latter ratio keeps its KSS value $\eta/s=1/4\pi$. This violation is conjectured to be present because of the viscoelastic properties of the material and a generalized ''KSS'' bound involving both the quantities (rigidity modulus and viscosity) has been conjectured in \cite{BaggioliViscosity}.
\end{itemize}
These results represent the first study of the viscoelastic properties of CFT dual to gravitational bulk theories and could potentially open a totally new direction into the holographic framework.\\[0.3cm]
\textbf{Holographic superconductors with momentum dissipation}\\[0.3cm]
Another natural question one should ask himself, once having at hand such toy models for momentum dissipation, is how the latter affects the superconducting phase transition. It is somehow straightforward to implement these generic massive gravity models along with the original holographic superconductor one as we did in \cite{BaggioliSC1} where we study all the effects of momentum dissipation to the SC background.\\
A more difficult, but more interesting as well, task would be to achieve at least via toy models the shape of the phase diagram of realistic High-Tc superconductors. Following previous succesful results we aimed at that target in \cite{BaggioliSC2} where we indeed obtained a dome shaped region of superconductivity surrounded by a metallic phase with finite DC conductivity as in the actual phase diagram nature exhibits. This surprising result pushed the holographic toy models scenario closer to the High-Tc reality and represents a nice starting point for further investigations.
\section{What can gauge-gravity duality teach us (more) about CM physics ?}
\textbf{Disclaimer:} from this point forward I will simply be thinking out loud. Do not take it too seriously!\\[0.3cm]
\textbf{Strange Metals and universal scalings}\\[0.3cm]
One of the most important condensed matter open questions refers to the so-called \textit{Strange Metals} \cite{SM1,SM2}: a large class of materials whose transport properties obey unusual temperature scalings which are not in agreement with the Fermi Liquid Theory (FL), one of the CM last century pillars.\\
\begin{figure}
\centering
\includegraphics[width=13cm]{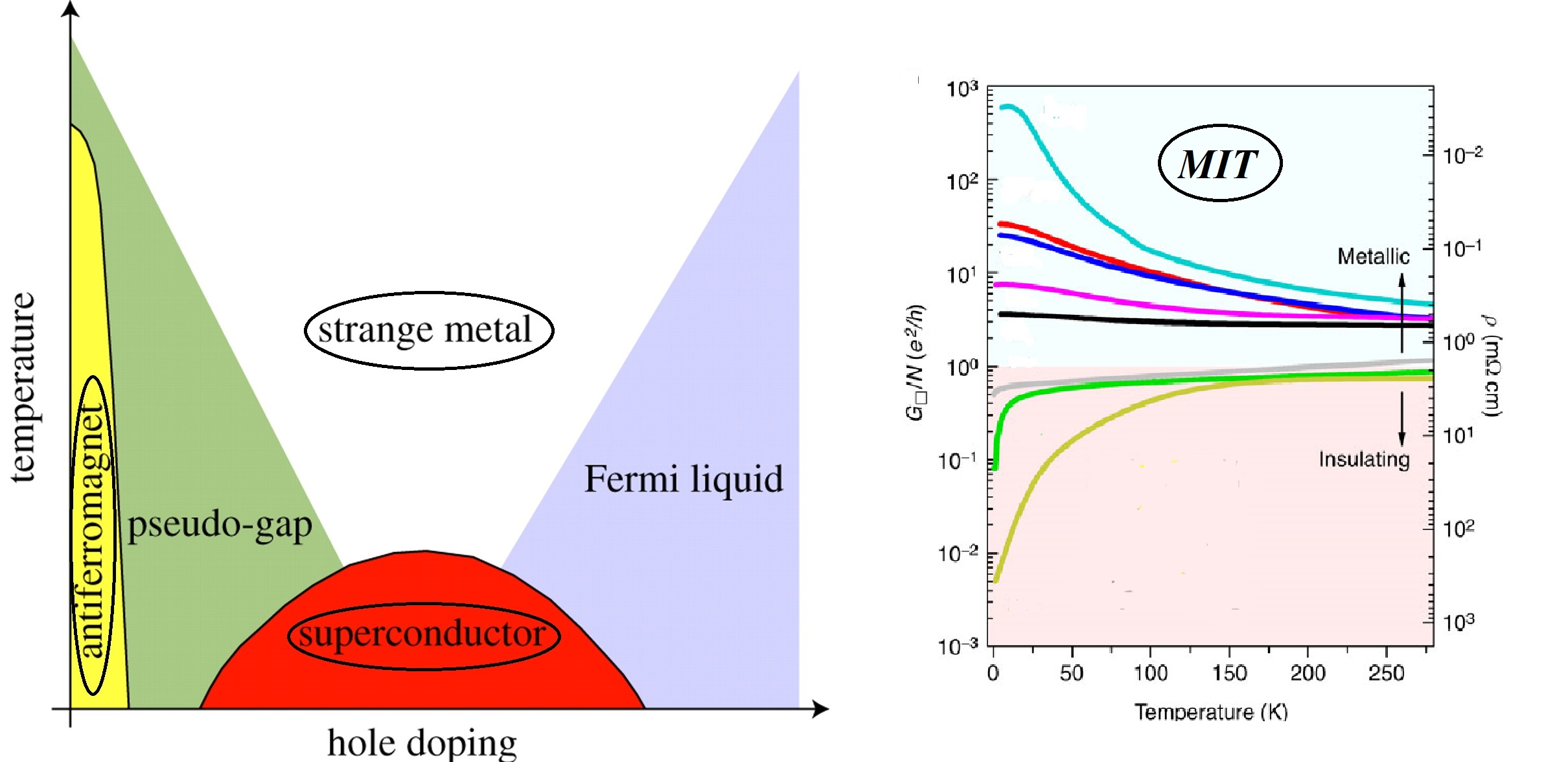}
\caption{Sketch of the various open issues in CM related which the AdS-CMT started to attack. Holographic effective theories for CM could be a new and suitable tool to adress these issues: Strange Metals scaling, nature of High Tc Superconductors, mechanism underlying Strongly Correlated Insulators, Many body localization.}
\label{CMissues}
\end{figure}
In short, Strange metals, which by the way realize the normal phase of most of the High-Tc superconductors materials we discussed previously (see fig.\ref{CMissues}), are characterized by the following features:
\begin{equation}
\rho\,\sim\,T\,(\neq\,T^2)\,,\qquad \theta_H\,\sim \,T^2\,(=\,T^2)\,.
\end{equation}
where $\rho$ and $\theta_H$ are respectively the electric resistivity and the Hall angle and the scalings in curved brakets represent the FL expectations.\\
A possible explanation, originally proposed by Anderson in \cite{AndersonSM}, refers to the possibility that the insertion of an effective spin-spin interaction leads to two different scattering rates for longitudinal and transverse modes which account for the different scalings $\sim\,T,\,T^2$ observed. Understanding and describing the Fermi Liquid scalings is one of the most pressing and early issue that the AdS-CMT program tried to adress. Importantly the achievement of such a task needs the introduction of a momentum dissipative sector into the game and it has been initiated in \cite{GouterauxDC,DavZaanen,BlakeHall}.\\
Realizing a linear in T resistivity turns out to be obtainable via the introduction of an opportune dilatonic field $\phi$ \cite{DavZaanen} in the context of holographic theories with broken translational symmetry. On the contrary accomodating both the linear in T resistivity and the Hall angle scaling seems to be an harder target. It \cite{BlakeHall} it was noticed that, in full generality, the conductivity and the Hall angle of an holographic system with momentum dissipation aquires the structure:
\begin{equation}
\sigma\,=\,\sigma_0\,+\,\sigma_{diss}\,,\qquad
\Theta_H\sim\,\frac{B}{\mathcal{Q}}\,\sigma_{diss}\,.
\end{equation}
where $\sigma_0$ and $\sigma_{diss}$ can have in general different temperature scalings.\\
It was therefore suggested that in order to accomodate the Stange Metals nature one needs the first term $\sigma_0$ (coinciding with the electric conductivity at zero heat current) to be the dominating one in the electric conductivity such that the scalings of dual theory would go like:
\begin{equation}
\rho\,\sim\,\sigma_0^{-1}\,,\qquad
\Theta_H\sim\,\sigma_{diss}\,.
\end{equation}
and the wanted phenomenology would be recovered whenever:
\begin{equation}
\sigma_0\,\sim\,1/T,\qquad
\sigma_{diss}\,\sim\,1/T^2\,.
\end{equation}
Unfortunately, it has been recently pointed out \cite{BaggioliGe} that, at least in the case where the charge density and the magnetic field are not relevant operators, this is not achievable.\\
The original dream of describing the Strange Metals phenomenology is still in the to-do list of holography \cite{Khveshchenko}. Several directions are worth to consider:
\begin{itemize}
\item Consider the cases with the charge $\mathcal{Q}$ or the magnetic field B being relevant operators of the dual CFT and having a strong impact on the IR physics.
\item Consider more generic system by adding a Chern-Simons term $\sim F\wedge *F$ and studying the effects of such a term to the transport properties of the dual material.
\item Introducing additional gauge fields as in \cite{Gursoy2,Cremonini} or with the idea of implementing holographically some kind of spin degrees of freedom with the aim of realizing the original Anderson's proposal (\cite{AndersonSM}) of two different scattering rates.\\[0.3cm]
\end{itemize}
\newpage
\textbf{High-Tc Superconductors}\\[0.3cm]
The understanding of the features of high-Tc superconductors, and in particular their microscopic origin, is very important from both theoretical and applicative persepctives. Understanding the phase diagram of such a materials is a primary question to adress. Despite the several efforts and attempt a resolution for such a puzzle has not yet been found.
\begin{figure}
\centering
\includegraphics[width=9cm]{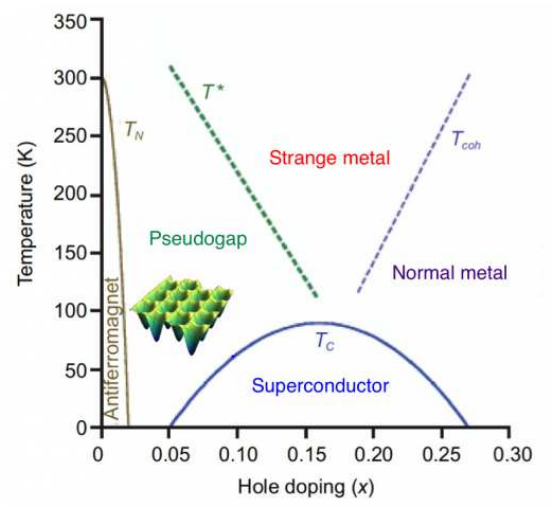}
\caption{Schematic phase diagram of the high-Tc cuprates. Figure taken from \cite{KiritsisLi}.}
\label{SCPD}
\end{figure}
The competition of various phases has been originally analyzed using the techniques of holography in \cite{KiritsisLi} and later on in \cite{Chen2}. Using a rather simple model an analogue of the real phase diagram \ref{SCPD} has been built and described. As an only shortcoming, momentum dissipation have not been introduced and as a consequence the normal phase of such a diagram is not a proper metallic state with a finite DC electric conductivity. The question whether such a diagram, resembling so much the experimental data, keeps its features in the presence of momentum dissipation is a very relevant question which can provide a step further towards the completion of a realistic phase diagram for unconventional superconductors.\\
As a first check, in \cite{BaggioliSC2} we studied whether the superconducting dome region survives in the presence of momentum dissipation and we got positive results. One can indeed build a SC dome region, as in \cite{KiritsisLi}, where the normal phase is realized by a proper metal with finite DC conductivity (unfortunately not a Strange Metal as in the actual critical region of the phase diagram). The question whether the other phases survive in the presence of translational symmetry breaking represents still an open problem which the holographic techniques could attach in a rather systematic way.\\[0.3cm]
\textbf{Mott Insulators}\\[0.3cm]
Strongly correlated materials are interesting because interactions play a very significant
role and therefore they are not easy to describe. One can distinguish 3 different mechanisms that can be responsible for the nontrivial (electrical) response: electron-phonon (e-ph), electron-disorder (e-dis), and electron-electron (e-e) interactions. Usually, Mott insulators refer to the materials that are dominated by the latter: charge-carrier self-interactions. The heuristic picture that summarizes the Mott behaviour (sometimes referred to as Mottness) is that of an electronic traffic jam: strong enough e-e interactions should, of course, prevent the available mobile charge
carriers to efficiently transport charge. Because the lack on controllable computational tools, it is worth and interesting to ask whether and how holography can consistently incorporate electron-electron interactions within its description.\\
The first positive results have been obtained in the context of probe fermions models \cite{gap1,gap2,gap3}, where the introduction of a dipole-interaction, was proved to lead to the dynamical formation of a gap in the Fermi surface which shares lots of features with the actual Mott insulators' nature.\\
Under a completely different perspective, it was recently showed \cite{BaggioliMott}, that introducing non-linearities in the bulk charge sector can provide an interesting phenomenology. In particular, using an effective model with generic non-linear self interactions for the gauge field $A_\mu$ of the form:
\begin{equation}
\sim\,\mathcal{K}\left(F_{\mu\nu}\,F^{\mu\nu}\right)
\end{equation}
it has been shown that insulating states, sharing several features with real Mott insulators, can be obtained. In addition, upon dialing the non-linearities of the system, representing the strenght of the ''electrons'' self-interactions, possible metal-insulator transitions could appear.\\
Despite the hunt for a dual for Mott insulators is still in progress, several interesting developments have been recently performed and the task is certainly at the horizon.\\[0.3cm]
\textbf{Localization}\\[0.3cm]
Strong disorder can induce insulating behaviours due to Localization mechanisms. The charged excitations in a strongly coupled medium can get localized (see fig.\ref{LOC}) making the correspondent conductivities drop down.\\
\begin{figure}
\centering
\includegraphics[width=9cm]{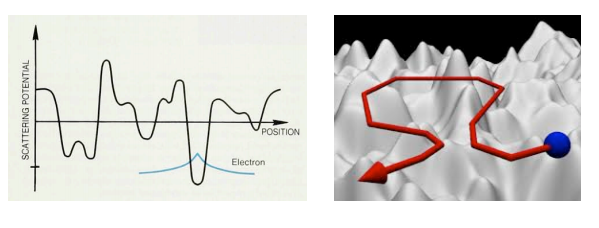}
\caption{Localization of an electron in a randomly distributed scattering potential.}
\label{LOC}
\end{figure}
The question whether holography can reproduce this behaviour and give some hints about its explanation has been recently considered in \cite{Grozdanov1}. It was claimed, that because the existence of a lower bound in the electric conductivity for generic ''simple'' holographic model, gauge-gravity duality is not able to reproduce Localization. On the contrary it seems that in the limit of strong disorder, holography just turns into a very incoherent metallic state.\\
The existence of a generic and universal lower bound for the electric conductivity has been proven to be generically incorrect in \cite{BaggioliDisorder,Gouterauxbound,Garcia1}, showing that more complicated holographic setups could reproduce insulating states driven by disorder. The nature of those insulating phases and whether they are connected to some sort of Localization mechanism is still a task in progress. It seems indeed pretty generic \cite{Grozdanov2} that an analogous lower bound on the thermal conductivity remains, hinting towards the absence of localized phases in holography.\\
It is not clear whether this shortcoming is due to the large N limit or most likely to the strong coupling regime. The presence and the search for localization mechanisms is an open and interesting question which more generic holographic effective models can adress. Probably, the homogeneous models (massive gravity, Q-lattices, helical lattices) are not enough to capture the effects of localization, and more complicated disordered setups (for example \cite{ha1,ha2}) may be the answer.\\[0.3cm]
\textbf{Metal-Insulator transitions, Mottness and the Hard Gap}\\[0.3cm]
In the last years, in the holographic community, many efforts and advances have been made in the direction of finding out possible insulating states and describing quantum phase transitions between the latter and proper metallic states, \textit{i.e.} MIT. The landscape of the different mechanisms leading to MITs is quite rich and several correlations between physical observable are present. The idea of exploring further this topic is of course intriguing with the aim of, not only identifying those mechanims, but also extracting physical results and maybe one day also possible predictions.
See for example \cite{CaiM} for an initial holographic study about the existence of colossal magnetoresistance at the metal-insulator transition.\\
Moreover, the presence of an \textit{hard gap} in the electric conductivity is a crucial feature of strongly coupled insulating phases. In particular, this is meant to say that such insulators have a conductivity dropping down exponentially at low frequencies in constrast with the so called \textit{soft-gapped} insulators where:
\begin{equation}
\sigma\left(\omega\rightarrow \infty\right)\,\sim\,\omega^p\,,\qquad p>0\,.
\end{equation}
It turns out that producing such an hard gap in the context of holography is a very diffucult problem. Despite few attempts (\cite{KiritsisRen})\footnote{The present model considers an hard-gapped geometry driven by a dilatonic dynamics that produces a discrete spectrum for all the modes of the system. It is not clear yet if producing such a cut into the geometry through the so-called hard-wall solution is the way to go.} the problem is still open and lot of people in the community start to believe that in presence of a smooth horizon such an exponential suppressed gap would never appear\footnote{A way to understand it refers to the fact that we are always dealing with (deformed) CFT states where all the correlation functions present power law decays. It seems therefore pretty unnatural, without tuned constructions, to achieve an exponential suppressed gap.}. In conclusion, whether and how holography could accomodate this class of hard-gapped materials stands as an important unanswered question.\\[0.3cm]
\textbf{Universal bounds}\\[0.3cm]
The presence of possible universal bounds on physical observable may be a solid hint for the emergence of a universal behaviour shared by many UV fixed points, which is of course of great interest.\\
Without doubts the most famous universal bound in the context of strongly coupled system is the well known KKS viscosity/entropy ratio $\eta/s\geq1/4\pi$. It has been recently claimed \cite{Hartnolleta,nicketa,BaggioliViscosity} that such a generic bound, which all the experimental checks respect so far, can be violated in particular theories which enjoy translational symmetry breaking. From the theoretical level, momentum dissipation does not lead directly to such a violation, but particular realizations (labelled as solid-type) do. The presence of an effective graviton mass for the transverse traceless helicity-2 mode provides indeed a violation of the bound which is arbitrary and can eventually drop down to a null value at $T=0$. The understanding of such a violation is still in progress and the possibility of a generalization of the KSS is being already considered but without particular success. All in all, this violation, which holographic theories seem to suggest, has still to be properly understood and it can be moreover verified in possible experiments with viscoelastic materials or solid materials in the future.\\
On a different, but somehow connected, line possible bounds on the diffusive constants of strongly coupled metals, known as incoherent metals, have been conjectured in \cite{Hartnollincoherent,Kovtunbound}. For simple holographic models, dealing with massive gravity theories, such a conjecture resulted to be incorrect \cite{genovabounds}. Nevertheless, recently, an unexpected connection between chaos (and possible bounds associated with it \cite{MaldacenaChaos}) has been analyzed in the framework of holographic theories with momentum dissipation and the results \cite{BBound} are pretty promising and worth of further investigation. Maybe, the search for universal quantities has to be pursued not at the level of the 2-point functions, such as the viscosity or the conductivity, but at the level of time scales. Nature and in particular its quantum character could indeed provide universal minimum time scales as recently proposed for example in \cite{BlakeB}.\\[0.3cm]
\textbf{What is the correct interpretation of MG AdS duals?}\\[0.3cm]
Is MG effectively encoding a lattice?\\[0.2cm]
Is MG an averaged version of disorder?\\[0.2cm]
Despite the widespread convictions that holographic massive gravity models provide a nice effective description of translational symmetry breaking mechanisms, the fundamental meaning and intepretation of these theories are still elusive and not understood.
What has been proven in \cite{MGlattice} is that, at least at linear order, an explicit holographic lattice gives rise to a mass for the graviton. Moreover, it is also clear that HMG do not share typical features of realistic lattices such as commensurability \cite{AndradeKrikun} meaning that the identification of \cite{MGlattice} is valid just at linear level and no more. There is certainly a connection between HMG and disorder. Indeed, disorder can relax momentum
via scatterings that involve only low-momentum ($k \sim 0$) processes, while in a lattice you relax momentum via high momentum ($k\sim k_L$) processes. MG is exactly realizing the first scenario and this is an important fact; however, it is only one of many features of disorder.
It would be extremely interesting and encouraging to test if there is any deeper connection between HMG theories and theories with explicit disorder, in particular up to the limit of strong disorder. Thinking of MG as an averaged, ''mean field version'', of translational symmetry breaking it would be important to check wether at that level one can distinguish between a periodic breaking (\textit{i.e. }lattice) or a disordered one (\textit{i.e.} impurities). This will also somehow adress the question of how many features of disordered systems MG can share or at least mimick. Perhaps a possible way would be a wise application of the replica trick method to holographic disordered models.\\[0.3cm]
\textbf{Phonons physics and elasticity}\\[0.3cm]
Can we reproduce phonons physics and elasticity through holography?\\
Can massive gravity be the path?\\
It certainly seems there is a sensible limit where the holographic description resembles closely the typical features of phonons (see \cite{BaggioliPRL,BaggioliSolid}). Unfortunately, following the field-operator mapping given by the AdS-CFT correspondence, it is not easy to get rid of the explicit breaking of translational symmetry which is always present in this homogeneous toy models via explicit sources for some bulk fields. Precisely identifying the holographic phonon modes, one could eventually re-build the theory of elasticity in the holographic context and improve the physical understanding of HMG theories. This is certainly something that the AdS-CMT program is still missing.\\
Inspiring works come from the description, in flat space, of the different spontaneous breaking patterns of Poincar\'e symmetry analyzed in \cite{LVEFT3,LVEFT4}. A first attempt of embedding such a framework into the holographic picture has been made in \cite{BaggioliSolid}; pursuing this path can be a promising direction to get a holographic low energy description of phonons through massive gravity theories. One, for example, could think about the possibility of gauging the ISO(d) group and construct the holographic dual of a real solid where such a symmetry is globally realized. Unfortunately, since such a group is not compact, the latter task appears not trivial. Similar suggestions have been made in \cite{Zaanen5} where potential connections between the theory of elasticity and linearized massive gravity have been speculated.
In conclusion, an optimistic way of thinking would suggest that such holographic MG theories could mimick phonons physics in the presence of disorder, where the usually massless goldstones (\textit{i.e} the phonons) acquire a mass gap directly proportional to the explicit breaking of the translational symmetry provided by disorder.
In view of the latter, a possible study of spontaneous and explicit symmetry breakings for translational symmetry could be pursued on the lines of \cite{Daniele} using the Ward identities in the Gauge-Gravity duality context and could shed light on the phonons physics underlying holographic massive gravity and the other holographic homogeneous toy models.\\[0.3cm]
\textbf{Non linear cross-correlations}\\[0.3cm]
An important improvement in the direction of giving physical predictions through holographic techniques would be to study the transport properties of the gravitational duals beyond the linear approximation with the target of finding cross-correlations between the linear and non-linear responses which could be eventually observed in experiments. A first step towards the achievement of this task has been initiated in \cite{BaggioliMott} where the non-linear electric response has been analzyed. Alternatively one can think of extending the computation of the elastic response done in \cite{BaggioliSolid,BaggioliViscosity} to non linear order (\cite{crossC}). It would be finally very interesting to try to generalize the methods (\cite{DonosDC}) for exctracting holographically the linear DC thermoelectric coefficients to non linear order going beyond the actual Ohm's law $J=\sigma E$. By now this could be done just via probe branes analysis where a non linear conductivity appears naturally.\\
In conclusion, peculiar features in the non-linear response, maybe linked with some linear features, could provide interesting physical predictions which could be even checked in real labs and give even more credit to the AdS-CMT program.\color{black}
\newpage
\section{Farewell}
\centering
\setstretch{2.0}
\textit{We conclude here our long journey.\\[0.3cm]
The take-home message is that the Gauge-Gravity duality provides a huge an unexplored playground where researchers from unbelievable distant fields could meet together and build up a common ground of ideas, questions and targets. It is an extraordinarly interdisciplinar setup which puts in contact the most fundamental questions and theoretical frameworks of physics with the most intriguing and exotic experimental results. As a matter of fact, such a connection may give in the future incredible outcomes both at the theoretical abstract level and at the ''real-world'' experimental one.}
\setstretch{1.0}
\cleardoublepage
\appendix
\cleardoublepage

\newpage\leavevmode\thispagestyle{empty}\newpage

\addcontentsline{toc}{chapter}{Bibliography}


\bibliography{Bibliography/myBib}
\bibliographystyle{ieeetr}
\cleardoublepage
\end{document}